\documentclass[aps,prx,twocolumn,superscriptaddress,nofootinbib,floatfix,longbibliography,10pt]{revtex4-2}
\usepackage{graphicx}
\usepackage{amsmath,amssymb}
\usepackage[colorlinks=true,citecolor=blue,linkcolor=blue,urlcolor=blue]{hyperref}
\makeatletter
\AtBeginDocument{\let\selectlanguage\@gobble}
\makeatother

\begin{document}

\title{In-orbit operation of a programmable quantum photonic processor}

\author{Simon Steiner}
\thanks{These two authors contributed equally}\affiliation{University of Vienna, Faculty of Physics, Vienna Center for Quantum Science and Technology (VCQ), Boltzmanngasse 5, Vienna 1090, Austria}\affiliation{University of Vienna, Faculty of Physics, Vienna Doctoral School in Physics (VDSP), Boltzmanngasse 5, Vienna 1090, Austria}

\author{Peter Schiansky}\thanks{These two authors contributed equally}
\affiliation{University of Vienna, Faculty of Physics, Vienna Center for Quantum Science and Technology (VCQ), Boltzmanngasse 5, Vienna 1090, Austria}\affiliation{University of Vienna, Faculty of Physics, Vienna Doctoral School in Physics (VDSP), Boltzmanngasse 5, Vienna 1090, Austria}\affiliation{QUBO Technology GmbH, Vienna 1090, Austria}

\author{Antonius Scherer}\affiliation{University of Vienna, Faculty of Physics, Vienna Center for Quantum Science and Technology (VCQ), Boltzmanngasse 5, Vienna 1090, Austria}\affiliation{University of Vienna, Faculty of Physics, Vienna Doctoral School in Physics (VDSP), Boltzmanngasse 5, Vienna 1090, Austria}\affiliation{Remote Sensing Technology Institute, German Aerospace Center (DLR), 82234 Wessling, Germany}

\author{Riccardo~Albiero}\affiliation{Istituto di Fotonica e Nanotecnologie, Consiglio Nazionale delle Ricerche (IFN-CNR), piazza L. Da Vinci 32, 20133 Milano, Italy}

\author{Mathias Dragosits}\affiliation{University of Vienna, Faculty of Physics, Vienna Center for Quantum
Science and Technology (VCQ), Boltzmanngasse 5, Vienna 1090, Austria}

\author{Zhenghao Yin}\affiliation{University of Vienna, Faculty of Physics, Vienna Center for Quantum Science and Technology (VCQ), Boltzmanngasse 5, Vienna 1090, Austria}\affiliation{University of Vienna, Faculty of Physics, Vienna Doctoral School in Physics (VDSP), Boltzmanngasse 5, Vienna 1090, Austria}

\author{Martin Mauser}\affiliation{University of Vienna, Faculty of Physics, Vienna Center for Quantum Science and Technology (VCQ), Boltzmanngasse 5, Vienna 1090, Austria}\affiliation{University of Vienna, Faculty of Physics, Vienna Doctoral School in Physics (VDSP), Boltzmanngasse 5, Vienna 1090, Austria}

\author{Patrik Zah\'{a}lka}\affiliation{University of Vienna, Faculty of Physics, Vienna Center for Quantum Science and Technology (VCQ), Boltzmanngasse 5, Vienna 1090, Austria}

\author{Raphael Pimenta}\affiliation{University of Vienna, Faculty of Physics, Vienna Doctoral School in Physics (VDSP), Boltzmanngasse 5, Vienna 1090, Austria}\affiliation{Departamento de F\'{i}sica, Universidade Federal de Santa Catarina, CEP 88040-900, Florian\'{o}plis, SC, Brazil}

\author{Crist\'obal Melo}\affiliation{Christian Doppler Laboratory for Photonic Quantum Computer, Faculty of Physics, University of Vienna, Vienna, Austria}\affiliation{Millennium Institute for Research in Optics, Universidad de Concepci\'on, Concepci\'on, 160-C, B\'io B\'io, Chile}

\author{C\'edric L\'eonard}\affiliation{Remote Sensing Technology Institute, German Aerospace Center (DLR), 82234 Wessling, Germany}

\author{Abhiram~Rajan}\affiliation{Istituto di Fotonica e Nanotecnologie, Consiglio Nazionale delle Ricerche (IFN-CNR), piazza L. Da Vinci 32, 20133 Milano, Italy}\affiliation{Dipartimento di Fisica, Politecnico di Milano, piazza L. Da Vinci 32, 20133 Milano, Italy}

\author{Niki~Di~Giano}\affiliation{Istituto di Fotonica e Nanotecnologie, Consiglio Nazionale delle Ricerche (IFN-CNR), piazza L. Da Vinci 32, 20133 Milano, Italy}\affiliation{Dipartimento di Fisica, Politecnico di Milano, piazza L. Da Vinci 32, 20133 Milano, Italy}

\author{Antonino~Caime}\affiliation{Istituto di Fotonica e Nanotecnologie, Consiglio Nazionale delle Ricerche (IFN-CNR), piazza L. Da Vinci 32, 20133 Milano, Italy}

\author{Roberto~Osellame}\affiliation{Istituto di Fotonica e Nanotecnologie, Consiglio Nazionale delle Ricerche (IFN-CNR), piazza L. Da Vinci 32, 20133 Milano, Italy}

\author{Francesco~Ceccarelli}\affiliation{Istituto di Fotonica e Nanotecnologie, Consiglio Nazionale delle Ricerche (IFN-CNR), piazza L. Da Vinci 32, 20133 Milano, Italy}

\author{Daniel Mart\'inez}\affiliation{University of Vienna, Faculty of Physics, Vienna Center for Quantum Science and Technology (VCQ), Boltzmanngasse 5, Vienna 1090, Austria}\affiliation{Christian Doppler Laboratory for Photonic Quantum Computer, Faculty of Physics, University of Vienna, Vienna, Austria}

\author{Tobias Guggemos}\affiliation{University of Vienna, Faculty of Physics, Vienna Center for Quantum Science and Technology (VCQ), Boltzmanngasse 5, Vienna 1090, Austria}\affiliation{Remote Sensing Technology Institute, German Aerospace Center (DLR), 82234 Wessling, Germany}

\author{Iris Agresti}\email[Corresponding author: ]{iris.agresti@univie.ac.at}\affiliation{University of Vienna, Faculty of Physics, Vienna Center for Quantum
Science and Technology (VCQ), Boltzmanngasse 5, Vienna 1090, Austria}

\author{Philip Walther}\email[Corresponding author: ]{philip.walther@univie.ac.at}\affiliation{University of Vienna, Faculty of Physics, Vienna Center for Quantum
Science and Technology (VCQ), Boltzmanngasse 5, Vienna 1090, Austria}\affiliation{QUBO Technology GmbH, Vienna 1090, Austria}\affiliation{Christian Doppler Laboratory for Photonic Quantum Computer, Faculty of Physics, University of Vienna, Vienna, Austria}\affiliation{University of Vienna, Faculty of Physics \& Research Network Quantum Aspects of Space Time (TURIS), Boltzmanngasse 5, 1090 Vienna, Austria}
\affiliation{Institute for Quantum Optics and Quantum Information Sciences (IQOQI), Austrian Academy of Sciences, Boltzmanngasse 3, Vienna 1090, Austria}

\begin{abstract}
Quantum technologies promise computational capabilities beyond the reach of classical systems. A forward-looking application lies in satellite missions, which increasingly depend on onboard computing under stringent constraints on size, weight and power. Quantum photonics is particularly attractive here: photon interference can enhance the machine-learning models needed to process large onboard data volumes, at fixed hardware resources. However, harnessing this interference requires more than generating single photons, as they must remain mutually indistinguishable: a fragile condition that is hard to maintain within the technically demanding framework of a space mission, which includes a rocket launch, strong thermal drifts, and radiation. This is why quantum states of light, though already generated and transmitted in orbit for secure communication and fundamental tests, have never been used as a computational resource. Here, we report a programmable quantum photonic platform operating on a nanosatellite, processing two photons in a six-mode universal integrated circuit. By programming distinct unitaries and tuning the photons into indistinguishability, we observe two-photon interference, establishing the on-board generation, manipulation and detection of non-classical light. This extends space-based technologies towards in-orbit quantum-assisted computing, for instance local encoding of Earth-observation data, or nodes in a distributed quantum network.
\end{abstract}

\maketitle

\noindent Over the past decade, photonic quantum technologies have moved from laboratory demonstrations to operation in space. Satellite platforms now support space-to-ground quantum key distribution, entanglement distribution and quantum
networks~\cite{liao2017satellite,liao2018satellite,yin2017satellite}. In addition, compact sources of correlated and entangled photon pairs have been qualified and operated on nanosatellites~\cite{villar2020entanglement,tang2016generation,tang2016photon}, as well as single-photon detectors, which have been characterized in the orbital radiation environment~\cite{yang2019spaceborne,wilson2021radiation}. However, using that light for computation rather than communication is a challenge that goes beyond these efforts, requiring quantum interference as a fundamental primitive and a reconfigurable platform for manipulating single photons; as such, it has not yet been addressed.

\begin{figure*}
	\centering
\includegraphics[width=\textwidth]{  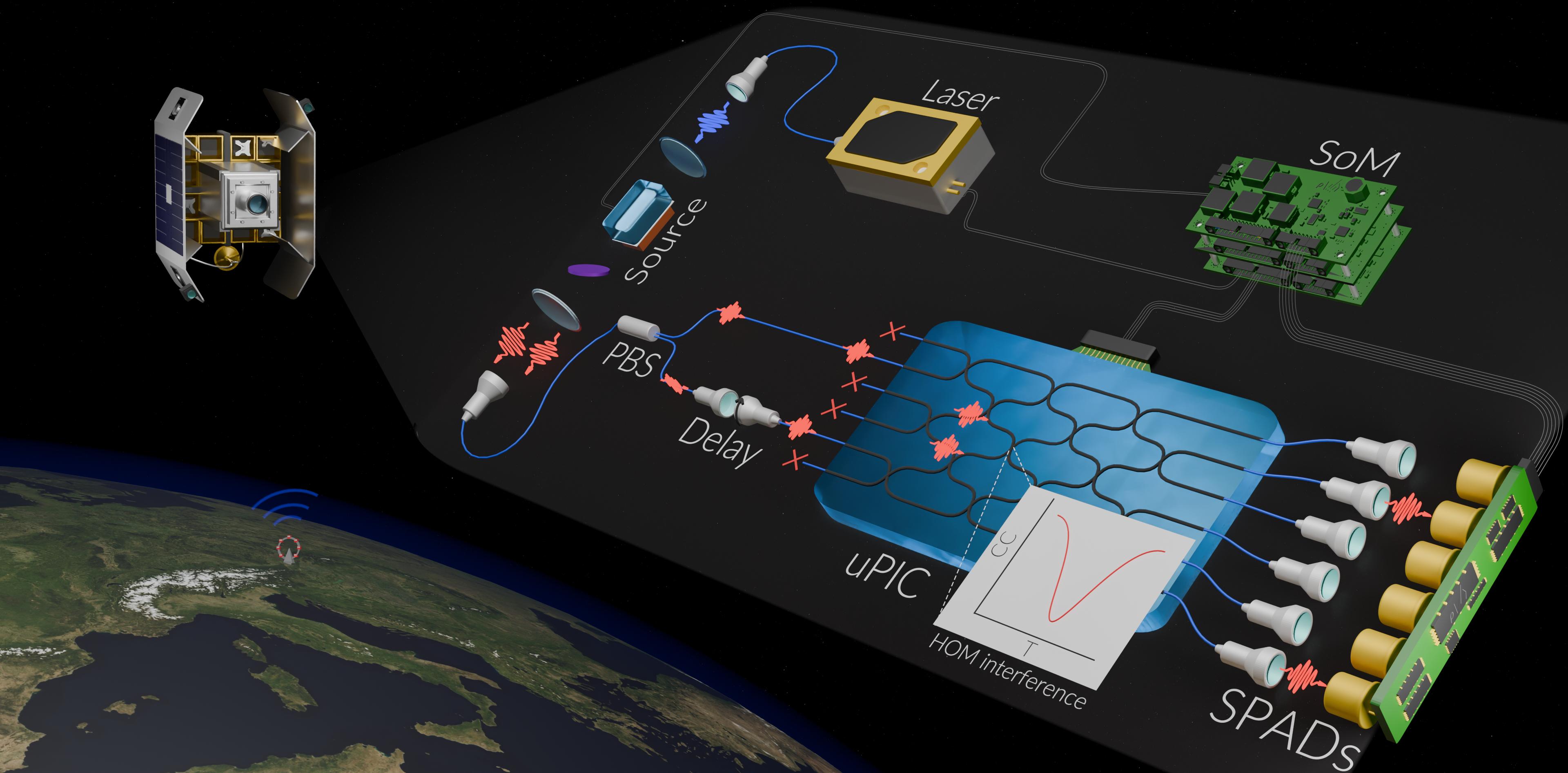}
	\caption[]{\textbf{Photonic computing platform in space.} We demonstrate a quantum computing unit operating in space, which comprises a source of indistinguishable photon pairs, a universal photonic integrated circuit (uPIC), and single-photon avalanche photodiodes (SPADs), controlled by a System on a Module (SoM). In more detail, the optical setup involves a type-II Spontaneous Parametric Down-Conversion (SPDC) unit based on a periodically poled potassium-titanyl phosphate (ppKTP) crystal, pumped by a $\ensuremath{405\,\mathrm{nm}}$ continuous-wave (CW)-laser. The two orthogonally polarized photons are deterministically split in path by a fibre-based polarizing beam splitter (PBS), whose polarization-maintaining (PM) output fibres are oriented so that both photons enter the circuit with the same polarization. A free-space delay line is used to synchronize the photons arrival time at the inputs of the uPIC. For clarity, this figure depicts the polarization matching as taking place at the delay line, through a proper orientation of the PM fibre. After undergoing the desired unitary, photons are detected at each of the six outputs of the uPIC by a SPAD.}
	\label{fig:concept}
\end{figure*}

The demand for computation in orbit, meanwhile, is growing sharply. Earth-observation constellations acquire data faster than it can be returned to the ground, so that even established real-time services, such as NASA's FIRMS wildfire tracking, deliver products up to an hour after acquisition~\cite{article}. Moving processing to the source reduces this latency and turns raw imagery into compact, actionable products before transmission~\cite{cao2020overview,denby2020orbital}. The challenge of meeting this demand within the tight size, weight and power budget of a spacecraft, while constantly being exposed to radiation and thermal cycling in high vacuum, has driven interest in computing hardware better matched to the space environment than conventional electronics~\cite{mateogarcia2023orbit}. Photonics is a compelling candidate: optical circuits perform linear transformations passively, through interference, so that the computational core of most inference algorithms is executed by propagation and dissipates no power along the optical path~\cite{hamerly2019large}. In addition, integrated implementations add a small footprint, low mass, mechanical and thermal robustness, and compatibility with semiconductor manufacturing~\cite{psiquantum2025manufacturable}. These properties have already motivated classical optical accelerators explicitly targeting spaceborne artificial intelligence~\cite{kubler2026concept}. The open question is what such hardware is ultimately capable of once it is in space.

\begin{figure*}
	\centering
	\includegraphics[width=\textwidth]{  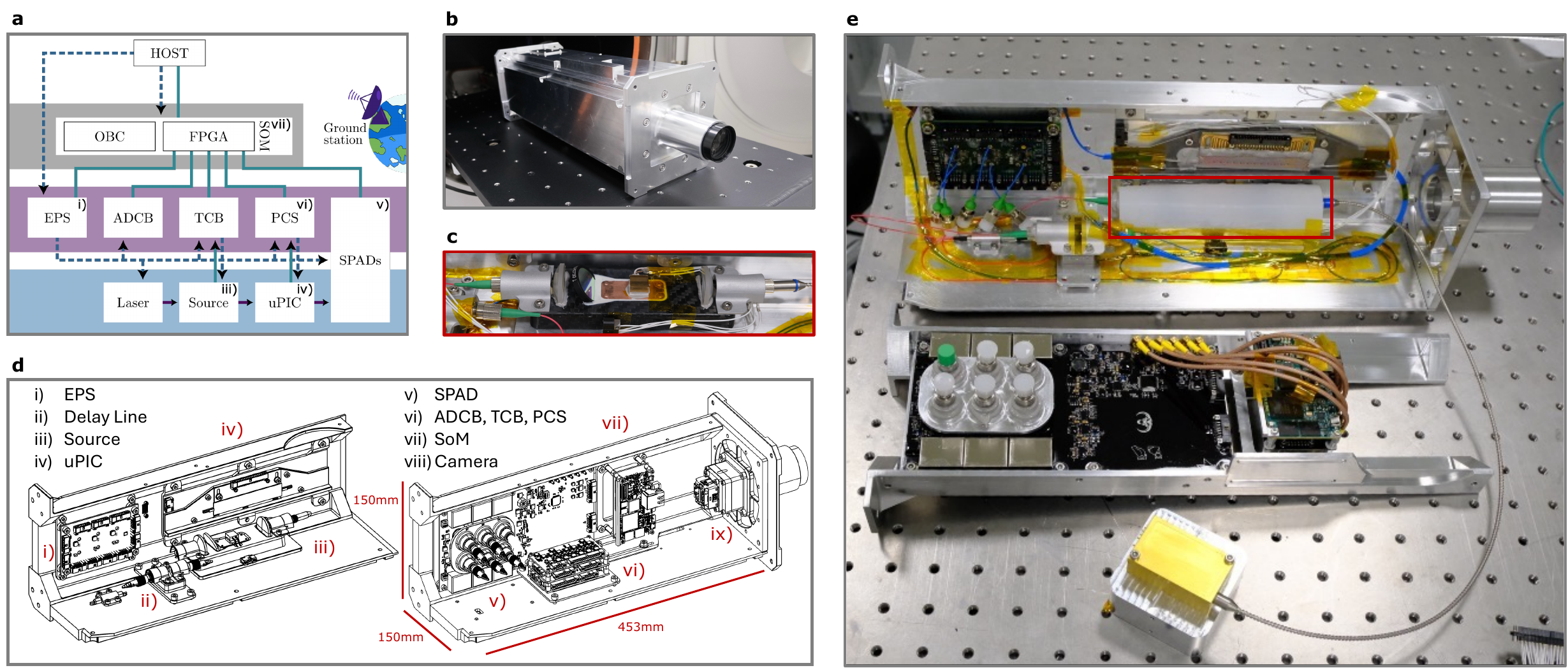}
	\caption[]{\textbf{Space-compatible quantum computing unit.}  The system fits within a $15\times15\times \ensuremath{45.3\,\mathrm{cm}}$ volume, has a total mass of $\ensuremath{9.8\,\mathrm{kg}}$, and an average power consumption of $\ensuremath{10\,\mathrm{W}}$. To withstand the mechanical shocks of launch, the components are mounted on purpose-built supports, while the optical fibres are secured to the frame with Kapton tape.
	(\textbf{a}) The payload is controlled by the System on a Module (SoM) containing an ARM Cortex-A9 processor and a Field-Programmable-Gate-Array (FPGA). The interplay between the electronics components and the optical payload is depicted by arrows for communication (green), power (blue, dashed) and light (purple). Power and communication with ground stations are provided by the host system. The electronic part is constituted by the Electrical Power System (EPS), Sensor Board (ADCB), Temperature Control Board (TCB), uPIC Current Source (PCS) and Detector Board (SPADs). A custom-made time-tagging program running on the SoM was used to detect coincidence events.
	(\textbf{b}) The assembled quantum computer before launch.
	(\textbf{c}) Zoomed-in, highlighted single-photon source, showing the components mounted on a carbon-fibre structure.
	(\textbf{d}) CAD model of the payload with highlighted components.
	(\textbf{e}) Interior view of the quantum computer, showing all components, including the integrated photonic circuit and the detection board hosting six SPADs. In this picture, the source is covered by polytetrafluoroethylene and highlighted in (\textbf{c}).}
	\label{fig:photos}
\end{figure*}

The answer is that the computational model implemented by an interferometric circuit is set not by its geometry, but by the state of light that evolves through it. When injected with classical light, a mesh of interferometers performs a linear map on the mode amplitudes, the vector-matrix product exploited by optical accelerators. By contrast, when the input is constituted by indistinguishable single photons, the very same unitary acts on the multi-photon Fock space, whose dimension grows combinatorially with the number of photons and modes. Moreover, the transition amplitudes are permanents of submatrices of that unitary, which makes sampling from the output distribution intractable for classical computers as the numbers of photons and modes grow, under standard complexity-theoretic assumptions~\cite{aaronson2011computational}. This is the origin of the computational advantages demonstrated with photonic hardware, from boson sampling~\cite{madsen_quantum_2022,zhong_quantum_2020} to quantum machine learning protocols, in which photon statistics enhance model expressivity at fixed circuit size~\cite{yin2025experimental,di2026time,paparelle2026experimental,cimini2026large,nerenberg2025photon,gan2022fock,monbroussou2025photonic,minati2026quantum,oh2024entanglement}. For a spacecraft, whose processor is fixed in size and power, the appeal is precisely that this larger computational space and complex transformations are obtained from the same optical platform, unlocking computational regimes that classical optical processing cannot attain. Whether the conditions for quantum interference, in terms of spectral degeneracy, phase stability and optical alignment, can be established and maintained in space is the prerequisite for everything above, and the problem we address here.

\begin{figure*}
	\centering
	\includegraphics[width=\textwidth]{  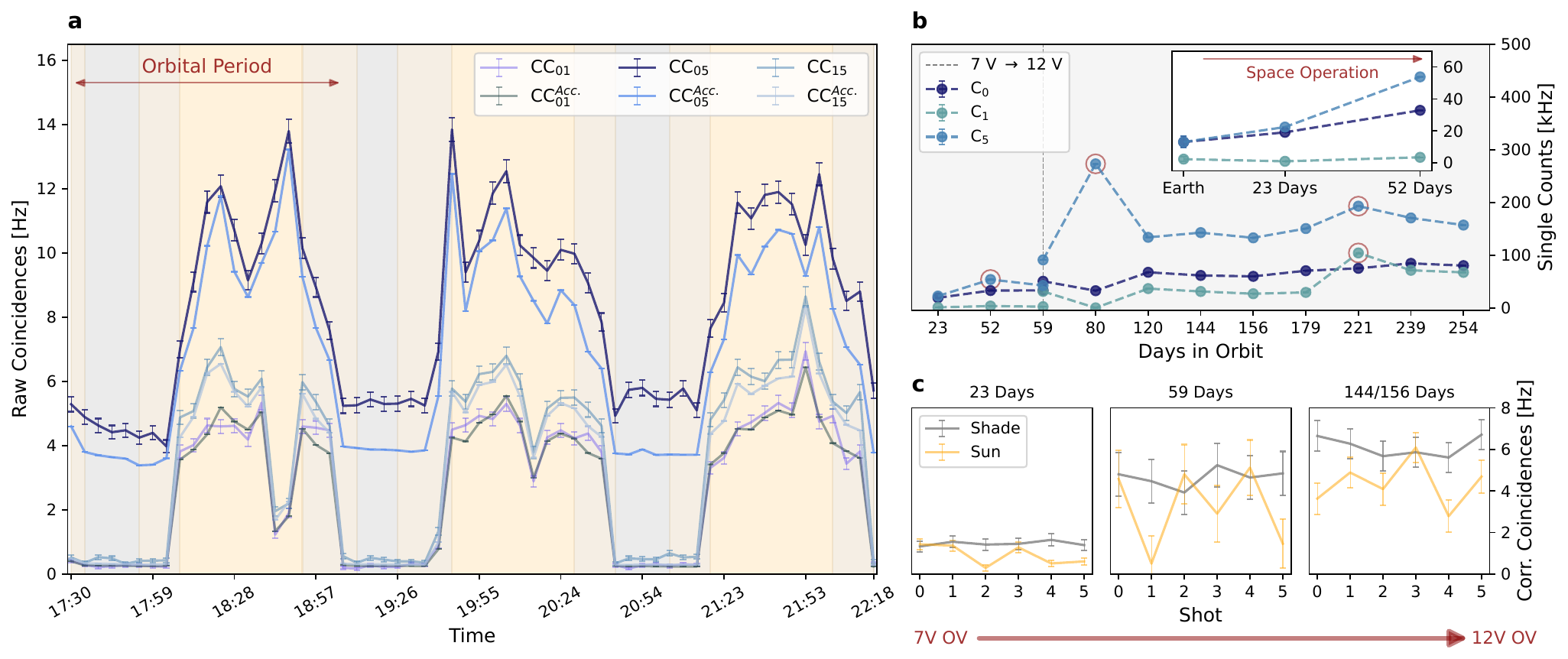}
	\caption[]{\textbf{In-orbit data acquisition.} (\textbf{a}) Raw coincidence counts CC$_{ij}$ between detectors C$_i$, C$_j$ and accidental noise floor CC$_{ij}^{Acc}$ for \ensuremath{7\,\mathrm{V}} SPAD overvoltage after $23$ days in orbit. The rate of coincidence counts and noise floor correlates with the payload position with respect to the sun: it is stable when the payload is shaded by the Earth (grey area) and it is strongly increased and fluctuating when facing the sun (yellow area). Given the \ensuremath{92\,\mathrm{min}} low-Earth orbital period (red arrow), low-noise measurements are restricted to \ensuremath{30\,\mathrm{min}} per orbit. For photons routed to detectors C$_0$ and C$_5$ (see Fig.~\ref{fig:unitary}a), the coincidence pair CC$_{05}$ (solid blue) remains clearly above the noise floor CC$_{05}^{Acc}$ (light blue). (\textbf{b}) Dark count rate (DCR) over time in Earth's shade, with SPADs at \ensuremath{-15\,{}^{\circ}\mathrm{c}}: a notable increase in DCR sets in after only $52$ days in orbit (small inset). SPAD overvoltage was raised from \ensuremath{7\,\mathrm{V}} to \ensuremath{12\,\mathrm{V}} (vertical dashed line) after $59$ days in orbit to recover signal, which also increased DCR. Detectors occasionally show transient DCR spikes (red circles) before restabilizing at a higher level. (\textbf{c}) Despite the increased DCR, genuine SPDC signal (coincidences corrected for accidentals), for instance for CC$_{15}$, could be improved significantly from $\sim$ \ensuremath{1.5\,\mathrm{Hz}} to \ensuremath{6\,\mathrm{Hz}} over the mission duration. Here, potential detector saturation during a sun measurement (yellow) is also reflected in a higher signal instability when compared to the shadow measurements (grey). Error bars are obtained via Gaussian error propagation and indicate one standard deviation.}
	\label{fig:countrate}
	\end{figure*}

In our platform we bring together the two building blocks of linear optical quantum computing: a programmable universal photonic integrated circuit (uPIC) and photon--photon interference in a single payload operated in low Earth orbit (LEO), at $510~\mathrm{km}$ altitude. The payload was launched on 23 June 2025, after two years of development time, on SpaceX's Transporter-14 rideshare mission and is hosted by D-Orbit's ION SCV orbital transfer vehicle in a Sun-synchronous orbit~\cite{SM}. Our quantum computing unit combines a source of indistinguishable photon pairs, a six-mode uPIC implementing arbitrary unitary transformations, and on-board single-photon avalanche photodiodes, within a payload frame of $15\times 15 \times 45.3~\mathrm{cm}$ (H $\times$ W $\times$ L), a mass of $9.8~\mathrm{kg}$, and an average power consumption of $10~\mathrm{W}$. Over the first eight months in orbit we programmed a set of distinct unitaries on the circuit, recovered their output distributions, and observed a Hong--Ou--Mandel (HOM) dip~\cite{hong1987measurement}, the signature of two-photon quantum interference. In doing so, we demonstrate the hardware primitive on which quantum-enhanced onboard processing rests. These results move photonic quantum technology in space from the distribution of quantum states towards their processing, and set the direction we intend to follow: encoding Earth-observation data directly into the operations of an orbiting quantum processor, and exploiting non-classical statistics for onboard inference, ultimately across the distributed, globally connected architecture that satellite constellations provide.

\begin{figure*}
	\centering
	\includegraphics[width=\textwidth]{  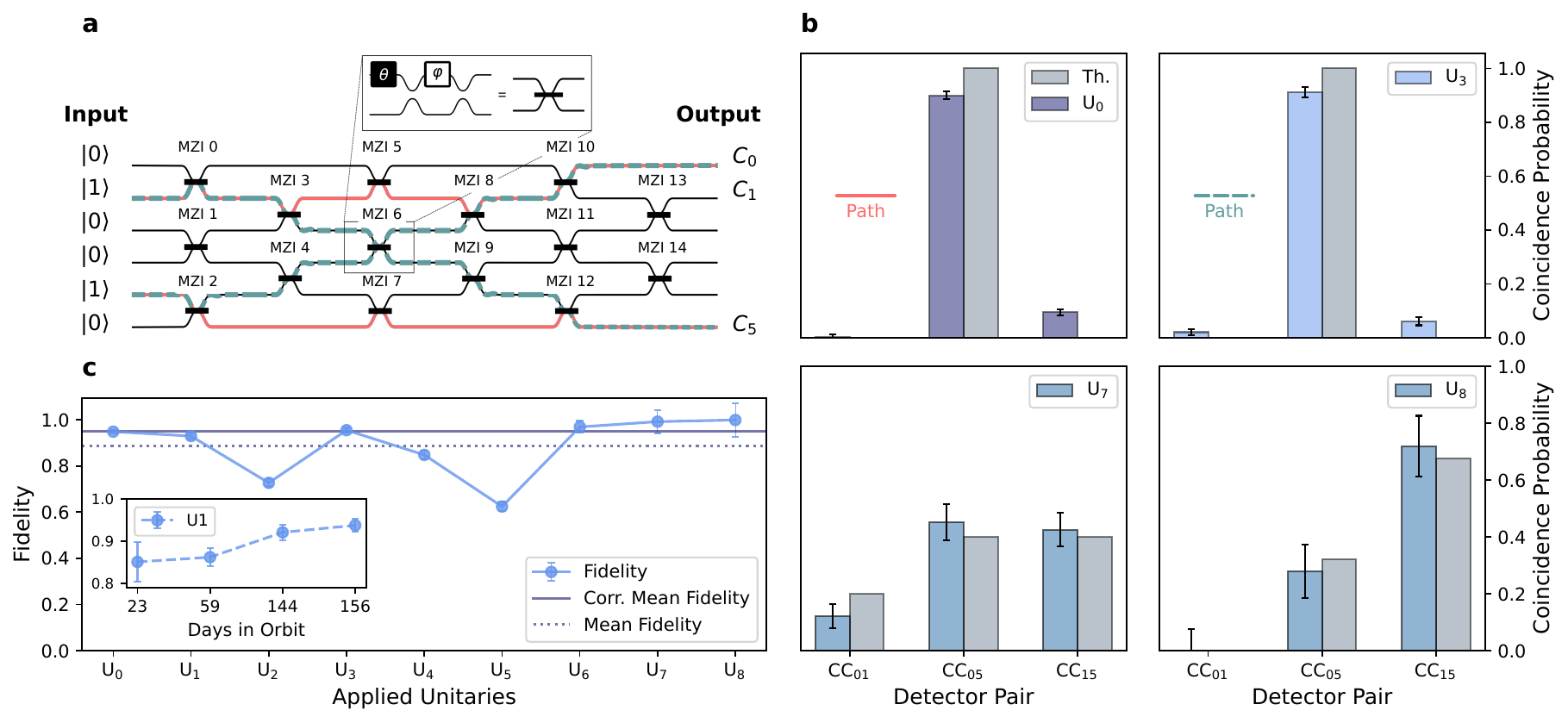}
	\caption[]{\textbf{On-board programmable unitary operations.} (\textbf{a}) Scheme of the uPIC that gives an overview of the $15$ MZIs and emphasizes the routing configuration for sending the two input photons to the detectors on modes $C_0$ and $C_5$. All MZIs have to be actively controlled to compensate for thermal crosstalk. (\textbf{b}) The top panels show the result of the two highlighted path configurations (solid red/ dashed green curve); $U_0$ corresponds to the case of no interference (solid red curve), while in $U_3$ both photons meet at the MZI$6$ (dashed green curve), which is set to a balanced beamsplitter. The outcome agrees well with theory (grey bars), and (\textbf{c}) compares the average fidelity across all measured unitaries for distinguishable photons. The majority have a fidelity higher than $0.9$, and some were even improved over the mission duration (small inset), while two significant outliers $U_2$, $U_5$ indicate a uPIC calibration issue, likely due to the absence of temperature stabilization. The bottom panels in (\textbf{b}) highlight that the uPIC is also capable of implementing more complex unitaries sending signal to all of the modes connected to the three working SPADs. Error bars are obtained via Gaussian error propagation and indicate one standard deviation.}
	\label{fig:unitary}
	\end{figure*}

\section{Photonic computing in space}
Our photonic computing platform is hybrid, combining a quantum optical front-end with conventional control electronics. The optical payload, shown in Fig.~\ref{fig:concept}, comprises three functional stages: input state preparation, programmable processing, and single-photon detection.

The input state is a two-photon Fock state $|010010\rangle$, generated by a collinear SPDC source, at $\ensuremath{810\,\mathrm{nm}}$. The process produces two orthogonally polarized photons in the same spatial mode, which are deterministically separated in path by a fibre-based polarizing beam splitter. Its two output fibres are polarization-maintaining (PM) and are oriented with their slow axis along the polarization of the beam they collect; mounted with the same orientation at the input of the circuit, they deliver the two photons with identical polarization. In these conditions, at the degeneracy temperature ($\sim32.5^\circ$C), the photons are indistinguishable in all degrees of freedom except arrival time. Thus, a free-space delay line compensates for fibre-length mismatch, ensuring simultaneous injection into the processing stage.

\begin{figure*}
	\centering
	\includegraphics[width=\textwidth]{  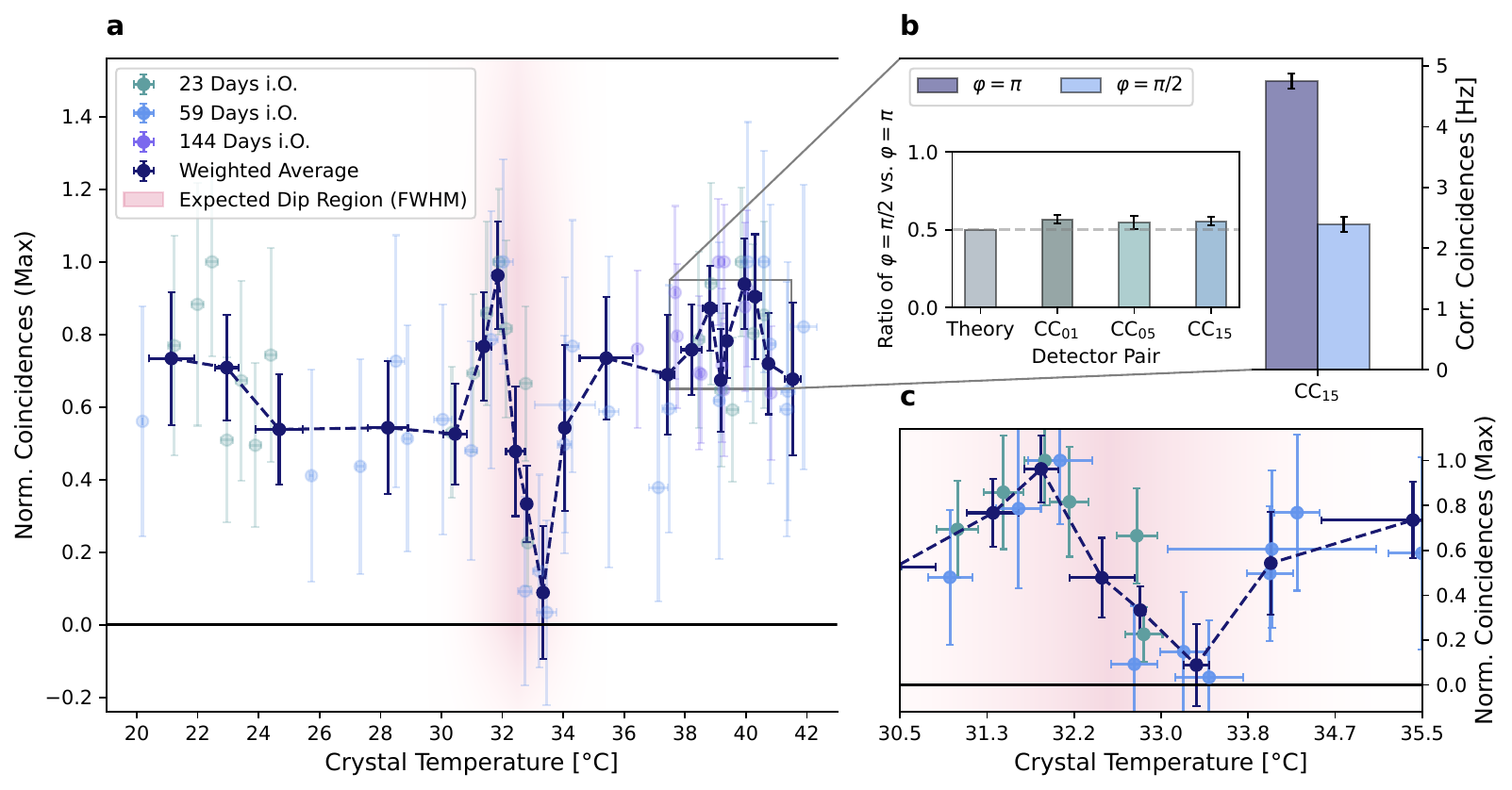}
	\caption[]{\textbf{Hong-Ou-Mandel interference in space.} (\textbf{a}) Two-photon interference scan with respect to the crystal temperature, including data from three different operation days. The uPIC is set such that two input photons (corresponding to the state $|010010\rangle$) are sent to MZI6, whose internal phase is set to $\varphi=\pi/2$ and whose outputs are then routed to detector pairs CC$_{05}$ (green) or CC$_{15}$ (light blue, violet). Individual data points of normalized coincidence counts for consecutive days in orbit (i.O.) are marked in the background while the inverse-variance weighted average of these points (dark blue) highlights the overall trend. Error bars indicate one standard deviation, obtained by Gaussian uncertainty propagation, and there is an explicit dip in coincidences around a crystal temperature of $\sim \ensuremath{32.5\,{}^{\circ}\mathrm{C}}$, which also corresponds to the value measured on Earth (red-shaded area). The peak in coincidences just before the dip is discussed in the Supplemental Material~\cite{SM} (Sec.~S9\,N). (\textbf{b}) The registered average coincidence rate of distinguishable photons when the internal phase of MZI6 is set to $\varphi = \pi/2$ (light blue), is half with respect to the setting of $\varphi = \pi$ (violet), as can be seen exemplarily for detector pair CC$_{15}$. This feature served as a reference measurement for the HOM scan. It was consistently observed when routing photons from MZI6 across all detector pairs (inset) and shows good agreement with theoretical predictions (grey)~\cite{hong1987measurement}. (\textbf{c}) Close-up of the HOM dip around the expected degeneracy temperature (red-shaded area) highlights the explicit drop of coincidence counts for the individual operation days (light blue, green) and the corresponding weighted average (dark blue).}
	\label{fig:hom}
	\end{figure*}

This stage is made of a uPIC with six input/output modes, composed of a mesh of Mach--Zehnder interferometers (MZIs), capable of realizing arbitrary unitary transformations~\cite{clements2016optimal}. The optical circuit is written by femtosecond laser waveguide writing in glass and reconfigurability is achieved via thermo-optic micro-heaters that locally modify the refractive index of the waveguides, inducing a controlled phase shift.
Each output mode is then connected to the detection stage, featuring single-photon avalanche photodiodes (SPADs).

The control electronics, whose scheme is shown in Fig.~\ref{fig:photos}a, include a Peltier-based temperature controller with proportional-integral-derivative (PID) feedback to maintain frequency degeneracy of the produced photons, with a precision of $0.1\,\mathrm{K}$. It also features a multichannel current driver to program the desired unitary on the uPIC, a passive quenching circuit, an electrical power system, and an on-board-computer for experiment management.

Deploying such a system in space required addressing several environmental hazards. The mechanical stress of a rocket launch, involving broadband vibrations and high-frequency shocks, demands a rigidly bonded optomechanical assembly tolerant to strong transient loads. Once in orbit, periodic transitions between sunlight and Earth's shadow over a $\sim90$-minute period drive a continuous thermal cycling, which can induce severe fatigue and consequent irreversible damage to the adopted materials~\cite{carvelli2015fatigue}. In addition, in the absence of a convective atmosphere, thermal management relies solely on conduction and radiation. High vacuum further constrains material selection, as outgassing (i.e., gas produced by the sublimation of organic matter) would contaminate and degrade optical surfaces~\cite{jiao2019outgassing}. Finally, at an altitude of \ensuremath{510\,\mathrm{km}} and \ensuremath{97.5^{\circ}} inclination to the equator, the payload is exposed to electrons and protons trapped in the radiation belts, which are the dominant cause of permanent deterioration to silicon SPADs, causing displacement damages, leading to an increase of the dark counts~\cite{wilson2021radiation}, and fibre darkening~\cite{Friebele5138}, reducing the signal rate.

To mitigate these effects, we took several countermeasures. First, the payload is enclosed in a $1~\mathrm{cm}$-thick aluminium shield (see Fig.~\ref{fig:photos}b and \ref{fig:photos}e), which effectively attenuates electrons and Bremsstrahlung~\cite{SM}. To further reduce the dark count rate (also radiation-induced), SPADs are cooled down through an internal temperature controller~\cite{yang2019spaceborne}.
To avoid mechanical shock-induced disruptions, all optical components are mounted on titanium baseplates, which act also as heatsinks~\cite{miao2021spacecraft}, and free-space components (such as the source and delay line) are additionally bonded to custom supports. The processor is fibre-pigtailed at both input and output facets, avoiding free-space alignments that would be difficult to maintain. This fibre--chip interface was a particularly delicate junction, requiring additional adhesive reinforcement to withstand launch vibrations~\cite{SM}. All the fibre-based devices use PM fibres, which were secured directly to the payload frame with strong adhesive tape and all optical components underwent mechanical stress and shock tests~\cite{SM}.
Fig.~\ref{fig:photos} shows the assembled payload, with all of the main optical and electronic components highlighted.

\section{In-orbit quantum information processing}

The first in-orbit measurements were aimed at verifying the correct functioning of the system; in particular, that the temperature of the source nonlinear crystal was stabilized, and that the SPADs were operated at the optimal overvoltage for the best signal-to-noise ratio.

Despite all the efforts to suppress external noise sources, a careful analysis of the photon count rate was required to isolate the genuine SPDC signal from the noise floor, set by dark counts and by sunlight-induced accidental coincidences. The rate of those was evaluated using the following formula: $S_1 \times S_2 \times \tau$, where $S_{1,2}$ indicate the rates of single counts of the two considered detectors and $\tau$ is the coincidence window. This estimate holds for the underlying Poissonian counting statistics~\cite{castelletto2003quantum,kim_low-cost_2005,bjurlin_versatile_2024} in the presence of uncorrelated noise (Supplemental Material~\cite{SM}, Secs.~S9\,A and~S9\,O).

Throughout the operation, sunlight manifested itself as a strong periodic noise component that could saturate the detectors and, as shown in Fig.~\ref{fig:countrate}a, count rates increased sharply within a 92-minute periodicity, matching our orbital period. During these intervals, the signal-to-accidentals contrast of, for example, target detector pair CC$_{05}$ (dark blue vs light blue in Fig.~\ref{fig:countrate}a), was markedly reduced, and, overall, fluctuations were considerably larger. We therefore restricted our measurements to the shaded portions of the orbit to ensure reliable estimates. As shown in Fig.~\ref{fig:countrate}b, the DCR of our SPADs also increased steadily over the operation days, with occasional sharp spikes likely caused by individual, severe radiation events.

The overall count rate is another challenge, when comparing with the results measured on Earth, mainly caused by a progressively reduced laser power (due to internal outgassing of the device~\cite{SM}) and by radiation damages of the detectors~\cite{ceccarelli2021recent}, lowering their efficiency. This significantly affected the statistics and thus  was crucial to find the optimal SPAD overvoltage for the signal-to-noise ratio throughout the mission ~\cite{SM}. Fig.~\ref{fig:countrate}c shows that the best results were obtained by selecting $\ensuremath{12\,\mathrm{V}}$, with an integration time of $\ensuremath{60\,\mathrm{s}}$ per measurement.

With these strategies in place, we measured the output distributions of several distinct unitary transformations applied to the input state to test the reconfigurability of the circuit. These measurements were performed with distinguishable photons, at temperatures far from the photon degeneracy regime. As only three SPADs remained functional after launch (C$_0$, C$_1$ and C$_5$)~\cite{SM}, we restricted our analysis to $3\times3$  sub-unitaries, using the full $6\times6$  circuit to reroute the relevant output modes to working detectors (see Fig.~\ref{fig:unitary}a and \ref{fig:unitary}b). We quantify the agreement between the measured coincidence distribution $\bar{p}_{ij}$ and the theoretical distribution $t_{ij}$ with the classical fidelity $F=\sum_{ij}\sqrt{t_{ij}\,\bar{p}_{ij}}$~\cite{SM}. Across nine measured unitaries, we obtained an average fidelity of $F= 0.888 \pm 0.011$, rising to $F= 0.949 \pm 0.014$ when excluding two significant outliers (see Fig.~\ref{fig:unitary}c). This shows that the photonic processor and the optical components remain operational in the uncontrolled and harsh environment of space.

To observe two-photon interference, we performed a scan of the crystal temperature, between $\ensuremath{20\,{}^{\circ}\mathrm{C}}$ and $\ensuremath{40\,{}^{\circ}\mathrm{C}}$. For this measurement, the circuit routed both input photons to a single MZI (MZI6 in Fig.~\ref{fig:unitary}a), whose internal phase was set to $\varphi=\pi/2$ to realize a balanced beamsplitter, with the outputs directed to a pair of working detectors.

As shown in Fig.~\ref{fig:hom}a, a clear drop in coincidence counts, showing a genuine quantum interference signature, appears close to the expected degeneracy temperature of $\sim \ensuremath{32.5\,{}^{\circ}\mathrm{C}}$, % TODO: check against Table~\ref{tab:binning_visibilities} in the supplement (see note there)
with an interference visibility $\mathcal{V}= 0.908 \pm 0.191$, which lies 2.14 standard deviations above the classical bound of 0.5~\cite{li_experimental_2022}. Only data acquired in the relatively low-noise shadow regions are considered here, as sunlight introduces a strong noise component (see Fig.~\ref{fig:countrate}a). This restriction limits the measurement time available per orbit, so that the results reported here combine experiments performed on three separate operation days. The individual datasets carry substantially different uncertainties, and were therefore merged into a weighted average. These differences arise from the ambient conditions of each day, such as background noise and environmental temperature, and from the measurement parameters used, such as SPAD overvoltage, integration time and target detector pair. Notably, the drop in the coincidence rate was independently reproduced on two of these days.
To rule out that the reduction in the coincidence counts was due to statistical fluctuations, we additionally measured, at each temperature setpoint, the coincidence rate with the MZI6 internal phase set to $\varphi=\pi$, a configuration in which no interference is expected. In this case, outside the dip region, the coincidence rate was twice that observed at $\varphi=\pi/2$ (see Fig.~\ref{fig:hom}b), while within the dip region, no corresponding decrease in the coincidence counts was observed (Supplemental Material~\cite{SM}, Sec.~S9).

\section{From communication to spaceborne quantum processing}

In this work, we report the operation of a photonic quantum processor in orbit. The payload comprises an SPDC photon-pair source, a uPIC, SPADs, and custom electronics, each engineered to survive the harsh conditions of space. Across the measured unitary evolutions of our two-photon input state, we recovered the output distributions with an average fidelity of $F= 0.888 \pm 0.011$, over the three-mode subspaces accessible with the detectors that remained operational after launch. This value rises to $F= 0.949 \pm 0.014$ once two outliers attributable to a circuit calibration issue are excluded. Scanning the crystal temperature, we then observed a HOM dip with a visibility of $\mathcal{V}= 0.908 \pm 0.191$. To benchmark the robustness of this result, we combined measurements taken independently across multiple operation days, separated by extended intermissions, and found consistent signatures of two-photon interference.

Our demonstration carries direct implications for photonic quantum computing. Interference between indistinguishable photons is the non-classical resource that underpins the computational advantages accessible to linear optical hardware, from boson sampling~\cite{aaronson2011computational} to the enhancements in model expressivity that allow quantum systems to match or surpass classical algorithms with fewer resources, for instance in machine learning~\cite{nerenberg2025photon,yin2025experimental,di2026time}. Generating and controlling this resource in the space environment is therefore a prerequisite for satellite-based edge quantum computing. It is also significant that this experiment was carried out on an integrated platform, as state-of-the-art schemes rely predominantly on photonic circuits of this kind, valued for their stability, compactness and scalability~\cite{psiquantum2025manufacturable,maring2024versatile,monbroussou2025photonic,spagnolo2022experimental,saggio2021experimental,madsen_quantum_2022}.

Beyond the demonstration reported here, the same architecture is in principle capable of executing more complex algorithms, including quantum machine learning protocols~\cite{yin2025experimental,monbroussou2025photonic,di2026time,mauser2026experimental,saggio2021experimental} and foundational quantum optics experiments~\cite{lombardi2002teleportation,kun2025direct,kun2026testing}. The next step is to close the loop between sensor and processor, encoding Earth-observation data directly into the unitary programmed on the circuit.
This long-term goal calls for progress on two fronts. The first is an
engineering one. The results reported here show that quantum interference can be established and controlled in orbit. What remains is to keep it stable over the long acquisition
times that an inference task demands, as the gradual degradation of the
components reduces the coincidence rate. Our own mission makes the terms
concrete: radiation damage eroded detector efficiency while raising dark
counts, and the pump power declined over time. Part of the response is
already available, in temperature stabilization of the source, detector
shielding, radiation-tolerant packaging and in-orbit
annealing~\cite{lim2017laser,anisimova2017mitigating,ceccarelli2021recent}; other
parts still have to be developed, from maintaining spectral degeneracy across
months of operation to recalibrating the circuit autonomously as thermal
conditions drift.

The second front is theoretical. An effective protocol must recognize which features carry the
information relevant to a given class, and commit only those to the available modes. Also in this case, the ingredients exist, from quantum kernel methods to multiphoton neurons for image classification and photonic convolutional
architectures, all demonstrated on the
ground~\cite{yin2025experimental,minati2026quantum,monbroussou2025photonic}, while an encoding that performs this selection at the scale of a compact orbiting processor has yet to be formulated.

The results reported here are the first step along this path we are pursuing, to allow for classification and compression on the photonic hardware itself, while sending only the inference product to the ground. This progress would make orbiting processors the natural counterpart to the quantum links already established between satellites and the ground~\cite{liao2017satellite,liao2018satellite,yin2017satellite}: nodes and channels together define a distributed architecture with global coverage and low-latency access, in which quantum computation is no longer confined to controlled laboratory environments but deployed at scale across the network.

\section*{Acknowledgements}
The authors wish to thank the staff of the German Aerospace Center (DLR), in particular the group RC3, with special mention to Hannes Brandt, David Freiknecht, and David Kleeman, for hosting the payload assembly at their facility and for offering their TVAC testing station. The authors also thank Steffen Babben, who performed the outbaking of the components.
A special thanks to Hans K\"uffner-McCauley for the valuable discussions, the technical support in the design of the mechanical structure of the payload, and the incredible help in organizing and managing the project.
The authors also wish to thank Alex Ling, Mirela Selimović, Hubertus Lauterbacher, Paul Kötter, Johann Kötter, Jakob Mayer, André Radloff, Ankush Sharma, Roland Blach, Mark Pruckner,  and Liam Ramsey for the valuable support and fruitful discussions.
N.D. and F.C. wish to thank Ciro Pentangelo for the help with the calibration of the uPIC.
The uPIC was partially fabricated at PoliFAB (www.polifab.polimi.it), the micro- and nanofabrication facility of Politecnico di Milano. A.R., R.O., F.C., T.G., and I.A. wish to thank the PoliFAB staff for the valuable technical support and for hosting the payload after assembly.
The authors acknowledge ``NASA Earth Observatory'' for the Earth surface images used in Fig.~\ref{fig:concept}. \\ This research was funded in whole, or in part, by the \"Osterreichische Forschungs\-f\"orderungs\-gesellschaft (FFG) grant agreement No FO999921413 (SPACE), by the Austrian Science Fund (FWF) through 10.55776/ESP205 (PREQUrSOR), and through 10.55776/F71 (BeyondC) and by the European Union’s Horizon 2020 and Horizon Europe research and innovation programme under grant agreements No 899368 (EPIQUS) and No 101135288 (EPIQUE), by the Marie S\l{}odowska-Curie grant agreement No 956071 (AppQInfo) and by the QuantERA program [10.55776/I6002] (PhoMemtor).
D.M. and C.M. acknowledge the financial support by the Austrian Federal Ministry of Labour and Economy, the National Foundation for Research, Technology and Development and the Christian Doppler Research Association.
F.C. acknowledges the financial support from the National Research Council of Italy (CNR) under the Short Term Mobility Program 2024, while R.A. and  F.C. acknowledge funding from the project HI-LIGHT (Hybrid Integration of Laser-written Interferometers and sinGle pHoton deTectors), grant n. 2022JRSST2, CUP B53D23002690006, funded by the Italian Ministry of University and Research (MUR) through the PRIN 2022 program (D.D. n. 104, 02/02/2022) and by the European Union – NextGenerationEU (NRRP Mission 4 Component 2 Investment 1.1).

\section*{Author contributions}
I.A., T.G. and A.S. conceived the project. S.S., M.D., I.A. and P.S. developed the photonic source. P.S., D.M. and P.Z. developed the PCBs. S.S. made the SPAD pigtailing. R.A., A.R., N.D., A.C., R.O., and F.C. fabricated and calibrated the uPIC. S.S., R.A., and F.C. devised and performed the space-compatible packaging of the uPIC. D.M., T.G., A.S., Z.Y., C.M., and C.L. developed the FPGA and SoM architecture. T.G., D.M., A.S., P.S., I.A., Z.Y., S.S., M.M., and R.P. developed the payload drivers and scripts. T.G. and Z.Y. developed the OBCP scripts to operate the payload in-orbit. T.G., I.A., S.S., A.S., D.M. and P.S. verified the scripts and operated the payload in-orbit,  N.D., F.C., and T.G. performed the TVAC testing. I.A., S.S., P.S., F.C., R.A., A.S. and M.D. performed the environmental testing of the payload.  S.S. and I.A. conducted the data analysis. S.S., P.S., D.M., A.S., N.D., F.C., R.P., T.G., and I.A. participated in the payload assembly.  T.G., I.A., and P.W. managed the project. All authors contributed to the revision of the manuscript.

\vspace{0.8cm}

\section*{Competing interests}
I.A., M.D., S.S., P.S., and P.W. are employees of the University of Vienna, which has applied for a patent (EP25193461) for a fibre-coupled single-photon source for space-applications and a method for a fibre-coupled single-photon source with I.A., M.D., S.S., P.S., and P.W. listed as inventors. The remaining authors declare no competing interests.

\section*{Data availability}
All data used in this work, together with data analysis scripts, are openly available in Zenodo. EDA projects of the in-house made electronic components are openly available in Zenodo.

\end{document}

\title{Supplemental Material:\\ In-orbit operation of a programmable quantum photonic processor}
\author{Simon Steiner}
\author{Peter Schiansky}
\author{Antonius Scherer}
\author{Riccardo Albiero}
\author{Mathias Dragosits}
\author{Zhenghao Yin}
\author{Martin Mauser}
\author{Patrik Zah\'{a}lka}
\author{Raphael Pimenta}
\author{Crist\'obal Melo}
\author{C\'edric L\'eonard}
\author{Abhiram Rajan}
\author{Niki Di Giano}
\author{Antonino Caime}
\author{Roberto Osellame}
\author{Francesco Ceccarelli}
\author{Daniel Mart\'inez}
\author{Tobias Guggemos}
\author{Iris Agresti}
\author{Philip Walther}
\affiliation{See the main text for affiliations.}
\date{\today}
\maketitle
\tableofcontents
\part*{Materials and Methods}

    \section{Host satellite and launch details}\label{sec:host}
After two years of project development, the experimental setup reported in this manuscript was launched on a Falcon 9 rocket as part of SpaceX’s Transporter-14 rideshare mission. The launch happened on June 23rd 2025 from Space Launch Complex 4E (SLC-4E) at Vandenberg Space Force Base, California, at 14:25 PT (21:25 UTC). Our experiment is a 3U payload hosted on the ION SCV Passionate Paula (Skytrail mission): an orbital transfer vehicle designed by the European company D-Orbit, whose main goal is to transport and deploy satellites into their target orbits. The satellite has been since June 2025 in a Sun-synchronous orbit at an altitude of approximately \ensuremath{510\,\mathrm{km}} and is expected to be deorbited within a few years after decommissioning through natural atmospheric drag, in line with D-Orbit's end-of-life disposal procedure \cite{dorbit_skytrail_2025}.
\section{Photonic hardware}
\label{sec:experimental_apparatus}
\subsection{Single-photon source}
The source is based on type-II SPDC in a collinear single-pass configuration using a 1 cm-long ppKTP crystal, pumped at $\ensuremath{405\,\mathrm{nm}}$ and generating photon pairs at $\ensuremath{810\,\mathrm{nm}}$.
The mechanical assembly employs a structure based on carbon-fibre reinforced polymer (T800/M55J) and titanium (Ti-6Al-4V), with optical components bonded into precision-machined negative pockets using a low-outgassing two-component epoxy. This integration strategy combines mechanical robustness with the flexibility of using commercial off-the-shelf (COTS) optical components. The main limitations are the reduced degrees of freedom available for crystal alignment and the absence of waveplates. Hence, polarization control is achieved through mechanical rotation of the fibre collimators and the use of PM fibres.
Environmental qualification tests confirmed the platform's mechanical robustness. The prototypes successfully passed sinusoidal/random vibration and shock tests, with no measurable degradation in either coincidence rate or optical insertion loss.
Finally, two pre-launch HOM interference measurements were performed~\cite{hong1987measurement}. Before the flight-acceptance test, a commercial temperature controller was used to scan the ppKTP crystal temperature, and a visibility of $V= 0.924 \pm 0.003$ was obtained at an optimal crystal temperature of $\sim \ensuremath{40\,{}^{\circ}\mathrm{C}}$.
Following the flight-acceptance test, the complete flight optics and electronics chain, including the fixed optical delay line, custom temperature controller, and SPADs acquisition board, yielded a visibility of $V= 0.788 \pm 0.078$ at a crystal temperature of $\sim \ensuremath{32.5\,{}^{\circ}\mathrm{C}}$ (see section~\ref{sec:two_photon_interference} for more details on the temperature calibration). The optical delay line is of a similar design to the single-photon source, consisting of a titanium baseplate on which two collimators are bonded at a fixed distance to synchronize the photons in their path and polarization degree of freedom. The reduced visibility after the flight-acceptance test is primarily due to the temperature controller, which, at this stage of the project, had not yet been fully optimized.
\subsection{Universal photonic integrated circuit}
\label{sec:photonic_processor}
Input state processing is performed using a uPIC. The circuit implements a rectangular mesh architecture~\cite{clements2016optimal} supporting six spatial modes. It comprises 15 MZIs, each equipped with two thermo-optic phase shifters (save 3 unobservable external phases) to enable full reconfigurability, occupying a footprint of \ensuremath{0.635 \times 60\,\mathrm{mm}}. Single-mode optical waveguides operating at \ensuremath{810\,\mathrm{nm}} are fabricated by femtosecond laser writing on a Corning EAGLE XG borosilicate glass substrate~\cite{corrielli2021femtosecond}. Following waveguide inscription, the MZI arms were fully suspended via laser ablation to improve thermal isolation and minimize power consumption~\cite{albiero2022toward}. The resulting waveguide bridges were maintained at a width of \ensuremath{50\,\mathrm{\mu{}m}} to provide sufficient mechanical strength. Chromium micro-heaters and copper interconnects were then patterned using standard photolithography, metal evaporation and lift-off, followed by a post-annealing step and $\text{SiO}_2$ passivation as reported in Ref.~\cite{albiero2022toward}. Ultrafast laser writing of waveguides in glass produces photonic circuits whose optical performances are particularly robust to the radiation dose expected in LEO missions \cite{piacentini2021}.

To ensure structural integrity under launch conditions, the photonic chip was mounted on a titanium baseplate. For optical integration, fibre arrays were precisely aligned and bonded to the input and output facets using an index-matched UV-curable adhesive. To mitigate the intrinsic mechanical fragility of these optical interfaces, the baseplate served as a rigid support: a two-component epoxy was injected above and beneath the V-grooves, fully encapsulating both coupling regions (see section~\ref{sec:pigtailing}). Finally, electrical interconnections between the phase shifters and custom printed circuit boards (PCBs) were established using electrically conductive adhesive, additionally covered with the same epoxy to guarantee enhanced mechanical robustness.

Following fabrication and packaging, the uPIC was characterized on the ground at \ensuremath{30\,{}^{\circ}\mathrm{C}} and \ensuremath{7\times10^{-6}\,\mathrm{mbar}} to implement arbitrary unitary transformations across the spatial modes. Thanks to the implemented isolation strategies, the $2\pi$ power dissipation was reduced to less than \ensuremath{11.5\,\mathrm{mW}} for all the shifters, while the thermal crosstalk between adjacent MZIs remained below 6\%. Benchmarking the device over Haar-random unitaries yielded reconstruction fidelities exceeding 0.996 on average, confirming its accurate operation as a universal processor. The chosen fidelity is the same as reported in \cite{pentangelo2024high} based on classical intensity measurements, while in the main text the fidelity is based on the ``Bhattacharyya coefficient,'' which quantifies the ``closeness'' of two random statistical ensembles~\cite{dodge_oxford_2003}, defined as $F=\sum_{ij} \sqrt{t_{ij}\cdot \bar{p}_{ij}}$. Here, $\bar{p}_{ij}$ is the experimental detection frequency for the $ij$th output, while $t_{ij}$ corresponds to the theoretical detection frequency for the $ij$th output~\cite{yin2025experimental}.
\subsection{Detectors}
\label{sec:detectors}
The detection circuit is built around commercially available Si-SPADs, SAP500, from Laser Components, in a passive quenching configuration based on~\cite{Stipcevic2010}. Here, an avalanche is detected by AC-coupling the anode voltage to a comparator (LT1719, Analog Devices) that acts as a constant-level discriminator. After pulse shaping through a monostable multivibrator, the signal is routed to coaxial connectors, which carry the pulses to the time-tagger.

Passively quenched circuits need to balance the trade-off between maximal achievable overvoltage $V_{\text{OV}}$, with the time constant of the avalanche recovery process $\tau_R$, as both of these quantities are dependent on the resistance of the quenching resistor $R_S$. The deployed $R_S=\ensuremath{402\,\mathrm{k\Omega}}$ yields a minimal avalanche recovery time constant of $\tau_R = R_S(C_{\text{SPAD}}+C_p) \approx \ensuremath{650\,\mathrm{ns}}$, where $C_{\text{SPAD}}$ is the biased diodes' capacitance, and $C_p$ is stray capacitance (assumed to be 0 for estimating the minimal time constant). Meanwhile, it still allows for an appreciable maximum overvoltage of $V_{\text{OV}}=\ensuremath{20\,\mathrm{V}}$ (the maximal voltage satisfying $I_{\text{latch}} < V_{\text{OV}}/R_S$, with a latch current of $I_{\text{latch}}\approx \ensuremath{50\,\mathrm{\mu{}A}}$).

The reverse bias voltage is generated by high-voltage DC-DC converters (TZ series, Matsusada Precision), one per diode, and is individually controllable by the System-on-a-Module (SoM) used for payload control via 16-bit digital-analog-converters, DACs, (AD5669, Analog Devices). Each diode is mounted on a dual-stage thermo-electric cooler (TEC), which is powered by its own buck converter (LT8613, Analog Devices) and controlled by a full-bridge pulse-width modulation (PWM) driver with PID control (LTC1923, Analog Devices). Temperature setpoints are individually controllable per diode by the SoM via another DAC, and actual temperatures are read by an analog-digital converter, ADC, (ADS1278, Texas Instruments) via negative temperature coefficients (NTCs) internal to the SPAD package. This setup allows us to reach temperatures between, depending on the detector, ($\ensuremath{-76\,{}^{\circ}\mathrm{C}}$ to $\ensuremath{-66\,{}^{\circ}\mathrm{C}}$) and ($\ensuremath{71\,{}^{\circ}\mathrm{C}}$ to $\ensuremath{74\,{}^{\circ}\mathrm{C}}$) during in-orbit operation.

This circuitry is combined on a single home-made electronic board, which, after mounting the detectors, receives the same post-processing treatment as most other electronic boards (detailed in section~\ref{sec:electronics_postprocessing}). Interfacing the fibre-coupled uPIC output to the free space detector surface is facilitated by bonding a collimation package and a focusing lens into an aluminium structure sitting on top of the SPAD packages.

Out of the six deployed detectors, three still produced a reliable signal after launch. While one detector was already defective pre-launch, two more detectors showed signals unrelated to SPDC-produced photon counts. However, since they still produce dark counts, we suspect these two failures are due to fibres breaking during launch.
\section{Electronics}\label{sec:electronics}
To ensure sustainable, long-term operation of the payload, all instruments and components must be capable of withstanding the harsh space environment, high vacuum, radiation, temperature fluctuations, as well as the mechanical stress of the rocket launch. The experimental apparatus footprint must also stay below \ensuremath{10\,\mathrm{L}} and \ensuremath{10\,\mathrm{kg}}, with power consumption under \ensuremath{30\,\mathrm{W}}. To meet these requirements, most of the instruments in the payload were self-designed and comprise the following boards: EPS, ADCB, TCB, PCS and SPADs (already detailed in section~\ref{sec:detectors}). All instruments were tested for space conditions, with mechanical testing playing a particularly important role in the mission's development (see section~\ref{sec:mechanical_testing}).
The payload is controlled by a space-proven single-board computer, the SoM, which is a commercially available Xiphos Q7S, built around a Xilinx Zynq-7000 system-on-chip, integrating a 32-bit ARM Cortex-A9 processor with tightly coupled FPGA logic. The FPGA runs the electronic signal sequences that drive the instruments. A self-designed time-tagger FPGA module computes correlations from photon-pair detections. The light source is a COTS fibre-coupled CW laser (Integrated Optics) operating at \ensuremath{405\,\mathrm{nm}}. The SoM controls the laser (power cycling) module throughout the experiment.

At the power-hardware level, the SoM manages the power distribution network by operating the EPS, enabling controlled activation of the other on-board instruments. It regulates and delivers independent voltage rails to each board, avoiding electrical interference and ensuring each board operates independently in the event of a failure.
Precise temperature stabilization is crucial for the optimal operation of the ppKTP crystal. The TCB consists of a dual-PID-loop current driver that monitors and regulates the temperature of the ppKTP crystal, achieving a stability of better than \ensuremath{0.1\,\mathrm{K}} (see Fig.~\ref{fig:temperature-control}).

A remarkable property of the uPIC is its reconfigurability, enabled by a set of 27 thermal phase shifters. The payload includes a programmable 28-channel electrical current source to fully exploit this reconfigurability. The PCS (Qontrol) has 16-bit precision and can drive more than a full phase rotation on all 27 shifters. It can be dynamically programmed to prepare an arbitrary 6$\times$6 unitary on the uPIC.
To perform quantum measurements, we use the architecture described in section~\ref{sec:detectors}. Prior to every operation, a SPAD breakdown-voltage calibration routine runs alongside SPAD temperature stabilization. The time-tagger architecture consists of a delay line based on sequential CARRY4 chains \cite{Adamic_zynqTDC}, enabling an average precision of $\sim$\ensuremath{27\,\mathrm{ps}}. The time-tagger continuously acquires single-photon detection events and computes coincidence statistics among all detector pairs in real time.
Finally, ADCB collects temperature, voltage and radiation-level measurements via 16 independent analogue channels. These measurements serve as \textit{health checks} for the system, since they are recorded during every experiment execution and provide information about the payload's general environment and system degradation.
\subsection{Staking and conformal coating}\label{sec:electronics_postprocessing}
Conformal coating is a protective layer applied to components to shield them from environmental conditions. In space, this protection is two-fold. Firstly, it prevents outgassing, protecting both the coated device from damage due to material loss and the remaining payload from contamination by the lost material. Secondly, conformal coating delays the onset of metal whisker growth and delays or prevents electrical shorts caused by these whiskers. This is particularly vital for COTS components, whose outgassing parameters are often undocumented or subject to batch-to-batch variation.

PCS, ADCB, TCB, the SoM break-out-board (SoMBOB), SPAD and EPS underwent the following procedure: Cleaning with an electrostatic discharged (ESD) brush under running deionized water. Ultrasonic cleaning in 2-propanol. Submersion cleaning in ethanol. Air drying. Oven drying at \ensuremath{70\,{}^{\circ}\mathrm{C}} for at least \ensuremath{6\,\mathrm{h}}.
Physically large components (predominantly inductors, tantalum capacitors and SPADs) were staked to the PCB using a low-outgassing, low-stress, thermally conductive epoxy (Loctite Ablestik 285 / CAT 9), both to decrease the risk of mechanical detachment of these components during launch, and to decrease the thermal resistance of the SPADs to the PCB. Staked assemblies were oven-cured at \ensuremath{70\,{}^{\circ}\mathrm{C}} for at least \ensuremath{6\,\mathrm{h}} before conformal coating.
Connectors were covered with polyimide tape, SPAD windows with optical tissue, and polyimide tape. A dual-cure urethane acrylate (Loctite Stycast PC40-UMF) was thinned with toluene and applied to the boards using an airbrush (Fengda FE-183K, nozzle diameter \ensuremath{0.5\,\mathrm{mm}}) operated with compressed nitrogen. To maintain thermal conductivity between PCBs and frame, the acrylate was removed from plated mounting holes using toluene and lint-free wipes. Initial UV cure was $\ensuremath{20\,\mathrm{s}}$ per side in a UV oven ($\ensuremath{365\,\mathrm{nm}}$, $\approx\ensuremath{300\,\mathrm{mW/cm^{2}}}$), followed by a moisture cure (ambient lab) for $\ge\ensuremath{24\,\mathrm{h}}$. PCBs were then individually packaged in ESD bags together with silica gel desiccant.
For each coating run, a dummy underwent the same treatment and was stored for reference.
After outbaking (PCBs, cables, fibres), all components were individually vacuum-sealed together with silica gel.

\section{Payload operation}
\label{sec:operation}
The on-board computer (OBC) is a Xiphos Q7S radiation-tolerant single-board computer based on a Xilinx Zynq-7000 system-on-chip, integrating a 32-bit ARM Cortex-A9 processor with tightly coupled FPGA logic. The payload runs a custom Linux distribution built with Yocto and extended with Python 3, NumPy, and OpenCV. Due to limited internal flash capacity, the user-space filesystem is hosted on an external SD card mounted at boot.
All subsystem control, including the single-photon detectors, time tagger, phase shifters, and supporting electronics, is exposed via a lightweight Python interface directly accessing memory-mapped Zynq registers. Higher-level experiment logic, including acquisition and control sequences, is implemented in Python on top of this driver layer.

The payload is operated through the host platform and is not directly commanded from the ground. All interactions are mediated by the ION SCV flight payload computer, which arbitrates power, timing, file transfer, and command execution via On-Board Control Procedures (OBCPs) over an RS-422 link. Three OBCPs define the full interface: OBC power control, file transfer between the platform staging area and payload filesystem, and bounded-time command execution.

Operational days are structured as OBCP sequences in which the OBC is powered on, the payload image is uploaded, a schedule script is executed, telemetry and data are retrieved, and the system is powered down within the allocated window. Operation windows last typically 8 hours within predefined UTC intervals. Within each window, the schedule coordinates on-board operations and on-chip experiments.

Software development is performed in a self-hosted browser-based workspace environment providing reproducible toolchains for each subsystem, including OBC image builds (Yocto), OBCP compilation, photonic calibration, and back-end simulation.
On the ground segment, the bus provider SDK mirrors the flight payload computer and exposes an identical OBCP interface, ensuring full equivalence between laboratory and in-orbit operations. An Engineering Model (EM) is maintained in functional parity with the Flight Model (FM), sharing OS image, drivers, OBCPs, and calibration procedures, differing only in device-specific calibration constants.

The EM hardware (OBC, photonic chip, detectors, TEC boards, and instrumentation) is integrated via a dedicated Ethernet network, with USB/serial peripherals re-exported through single-board computers as TCP endpoints and mounted as virtual devices in the workspace environment.
All flight software and OBCPs are first validated on the EM before FM deployment. Only after successful EM execution are updates integrated into the flight operation pipeline and uplinked to the FM.

\clearpage

\section{Space qualification of optical and electronic components}
\label{sec:space_as_environment}
The demand for a quantum signature, i.e., two-photon interference, in the uncontrollable and harsh environment of space imposes stringent requirements on the optical hardware. More precisely, the single-photon source (SPS), universal photonic processor (uPIC), and single-photon detectors (SPADs) must deliver, process, and register indistinguishable photons with high efficiency while remaining compact enough to fit within a 3U CubeSat framework and comply with its resource limits. Beyond these functional demands, a space mission begins with a rocket launch in which strong mechanical forces act on the sensitive alignment of the optical framework; the optomechanical design must therefore withstand short-term mechanical stress such as vibration loads and pyrotechnic shocks \cite{anandito2019dynamic,secretariat2013spacecraft}.

Beyond the short-term hazards of launch, the space environment itself, particularly in sun-synchronous orbits (SSO) at an altitude of $510~\mathrm{km}$, poses additional risks to optical and electronic components through periodic temperature cycling, high vacuum, and radiation. Temperature drifts are unavoidable in SSO, as the satellite alternates between facing the sun and the Earth's shadow (umbra), producing periodic heating and cooling of the payload; see Fig.~3a of the main text. Since a full SSO orbit lasts approximately 90 minutes\footnote{In the case of a typical SSO the satellite faces the sun for $\sim \ensuremath{60\,\mathrm{min}}$, while it passes in the shade of the Earth for $\sim \ensuremath{30\,\mathrm{min}}$ \cite{seon_brief_2000}.}, these thermal cycles are relatively fast and can cause material damage through thermally induced fatigue \cite{boain2004ab,carvelli2015fatigue}. Compounding this, the near-absence of atmosphere in low Earth orbit limits heat transfer to conduction and radiation alone, so temperature-sensitive components require dedicated heatsinks or radiators for efficient heat distribution \cite{miao2021spacecraft}. The vacuum environment also restricts material choice: components with high total mass loss (TML) undergo outgassing during the transition from atmospheric pressure to vacuum, and the resulting sublimated organic matter can contaminate and degrade nearby optical elements, for instance, by increasing the insertion loss of a lens through surface adsorption \cite{jiao2019outgassing,fisher1971compilation,calders2018modeling}.

A further major challenge for space missions is the abundance of energetic particles trapped in Earth's magnetic field. Particle density depends strongly on satellite altitude and inclination; at $510~\mathrm{km}$, the payload primarily encounters protons and electrons from the inner Van Allen radiation belt \cite{calders2018modeling}. Radiation damage to spacecraft components arises from the deposition of energy from these particles, which can be classified as ionizing or non-ionizing, both with significant impact on semiconductor performance and lifetime.

Semiconductor devices generally respond to radiation with two distinctive mechanisms. For one, radiation may slowly degrade electrical parameters until they fall out of specification. This behaviour is characterized by the acceptable total ionizing dose (TID) (even though the response generally depends on the timeframe over which the dose was acquired). Furthermore, high-energy particles may disrupt the device immediately upon deposition. These single-event effects (SEE) pose a risk of abruptly creating shorts in the circuit, leading to localized overcurrents and the potential destruction of the biased IC within a short timeframe. A device's resilience to this failure mode is characterized by the linear energy transfer (LET) immunity threshold.
While radiation-hard or radiation-resistant parts that cover most of the typical functions of semiconductor devices exist, acquiring them can be challenging for an academic project (due to economic constraints, project-timeline incompatible lead times, or export controls). Confidence in a commercial-off-the-shelf (COTS) part's reliability in the space environment can alternatively be increased either by after-market radiation testing or by anecdotal survivability in previous missions ("space heritage").
All ICs and connectors deployed in this mission are collected in Table~\ref{tab:sup_component_qualification_level}, together with their respective space qualification level. All passives were either high-reliability COTS components (tantalum caps) or AEC-Q qualified (all other passives).

Apart from the gradually increasing dark count rate of the SPADs, no failure of any of these components was noticed until the loss of functionality of one of the eleven buck-converters (EPS, 5V rail) occurred after about 8.5 months in orbit, with less than $\ensuremath{136\,\mathrm{h}}$ of operation. While we are certainly not sensitive to all possible degradation modes of these devices, we still believe that Table~\ref{tab:sup_component_qualification_level} can prove useful as one (more) data point of heritage for these ICs.

    \begin{table}
        \centering
\caption[]{\textbf{Component Qualification Levels.} Overview of the electronic and optoelectronic components used in the payload, together with their space qualification status. ``Heritage'' denotes components with a documented flight history, ``COTS screened'' refers to commercial off-the-shelf parts subjected to, and passing, radiation tests (TID, SEL, or both), and ``RH sister part'' indicates the existence of a radiation-hardened variant of the listed part. Components listed as ``none'' underwent no dedicated space qualification and were selected based on datasheet specifications.}
\label{tab:sup_component_qualification_level}
        \begin{tabular}{|c c c c|}
            \hline
            Part & Type & Qualification & Ref \\ [0.5ex]
            \hline\hline
            AD5669 & DAC & none & N.A. \\
            \hline
            ADS1278 & ADC & RH sister part & N.A. \\
            \hline
            DMC3016 & MOSFET & none & N.A. \\
            \hline
            DMC3025 & MOSFET & none & N.A. \\
            \hline
            LMP7704 & OpAmp & RH sister part & N.A. \\
            \hline
            LT1498 & OpAmp & COTS screened & \cite{Tscherne2020,Wind2022} \\
            \hline
            LT1719 & Comparator & heritage & \cite{Chandrasekara2017,Bedington2015} \\
            \hline
            LT8613 & DC-DC Buck & heritage, COTS screened & \cite{Kobayashi2019} \\
            \hline
            LTC1923 & TEC Controller & none & N.A. \\
            \hline
            LTC2991 & V, I, T Monitor & none & N.A. \\
            \hline
            LTC6240 & OpAmp & COTS screened & \cite{Tscherne2020,Wind2022,Wind2023} \\
            \hline
            tz-0.2p-5 & HV PSU & heritage sister series & \cite{speqs1optica} \\
            \hline
            RM-VT01-A & radiation sensor & N.A. & N.A. \\
            \hline
            SAP500 & SPAD & heritage & \cite{Perumangatt2021} \\
            \hline
            SN74LVC74 & FlipFlop & COTS, 74LVC screened & \cite{Spiezia2013} \\
            \hline
            THS4541Q & Differential Amplifier & RH sister part & \cite{Narayanan2019} \\
            \hline
            TPS79318-EP & LDO & COTS high-rel & N.A. \\
            \hline
            Zynq-7020 & SoC & heritage & \cite{Xiphos} \\
            \hline
            Harwin M80 & connector & heritage & various, e.g. \cite{Ubbels2008} \\
            \hline
            SSMC & coaxial connector & none & N.A. \\
            [1ex]
            \hline
        \end{tabular}

    \end{table}

\subsection{Shielding against radiation}\label{sec:shielding}
\begin{figure}[!tbp]
    \centering
    \includegraphics[width=1\textwidth,height=0.63\textheight,keepaspectratio]{ 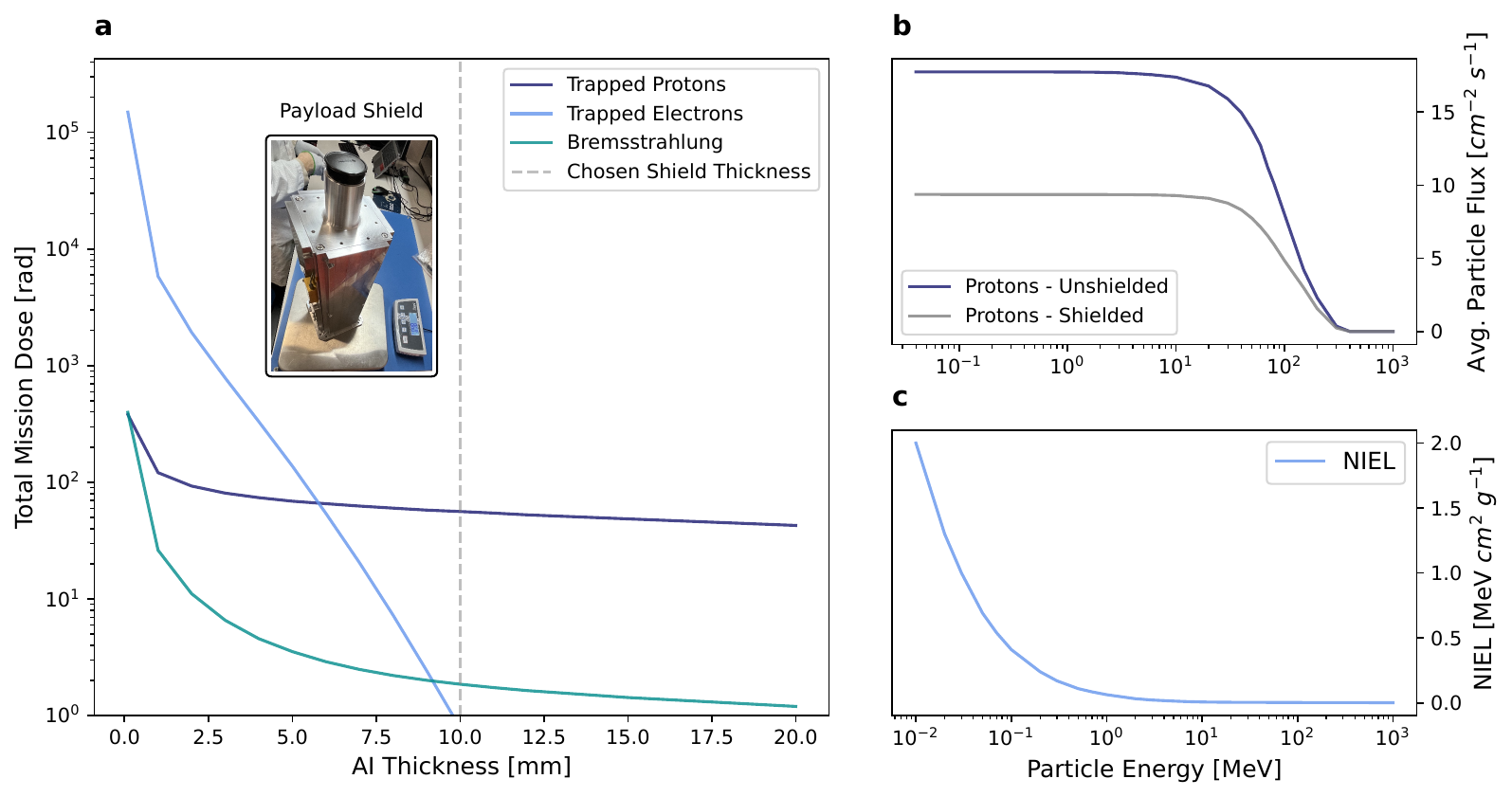}
    \caption[]{\textbf{Radiation Profile of SSO and Shielding.}  Simulated total radiation dose vs. shielding thickness (\textbf{a}), average particle flux (\textbf{b}), and displacement damage (\textbf{c}) for a one-year SSO mission at an altitude of $510~\mathrm{km}$ and inclination of $97.5^\circ$ \cite{calders2018modeling}. The shielding material under consideration is the Aluminium alloy AA7075, commonly used in space missions for its high strength and fatigue resistance \cite{couteau2018spontaneous}. \textbf{a} Shield effectiveness is shown for energetic particles and their corresponding Bremsstrahlung (solid lines), with the chosen Aluminium thickness marked as a vertical dashed line. The total mission proton dose changes little for shielding thicker than $1~\mathrm{cm}$, whereas electrons and their Bremsstrahlung are significantly attenuated; the optimal thickness was therefore chosen to be $1~\mathrm{cm}$. The inset shows the payload with the corresponding Aluminium mass relative to its total weight of nearly $10~\mathrm{kg}$. \textbf{b} The average particle flux is compared for $1~\mathrm{cm}$ shielding (grey) versus no shielding (blue), showing a marked reduction. In both cases, low-energy protons dominate the particle population and, as shown in \textbf{c}, contribute most strongly to non-ionizing displacement damage. Shielding thus offers substantial protection overall, but effectively shielding the payload, particularly the SPADs, from low-energy protons remains inherently difficult.}
    \label{fig:00}
\end{figure}
For our experiment, radiation damage primarily affects the performance of single-photon detectors (SPADs), as ionizing events can trigger an avalanche response, producing a false count unrelated to an actual photon; such events increase the effective dark count rate (DCR) without damaging the underlying crystal structure. Non-ionizing events have more severe consequences, as an energetic particle can induce displacement damage in the semiconductor lattice. The resulting lattice defects create new energy levels within the bandgap, leading to a permanent increase in DCR and pushing the SPAD closer to saturation. The non-ionizing energy loss (NIEL) provides a quantitative measure of this displacement damage and can be used to assess its severity \cite{wilson2021radiation}.

Since low DCR is essential for a favorable signal-to-noise ratio in quantum optics experiments, mitigating radiation-induced damage is critical. A straightforward strategy is to shield the payload from incoming energetic particles, and Fig.~\ref{fig:00} reports the simulated radiation dose (in rads) for a one-year SSO mission, estimating the required Aluminium shielding thickness and the corresponding expected displacement damage. Such numerical results were obtained using SPENVIS (SPace ENVironment Information System), which is a web-based tool developed by ESA which adopts recognized models of the radiation environment to estimate practical effects on hardware \cite{calders2018modeling}. The simulations show that a $1~\mathrm{cm}$ shield effectively damps electrons and their Bremsstrahlung; protons, however, are considerably harder to attenuate, as the total radiation dose remains largely unchanged even for shielding thicker than $1~\mathrm{cm}$. Given the additional constraint of the $10~\mathrm{kg}$ mass allocation for the payload, a $1~\mathrm{cm}$ shield of aerospace-grade Aluminium (AA7075) was chosen as a practical trade-off, reducing the average proton flux to roughly half of the unshielded value.

\subsection{Photonic circuit and SPAD board pigtailing procedure}\label{sec:pigtailing}
The universal photonic processor (uPIC) and the optical fibre connections of the detection module also had to be adapted for a space mission. As the underlying mechanical challenges are the same for both the source and the delay line (see section~\ref{sec:experimental_apparatus}), we bonded optical elements onto a robust framework using strong epoxy adhesive.

The uPIC consists of optical waveguides inscribed on a borosilicate glass substrate via femtosecond laser writing, arranged such that a total of $15$ Mach-Zehnder interferometers (MZIs) are imprinted on the substrate. Each interferometer is equipped with two thermal phase shifters for reflectivity and phase tuning, and the resulting MZI mesh is accessed via six fibre-coupled input and output modes. This arrangement enables the application of any $6 \times 6$ unitary transformation to the input photon state, providing full universality of the photonic processor \cite{ceccarelli2020low,clements2016optimal}.

\begin{figure}[!tbp]
    \centering
    \includegraphics[width=1\textwidth,height=0.67\textheight,keepaspectratio]{ 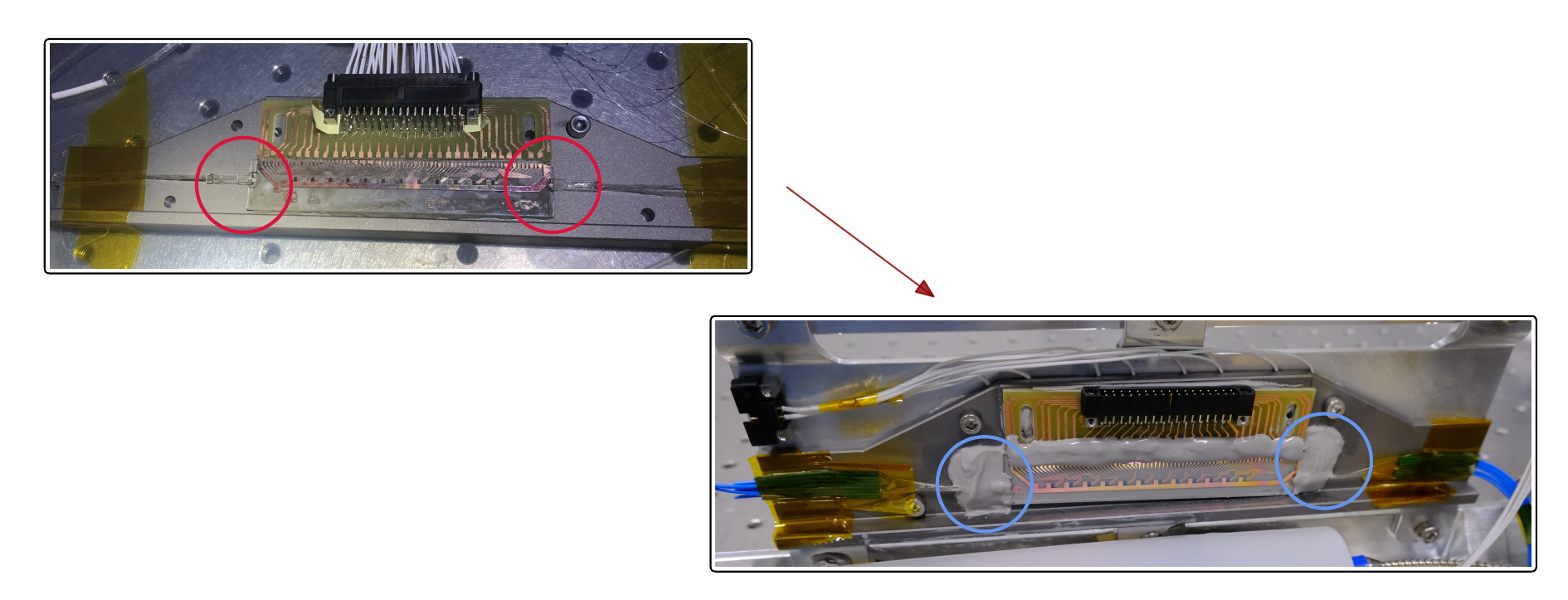}
    \caption[]{\textbf{Mechanical adaptation and flight model version of the uPIC.} The photonic processor is shown in its standard lab-model configuration (upper left) and flight-model configuration (lower right). In both cases, the borosilicate glass substrate is bonded to the titanium plate via strong epoxy adhesive (grey film), and the thermal phase shifters (copper lines) on the glass are clearly visible. Adjacent to the glass substrate sits the electronic board (green) with a connector, whose wires are linked to the phase shifters via gold wires (lab model) or conductive adhesive (flight model). The red circles in the upper-left indicate the V-groove fibre array and the locations of the gold wires. Both components are highly sensitive to mechanical stress, and the fibre array in particular can easily be misaligned by slight pulling or pushing motions. For this reason, the flight-model adaptation includes an additional layer of epoxy adhesive (grey film) around the V-groove array and the wire bonding (blue circles), an approach that has proven successful in protecting the uPIC from mechanical stress.}
    \label{fig05}
\end{figure}

The most mechanically delicate part of the uPIC is the input/output fibre-coupling section, consisting of a V-groove fibre array precisely aligned with the input/output facet of the optical chip. In its standard laboratory configuration, the fibre array is bonded to the facet with a UV-curing optical adhesive, which has the advantage of a refractive index close to that of glass but a bonding strength considerably lower than that of the two-stage epoxy adhesive used for other optical components (see section~\ref{sec:experimental_apparatus}). To mechanically harden the optical chip, the complete glass substrate, including fibre arrays, is bonded to a robust titanium baseplate: the glass substrate is first bonded to the baseplate, after which the fibre arrays are covered with epoxy adhesive. In this assembly procedure, the fibre arrays are already aligned and bonded to the facet with the standard optical adhesive, and the baseplate is designed to leave additional space around the V-groove array. Epoxy adhesive is then applied below and on top of the fibre array, fully covering both input and output arrays. Fig.~\ref{fig05} shows the evolution of this mechanical modification; the facets of the flight-ready photonic processor (lower right) are nearly completely covered in epoxy. In addition to the fragile V-groove fibre arrays, the electronic connections between the circuit board and the phase shifters are likewise protected by a thick adhesive layer.

\begin{figure}[!tbp]
    \centering
    \includegraphics[width=0.8\textwidth,height=0.75\textheight,keepaspectratio]{ 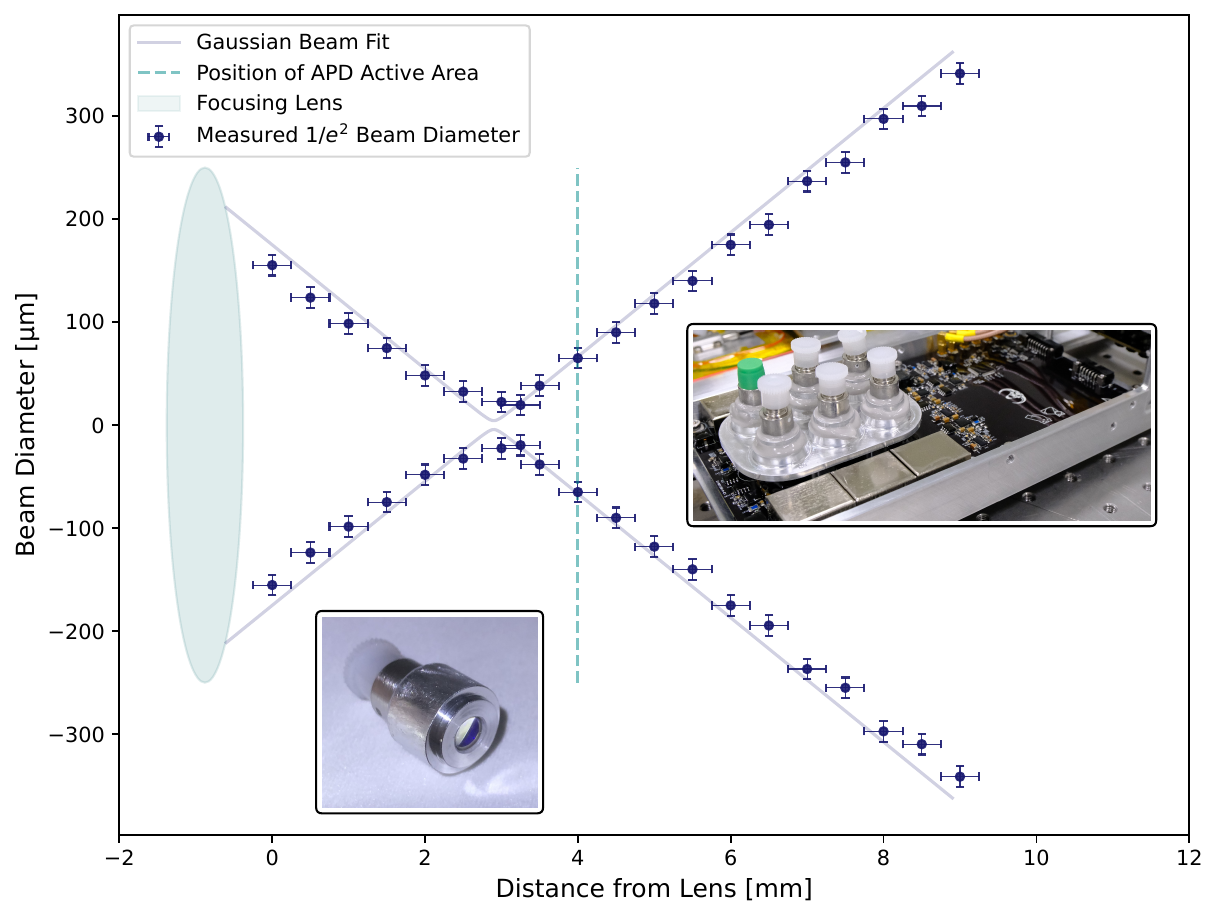}
    \caption[]{\textbf{Fibre Pigtailing of the SPAD Board.} Measured Gaussian beam evolution for the fibre pigtailing of the SPAD board. In this setup, the beam from a commercial collimator is focused by a small convex lens (blue ellipse) with focal length $f \sim 6\mathrm{mm}$. The convex lens is bonded to the collimator via a collimator sleeve (small inset, left), and this package is then aligned to the active area of the APD (green dashed line). The collimator is positioned slightly off focus to allow greater flexibility in mounting and alignment. Once optimal alignment is achieved, the collimator package is bonded into the Aluminium bracket; the inset on the right shows the SPAD board with Aluminium bracket (grey collimator host) and complete pigtailing.}
    \label{fig03}
\end{figure}

The bare SPAD board consists of six APDs arranged in a tight $2 \times 3$ geometry. The photosensitive area of each APD is centered within the circular TO-8 can and limited to $500\,\mathrm{\mu m}$ in diameter, requiring incoming photons to be precisely aligned and focused onto this active area. To host a fibre pigtail, the SPAD board is extended with a robust AA7075 Aluminium bracket featuring holes centered on each APD facet. Collimators with additional focusing optics are aligned to the active area and bonded symmetrically to the bracket. To achieve a sufficiently small photon spot size, a commercial collimator is extended with a collimator sleeve housing a small convex lens, yielding a spot size of \ensuremath{70\,\mathrm{\mu{}m}} on the photosensitive area of the APD. Fig.~\ref{fig03} shows the measured beam evolution for this setup and the resulting pigtailing of the SPAD board.
\clearpage
\subsection{Mechanical testing}\label{sec:mechanical_testing}
To avoid damage from the strong mechanical loads during a rocket launch, we mechanically hardened optical systems and validated them against loads of the same intensity. For this reason, prototypes of the optical components, i.e., SPS, DL, and uPIC, were tested for mechanical vibration loads and pyrotechnic shocks. If in resonance with the device under test (DUT), these mechanical stresses can cause partial damage or complete failure (e.g., optical misalignment); each optical component therefore underwent mechanical testing at a designated aerospace testing facility. Each DUT was tested under sinusoidal/random vibrations, comprising frequency sweeps in the range $0 < \omega < 2~\mathrm{kHz}$, as well as pyrotechnic shock tests in which high-frequency shocks around $\omega \sim 10~\mathrm{kHz}$ are imparted on the device.

\begin{figure}[!tbp]
    \centering
    \includegraphics[width=1\textwidth,height=0.61\textheight,keepaspectratio]{ 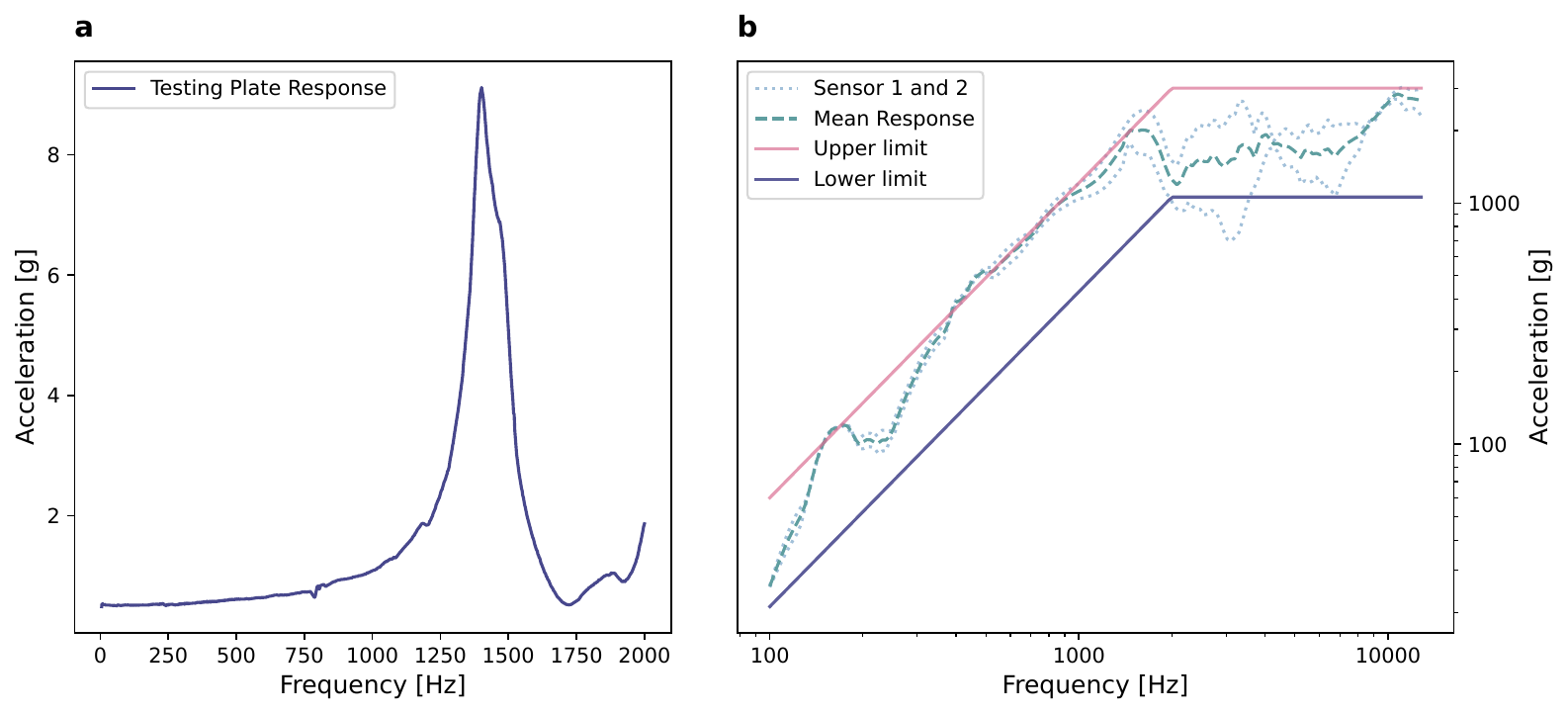}
    \caption[]{\textbf{Vibration and Shock Tests of Source Prototype.} Measured test results for the sinusoidal vibration test \textbf{(a)} and the shock response spectrum \textbf{(b)} of the source prototype. Both panels show data for a single acceleration direction. The vibration test in \textbf{(a)} includes both the source and delay-line prototypes mounted together on an Aluminium plate (inset, left), while in the shock test, the two DUTs are tested independently, mounted on a smaller Aluminium plate of similar form factor (inset, lower right). Both figures show the resulting forces in g as a function of applied frequency \textbf{(a)} and natural frequency \textbf{(b)}. In the sinusoidal vibration test, the driving frequency is swept from $0 < \omega < 2 \mathrm{kHz}$. The natural frequency of the full test assembly—Aluminium plate with mounted DUTs—is found at $\omega \sim 1.3 \mathrm{kHz}$, where resonance with the driving frequency amplifies the measured acceleration up to 9 g. In \textbf{(b)}, the shock response spectrum (SRS) for the SPS prototype predicts the peak acceleration the system must withstand as a function of natural frequency \cite{irvine2002introduction}. The upper (red) and lower (blue) limits denote numerical bounds, while the dashed lines show the transformed measured sensor data. At its natural frequency of $\omega = 1.3 \mathrm{kHz}$, the SPS would need to withstand accelerations of up to 1500 g. Note that the SPS and DL were tested on a smaller Aluminium plate of similar form factor, so their natural frequencies may differ from those in \textbf{(a)}.}
    \label{fig04}
\end{figure}

For the vibration tests, the DUT is mounted on an electrodynamic shaker, and its resonance frequency is identified by sweeping the drive frequency in both sinusoidal and random fashion. At resonance, the DUT experiences strong acceleration forces, allowing this test to simulate the dynamic vibration loads of a rocket launch. Fig.~\ref{fig04}a shows the measured acceleration response to a sinusoidal drive frequency for a typical ``testing plate'', on which optical- and/or circuit board components were mounted for testing.

The shock test is performed using an impact hammer dropped from a specific height in close proximity to the DUT. The mechanical response during a shock test is characterized in terms of a shock-response spectrum (SRS), as shown in Fig.\ref{fig04}b. An SRS transforms the short-lived, complex shock pulse into the domain of the system's natural frequencies, allowing the shock response to be extracted in terms of the system's eigenfrequency and enabling a more realistic characterization of the response \cite{irvine2002introduction}. Accordingly, the DUT in Fig \ref{fig04}b, with a resonance frequency of $\omega \sim 1.3~\mathrm{kHz}$, had to withstand short-term accelerations of up to $1500~\mathrm{g}$.

The performance of the optical devices, in terms of two-photon rate and insertion loss, was measured before and after mechanical testing, and no signs of degradation from mechanical stress were observed.

\section{Thermal Vacuum Testing and Laser Damage}\label{sec:tvac}
The Engineering Model (EM), which is a near identical copy of the one sent to space (Flight Model FM), was tested in thermal vacuum (TVAC) in a designated facility at the German Aerospace Center (DLR) group RC3 with a TVAC system (see e.g. top right of Fig.~2 in the main text) that is capable of reaching $\sim \ensuremath{-20\,{}^{\circ}\mathrm{C}}$ to $\ensuremath{70\,{}^{\circ}\mathrm{C}}$. This temperature range is also considered the maximum survival temperature range in Low Earth Orbit under D-Orbit Standards. After oral communication with the host provider, the nominal cycling temperature was chosen to be $\sim \ensuremath{-20\,{}^{\circ}\mathrm{C}}$ to $\ensuremath{0\,{}^{\circ}\mathrm{C}}$ as this was found to be the typical environmental temperature range on prior D-orbit ION SCV missions. The cycle duration was set to match the actual orbit duration of $\sim \ensuremath{92\,\mathrm{min}}$ in a sun synchronous orbit at $\sim \ensuremath{510\,\mathrm{km}}$ altitude and the payload was kept at a fixed pressure of $\sim \ensuremath{1\times10^{-5}\,\mathrm{mbar}}$ for three consecutive days. During this TVAC test, the system health in terms of single counts, coincidence counts, TEC performance, and power consumption was monitored.

The only component that took damage from the TVAC test is the fibre-coupled commercial CW ``Matchbox laser'' by Integrated Optics. For this mission, the laser was modified to meet the challenging requirements of a space mission, and the modifications included changes to the laser housing materials. In addition to hardware hardening, this specific line of Matchbox laser modules was also tested under mechanical loads (vibration/shock). The most prominent change with respect to a standard module is the use of a Kovar housing, which offers lower thermal expansion compared to standard metal alloys \cite{ardebili_encapsulation_2018}. The internal components will thus experience less mechanical strain during thermocycling. The Kovar body is also gilded, which lowers the module's emissivity and thus its radiative heat loss, making it easier for the module to reach and maintain its optimal operating temperature (internal TEC) of $\sim \ensuremath{20\,{}^{\circ}\mathrm{C}}$ \cite{langley_gold_1971}.

\begin{figure}[!tbp]
            \centering
            \includegraphics[width=1\textwidth,height=0.63\textheight,keepaspectratio]{ 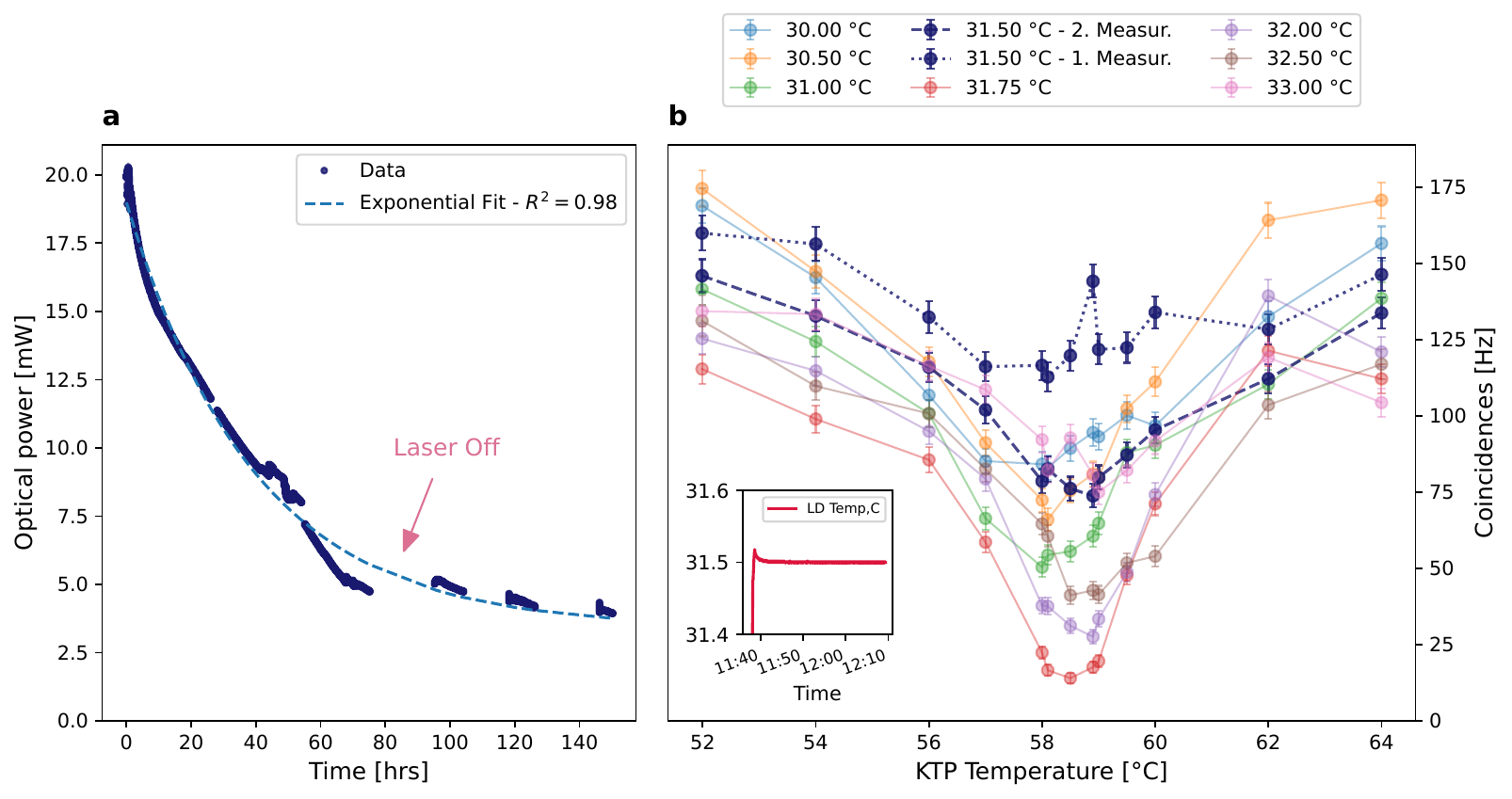}
            \caption[]{\textbf{Outgassing induced power- and coherence loss.} \textbf{a} Power loss of a Matchbox laser after $140$ hrs in vacuum due to severe internal outgassing of an adhesive; every internal optical component is contaminated by a layer of outgassed matter, effectively reducing the transmission of each component (e.g., lens). Start of measurement is at atmospheric pressure, and the final reached pressure (after $1$ hour) is at $\sim \ensuremath{1\times10^{-6}\,\mathrm{mbar}}$; final laser output power is at $\sim \ensuremath{20\,\%}$ of the initial value and power degradation only occurs during lasing (red arrow). \textbf{b} Hong-Ou Mandel scan with damaged EM laser for different laser diode temperature shows how e.g. outgassing and/or thermal distortion can affect the interference visibility; expected dip in coincidences is at $\sim \ensuremath{58.5\,{}^{\circ}\mathrm{C}}$ (healthy laser) with factory set diode temperature (LD Temp, small inset) of $\ensuremath{31.5\,{}^{\circ}\mathrm{C}}$. However, when scanning the KTP temperature around the expected value, no dip in coincidences is observed (dotted dark blue). Only when starting to scan the diode temperature (light blue - pink), the signature of two-photon interference sets in gradually; highest visibility is at a diode temperature of $\ensuremath{31.75\,{}^{\circ}\mathrm{C}}$ (red), while with the factory set value (dotted dark blue, dashed dark blue) interference is strongly suppressed.
            \label{fig:laser_power_vac}
            }
        \end{figure}

Despite these efforts to mechanically and thermally harden the laser module for a space mission, it was found that the laser incorporated a specific adhesive that did not meet low outgassing requirements. Moreover, this adhesive was used in close proximity to critical optical components, including the Volume-Bragg grating, collimation lens, and polarizing beam splitter. During vacuum operation, the outgassing of the adhesive leads to severe contamination (i.e., polymers from the epoxy adhesive) of surrounding optical components, effectively creating a layer of carbonaceous deposits. It is known that this deposit reduces the transmission of an optical component \cite{wagner_laser_2020} and Fig.~\ref{fig:laser_power_vac} shows the gradual decrease in output power of the EM Matchbox laser during a one-week vacuum ($\sim \ensuremath{1\times10^{-6}\,\mathrm{mbar}}$) measurement. The decay is exponential, starting from the factory set power of $\ensuremath{20\,\mathrm{mW}}$ and slowly converging to approx. $\ensuremath{4\,\mathrm{mW}}$, which suggests that the adhesive had largely outgassed by the end of the measurement.

It is important to note that this problem of laser-induced molecular contamination \cite{wagner_laser_2020} was identified during a project stage in which the FM was already integrated into the Falcon 9 rocket, so nothing could have been done to change the adhesive and/or clean the optical components of the outgassing deposit. The final on-earth vacuum testing of the FM also showed a significant coincidence count drop from $\sim \ensuremath{220\,\mathrm{Hz}}$ (atmospheric pressure, full laser power) to $\sim \ensuremath{50\,\mathrm{Hz}}$ ($\sim \ensuremath{1\times10^{-3}\,\mathrm{mbar}}$, low laser power) at $\ensuremath{7\,\mathrm{V}}$ SPAD overvoltage. As highlighted by the red arrow in Fig.~\ref{fig:laser_power_vac}a this decrease in optical power only happens during lasing and the final measured value of $\sim \ensuremath{50\,\mathrm{Hz}}$ coincidence counts was recorded after $\sim \ensuremath{70\,\mathrm{h}}$ of non-continuous in-vacuum laser operation (random on/off operation). In this sense, the laser was not yet fully ``outgassed'' during final on-earth testing and continued to outgas during in-orbit operation. Together with the increased DCR and damage in SPADs due to radiation (see Sec.~\ref{sec:space_as_environment}), the problem of laser outgassing explains the low in-orbit coincidence rate described in Fig.~3c of the main text, and it was crucial to tune the SPAD overvoltage to regain some of the signal.

In addition to the already severe complications of reduced output power, internal outgassing affected the laser's single longitudinal mode (SLM) performance. The Matchbox module itself consists of three alignment-sensitive components: the laser diode, the collimation lens, and the volume Bragg grating (VBG). The latter ensures that the laser is locked at a single frequency, i.e., enables the SLM property. If slightly misaligned ($\sim \ensuremath{\mathrm{\mu{}rad}}$ angle), SLM properties can be lost, and these misalignments may happen due to thermal distortion, vibration/mechanical shock, or contamination of optics \footnote{Described working principle and laser-specific details are based on oral communication with the laser company Integrated Optics.}. The first two contributions are inevitable in a space mission, and thermal distortion will eventually affect the laser during in-orbit thermal cycling. Measurements of the payload frame temperature and the host satellite temperature are shown in Fig.~\ref{fig:temperature-control} and underscore that the temperature drifts are significant. The frame on which the laser is mounted (passive heatsink), experiences a temperature drift of $\sim \ensuremath{7.5\,{}^{\circ}\mathrm{C}}$ over $\sim 8$ hours, and despite being assembled with a Kovar housing, the laser and its internal parts are exposed to strong periodic temperature changes.

The effect of contamination is more subtle, and in a typical optical configuration, the multimode (frequency) output of the laser diode is recollimated by a collimation lens and guided on a VBG. The VBG reflects only a narrow bandwidth back into the laser diode (optical feedback), thereby forcing the diode to emit only at this narrow frequency range. As the spectrum and center wavelength of a laser diode output also change with diode temperature, it is critical to keep the diode body at a specific temperature for optimal wavelength locking \cite{havermeyer_volume_2000}. This temperature is set by an internal Peltier module that, in the case of the FM laser, keeps the diode at $\ensuremath{32\,{}^{\circ}\mathrm{C}}$ to ensure optimal and stable SLM performance.

In the case where the critical optical components are layered with a carbonaceous deposit, the light emitted by the laser diode is scattered randomly, thus unwanted frequency modes (broadband) may back-reflect into the diode and worsen the side-mode suppression ratio (SMSR). The latter is an important measure as it specifies how strongly the side modes are suppressed relative to the main mode, and typical SLM lasers are considered to have values of at least $\ensuremath{40\,\mathrm{dB}}$ \cite{moeyersoon_degradation_2004}. In the worst case scenario, the amount of broadband parasitic back reflections overcomes the narrow-band feedback of the VBG, hence the laser loses its SLM feature, or in other words, the spectrum of the laser diode becomes broadband\footnote{Details on loss of coherence based on oral communication with the laser company Integrated Optics.}.

This phenomenon of coherence loss was observed for the two EM lasers (EM$_1$, EM$_2$) used and tested on Earth, and we attribute the behaviour of the FM laser in space to the same mechanism (see Sec.~\ref{sec:two_photon_interference}). The direct and drastic consequence of the loss of the laser's SLM property is the vanishing of two-photon interference capability when using the damaged laser. This is shown in Fig.~\ref{fig:laser_power_vac}b, and here EM$_2$ is used in a two-photon interference experiment after it lost its SLM property. In the given case, the experimental setup is the same as described in Fig.~1 of the main text, and the expected degeneracy temperature for this specific laser module and KTP crystal (EM source model) is at $\sim \ensuremath{58.5\,{}^{\circ}\mathrm{C}}$. It was found in testing that only when the diode temperature is swept in a specific fashion (from below the factory-set temperature to above), the laser's coherence can be restored deterministically, and Hong-Ou Mandel interference is enabled. It is important to note that the ``new'' optimal diode temperature is shifted, while at the ``old'' one, two-photon interference is strongly suppressed. This behaviour could be shown consistently for the two EM lasers\footnote{Note that one of the EM lasers underwent the TVAC test and lost its SLM property after $\sim 7$ months, while the other lost it randomly without any external influence. The latter is shown in Fig.~\ref{fig:laser_power_vac}b.} on Earth and in both cases, the sweeping process described in Fig.~\ref{fig:laser_power_vac}b was crucial to regain two-photon interference capabilities.

\begin{figure}[!tbp]
    \centering
    \includegraphics[width=1\textwidth,height=0.70\textheight,keepaspectratio]{ 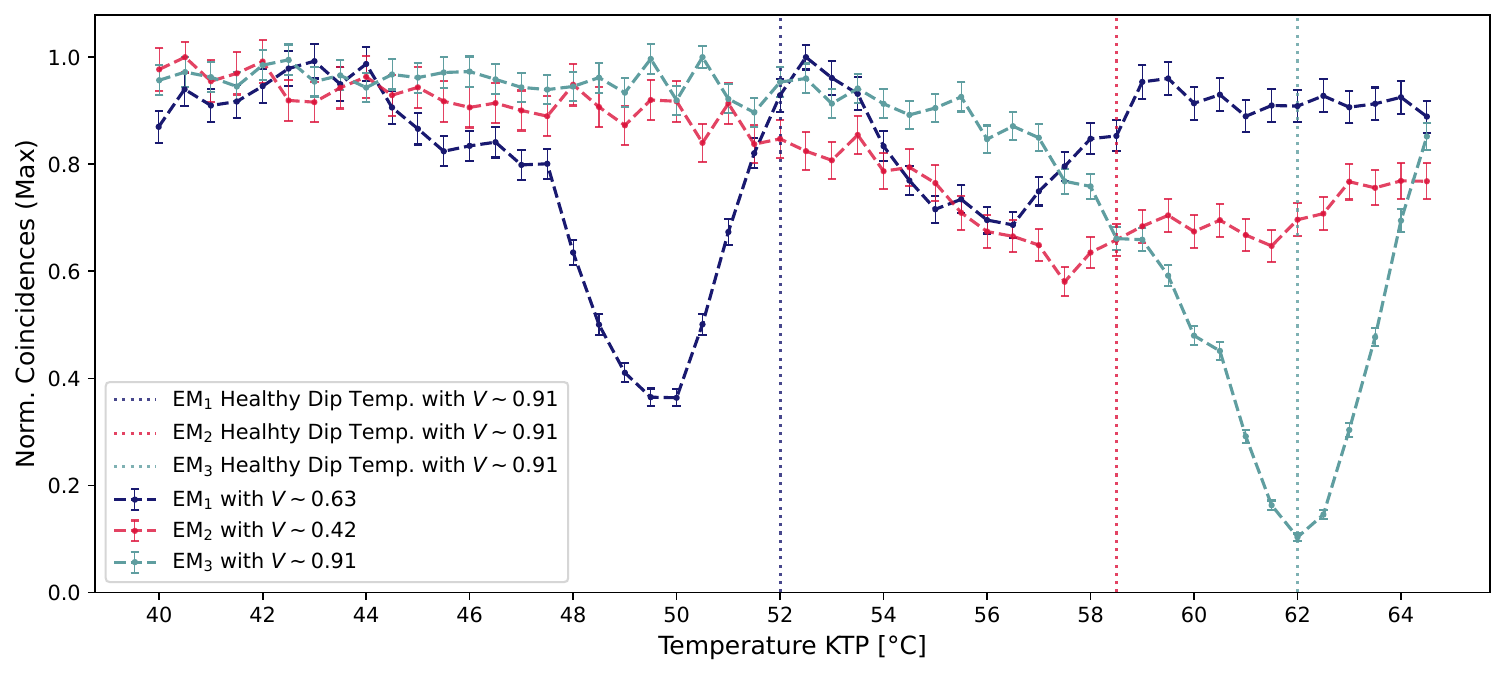}
    \caption[]{\textbf{Mode hopping and SLM damage.} Hong-Ou Mandel scan with respect to KTP temperature of the EM model is shown for three EM lasers, EM$_1$, EM$_2$, and EM$_3$. EM$_1$ (dark blue) was damaged in the TVAC test and lost its coherence $\sim 7$ months after, while EM$_2$ (red) lost its SLM feature without any external influence; vertical dashed lines correspond to the maximal dip KTP temperature of the corresponding healthy lasers (out of the package). Here, EM$_1$ shows that even at its factory-set diode temperature, it exhibits quantum interference, with reduced visibility $V$ and a shifted optimal KTP temperature. This phenomenon is not deterministic and only happens randomly after power cycling the laser (frequency mode-hopping). HOM scan with EM$_2$ shows a complete absence of interference at its original optimal HOM dip temperature, but can be regained via the diode temperature sweep described in Fig.~\ref{fig:laser_power_vac}. EM$_3$ (light blue) is a healthy laser shown for reference.}
    \label{fig:mode_hopping}
\end{figure}

It is important to note that even after the TVAC test, the damaged laser EM$_1$ maintained its original coherence at a factory-set laser-diode temperature for $\sim 7$ months. During this episode, the EM laser was (unintentionally) improperly mounted on a heatsink and therefore overheated regularly. In this sense, this module experienced additional thermal distortion during EM payload operation and can therefore be considered a reasonable proxy for the FM laser, as both experienced outgassing damage and thermal distortion. This damage eventually led to the problematic outcome that approx. $7$ months after the TVAC testing, the EM laser lost its SLM feature, i.e., no more interference could be observed in a Hong-Ou Mandel scan. However, it could be shown that both lasers sometimes randomly ``hop back'' into a sufficient SLM mode at their original diode temperature after power cycling. An example for this is shown for EM$_1$ in Fig.~\ref{fig:mode_hopping}. Here, a HOM dip could be measured at the original diode temperature, but at a shifted degeneracy temperature and with worsened visibility. This means that the center wavelength of the ``new'' SLM mode is shifted with respect to the ``old'' one. In comparison to the diode temperature-tuning procedure in Fig.~\ref{fig:laser_power_vac}, this effect was not deterministic and occurred only randomly after power cycling the laser, although it might explain the interference scenario in Op. Day $12$ in Sec.~\ref{sec:operation_day_12}.

The procedure of laser diode temperature tuning is very delicate, as in addition to polarization, path, and frequency, the laser diode temperature enters as a degree of freedom in a Hong-Ou Mandel experiment. In addition, controlling the diode temperature requires UART serial communication, in addition to powering the module on or off. In the case of the FM, only the latter (powered via the $\ensuremath{5\,\mathrm{V}}$ rail) with a fixed factory-set diode temperature of $\ensuremath{32\,{}^{\circ}\mathrm{C}}$ was available. The critical communication channel was potentially damaged during the launch or in space operation, and this circumstance also explains why two-photon interference (see Sec.~\ref{sec:two_photon_interference} for further details) could not be regained by the procedure described in Fig.~\ref{fig:laser_power_vac}b if once lost due to thermal fatigue and/or laser-induced molecular contamination (outgassing).
\clearpage

\section{Two-photon interference on Earth and TEC Calibration}
\label{sec:two_photon_interference}
\begin{figure}[!tbp]
    \centering
    \includegraphics[width=1\textwidth,height=0.67\textheight,keepaspectratio]{ 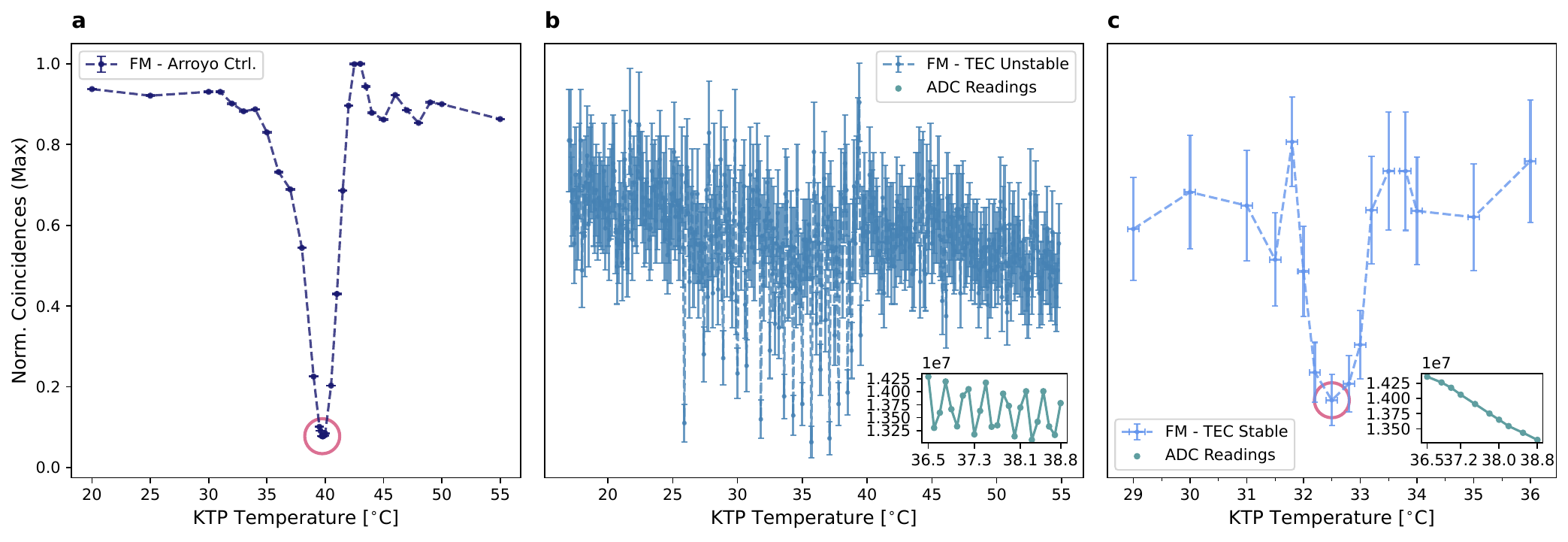}
    \caption[]{\textbf{Two-photon Interference on Earth.} \textbf{a} Two-photon interference of FM measured with commercial devices (temperature control, detection module) for calibration reasons; maximal interference visibility is at a read temperature of $T_A\sim\ensuremath{40\,{}^{\circ}\mathrm{C}}$ with visibility $\mathcal{V}= 0.924 \pm 0.003$. \textbf{b} HOM dip trial after flight-acceptance test (full integration) shows that TEC board PID parameters were not yet optimised. Strong fluctuations in temperature during measurement are visible in the strongly fluctuating ADC readings (temperature readings, small inset) and hinder proper interference measurement. Despite severe fluctuations, signs of two-photon interference occur between $\sim \ensuremath{20\,{}^{\circ}\mathrm{C}}$ and $\sim\ensuremath{40\,{}^{\circ}\mathrm{C}}$ of KTP (set) temperature. \textbf{c} Improvements in TEC stability (stable ADC readings, small inset) show a genuine sign of two-photon interference with visibility $\mathcal{V}= 0.788 \pm 0.078$. Larger errorbars with respect to the case in \textbf{a} are attributed to lower integration time and lower rate due to laser outgassing (see Fig.~\ref{fig:laser_power_vac}); KTP temperature values of TEC board could be extracted from calibration mapping in Fig.~\ref{fig:arroyo_vs_tec_board} and show maximal interference at $T_{TEC}\sim\ensuremath{32.5\,{}^{\circ}\mathrm{C}}$.}
    \label{fig:hom_interference_earth}
\end{figure}
The two-photon interference capability of the FM was tested prior to launch for two different scenarios. In Fig.~\ref{fig:hom_interference_earth}a, a commercial KTP temperature control\footnote{Arroyo 5240 Series TECSource with preset PID values. Setting for this setup: Gain 1.} and off-the-shelf detector modules were used to align the delay line (path- and polarization synchronization) before final integration of the source and delayline into the payload. Here, at a ppKTP temperature of $T_{\text{opt}} \sim 40^{\circ}$C the two-photon interference visibility is maximal (red circle in Fig.~\ref{fig:hom_interference_earth}) with $\mathcal{V}= 0.924 \pm 0.003$ and a signature of spectrum mismatch (beating) outside the dip region is clearly visible (see \cite{zhou2019second}). The measurement with commercial tools served as a calibration benchmark, and after SpaceX's Flight Acceptance Test (full integration), the two-photon interference experiment was repeated in vacuum with the full experimental pipeline, including TEC- and SPAD board. The evolution in interference quality from Fig.~\ref{fig:hom_interference_earth}b-c shows that, during this project stage, the PID parameters and ADC readings of the TEC board were not fully developed. This had a significant influence on the interference visibility due to strong temperature fluctuations (see e.g. Fig.~\ref{fig:hom_interference_earth}b). Despite these complications, genuine Hong-Ou Mandel interference could be shown prior to launch with improved TEC stability in Fig.~\ref{fig:hom_interference_earth}c with an obtained visibility of $\mathcal{V}= 0.788 \pm 0.078$.

\begin{figure}[!tbp]
    \centering
    \includegraphics[width=1\textwidth,height=0.75\textheight,keepaspectratio]{ 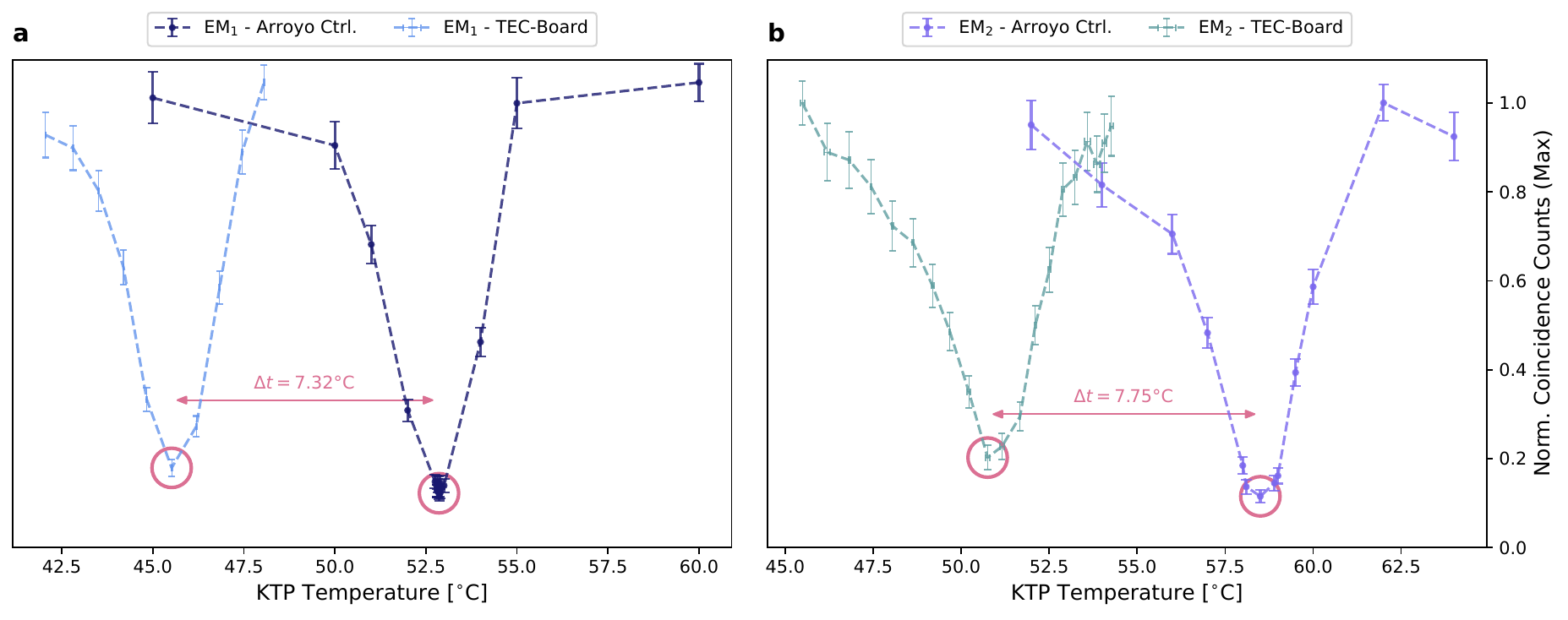}
    \caption[]{\textbf{Temperature Calibration.} \textbf{a} Two-photon interference measurement for the EM laser (EM$_1$) shows a difference of $\Delta T \sim \ensuremath{7.3\,{}^{\circ}\mathrm{C}}$ in KTP reading temperature from the commercial temperature controller (dark blue) and the home-made TEC board (light blue). This behaviour is also verified in \textbf{b} with an alternative EM laser module (EM$_2$) and the resulting temperature mapping between Arroyo Ctrl. and the TEC board is approx. $T_{TEC}\sim T_A-7.5\ensuremath{{}^{\circ}\mathrm{C}}$. Note that both modules come from the same production batch; hence, they are nearly identical. The only difference lies in a slightly different center wavelength; thus, the Hong-Ou Mandel interference temperature also changes for the same KTP crystal (EM source model).}
    \label{fig:arroyo_vs_tec_board}
\end{figure}

As mentioned, during this project stage (prior launch) there was no clear correspondence between the ADC readings (small insets in Fig.~\ref{fig:hom_interference_earth}b-c) and the actual, physical temperature. Similarly, the mapping between the set KTP temperature and the actual KTP reading temperature was not yet defined; however, with the input of the near-identical EM, a proper temperature calibration could be established (after launch). Fig.~\ref{fig:arroyo_vs_tec_board} shows the correspondence in temperature calibration between the commercial temperature control (Arroyo Ctrl.) and the TEC-board. The dip in a Hong-Ou Mandel (HOM) experiment is used as a calibration measure, and the two EM lasers (EM$_1$, EM$_2$) indicate that the read temperature of the TEC board is approx. $\ensuremath{7.5\,{}^{\circ}\mathrm{C}}$ lower than the reading value of the commercial product\footnote{This behaviour is also verified by two additional Matchbox lasers (not shown), that come from the same production batch but have slightly different center wavelengths. The latter is responsible for also slightly different HOM dip KTP temperatures between the lasers.}. This means that if a maximal HOM dip is measured at $T_A \ensuremath{{}^{\circ}\mathrm{C}}$ by the commercial device (Arroyo Ctrl.), then the maximal dip with the TEC board will be $T_{TEC}\sim T_A-7.5\ensuremath{{}^{\circ}\mathrm{C}}$. This result then allows for a meaningful labeling of the KTP temperature in Fig.~\ref{fig:hom_interference_earth}c and the genuine HOM dip occurred at $T_{TEC}\sim T_A-7.5\ensuremath{{}^{\circ}\mathrm{C}} \sim (40-7.5)\ensuremath{{}^{\circ}\mathrm{C}} \sim \ensuremath{32.5\,{}^{\circ}\mathrm{C}}$. This is also the temperature at which we expect to observe two-photon interference in space.

\section{Overvoltage calibration of single photon detectors}\label{sec:spad_calibration}
\begin{figure}[!tbp]
           \centering
            \includegraphics[width=0.95\linewidth,height=0.80\textheight,keepaspectratio]{ 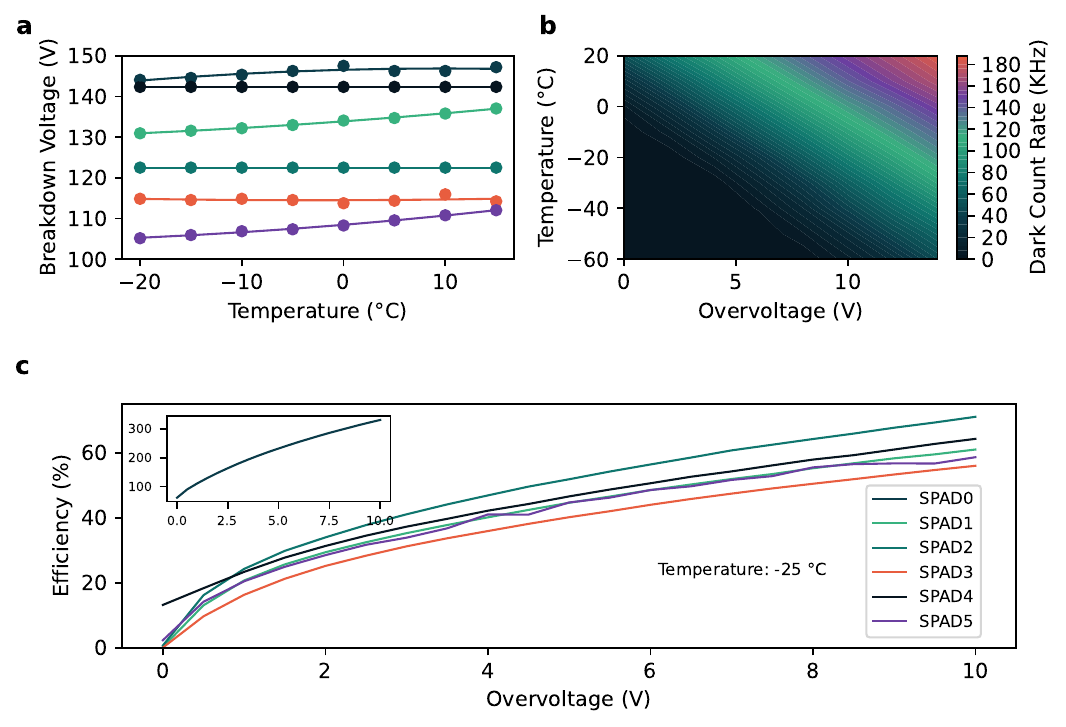}
            \caption[]{\textbf{SPAD characterization.} a) Breakdown-voltage calibration vs. temperature measured during flight operation. b) Dark-count rate as a function of temperature and overvoltage above breakdown measured during flight operation. c) Calibrated detection efficiency under laboratory conditions for a temperature of $-25^\circ$C for the EM. Inset shows C$_0$ relative efficiency. The discrepancy is caused by electronic noise and afterpulsing detection.}
            \label{fig:spads}
        \end{figure}

%In order to enable and tune the quantum detection efficiency and signal-to-noise ratio of the SPADs the temperature and high voltage applied to the diode terminals are controlled by the SoM \cite{Ursin_SAP500}. To find the voltage for which the SPADs photon detection state is enabled, a routine was developed that determines the operation point of the SPADs for a given temperature and a given voltage. This routine also works as a \textit{health check} of the SPADs board. For each SPAD a temperature is fixed and with a binary tree search the breakdown voltage is found. The script continues to the next temperature settings and finds the corresponding voltages as it is shown in \ref{fig:spads}a. Once the breakdown voltage is found, an Over voltage is applied on top of the breakdown voltage to set the quantum detection efficiency of each individual SPAD. The over voltage has a deep impact on the Dark Count Rate and efficiency, as shown in figure \ref{fig:spads}b and c. In order to calibrate the efficiency of the SPADs measurements were taken and compared to a pre calibrated detection system (\textcolor{red}{Excelitas}), demonstrating an overall detection efficiency of \textcolor{red}{10$\%$}.
%During the space operation days we a range of over voltages from 7 to 15 V to maximize the signal to noise ratio and extract a faithful quantum signature of our experiment by measuring photon pair coincidences with a coincidence window of 5 ns.
Tuning the quantum detection efficiency and signal-to-noise ratio of the SPADs is done by controlling the temperature and high voltage applied to the diode terminals \cite{Ursin_SAP500}. To find the voltage at which each SPAD photon-detection state is enabled, we developed a routine that determines the operating point of the SPADs for a given temperature and voltage. This routine also serves as a \textit{health check} of the SPADs board. For each SPAD, the temperature is fixed and a binary tree search is used to find the breakdown voltage. The script then proceeds to the next temperature setting and finds the corresponding voltage, as shown in Figure \ref{fig:spads}a. Once the breakdown voltage is found, an overvoltage is applied on top of the breakdown voltage to set the quantum detection efficiency of each individual SPAD. This overvoltage has a strong impact on both the dark count rate and the detection efficiency, as shown in Figure \ref{fig:spads}b and c. To calibrate the efficiency of the SPADs, pre-launch measurements were taken and compared against a calibrated reference detection system (SPCM-800-12-FC, by Excelitas), demonstrating an overall detection efficiency of $\sim40$\% when operating at 7 V of overvoltage. Figure \ref{fig:spads}c corresponds to the calibration made on the Engineering Model (EM), where a special case arose with SPAD C$_0$, which has a higher noise floor and exhibits afterpulsing detection signals. We suspect that a similar case occurred with SPAD C$_4$ in the Flight Model (FM).
During the space operation days, we used a range of overvoltages from 7 to 15 V to maximize the signal-to-noise ratio and extract a faithful quantum signature of our experiment, measuring photon-pair coincidences with a coincidence window of $\ensuremath{4.6\,\mathrm{ns}}$.

\clearpage

\clearpage
\clearpage
\part*{Supplementary Text}

\section{Experimental Results of Space Operation}
\label{sec:space_operation}
This section presents the complete data analysis from the payload's in-orbit operation and provides a detailed explanation of how the experimental results presented in the main text were achieved. Due to its extensiveness, it is divided into subsections that address the outcome of each operation day individually, and the main results are summarized in a concluding chapter in Sec.~\ref{sec:main_figures_information}.
\subsection{Operation Day 1 - 23 Days in Orbit}
\label{sec:operation_day_1}
\begin{figure}[!tbp]
    \centering
    \includegraphics[width=1\textwidth,height=0.71\textheight,keepaspectratio]{ 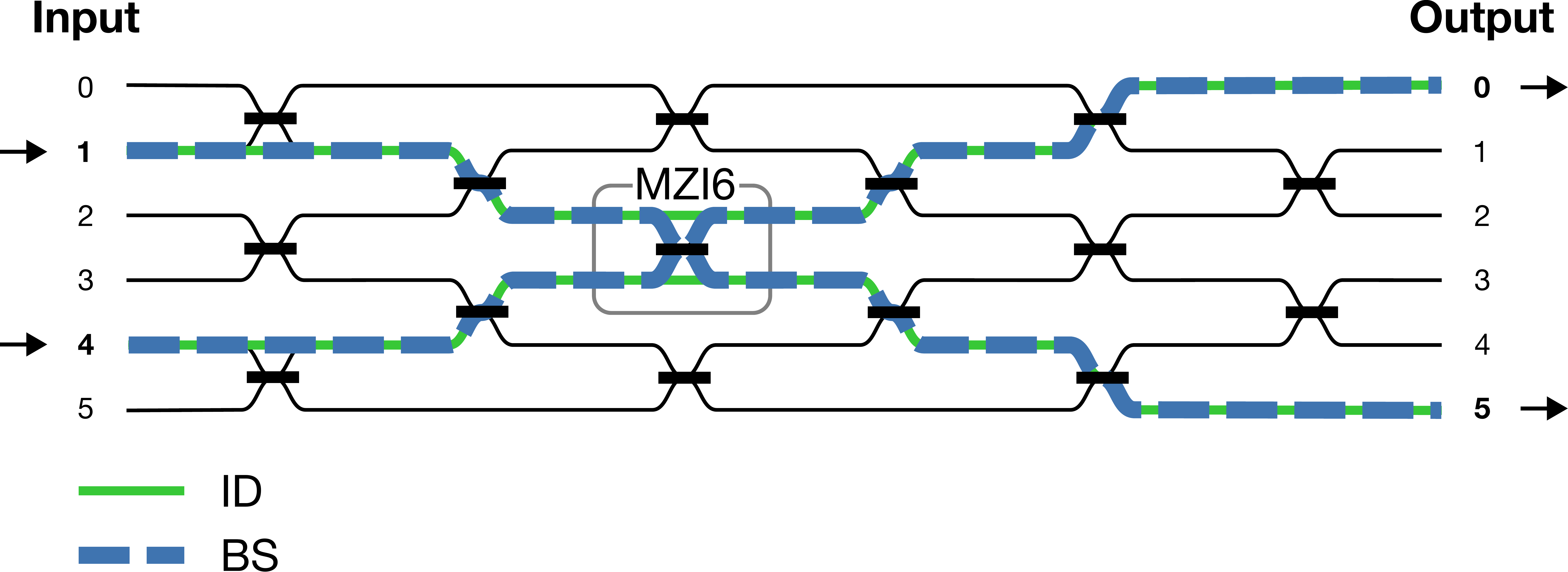}
    \caption[]{\textbf{Conceptual scheme of the photonic processor.} Two photons are injected into modes 1 and 4 (black arrows) of the 6$\times$6 universal circuit and routed along the paths indicated by the green and blue curves. To probe photon-photon
interference we implemented the same two unitaries on every operation day:
the identity (ID, green solid curve) and a balanced beam splitter (BS, blue dashed one). In both cases the
photons are directed to the same interferometer, MZI6, whose internal phase
is set to $\pi$ for ID and $\pi/2$ for BS, and its two outputs are then
routed to a pair of working detectors (e.g. modes 0 and 5, feeding $C_0$ and
$C_5$). For ID the photons do not interfere and the coincidence rate is
always maximal. For BS, indistinguishable photons bunch at the degeneracy
temperature, suppressing coincidences, whereas away from degeneracy they
behave classically and the coincidence rate is half that of ID (see
Eqs.~\ref{eq:dist_mzi_outcome} and~\ref{eq:indist_mzi_outcome}).}
    \label{fig:in_out_chip}
\end{figure}

The main goal of the first operation day was to verify that all optical and electronic components survived the rocket launch and could withstand the environment of low Earth orbit (LEO). The experimental protocol was designed to send the input photons, entering the chip from modes 1 and 4 to detector pair C$_0$, C$_5$ (see Fig.~\ref{fig:in_out_chip}). In this configuration, we performed a broad ppKTP temperature scan, looking for quantum interference. The covered range was from $\ensuremath{20\,{}^{\circ}\mathrm{C}}$ to $\ensuremath{50\,{}^{\circ}\mathrm{C}}$ with a  stepsize $\ensuremath{0.5\,{}^{\circ}\mathrm{C}}$. During this scan, the circuit was reconfigured, alternatively, to two unitary operations. We identify the first as ``Identity'' in which the photons are sent to C$_0$, C$_5$ without interfering, while for unitary ``Beamsplitter'' the photons meet at a balanced beamsplitter (see scheme in Fig.~4a of the main text) and, when in the indistinguishability regime, they interfere. This allows us to use Identity as a reference measurement, as here we should see double the coincidence rate (distinguishable photons) with respect to Beamsplitter. For this reason, we repeated this strategy for every operation day. This phenomenon follows directly from the treatment of the problem within the framework of second quantization. In the case of Identity, we send two photons directly to modes C$_0$, C$_5$, and these photons can, in general, be discriminated by a feature $m \neq n$, which renders them distinguishable (i.e., their frequency). The corresponding state in terms of the Fock basis is given by
\begin{equation}
    \ket{\psi_{ID}}=\hat{a}_{m}^{\dagger} \hat{b}_{n}^{\dagger} \ket{0}_{C_0} \ket{0}_{C_5} = \ket{1,m}_{C_0} \ket{1, n}_{C_5},
    \label{beamsplitterinput}
\end{equation}
and will lead to a coincidence event CC$_{05}$ between detectors C$_0$ and C$_5$, i.e. a simultaneous click of both detectors. If a balanced beamsplitter is included, we will have to transform our input photon operators accordingly via
\begin{equation}
    \begin{aligned}
        \hat{a}^{\dagger}_m &\rightarrow \frac{1}{\sqrt{2}} \left( \hat{a}^{\dagger}_m + \hat{b}^{\dagger}_m \right) \\
        \hat{b}^{\dagger}_n &\rightarrow \frac{1}{\sqrt{2}} \left( \hat{a}^{\dagger}_n - \hat{b}^{\dagger}_n \right).
        \label{eq:bs_trafo}
    \end{aligned}
\end{equation}

Inserting these transformations back into Eq. \ref{beamsplitterinput} will lead to the more complex output
\begin{equation}
    \ket{\psi_{BS}} =\frac{1}{2} \left(  \hat{a}^{\dagger}_m \hat{a}^{\dagger}_n - \hat{a}^{\dagger}_m   \hat{b}^{\dagger}_n +  \hat{a}^{\dagger}_n \hat{b}^{\dagger}_m - \hat{b}^{\dagger}_m \hat{b}^{\dagger}_n  \right) \ket{0}_{C_0} \ket{0}_{C_5},
    \label{eq:single_bs_output}
\end{equation}
in which the first and last term correspond to the loss of both photons into one detector mode. As our setup does not include number-resolving single-photon detectors, the difference between one- or multi-photon detection events cannot be resolved and is thus ignored. The middle terms, on the other hand, will also lead to a coincidence event CC$_{05}$, although with a significant difference relative to Eq. \ref{beamsplitterinput}, as the present prefactor reduces the coincidence probability by one half. In the special case of indistinguishable photons ($m=n$), the coincidence probability drops to zero, and this phenomenon corresponds to the characteristic Hong-Ou Mandel dip in coincidences \cite{branczyk_hong-ou-mandel_2024}. In terms of the effective measured coincidence rate, distinguishable photons in a Beamsplitter setting will result in half of the Identity rate, while indistinguishable photons will lead to no measurable coincidences.

In this spirit, applying the Identity unitary for every temperature step $T_{TEC}$, helps to identify potential artifacts in a Hong-Ou Mandel scan (HOM scan), e.g., if both Identity and Beamsplitter show zero coincidences for some KTP temperature, then this behaviour is not physical and must be excluded from the data analysis.

\begin{figure}[!tbp]
    \centering
    \includegraphics[width=1\textwidth,height=0.71\textheight,keepaspectratio]{ 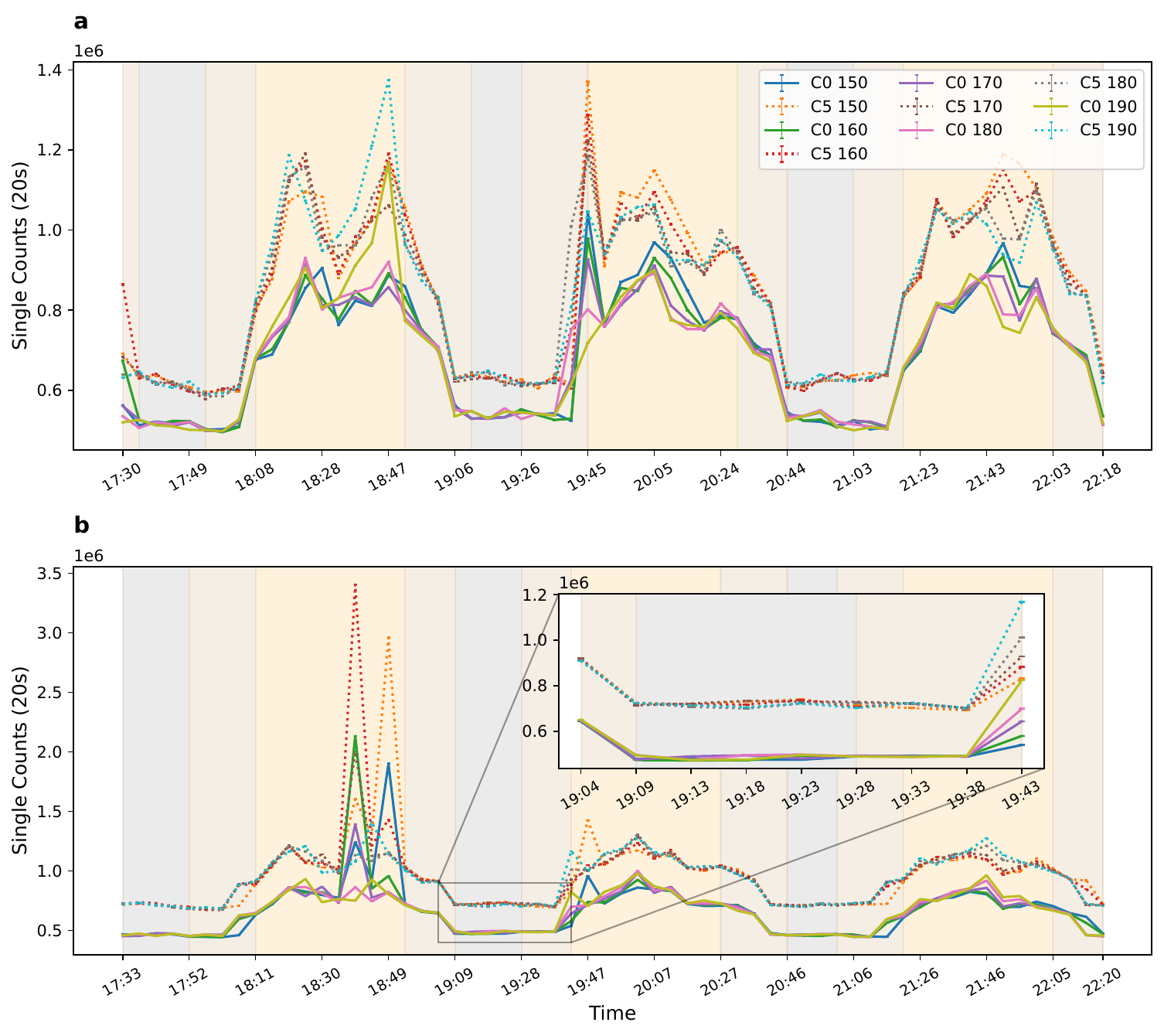}
    \caption[]{\textbf{Op. Day 1 - Single counts and noise pattern.} \textbf{a} Single counts for Identity unitary and five different measured coincidence windows $\tau$; target detectors C$_0$ (solid), C$_5$ (dashed) show significant periodic noise pattern over $6$ hrs. Background coloring is extracted from publicly available TLE data, which correlates measurement time with the host satellite's position. Spikes together with an increased number in single counts occur when the host satellite ION is facing the sun (yellow area), while single counts are lower and relatively stable when ION is in the shadow of the Earth (grey area). \textbf{b} Unitary Beamsplitter single counts reveal a similar scenario as in the case of \textbf{a}. Here, even stronger spikes in counts are visible (first yellow area) as the rate is increased by a factor of $6$. The small inset shows that during umbra/penumbra the countrate of $C_0$ and $C_5$ is stable and rises sharply once the satellite faces the sun again.}
    \label{fig:op_day_1_singles}
\end{figure}

Besides the cyclic switch between Identity and Beamsplitter, the first operation day also included a scan of the coincidence window $\tau$, for each set temperature $T_{TEC}$. This procedure had two reasons: first, to find the most optimal window in terms of Signal to Noise, and second, in terms of Hong-Ou Mandel interference visibility \footnote{Choosing an appropriate coincidence window in a HOM scan was found to be relevant for optimal interference visibility. Preliminary alignment measurements with a commercial Timetagger (UQD Devices - Logic 16) and commercial detectors (Excelitas SPCM-NIR) showed this feature, as well as measurements on the Engineering model (EM).}. The range of $\tau$ includes values of $\{4.05,\allowbreak 4.32,\allowbreak 4.59,\allowbreak 4.86,\allowbreak 5.13\}$ $\ensuremath{\mathrm{ns}}$ and each datapoint consists of the sum of two individual data acquisition chunks of $\ensuremath{10\,\mathrm{s}}$. This implies that each window $\tau$ has an effective integration time of $\ensuremath{20\,\mathrm{s}}$. As a first step towards a complete analysis, Fig.~\ref{fig:op_day_1_singles} gives an overview of the single counts C$_0$, C$_5$ for Identity and Beamsplitter and here the dramatic periodic noise pattern for each measured window $\tau$\footnote{Note that here the technical labeling for the coincidence windows is $\{150,160,170,180,190\}=\{4.05,\allowbreak 4.32,\allowbreak 4.59,\allowbreak 4.86,\allowbreak 5.13\} \ \ensuremath{\mathrm{ns}}$.} reveals the first drastic difficulty of this space mission.

The colored shaded areas in the background mark the current position of the ION satellite relative to the sun. This information was extracted from publicly available two-line element set (TLE) data. The latter is used as a standard in satellite tracking (position, velocity) \cite{vallado_revisiting_2006} and here grey areas indicate that the ION satellite was in the shadow of the earth (umbra), while the yellow areas indicate that the host satellite faced the sun. The stage between sun and shadow is also called penumbra and can be identified as the overlap between yellow and grey. According to this result (and all subsequent results from the other operation days), there is a strong correlation between the host satellite facing the sun (yellow area) and the detectors showing highly fluctuating, increased counts. In this regime, the SPADs or the quenching circuit readout may saturate (see e.g. spikes in Fig.~\ref{fig:op_day_1_singles}b), while in the regime of penumbra/umbra the countrate is relatively stable and reduced.

\begin{figure}[!tbp]
    \centering
    \includegraphics[width=0.8\textwidth,height=0.80\textheight,keepaspectratio]{ 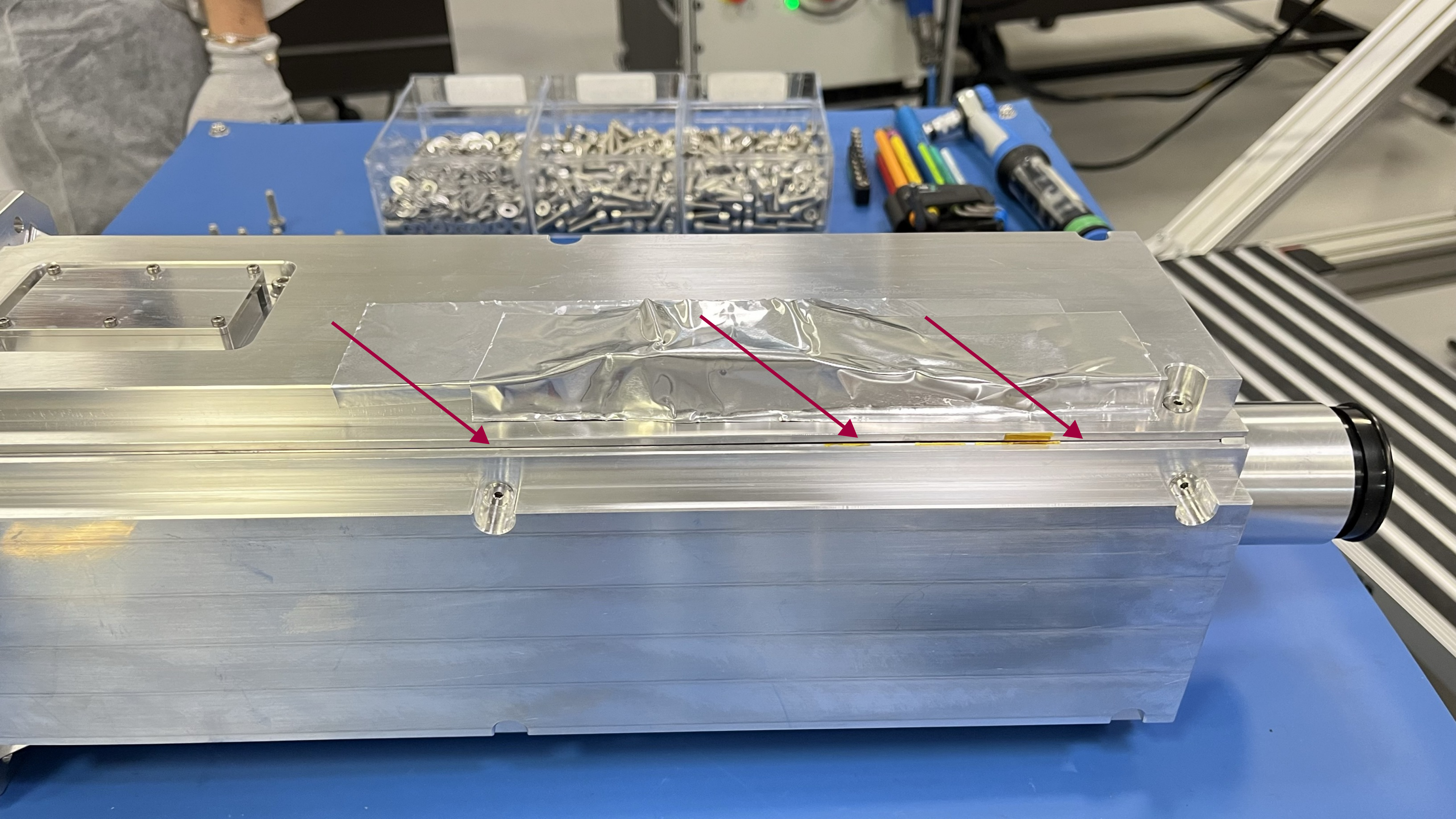}
    \caption[]{\textbf{Potential leak in payload.} Red arrows mark a small gap that could be responsible for allowing the parasitic photons of the sun to enter the payload interior. While other larger gaps were covered by Aluminium tape (e.g. Aluminium tape on the laser cap), some gaps were not visible by eye and thus not additionally covered. This might explain the strong noise contributions that occur when the payload faces the sun.}
    \label{fig:potential_gap}
\end{figure}

A potential source of this noise is a small gap in our payload frame, which allows parasitic photons from the sun to enter the payload. One of the responsible gaps is indicated with red markers in Fig.~\ref{fig:potential_gap}. Furthermore, as can be seen from Fig.~2b and 2c, in the main text, the SPADs are not completely covered, so they are sensitive to visible noise photons entering our payload, despite the collimator housing acting as a good light block. To protect them from stray pump light ($\ensuremath{405\,\mathrm{nm}}$), a relatively thick layer of Kapton tape was wrapped around the single SPAD modules as the latter was found to be efficient in blocking ultraviolet light. While Kapton is effective at blocking the pump laser, it does not prevent broadband photons from entering the SPAD active area.

\begin{figure}[!tbp]
    \centering
    \includegraphics[width=0.9\textwidth,height=0.63\textheight,keepaspectratio]{ 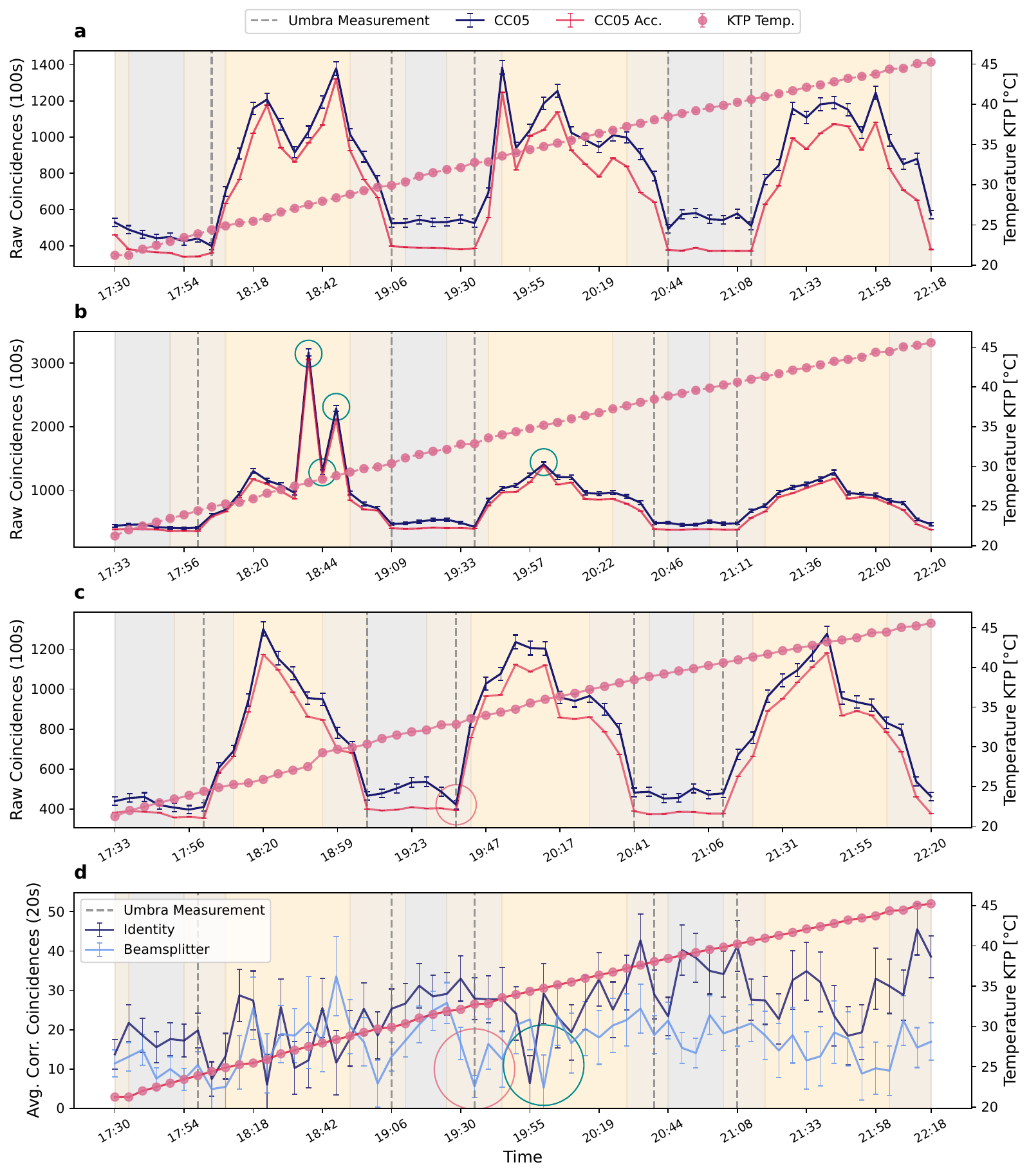}
    \caption[]{\textbf{Op. Day 1 - Coincidences overview.} \textbf{a} HOM scan that targets Identity unitary which sends both photons to detector modes C$_0$, C$_5$ for different KTP temperatures (red points, secondary $y$-axis); clear signal (dark blue) on top of noise floor (red) reveals that the payload survived the rocket launch and here the measurements in the shade (vertical dashed lines) indicate a stable countrate with respect to the measurements in the sun. \textbf{b} Similar to \textbf{a}, but here the target unitary is Beamsplitter, and strong spikes in the noise floor (green circles) that regularly wash out the signal indicate that parasitic photons from the sun are capable of saturating the detectors/quenching circuit. \textbf{c} After removing outliers (green circles) in \textbf{b}, the Beamsplitter unitary countrate is easier to analyze, and a dip in coincidences (red circle) in the shade is visible. \textbf{d} Average corrected coincidences for Identity (dark blue) and Beamsplitter (light blue) for all five coincidence windows $\tau$ in Eq. \ref{eq:corrected_signal}. Here, sun measurements also show generally more fluctuations and have larger error in countrate, and even Identity unitary could mimic a Hom dip in coincidences in the sun (green circle). For these reasons, data acquisition during a sun measurement is not reliable for a HOM scan. Total measurement time is $\sim 6$ hrs.}
    \label{fig:op_day_1_counts}
\end{figure}
As SPDC is a heralded single-photon generation process, analyzing single counts does not provide sufficient information about the experimental outcome. We need to deepen our understanding by evaluating coincidence counts between the corresponding detector pairs. As mentioned, the first operation day considered C$_0$ and C$_5$ as the target detector modes for the two partner photons. The raw timetagger signal of coincidences for both the Identity and Beamsplitter unitary is shown in Fig.~\ref{fig:op_day_1_counts}a-c and here the HOM dip scan is presented with respect to time on the $x$-axis, while the KTP temperature data is projected on the secondary $y$- axis. Due to the low signal rate (approx. $\ensuremath{1.5\,\mathrm{Hz}}$), the raw measurements of the five different coincidence windows are summed up to improve the overall statistics. This gives a good overview on the overall (raw) countrate stability and as the individual windows are similar to each other, this approach is also meaningful.

The measured coincidence signal CC$_{ij}$ corresponds to the sum of the data collected in several measurements for a timetagger channel $ij$\footnote{For our working detector pairs, this would imply that $ij=\{01,05,15\}$.}. Each measurement was performed with a different coincidence window $s_\tau$. With its error CC$_{ij,\text{Err.}}$ this sum reads
\begin{equation}
\text{CC}_{ij}=\sum_\tau s_\tau ,\quad  \text{CC}_{ij,\text{Err}}=\sqrt{\sum_\tau \sigma_\tau^2} \quad \text{for} \  \tau \in \{4.05,\allowbreak 4.32,\allowbreak 4.59,\allowbreak 4.86,\allowbreak 5.13\} \ \ensuremath{\mathrm{ns}}.
\label{rawsignal}
\end{equation}
Here, we can assume Poissonian counting statistics and the error of each individual window measurement is thus $\sigma_\tau=\sqrt{s_\tau}$ \cite{c_data_1972}. The underlying Poissonian counting statistics has also strong implications on the estimate of the noise floor $\text{CC}_{ij,\text{Acc}, \tau}$ and here we make use of the canonical accidentals model with related error $\sigma_{\text{Acc},\tau}$ \cite{kim_low-cost_2005}, \cite{bjurlin_versatile_2024}
\begin{equation}
    \begin{gathered}
    \text{CC}_{ij,\text{Acc}, \tau} =C_i(\tau)\cdot C_j(\tau) \cdot \tau, \quad \sigma_{\text{Acc},\tau}=\tau\sqrt{C_i^2(\tau) \cdot \sigma^2_j(\tau)+C^2_j(\tau) \cdot \sigma^2_i(\tau)},\\
    \text{and} \quad \sigma_{i,j}(\tau)=\sqrt{C_{i,j}(\tau)}.
    \end{gathered}
    \label{accidentals}
\end{equation}
The accidental noise floor is therefore evaluated by multiplying the singles C$_i(\tau)$, C$_j(\tau)$ of a given detector pair with the chosen coincidence window $\tau$. For each window and coincidence pair CC$_{ij}$, the accidentals are accordingly estimated, and in Fig.~\ref{fig:noise_analysis}a, the model in Eq. \ref{accidentals} is also experimentally verified. The total noise floor $\text{CC}_{ij} \ \text{Acc.}$ in Fig.~\ref{fig:op_day_1_counts}a-c (red) is calculated in accordance with the sum of the raw signal in Eq. \ref{rawsignal} and given with its error $\text{CC}_{ij,\text{Err.}} \ \text{Acc.}$ as
\begin{equation}
\begin{gathered}
\text{CC}_{ij} \ \text{Acc.}=\sum_{\tau} C_i(\tau)\cdot C_j(\tau) \cdot \tau ,\quad  \text{CC}_{ij,\text{Err.}} \ \text{Acc.}=\sqrt{\sum_\tau \sigma^2_{\text{Acc.},\tau}}\\
\text{for} \  \tau \in \{4.05,4.32,4.59,4.86,5.13\} \ \ensuremath{\mathrm{ns}}.
\end{gathered}
\label{eq:rawsignal_acc}
\end{equation}
 To estimate the corrected signal, reported in Fig.~\ref{fig:op_day_1_counts}d, we subtract the predicted accidental counts from the raw timetagger data $s_\tau$ for window $\tau$. The resulting signal $\tilde{s_\tau}$ then amounts to the following:
\begin{equation}
\tilde{s}_{\tau} = s_\tau-C_i(\tau) \cdot C_j(\tau) \cdot \tau ,\quad  \tilde{\sigma_{\tau}}=\sqrt{\sigma_{\tau}^2+\sigma^2_{\text{Acc},\tau}}.
\label{eq:signal_extraction}
\end{equation}
The given routine holds for all operation days where a fixed coincidence window $\tau$ is chosen. The only exception is operation day $1$, where we carried out a more refined analysis not to overshoot the accidental estimation (as detailed in Sec.~\ref{sec:noise_model}). %These five window measurements, with a $\ensuremath{20\,\mathrm{s}}$ acquisition time, target, in principle, the same physical quantity (e.g., Beamsplitter at a given temperature $T_{TEC}$), but they have slightly different accidental noise floors due to varying $\tau$. For this reason, we take the mean from the five windows in Eq. \ref{eq:signal_extraction} and a more detailed error analysis in Sec.~\ref{sec:noise_model} reveals that within the shadow of the Earth, the error estimation in Eq. \ref{eq:signal_extraction} would overshoot the data from the five coincidence windows.
To faithfully describe the observed data distribution, the correct error follows a Poissonian distribution around the mean value of the five coincidence windows measurements, and therefore, we have:
\begin{equation}
\text{CC}_{ij,\text{Corr}}= \frac{\sum_\tau  \tilde{s}_\tau}{N_\tau} ,\quad  \text{CC}_{ij,\text{Corr},\text{Err}}=\sqrt{\frac{\sum_\tau  \tilde{s}_\tau}{N_\tau}} \quad \text{for} \  \tau \in \{4.05,\allowbreak 4.32,\allowbreak 4.59,\allowbreak 4.86,\allowbreak 5.13\} \ \ensuremath{\mathrm{ns}}.
\label{eq:corrected_signal}
\end{equation}
This analysis was only done for measurements taken in the shadow regions. Indeed, if we compare the accidental noise floor for the two different scenarios in Fig.~\ref{fig:op_day_1_counts}a-c, the accidental noise floor is highly stable only in these regions. For all the other operation days, the coincidence window is fixed at an optimal $\tau=170=\ensuremath{4.59\,\mathrm{ns}}$ and Eq. \ref{eq:signal_extraction} is used to extract the true signal with its error. Note that in the following figures regarding experimental results, the technical labeling of the coincidence window $\tau$ is mapped as follows
\begin{equation}
     \tau = \{150,160,170,180,190\}= \{4.05,\allowbreak 4.32,\allowbreak 4.59,\allowbreak 4.86,\allowbreak 5.13\} \ \ensuremath{\mathrm{ns}}
     \label{window_labeling},
\end{equation}
and the errors in all figures indicate one standard deviation ($1 \sigma$).

Within this framework, the results of Fig.~\ref{fig:op_day_1_counts}a-c show that there is a clear signal on top of the noise floor for target detectors C$_0$, C$_5$, and this is also reflected in the average corrected coincidences in Fig.~\ref{fig:op_day_1_counts}d. In more detail, similarly to the scenario for the single counts in Fig.~\ref{fig:op_day_1_singles}, the coincidences also reflect the strong periodic noise contribution when the payload faces the sun. This is clearly shown when comparing the accidental noise floor during shadow (grey) with that during sun (yellow). In the former, the accidentals are stable over a measurement duration of $\sim \ensuremath{30\,\mathrm{min}}$, while in the latter, the additional parasitic photons from the sun strongly increase the amount of accidentals and disturb the measurement over $\sim \ensuremath{60\,\mathrm{min}}$ by introducing severe count fluctuations. This is brought to an extreme in, e.g., Fig.~\ref{fig:op_day_1_counts}b, where short-term detector saturation occurs, as here the counts sharply increase, and the accidentals regularly wash out the signal (i.e., green circles). Comparing the behaviour of the singles in Fig.~\ref{fig:op_day_1_singles} with the behaviour of the accidentals in Fig.~\ref{fig:op_day_1_counts}a-c allows us to choose the accidental noise floor as a proper candidate to validate the general quality of a measurement\footnote{This follows directly from Eq. \ref{accidentals} and is also meaningful as the accidentals are directly extracted from the singles.}.

In this spirit, we divide a full operation day into individual chunks of sun/shadow measurements by analyzing the behaviour of the accidentals (singles). The complete measurement of Fig.~\ref{fig:op_day_1_counts} can be divided into three shadow measurements (vertical dashed lines) where the accidentals are stable, and a clear coincidence signal can be extracted. Moreover, in these regions, a sort of baseline in countrate can be identified in which accidentals are not only stable over a longer timeframe, but also minimal. The latter implies that the amount of external noise is lowest here, and, in this sense, we can rely on the accidentals to understand the payload's noise environment more precisely. This feature, together with the TLE position data for the host satellite, provides a good indication of when parasitic photons enter our payload.

\begin{figure}[!tbp]
    \centering
    \includegraphics[width=1\textwidth,height=0.69\textheight,keepaspectratio]{ 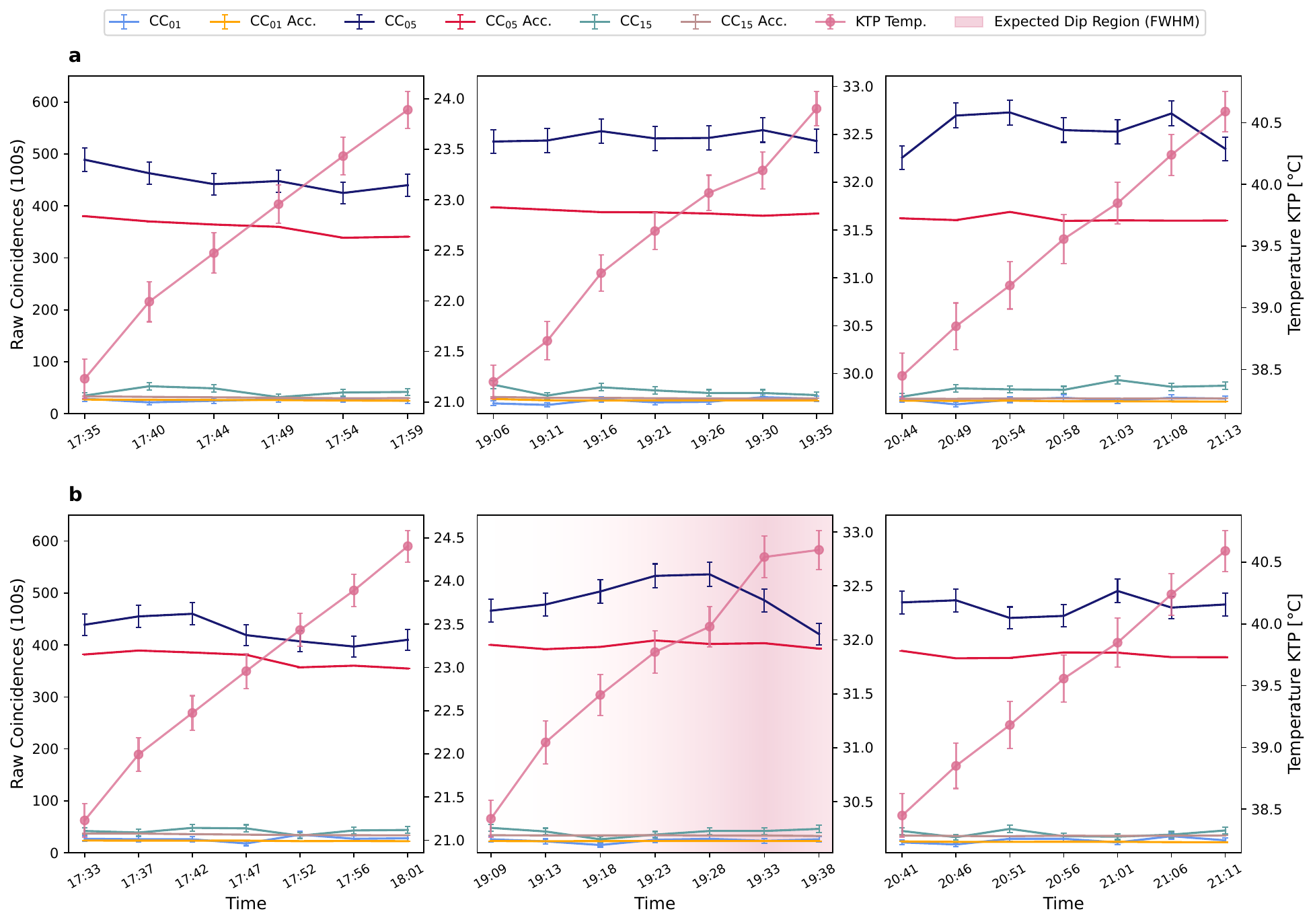}
    \caption[]{\textbf{Op. Day 1 - Coincidences in shade.} \textbf{a (row)} Extract of raw coincidence counts (Identity) from Fig.~\ref{fig:op_day_1_counts}a shows only the part where payload is in the shadow of the Earth, hence coincidence counts CC$_{ij}$ and noise floor CC$_{ij}$ Acc. are more reliable; target detector pair CC$_{05}$ (dark blue) shows clear and relatively stable signal above noise floor (red), while unused detector pairs CC$_{01}$, CC$_{15}$ are compatible with noise floor (as expected). KTP temperature is given on the secondary $y$-axis and shows sufficiently stable TEC stabilisation performance. \textbf{b (row)} Same measurement parameters as in \textbf{a} but with Beamsplitter unitary, thus interference between both photons is enabled. In the regime of an expected HOM dip ($T_{TEC} \sim \ensuremath{32.5\,{}^{\circ}\mathrm{C}}$ - red-shaded area), coincidences start to drop and nearly reach the noise floor. Timeframe of stable shadow measurements is $\sim \ensuremath{30\,\mathrm{min}}$, and this is also aligned with the expected duration of a $\sim \ensuremath{90\,\mathrm{min}}$ orbit in SSO (see Sec.~\ref{sec:space_as_environment}.)}
    \label{fig:op_day_1_umbra_first_look}
\end{figure}
The approach of considering only data from a stable, relatively low-noise measurement regime is essential for high-precision measurements, such as a Hong-Ou Mandel scan. This goes by the reason that the strong fluctuations or short-term detector saturation, i.e., when noise washes out the real signal, can mimic a dip in coincidences (see e.g., dip in Identity unitary in green circle in Fig.~\ref{fig:op_day_1_counts}d). The latter is the characteristic signature of a two-photon interference measurement, and it is therefore crucial to restrict Hong-Ou Mandel data acquisition to measurements when the payload is in the shadow of the Earth, so no artifacts are included in the analysis.

\begin{figure}[!tbp]
    \centering
    \includegraphics[width=1\textwidth,height=0.73\textheight,keepaspectratio]{ 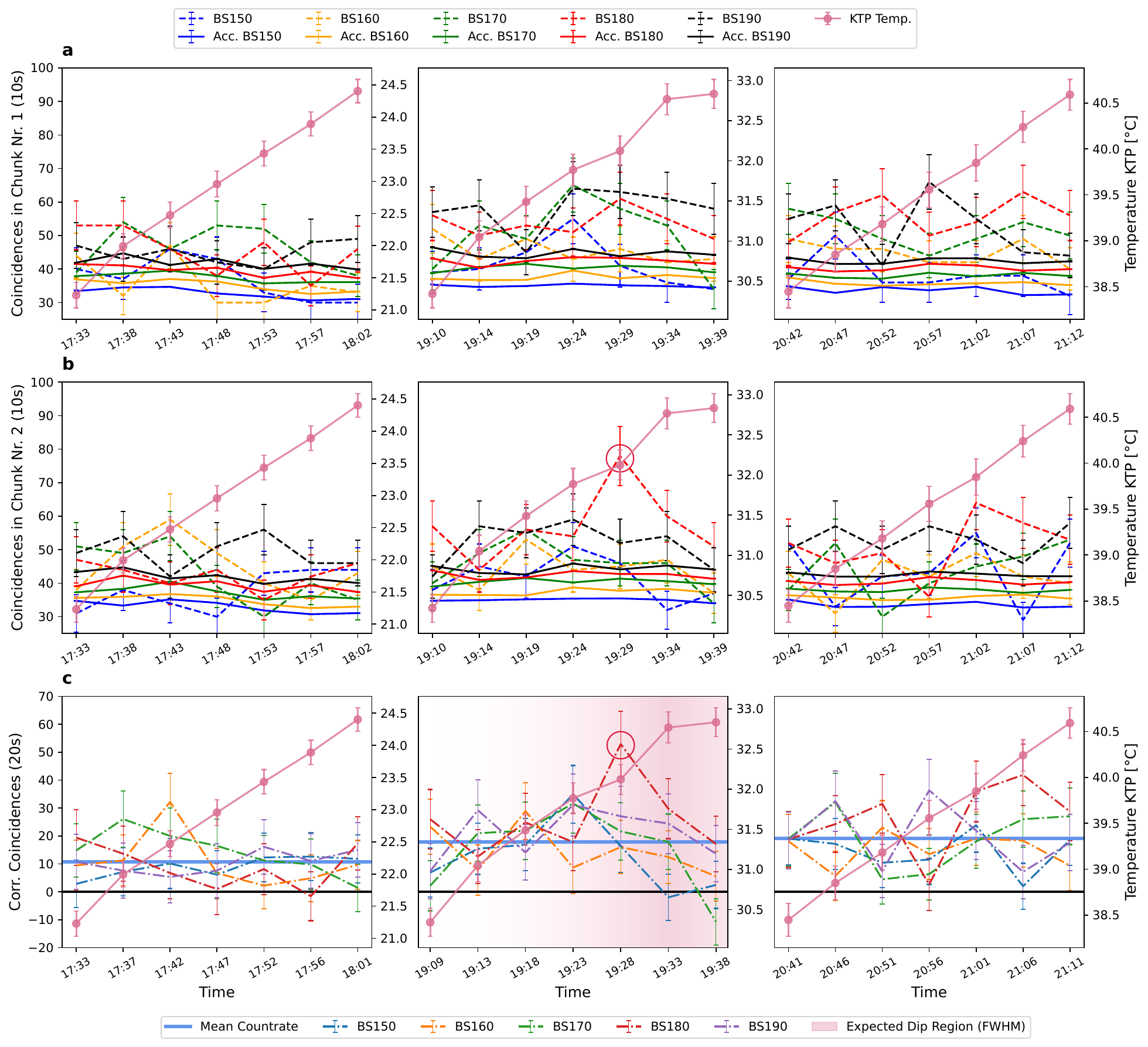}
    \caption[]{\textbf{Op. Day 1 - Beamsplitter outlier analysis.} \textbf{a (row)} Raw Beamsplitter coincidence counts (dashed) for all coincidence windows $\tau$ (refer to labeling in Eq. \ref{window_labeling}) of first $\ensuremath{10\,\mathrm{s}}$ measurement chunk and target detector pair CC$_{05}$ in shadow areas from Fig.~\ref{fig:op_day_1_counts}b (row). No timetagging anomaly of raw signal with respect to noise floor (solid) can be identified (compare with Fig.~\ref{fig:op_day_9_outlier_anomaly}) as the raw signal is either above or compatible with the noise floor. \textbf{b (row)} Same condition as in \textbf{a} but second chunk of $\ensuremath{10\,\mathrm{s}}$ measurement. Also, here, no strong outlier below the noise floor can be detected. \textbf{c (row)} Corrected coincidence counts for all windows $\tau$ via Eq. \ref{eq:signal_extraction} effectively shows the sum of the measurements in \textbf{a,b} (columnwise). Here, the only significant outlier is maximal with the spike in coincidences (red circle) for window $\tau = 180=\ensuremath{4.86\,\mathrm{ns}}$.}
    \label{fig:op_day_1_chunks}
\end{figure}
In this manner, Fig.~\ref{fig:op_day_1_umbra_first_look} gives a first overview of the three stable shadow measurements of the first operation day and also includes the other working detector pairs CC$_{01}$ and CC$_{15}$ for comparison. As expected, the measured raw signals from these two unused channels overlap well with the noise floor for CC$_{01}$ Acc. and CC$_{15}$ Acc., respectively. In stark contrast, the target detector pair CC$_{05}$ (dark blue) shows a clear signal on top of the noise floor (red) and the latter is also highly stable over time. Moreover, in Fig.~\ref{fig:op_day_1_umbra_first_look}b (middle) the Beamsplitter unitary shows a drop in coincidences around the expected interference regime of $T_{TEC} \sim \ensuremath{32.5\,{}^{\circ}\mathrm{C}}$ (red-shaded area) and the observed dip happens close to $T_{TEC} \sim \ensuremath{32.8\,{}^{\circ}\mathrm{C}}$. Most importantly, this drop in coincidences does not occur in the case of Identity in \ref{fig:op_day_1_umbra_first_look}a (center) for the same KTP temperature and is therefore a strong first sign of two-photon interference.

\begin{figure}[!tbp]
    \centering
    \includegraphics[width=1\textwidth,height=0.67\textheight,keepaspectratio]{ 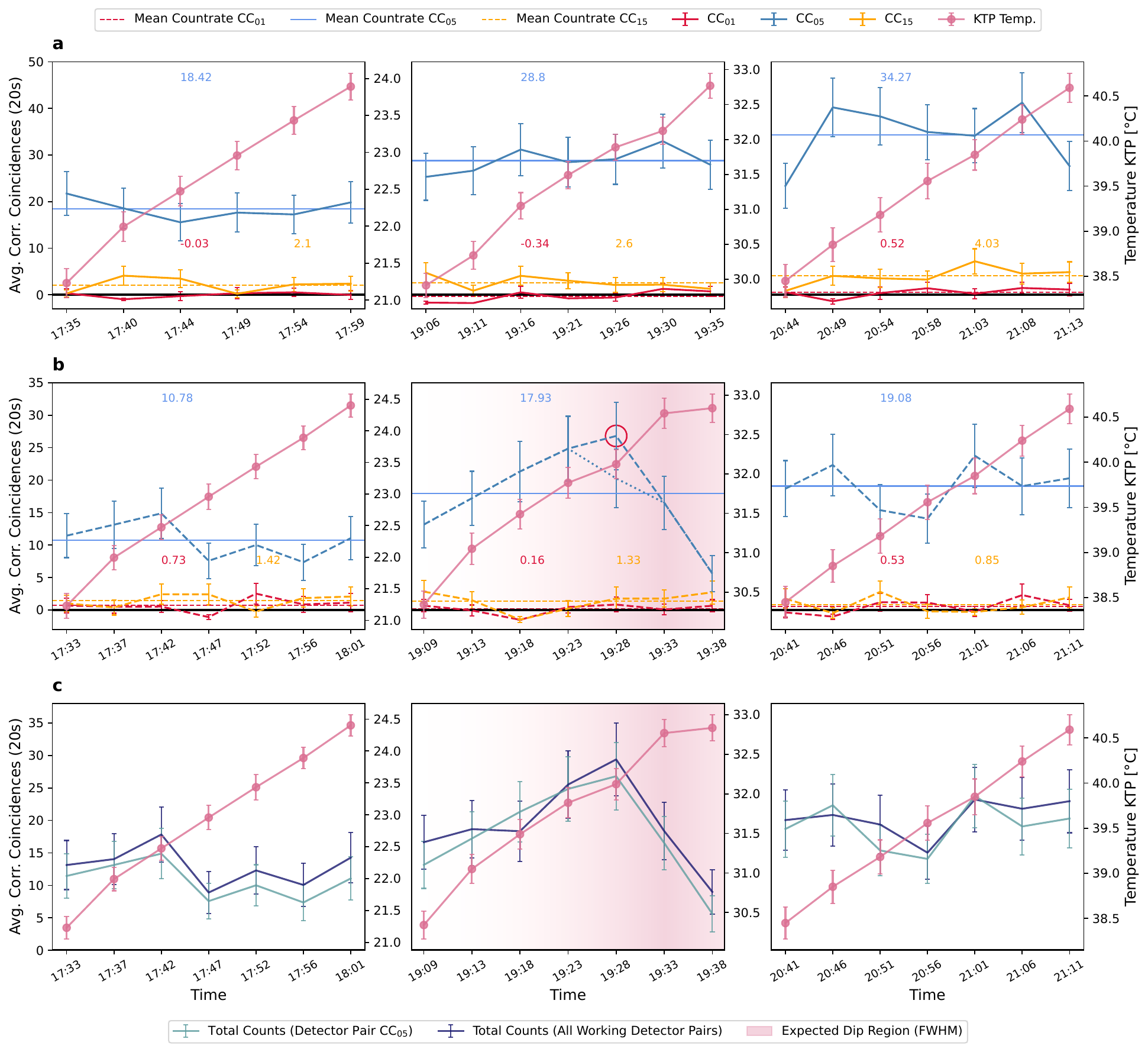}
    \caption[]{\textbf{Op. Day 1 - Total corrected coincidences.} \textbf{a (row)} Corrected coincidence counts (Identity) for all working detector pairs CC$_{ij}$ for shadow regimes show a clear and stable signal for target detector pair CC$_{05}$ (blue); mean countrate (horizontal blue line) increases over time due to thermalization of the payload. Spurious counts in CC$_{15}$ suggest slight uPIC miscalibration as some photons are misrouted to other detectors. \textbf{b (row)} Same condition as in \textbf{a}, but here Beamsplitter is applied and mean countrate also increases gradually over time; strong spike (red circle) can be attributed to the outlier measurement of coincidence window $\tau = 180=\ensuremath{4.86\,\mathrm{ns}}$ in Fig.~\ref{fig:op_day_1_chunks}c (center). Significant drop in coincidences is measured around the degeneracy temperature (red-shaded area) and suggests two-photon interference at $T_{TEC} \sim \ensuremath{32.8\,{}^{\circ}\mathrm{C}}$. \textbf{c (row)} To further rule out any artefact in the HOM dip regime, the sum of all detector pairs CC$_{ij}$ (dark blue) is compared with the target pair CC$_{05}$ (light blue) and underlines that dip in coincidences can not be attributed to falsely routed photons to CC$_{01}$ and CC$_{15}$.}
\label{fig:op_day_1_total_counts_anomaly_check}
\end{figure}
To further deepen our knowledge about this drop in coincidences and rule out any possible artifact, the raw coincidence counts for the Beamsplitter unitary are further dissected in their individual coincidence window $\tau$ and measurement chunks of $\ensuremath{10\,\mathrm{s}}$ integration time. This allows us to identify potential outliers that may result in an artifact. In Fig.~\ref{fig:op_day_1_chunks}a-b (row), the raw Beamsplitter coincidence counts for all coincidence windows and measurement chunks of $\ensuremath{10\,\mathrm{s}}$ reveal that there is no outlier that might mimic a dip in coincidences. This analysis is important, as on a later operation day in Sec.~\ref{sec:operation_day_9}, the detector pair CC$_{01}$ shows a timetagger anomaly/SPAD board saturation that could mimic a dip in coincidences. This anomaly reveals itself as a strong deviation from the measured signal below the predicted noisefloor (more than $2\sigma$) as shown in Fig.~\ref{fig:op_day_9_outlier_anomaly}. In contrast, the coincidences for each window (dashed) in Fig.~\ref{fig:op_day_1_chunks}a-b (row) of the first operation day do not drop below the predicted noisefloor (solid). The only identified potential outlier is a maximum in coincidences of window $\tau = 180=\ensuremath{4.86\,\mathrm{ns}}$ in Fig.~\ref{fig:op_day_1_chunks}c (red circle) that strongly overshoots the other counts. In this aspect, any timetagger anomaly that would mimic a dip in total coincidences, such as the one in Fig.~\ref{fig:op_day_9_outlier_anomaly}, can be safely excluded.

\begin{figure}[!tbp]
    \centering
    \includegraphics[width=1\textwidth,height=0.65\textheight,keepaspectratio]{ 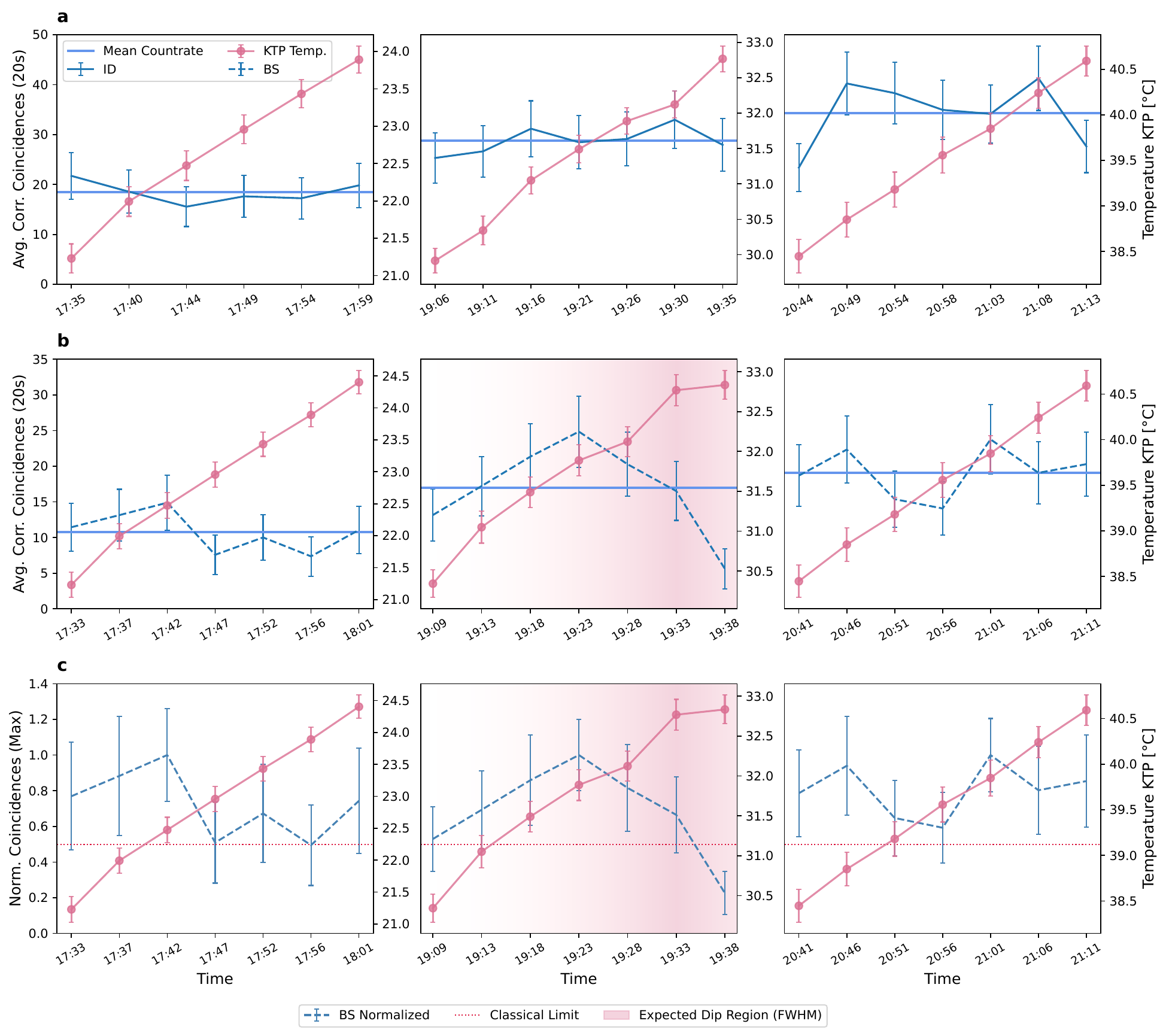}
    \caption[]{\textbf{Op. Day 1 - Average corrected Coincidences and HOM Interference.} \textbf{a (row)} Average corrected coincidence counts (Identity) for target detector pair CC$_{05}$ shows again gradual rate improvement over time from $\sim \ensuremath{1\,\mathrm{Hz}}$ to $\ensuremath{1.7\,\mathrm{Hz}}$ and together with similar rate improvement for Beamsplitter in \textbf{b (row)} explains that each shadow measurement has to be considered an individual measurement. Beamsplitter coincidence rate in \textbf{b (row)} is approx. half of the one in Identity, however, center measurement contains datapoints (middle) that are nearly identical to Identity rate even after removing outlier datapoint for window $\tau = 180=\ensuremath{4.86\,\mathrm{ns}}$. Deviation from theory is consistent with a jitter in MZI6 internal phaseshifter (see Fig.~\ref{fig:op_day_1_mzi_jitter}) that is closer to $\varphi=\{\pi, 0 \}$ (Identity) instead of target value $\varphi=\pi/2$ (Beamsplitter); offset in internal phase shifter from target value cannot explain dip in coincidences in expected KTP temperature regime around $T_{TEC}\sim \ensuremath{32.5\,{}^{\circ}\mathrm{C}}$ (red-shaded area). \textbf{c (row)} Normalized coincidence counts with respect to individual maximum are given for all shadow measurements; Dip in coincidences in the expected KTP temperature regime falls below the classical limit of $p=0.5$, consistent with quantum interference.}
    \label{fig:op_day_1_bs_normalized}
\end{figure}
Taking the mean and its Poissonian error via Eq. \ref{eq:corrected_signal} of the five different coincidence windows in Fig.~\ref{fig:op_day_1_chunks}c (row), provides more insight into the experimental outcome, and as a next step Fig.~\ref{fig:op_day_1_total_counts_anomaly_check} gives an overview of the average corrected coincidences in the shadow of the Earth. Here, the first row shows a stable corrected signal for Identity, and it is now also evident that the mean count rate of target detector pair CC$_{05}$ improves over time (vertical blue line) from $\sim \ensuremath{0.92\,\mathrm{Hz}}$ to $\sim \ensuremath{1.71\,\mathrm{Hz}}$. This can be explained by the gradual warming of the payload over a complete operation day, as shown in Fig.~\ref{fig:temperature-control}. During operation, the electronic boards generate excessive heat, which slowly warms the surrounding material via radiation and conduction. This thermalization effect eventually improves the optical alignment over time, as the final frame temperature of $\sim \ensuremath{15\,{}^{\circ}\mathrm{C}}$ is closer to the original alignment and bonding temperature of the optical components.

Besides, the unused detector modes CC$_{01}$ and CC$_{15}$ show only spurious counts, which agrees well with our expectation, as they are not targeted. In Fig.~\ref{fig:op_day_1_total_counts_anomaly_check}b (row), the same scenario is shown for the Beamsplitter unitary, and here also a gradual improvement in mean count rate is visible. Most importantly, the corrected coincidences in the center figure show a significant drop in coincidences around the expected HOM dip KTP temperature of $T_{TEC} \sim \ensuremath{32.5\,{}^{\circ}\mathrm{C}}$ (red-shaded area), and this is not the case for the corrected Identity measurement. A strong spike in coincidence counts can be identified right before the dip starts (red circle), and this behaviour can be attributed to the outlier of coincidence window $\tau = 180=\ensuremath{4.86\,\mathrm{ns}}$ in Fig.~\ref{fig:op_day_1_chunks}c (center). The dotted line represents the data excluding the significant outlier, and the last row \textbf{c} in Fig.~\ref{fig:op_day_1_total_counts_anomaly_check} shows the sum of counts (dark blue) of all working detector pairs, i.e., CC$_{01}$, CC$_{05}$, and CC$_{15}$. This test is carried out to further rule out any potential HOM dip artifact that could occur when photons are erroneously routed to other detector pairs in the expected HOM dip temperature regime. Here, the counts of the sum of all detector pairs and the target detector pair CC$_{05}$ (light blue) itself are compatible, meaning that the photons are not lost to the unwanted modes CC$_{01}$ and/or CC$_{15}$.

The final shadow analysis in Fig.~\ref{fig:op_day_1_bs_normalized}a-b focuses again on the corrected countrate for the specific target detector CC$_{05}$ and further analysis reveals some additional information on general in-orbit programmable unitary performance. More specifically, the center plot in row \textbf{b} shows that even with removing the outlier of $\tau = 180=\ensuremath{4.86\,\mathrm{ns}}$, the countrate of the Beamsplitter unitary is very close to that of Identity. This is not aligned with the expected behaviour in Eq. \ref{eq:single_bs_output} where approx. half of the Identity rate should be measured for the Beamsplitter setting. This condition is, on average, true for the beginning of the measurement in Fig.~\ref{fig:op_day_1_bs_normalized}b (center) or also for the neighboring shadow measurements (left, right); however, the anomaly of the peak in the center result has to be understood.

\begin{figure}[!tbp]
    \centering
    \includegraphics[width=1\textwidth,height=0.55\textheight,keepaspectratio]{ 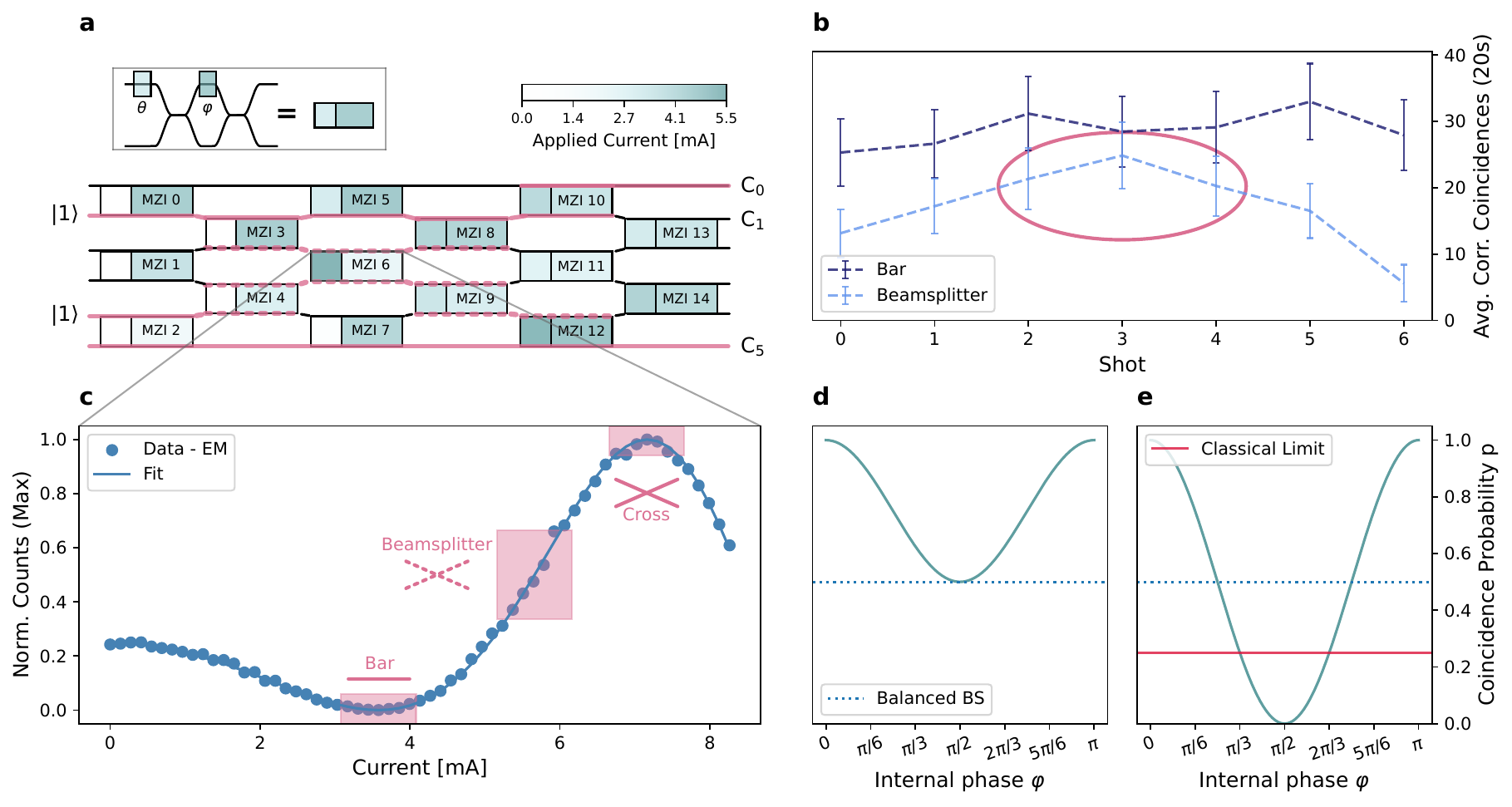}
    \caption[]{\textbf{Op. Day 1 - Phase shifter fluctuations.} \textbf{a} Scheme of uPIC shows path for Identity- (solid) and Beamsplitter unitary (dashed). Photon input is highlighted with Fock state input in uPIC mode $1,4$, and target detectors are C$_0$, C$_5$. MZI6 is responsible for the Beamsplitter configuration and inset in \textbf{c} shows MZI6 current sweep of EM model; a single photon is injected from mode $2$ and registered at modes $2,3$ respectively. Applied current on internal shifter $\varphi$ is transformed to heat and results in a phase shift that routes the photon from mode $2$ (Bar) to mode $3$ (Cross). Steepest point in current sweep relates to Beamsplitter configuration and red-shaded boxes (width) correspond to $\ensuremath{1\,\mathrm{mA}}$ current; Beamsplitter setting is much more sensitive to changes in applied current/heat with respect to Bar/Cross (box height), and this eventually explains the countrate behaviour of Beamsplitter unitary in \textbf{b}. Here, some data points (red ellipse) are close to Identity rate (e.g., Cross/Bar), and this potentially arises from the unstable payload temperature (see Fig.~\ref{fig:temperature-control}) that leads to shifting of the internal shifter closer to $\varphi=\{0,\pi\}$, respectively. \textbf{d} Coincidence probability at detector pair CC$_{05}$ is simulated for changing internal phase $\varphi$ in Beamsplitter unitary of scheme \textbf{a} (dashed). Minimal rate with respect to Identity is exactly one half for configuration $\varphi=\pi/2$, and any offset will lead to a higher coincidence rate. \textbf{e} Indistinguishable photons will only show perfect quantum interference ($p=0$) if MZI6 is a perfectly balanced beamsplitter with $\varphi=\pi/2$. Any offset will gradually decrease interference visibility and eventually vanish into the classical regime (horizontal red line) if, e.g., $\varphi=\pi/3$.}
    \label{fig:op_day_1_mzi_jitter}
\end{figure}

A general deviation from theory can be attributed to a jitter in phase shifter performance of the MZI that is set to a balanced Beamsplitter (see MZI6 in Fig.~4a of the main text or Fig.~\ref{fig:op_day_1_mzi_jitter}c for this specific unitary). In a formal sense, we can identify the responsible MZI and model the effect of a jitter in phase shifter performance. For this task, we consider the realistic scenario in which both photons enter the uPIC from modes $1,4$ and are subsequently routed to MZI6. The applied unitary is schematically shown in Fig.~\ref{fig:op_day_1_mzi_jitter}a and after potential interference at MZI6, both photons are sent to detector pair CC$_{05}$. In the framework of the Fock basis, we can write the two-photon input state as
\begin{equation}
    \ket{\psi_{MZI6}}=\hat{a}_{m}^{\dagger} \hat{b}_{n}^{\dagger} \ket{0}_{2} \ket{0}_{3} = \ket{1,m}_{2} \ket{1, n}_{3},
    \label{eq:mzi_input}
\end{equation}
and here the subscript $2,3$ corresponds to the uPIC spatial mode in which MZI6 is embedded. The first balanced beamsplitter of the MZI renders the input state according to Eq. \ref{eq:bs_trafo}
\begin{equation}
    \ket{\psi_{MZI6}} \rightarrow\frac{1}{2} \left(  \hat{a}^{\dagger}_m + \hat{b}^{\dagger}_m \right) \cdot \left(\hat{a}^{\dagger}_n-  \hat{b}^{\dagger}_n  \right) \ket{0}_{2} \ket{0}_{3}.
    \label{eq:first_bs_mzi}
\end{equation}
In this case, we omitted the external phase $\theta$ from mode $2$ as it will just evolve as a global phase; however, the internal phase $\varphi$ in mode $2$ will have a strong impact on the evolution of the photon path and is now formally included
\begin{equation}
    \ket{\psi_{MZI6}} \rightarrow\frac{1}{2} \left(  \hat{a}^{\dagger}_m \cdot e^{i\varphi} + \hat{b}^{\dagger}_m \right) \cdot \left(\hat{a}^{\dagger}_n\cdot e^{i\varphi} -  \hat{b}^{\dagger}_n  \right) \ket{0}_{2} \ket{0}_{3}.
    \label{eq:first_bs_mzi_including_internal}
\end{equation}
The second waveguide beamsplitter will further transform the state into
\begin{equation}
    \ket{\psi_{MZI6}} \rightarrow\frac{1}{4} \left(  (\hat{a}^{\dagger}_m+\hat{b}^{\dagger}_m) \cdot e^{i\varphi} + (\hat{a}^{\dagger}_m-\hat{b}^{\dagger}_m) \right) \cdot \left((\hat{a}^{\dagger}_n+\hat{b}^{\dagger}_n)\cdot e^{i\varphi} -  (\hat{a}^{\dagger}_n-\hat{b}^{\dagger}_n)  \right) \ket{0}_{2} \ket{0}_{3},
\label{eq:first_bs_mzi_including_internal_inclduing_2nd_bs}
\end{equation}
and there exist now three special outcomes for the internal shifter configuration
$\varphi$ which read
\begin{equation}
    \begin{aligned}
    \varphi=\pi &\rightarrow \hat{a}^{\dagger}_m \hat{b}^{\dagger}_n \ket{0}_{2} \ket{0}_{3} \\[0.5em]
    \varphi = 0 &\rightarrow \hat{b}^{\dagger}_n \hat{a}^{\dagger}_m\ket{0}_{2} \ket{0}_{3} \\[0.5em]
    \varphi=\frac{\pi}{2} &\rightarrow \frac{1}{2} e^{i \pi} \left(\hat{a}^{\dagger}_m \hat{a}^{\dagger}_n - \hat{a}^{\dagger}_m \hat{b}^{\dagger}_n + \hat{b}^{\dagger}_m \hat{a}^{\dagger}_n + \hat{b}^{\dagger}_m \hat{b}^{\dagger}_n \right)\ket{0}_{2} \ket{0}_{3}.
    \label{eq:mzi_output}
    \end{aligned}
\end{equation}
The first case of $\varphi=\pi$ corresponds to the MZI state ``Bar'' in which the two photons just travel straight through the interferometer. The second case $\varphi=0$ is the MZI ``Cross'' state in which the two input photons swap their paths. As both photons nevertheless end up at detector pair CC$_{05}$, these two configurations cannot be discriminated in coincidence counting as $\hat{a}^{\dagger}_m \hat{b}^{\dagger}_n$ and $\hat{b}^{\dagger}_n \hat{a}^{\dagger}_m$ are effectively identical. In this sense, the phase shifter setting $\varphi=\{0,\pi\}$ will lead to the same output as the Identity unitary (see Fig.~\ref{fig:op_day_1_mzi_jitter}a solid red) in which both photons do not interfere and are just routed to CC$_{05}$.

The third case of $\varphi=\pi/2$ differs from the previous configurations, and here the output state Eq. \ref{eq:mzi_output} is identical to the one of a single balanced beamsplitter in Eq. \ref{eq:single_bs_output}. Given this similarity, we expect to have half of the Identity coincidence rate for distinguishable photons or zero coincidences for indistinguishable photons. A jitter in internal phase shifter performance $\varphi \pm \Delta\varphi$ can be simulated by considering Eq. \ref{eq:first_bs_mzi_including_internal_inclduing_2nd_bs} and explicitly writing out the action of the Fock operators on the vacuum state  $\ket{0}_2\ket{0}_3$
\begin{equation}
    \begin{aligned}
    \ket{\psi_{MZI6}} \rightarrow\frac{1}{4} \big( &e^{i2\varphi} \cdot (\ket{1,m}_2 \ket{1,n}_2 +\ket{1,m}_2 \ket{1,n}_3+\ket{1,m}_3 \ket{1,n}_2+\ket{1,m}_3 \ket{1,n}_3)  \\[0.5ex]
    & - 2e^{i\varphi} \cdot (\ket{1,m}_2 \ket{1,n}_3-\ket{1,m}_3 \ket{1,n}_2) \\[0.5ex]
    & -\ket{1,m}_2 \ket{1,n}_2 + \ket{1,m}_2 \ket{1,n}_3 + \ket{1,m}_3 \ket{1,n}_2 - \ket{1,m}_3 \ket{1,n}_3 \big).
    \end{aligned}
    \label{eq:mzi_jitter_1}
\end{equation}
Photons that end up in the same mode, i.e., $\ket{1,n}_2 \ket{1,m}_2$, can be ignored, and so the only relevant terms are the ones that lead to a coincidence event. The former result therefore renders to
\begin{equation}
    \begin{aligned}
    \ket{\psi_{MZI6}} \rightarrow\frac{1}{4} \big( & e^{i2\varphi} \cdot (\ket{1,m}_2 \ket{1,n}_3+\ket{1,m}_3 \ket{1,n}_2) \\[0.5ex]
    & - 2e^{i\varphi} \cdot (\ket{1,m}_2 \ket{1,n}_3-\ket{1,m}_3 \ket{1,n}_2) \\[0.5ex]
    & +\ket{1,m}_2 \ket{1,n}_3 + \ket{1,m}_3 \ket{1,n}_2  \big).
    \end{aligned}
    \label{eq:mzi_jitter_2}
\end{equation}
This result can be simplified in terms of trigonometric identities
\begin{equation}
    \begin{aligned}
    \ket{\psi_{MZI6}} \rightarrow\frac{1}{2} e^{i \varphi} \big( & \cos{\varphi} \cdot (\ket{1,m}_2 \ket{1,n}_3+\ket{1,m}_3 \ket{1,n}_2) \\[0.5ex]
    & -  (\ket{1,m}_2 \ket{1,n}_3-\ket{1,m}_3 \ket{1,n}_2) \big),
    \end{aligned}
    \label{eq:mzi_jitter_3}
\end{equation}
and for distinguishable photons, i.e. $m\neq n$, Born's rule reveals the coincidence probability $p$ with respect to the internal phase $\varphi$ as follows
\begin{equation}
    p=\frac{1}{2} \big(1+\cos^2{\varphi} \big).
    \label{eq:dist_mzi_outcome}
\end{equation}
For indistinguishable photons, on the other hand, i.e., $m=n$, we expect the coincidence probability to follow
\begin{equation}
    p=\cos^2{\varphi}.
    \label{eq:indist_mzi_outcome}
\end{equation}
The coincidence probability of the first case is shown in Fig.~\ref{fig:op_day_1_mzi_jitter}d and it is now evident that only for a properly set internal phase of $\varphi=\pi/2$ the coincidence probability can be exactly one half of the Identity case. Any offset will lead to a higher probability $p>0.5$, and this is effectively reflected in a higher coincidence rate that overlaps with the one of Identity unitary if $\varphi=0$ or $\varphi=\pi$. The two scenarios for distinguishable or indistinguishable photons can be combined into a mixture
\begin{equation}
    \rho= \nu \cdot \cos^2{\varphi}  + (1-\nu) \cdot \frac{1}{2} \big(1+\cos^2{\varphi} \big),
    \label{eq:mixture}
\end{equation}
and here the degree of indistinguishability is encoded as $\nu$. Sweeping the internal phase from $0$ to $\pi$ as done in Fig.~\ref{fig:op_day_1_mzi_jitter}e reveals that it is crucial to have the internal phase set to precisely $\varphi=\pi/2$ in order to faithfully conduct a Hong-Ou Mandel experiment. In this simulation, both photons are considered to be perfectly indistinguishable, i.e. $\nu=1$, and a perfect HOM dip with $p=0$ can be found for $\varphi=\pi/2$. This is not true if the internal phase starts to shift from this optimal setting, and even a slight offset of $\Delta \varphi=\pi/6$ can push the HOM dip outside of the quantum regime. This can be seen for, e.g., $\varphi=\pi/3$, as here the coincidence probability already increased to $p\sim 0.5$\footnote{Note that in Fig. \ref{fig:op_day_1_mzi_jitter}e the counts are normalized with respect to MZI6 configuration $\varphi=\pi$, hence the classical threshold in the Figure is at $p \sim 0.25$ (horizontal red line). This treshhold corresponds to the case of $p=0.5$ in the main text.}. The latter is considered a classical limit (red line in Fig.~\ref{fig:op_day_1_mzi_jitter}e) and to claim true quantum interference, the coincidence probability has to be below this threshold \cite{li_experimental_2022}.

\begin{figure}[!tbp]
    \centering
    \includegraphics[width=1\textwidth,height=0.73\textheight,keepaspectratio]{ 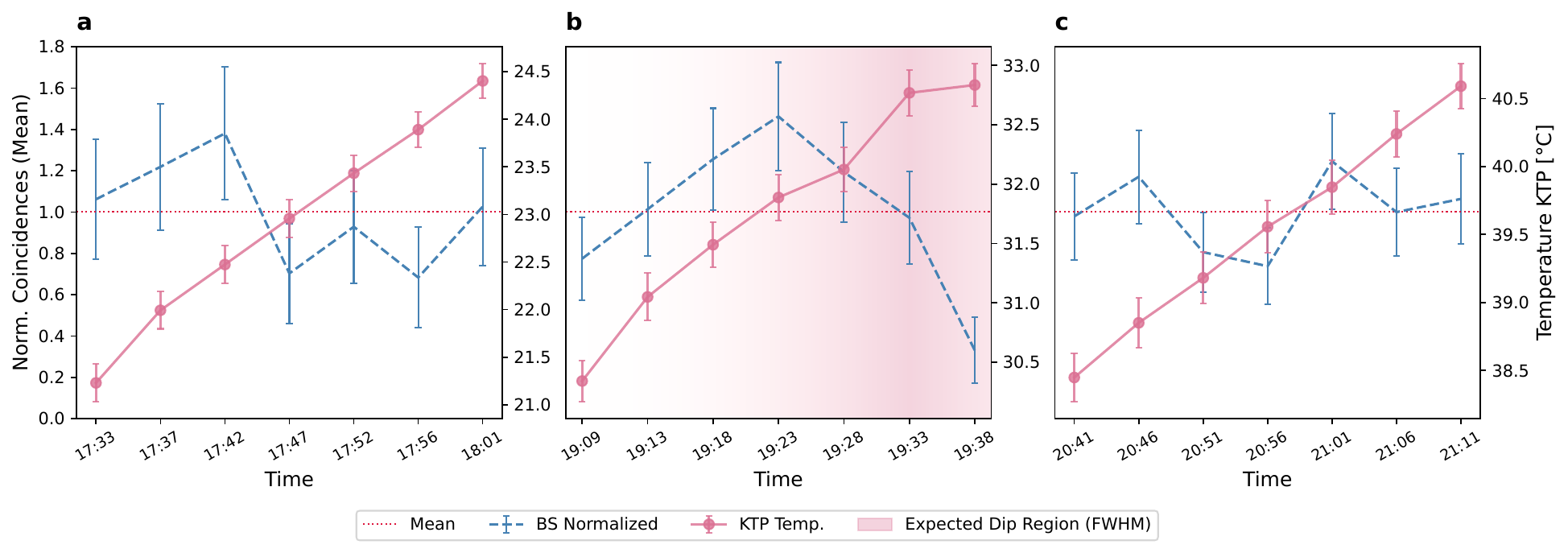}
    \caption[]{\textbf{Op. Day 1 - HOM Interference - Mean normalization.} \textbf{a-c} Identical to the measurement configuration of normalized Beamsplitter in Fig.~\ref{fig:op_day_1_bs_normalized}c, but here the raw data of the three shadow measurements is normalized with respect to the individual mean. This normalization routine helps in finding overall trends in noisy data and is also meaningful given the count rate fluctuations of the Beamsplitter unitary (see MZI jitter in Fig.~\ref{fig:op_day_1_mzi_jitter}). For a measurement with no Hong-Ou Mandel interference, the counts fluctuate around their mean (red dashed line), as shown in \textbf{a} or \textbf{c}. However, within the expected HOM dip regime (red-shaded area), the measured dip in coincidences is clearly below the mean, thereby manifesting the quantum interference signature of the first operation day.}
    \label{fig:op_day_1_bs_normalized_mean}
\end{figure}
\begin{figure}[!tbp]
    \centering
    \includegraphics[width=1\textwidth,height=0.65\textheight,keepaspectratio]{ 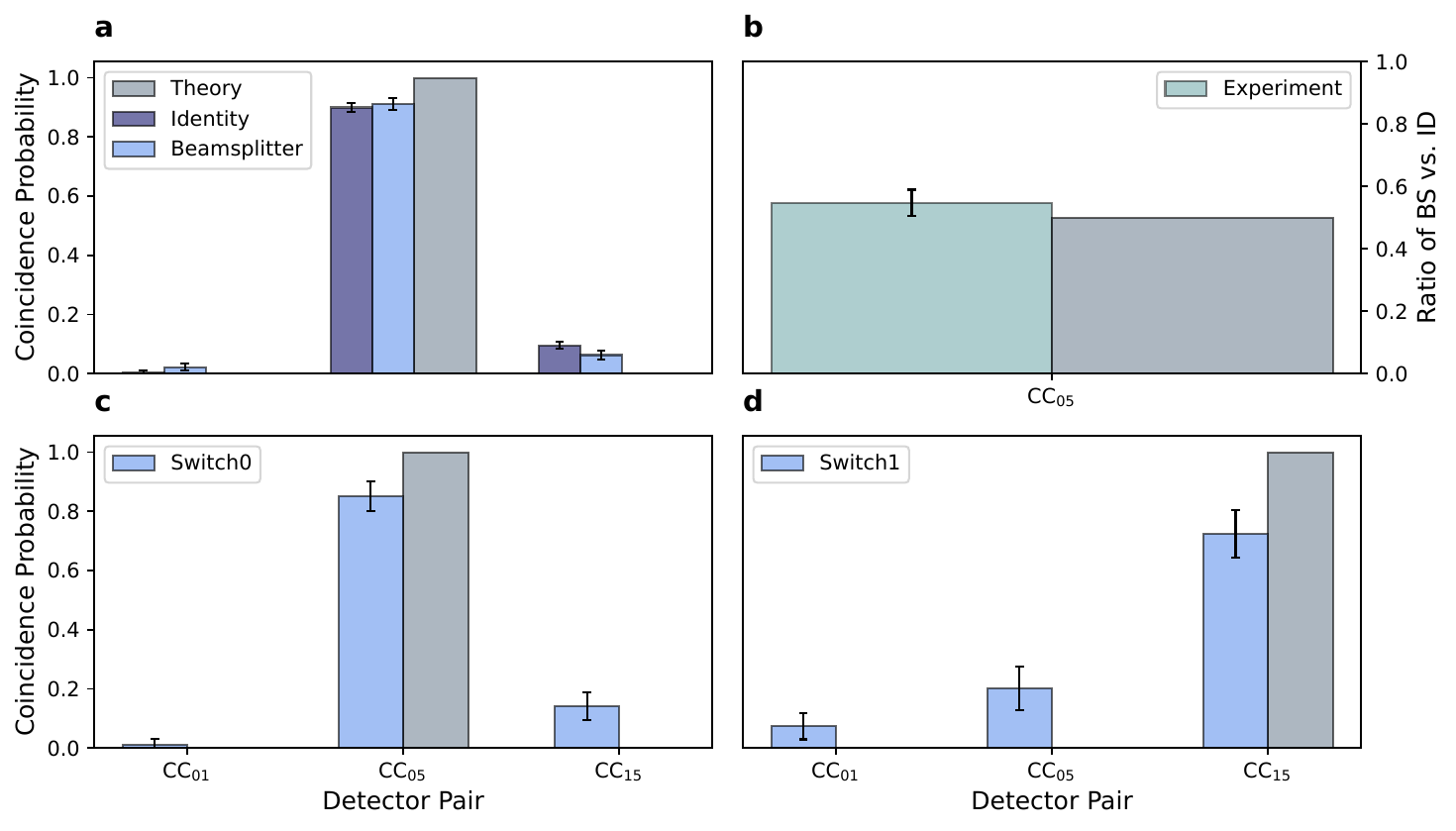}
    \caption[]{\textbf{Op. Day 1 - On-board programmable unitary performance and ratio.} \textbf{a} Average coincidence probability for Identity (dark blue) and Beamsplitter (light blue) unitary from the three shadow measurements are compared to theory (grey bars), and the measured output agrees well with the expectation. \textbf{b} Average ratio between Beamsplitter and Identity coincidence counts for distinguishable photons and target detector pair CC$_{05}$ is shown. Here, data from the first and last shadow measurements is considered, as well as the final section of the sun measurement; inclusion of the latter is justified for measurements after $\sim 4$ hrs. of operation (see Fig.~\ref{fig:op_day_1_counts}d), as here the rate is stabilised and more generally the ratio is more robust against outliers and noise. The second shadow regime is excluded because photons are partially indistinguishable here. The slightly higher average ratio can be explained by increased thermal sensitivity of the Beamsplitter unitary as shown in Fig.~\ref{fig:op_day_1_mzi_jitter}, which eventually increases the average Beamsplitter rate. \textbf{c,d} Health check of uPIC performance after HOM dip scan shows that also detector pair CC$_{15}$ is active when targeted (Switch1); fidelity of the latter could be improved in later operation days (see e.g. Sec.~\ref{sec:operation_day_8})}
    \label{fig:op_day_1_probabs}
\end{figure}
The simulation can now be concluded with an experimental test in Fig.~\ref{fig:op_day_1_mzi_jitter}c where data of the EM model shows the phase sweeping behaviour of the internal phase in MZI6. Here, single photons are injected into mode $2$ and registered at modes $2, 3$, respectively. As the physical phase shifters are thermal, a current has to be applied on the underlying resistor to effectively introduce a phase shift due to local heating of the waveguide (see also section~\ref{sec:photonic_processor} or Refs. \cite{ceccarelli2020low}, \cite{albiero2022toward}.). The applied current on the internal phase shifter $\varphi$ is swept from $\ensuremath{0\,\mathrm{mA}}$ to $\ensuremath{8\,\mathrm{mA}}$ and the generated heat will introduce a phase shift that routes the photon from mode $2$ (Bar) to mode $3$ (Cross). These MZI configurations can be seen as the red-shaded areas labeled in Fig.~\ref{fig:op_day_1_mzi_jitter}c, and the configuration for Beamsplitter ($\varphi=\pi/2$) is located at the steepest section of the curve fit. The total width of the red-shaded area corresponds to $\ensuremath{1\,\mathrm{mA}}$, and comparing the height of the red-shaded areas reveals that the Beamsplitter setting is much more sensitive to changes in current. In other words, slight changes in applied current or heat have a more drastic consequence on the Beamsplitter than on Cross/Bar, and together with the problematic non-stable thermal environment of the payload (see drift in payload temperature in Fig.~\ref{fig:temperature-control}), the Beamsplitter configuration might suffer from this fact of higher (thermal) sensitivity.

This additional thermal sensitivity of the Beamsplitter unitary eventually explains the momentarily high countrate\footnote{This problem is also encountered again in later operation days.} in Fig.~\ref{fig:op_day_1_mzi_jitter}b (red ellipse), where right before the dip in coincidences MZI6 was most likely closer to $\varphi=0$ or $\varphi=\pi$ instead of being properly set to $\varphi=\pi/2$. This jitter in the Beamsplitter unitary performance during a HOM scan further complicates a two-photon interference measurement, as it is essential that the internal phase shifter is very close to $\varphi=\pi/2$, otherwise no interference can be observed (compare Fig.~\ref{fig:op_day_1_mzi_jitter}e). Nevertheless, the significant dip in coincidences in Fig.~\ref{fig:op_day_1_bs_normalized}b (center) indicates that MZI6 was close to a balanced beamsplitter at that point, since a phase offset alone raises rather than lowers the coincidence rate.

In the final row of Fig.~\ref{fig:op_day_1_bs_normalized}c, the normalized coincidence counts $P_{i,max}$ from the three shadow measurements are shown. For easier readability, the mean of the five different windows in Eq. \ref{eq:signal_extraction} is written as $x_{i}$ with its error $\sigma_{x_{i}}$, and here the index $i$ refers to the KTP temperature. As a first step, the canonical maximum normalization is chosen in which $x_{i}$, with propagated error $\sigma_{x_{i}}$, is normalized as follows
\begin{equation}
    P_{i,max}=\frac{x_ {i}}{\max{(x_{i})}}, \quad \text{and} \quad\sigma_{P_i,max}=\sqrt{\frac{x_{i}^2\cdot \sigma^2_{\max{(x_{i})}}+\max^2{(x_{i})}\cdot \sigma^2_{x_{i}}}{{\max^4{(x_{i})}}}}.
    \label{eq:maximum_normalization}
\end{equation}

Here, it must be noted that each shadow measurement is treated as an individual measurement, as the count rate of each section is different\footnote{Note that to faithfully evaluate an errorbar for the maximum $P_{i=max(x_i)}$, the relative error $\sigma_{max(x_i)}/\max{(x_i)}$ is evaluated. This is necessary, as otherwise $P_{i=max(x_i)}$ would not have an assigned error.}. This behaviour is clearly illustrated in the rising count rate (vertical blue line) in Fig.~\ref{fig:op_day_1_bs_normalized}a-b.
In this sense, every shadow measurement in Fig.~\ref{fig:op_day_1_bs_normalized}c has its own maximal data point $\max{(x_{i})}$ in Eq. \ref{eq:maximum_normalization} and by evaluating $P_{i,max}$ for the three different data sets, it is now clear that the dip in coincidences in the expected dip regime (red-shaded area) already falls below the classical threshold of $p=0.5$ (red dotted line).

To faithfully interpret the Hong-Ou Mandel interference pattern, the given shadow measurements are also normalized to the mean. This is meaningful, as on one hand the underlying raw data is inherently noisy, and on the other hand, the apparent jitter in MZI performance in Fig.~\ref{fig:op_day_1_mzi_jitter} calls for a stricter normalization routine. As the Beamsplitter count rate is close to that of Identity for some data points, the maximum normalization might favor an otherwise non-significant dip in coincidences. For this reason, we have to further test the validity of the dip in Fig.~\ref{fig:op_day_1_bs_normalized}c and similar to Eq. \ref{eq:maximum_normalization}, the data $x_{i}$ with its error $\sigma_{x_{i}}$ is normalized with respect to the individual shadow mean $\sum_{i} x_{i}/N$ as follows
\begin{equation}
    P_{i,mean}=N\frac{x_{i}}{\sum x_{i}}, \quad \text{and} \quad\sigma_{P_i,mean}=N\sqrt{\frac{x_{i}^2\cdot \sum_j \sigma^2_{x_{j,j\neq i}}+\sigma^2_{x_{i}} \cdot (\sum_j x_{j,j \neq i})^2}{(\sum_i x_i)^4}}.
    \label{eq:mean_normalization}
\end{equation}

Here, $N$ represents the number of data points of a given shadow regime, and Fig.~\ref{fig:op_day_1_bs_normalized_mean}b illustrates that this more rigid normalization technique also shows a clear dip in coincidences in the expected HOM dip regime (red-shaded area). This outcome further supports quantum interference on the first operation day. In the summarizing chapter of the complete HOM dip analysis in Sec.~\ref{sec:main_figures_hom}, additional measurements from upcoming operation days are combined to complete the temperature scan and extract a faithful interference visibility.

The chapter concludes with additional insights into the on-board programmable unitary performance, and Fig.~\ref{fig:op_day_1_probabs} provides an overview of the coincidence probabilities of the applied unitaries. In particular, Fig.~\ref{fig:op_day_1_probabs}a shows the average coincidence probabilities obtained for Identity and Beamsplitter across the three shadow measurements. Here, the inverse-variance-weighted arithmetic mean is used to extract the average coincidence probability over a full operation day. The individual Identity or Beamsplitter data points for different KTP temperatures in Eq. \ref{eq:signal_extraction} are first summed up for all working detector pairs and form the quantity $v_{ij}$ with error $\sigma_{v_{ij}}$ for $ij=\{01,05,15\}$. To extract the coincidence probability $p_{ij}$ with  related error $\sigma_{p_{ij}}$, the total counts $v_{ij}$ for specific detector pair $ij$ are divided by the complete sum of all working detector pairs $ij$ via
\begin{equation}
    p_{ij}=\frac{v_{ij}}{\sum_{ij}v_{ij}}, \quad \text{and} \quad \sigma_{p_{ij}}=\sqrt{\frac{v_{ij}^2\cdot \sum_{mn} \sigma^2_{v_{mn,mn\neq ij}}+\sigma^2_{v_{ij}} \cdot (\sum_{mn} v_{mn,mn \neq ij})^2}{(\sum_{ij} v_{ij})^4}}
    \label{eq:coinc_probability}
\end{equation}
 This quantity is evaluated for every shadow period, and the weighting of all shadow regimes is then given by
\begin{equation}
    \bar{p}=\sum_i \tilde{w}_ip_i \quad \text{with} \quad \tilde{w}_i=\frac{w_i}{\sum_jw_j} \quad \text{and} \quad  w_i=\frac{1}{\sigma_{p_i}^2} \quad \text{with} \quad \sigma_{\bar{p}}=\sqrt{\frac{1}{\sum_i w_i}}.
    \label{eq:weighted_average}
\end{equation}
The natural approach of taking the weighted average of all individual summed shadow measurements (if available) is also standard for all upcoming operation days. The measured coincidence probability in Fig.~\ref{fig:op_day_1_probabs}a reflects good uPIC performance and calibration. This is also underscored by the fact that, for these unitaries to work, all $15$ embedded MZIs must be active to compensate for thermal crosstalk. This feature can be visually seen in Fig.~\ref{fig:op_day_1_mzi_jitter}a, where all individual MZIs are addressed with a specific current (green color code) to facilitate the Identity or Beamsplitter unitary.

The neighboring result in \textbf{b} of Fig.~\ref{fig:op_day_1_probabs} corresponds to the average ratio of Beamsplitter vs. Identity counts. Here, the summed measurements $v_{ij}$ for Beamsplitter and Identity, for a specific target detector (e.g., $ij=05$ on operation day $1$), are divided to extract the ratio. Due to the sum of many data points, the ratio is fundamentally more robust to outliers and noise (see also upcoming operation days, e.g., Sec.~\ref{sec:operation_day_9}); therefore, we can faithfully include measurements where the payload faces the sun. In more detail, the given result includes the first and third shadow measurements (compare sections in Fig.~\ref{fig:op_day_1_umbra_first_look}) as well as the complete data after the third shadow measurement. The latter can be justified by considering Fig.~\ref{fig:op_day_1_counts}d, where the coincidence rate after $\sim 4$ hrs has already improved significantly with respect to the first and second ``sun'' measurements. In this spirit, averaged measurements from this regime are more reliable and can be considered in the ratio data analysis. The second shadow measurement is excluded because the HOM dip is measured here, hence the photons are not fully distinguishable. Given these boundary conditions, the measured weighted-average ratio for the detector pair CC$_{05}$ in Fig.~\ref{fig:op_day_1_probabs}b is close to the expected value of $R_T=0.5$. The slightly higher value can be explained by the mentioned MZI performance jitter for the Beamsplitter unitary in Fig.~\ref{fig:op_day_1_mzi_jitter}. Due to the unstable thermal environment, it is harder for the uPIC to implement $\varphi = \pi/2$, and thus the rate is, on average, slightly closer to that of Identity.

The last two results in Fig.~\ref{fig:op_day_1_probabs}c-d represent a uPIC health test after the HOM dip scan. In this case, the two photons are routed to CC$_{05}$ and CC$_{15}$ to check the general uPIC response in a short timeframe. Here, the detector pair CC$_{15}$ also shows a clear response, and its unitary fidelity could be improved on later operating days.

Let us note that the position and velocity of the host satellite are, in principle, predictable from TLE data, which would imply that before each operation day, we could estimate the times when the satellite is in Earth's shadow and therefore conduct our experiments only within these time slots. This was however not possible in the organization of our operation days, so we measured continuously and separated the data into chunks afterwards. Given the limits of the operations agreement with the host provider, the periodic invasion of parasitic photons from the sun could not be avoided.

Additionally, the KTP calibration mapping result (see Fig.~\ref{fig:arroyo_vs_tec_board}) was obtained in a later project stage (around operation day $10$ in Sec.~\ref{sec:operation_day_10}), as the TEC board calibration was still not fully understood after launch. This also explains why on the first operation day, the KTP scan range was chosen to be broad, and until operation day $10$, it was believed that there exists a one-to-one mapping between Arroyo Ctrl. and TEC board, i.e. $T_A=T_{TEC}$. The latter was found to be wrong with the definitive measurements in Fig.~\ref{fig:arroyo_vs_tec_board} and this explains why the KTP setpoint in the upcoming operation days is, in general, not around the expected dip regime of $T_{TEC}\sim \ensuremath{32.5\,{}^{\circ}\mathrm{C}}$.
\vspace{6pt}
\clearpage
\subsection{Operation Day 2 - 35 Days in Orbit}
\label{sec:operation_day_2}
\begin{figure}[!tbp]
    \centering
    \includegraphics[width=1\textwidth,height=0.69\textheight,keepaspectratio]{ 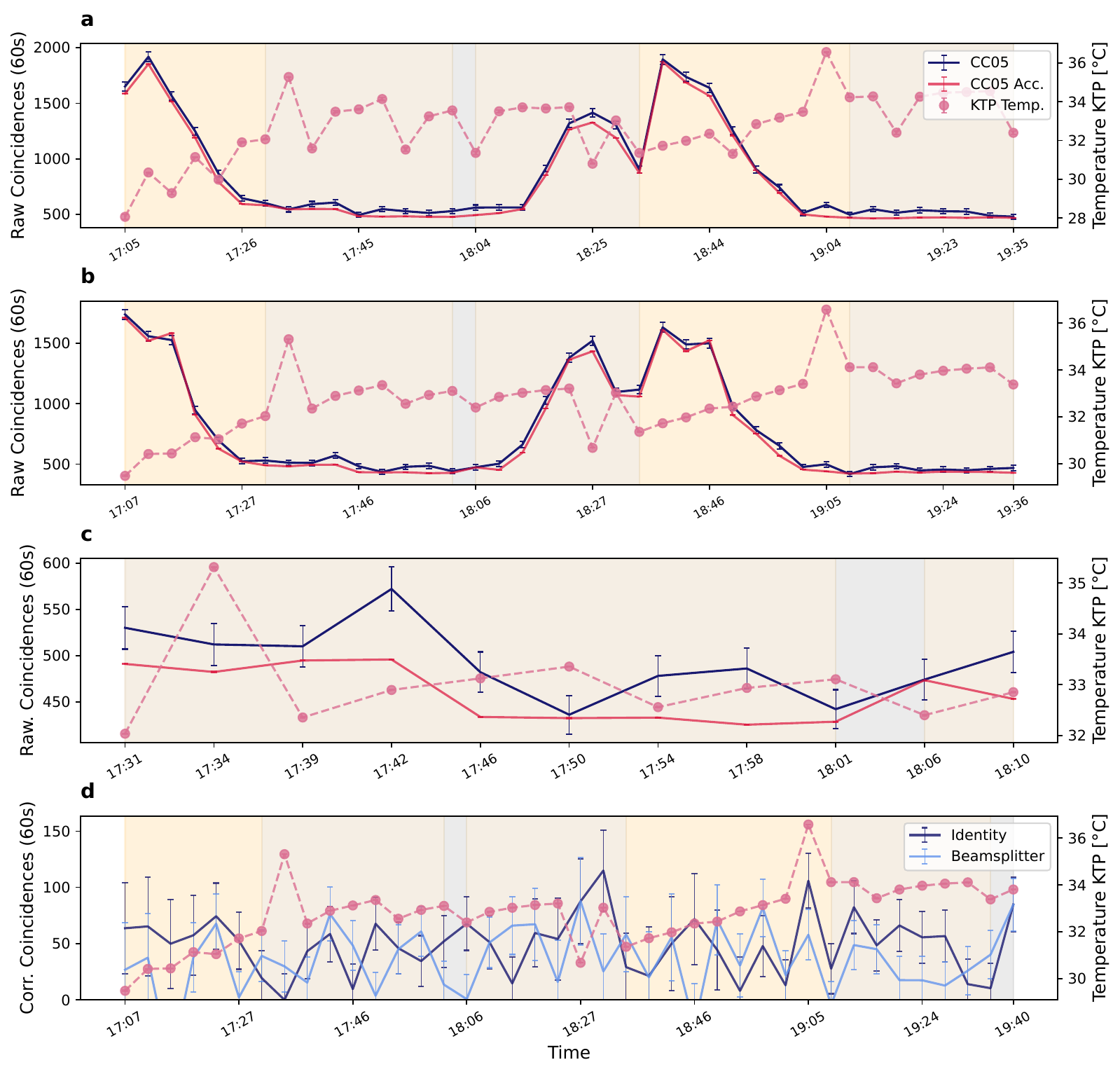}
    \caption[]{\textbf{Op. Day 2 - Coincidences overview.} \textbf{a} Raw coincidence counts (Identity) for target detector pair CC$_{05}$ (dark blue) shows a strong decline in measurement quality with respect to Fig.~\ref{fig:op_day_1_counts} of operation day $1$. This is reflected in increased instability of KTP temperature stabilisation (red points - secondary $y$-axis), increased instability in noise floor (red) in shadow regimes, and overall less clear distinction between noise and signal. \textbf{b} Identical setting as \textbf{a}, but Beamsplitter configuration shows similar fate in reduced measurement quality, and this is even further highlighted in \textbf{c}, where shadow measurement shows stronger noise fluctuations with respect to, e.g., Fig.~\ref{fig:op_day_1_umbra_first_look}; countrate close to zero can potentially be attributed to complete measurement being within KTP temperature regime of expected HOM interference. However, the reference measurement Identity (dark blue) is compatible with the Beamsplitter (light blue) in \textbf{d}; thus, artifacts cannot be excluded, and this effect renders the entire measurement unreliable.}
    \label{fig:op_day_2_counts}
\end{figure}

The task of the second operation day consisted of further HOM dip scans and additional system health checks; however, the outgassing induced power loss of the laser (see Fig.~\ref{fig:laser_power_vac}) and possible radiation-induced damage in SPADs (see Sec.~\ref{sec:space_as_environment}) had a significant impact on our payload performance. This can be directly seen in Fig.~\ref{fig:op_day_2_counts}, which is formally identical to that of the first operation day in Fig.~\ref{fig:op_day_1_counts}. There is, however, a drastic difference in KTP performance with respect to temperature stability and overall count rate. The latter is reflected, e.g., in the Beamsplitter setting of a shadow measurement in Fig.~\ref{fig:op_day_2_counts} \textbf{c}, where the raw coincidence counts are nearly overlapping with the noise floor. On the one hand, this might be explained by the KTP temperature being around the degeneracy temperature $T\sim \ensuremath{32.5\,{}^{\circ}\mathrm{C}}$ for nearly the entire measurement. In other words, this would mean that we are constantly measuring around the HOM dip regime; however, when looking at the corrected Identity rate (dark blue) for the same temperature in Fig.~\ref{fig:op_day_2_counts}d and comparing it to the corrected Beamsplitter rate (light blue), both configurations are essentially overlapping within their error. This means we cannot use the Identity measurement as a reference point to rule out potential artifacts, and, furthermore, the countrate appears to randomly overlap with the noise floor (i.e., countrate is zero) on many occasions throughout the measurement.

\begin{figure}[!tbp]
    \centering
    \includegraphics[width=1\textwidth,height=0.74\textheight,keepaspectratio]{ 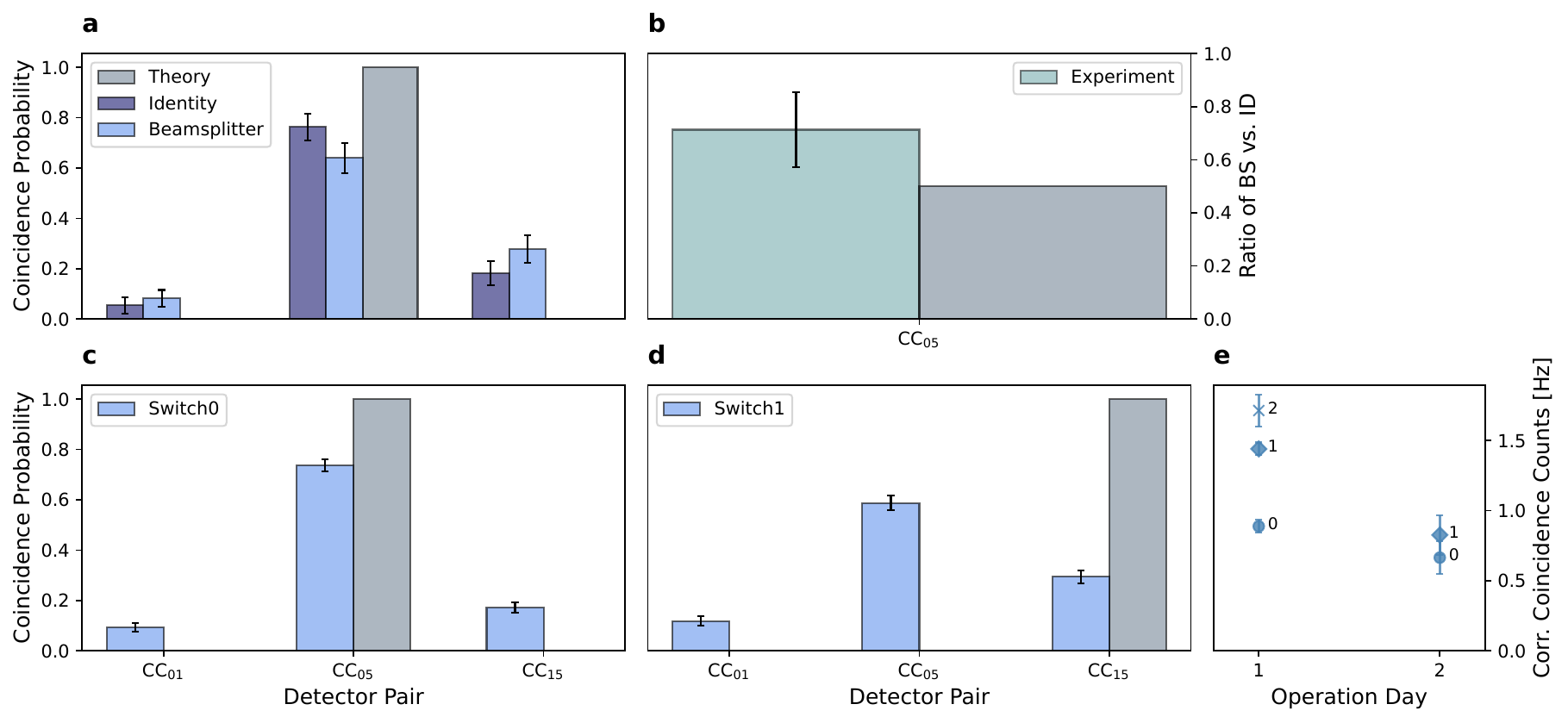}
    \caption[]{\textbf{Op. Day 2 - On-board programmable unitary performance and ratio.} \textbf{a} (Weighted) average of coincidence probability for Identity (dark blue) and Beamsplitter (light blue) unitary from shadow measurements also show a significant drop in unitary fidelity with respect to the same measured unitaries on operation day $1$ in Fig.~\ref{fig:op_day_1_probabs}. This decay in performance is also reflected in the ratio in \textbf{b} and in the uPIC system health test in \textbf{c,d}, where CC$_{15}$ does not respond as expected. \textbf{e} Shadow measurement corrected coincidence rate compares the first two operation days and highlights the significant decline in rate; small label indicates the general gradual improvement in count rate over an individual operating day, as label corresponds to individual shadow measurement.}
    \label{fig:op_day_2_probabs}
\end{figure}

This significant drop in performance essentially means that the measurements of this operation day cannot be considered valid, as no useful information can be extracted. The loss in count rate with respect to the first operation day in Sec.~\ref{sec:operation_day_1} is a critical factor, but another relevant contribution for the exclusion is the relatively unstable noise floor even in the shadow measurements, in, e.g., Fig.~\ref{fig:op_day_2_counts}c. The overall reduced measurement quality is also reflected in the averaged coincidence probability for Identity/Beamsplitter in Fig.~\ref{fig:op_day_2_probabs}a, or the uPIC health test in Fig.~\ref{fig:op_day_2_probabs}c-d.
\clearpage
\subsection{Operation Day 3 - 52 Days in Orbit}
\label{sec:operation_day_3}
\begin{figure}[!tbp]
    \centering
    \includegraphics[width=1\textwidth,height=0.78\textheight,keepaspectratio]{ 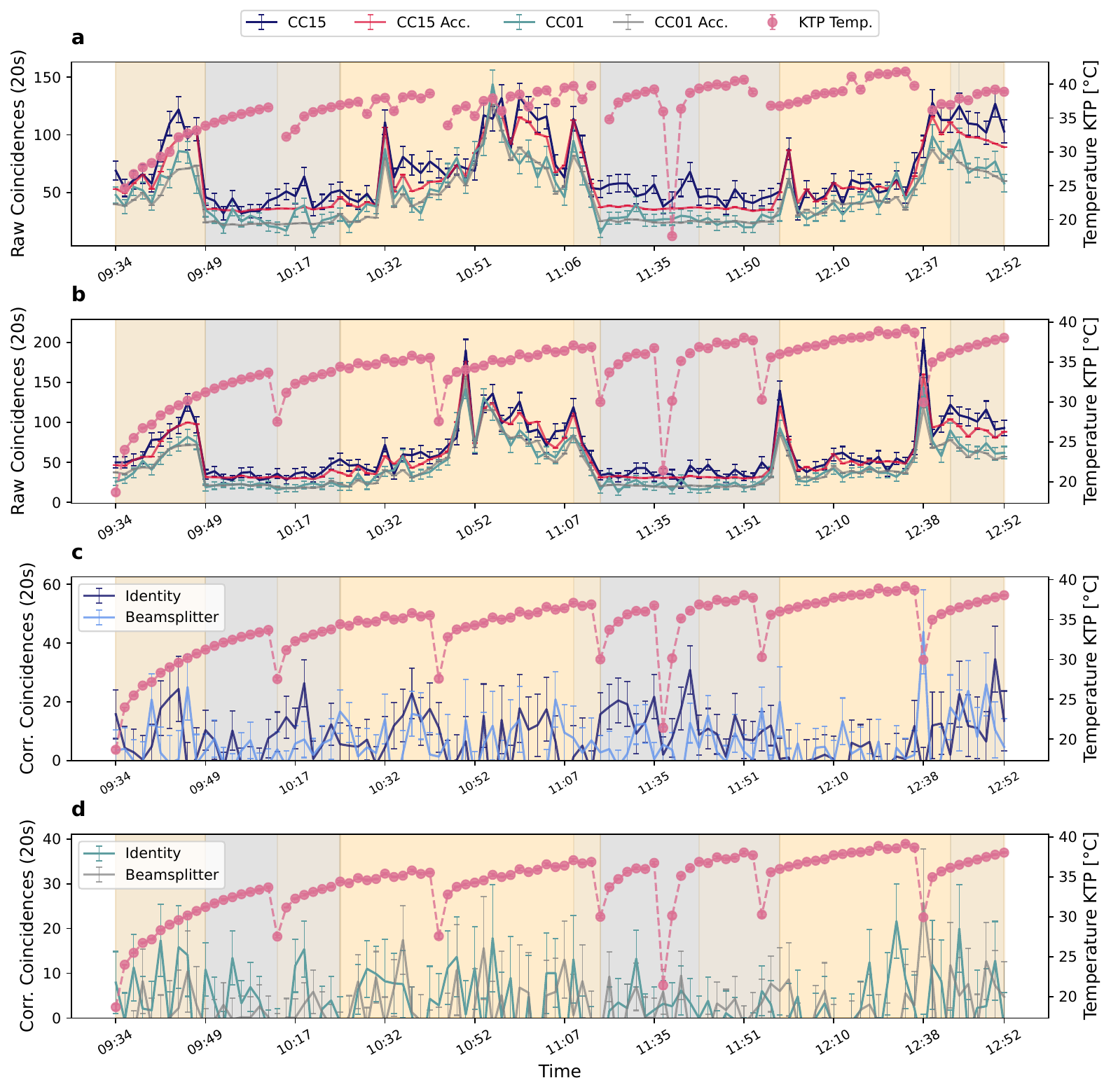}
    \caption[]{\textbf{Op. Day 3 - Coincidences overview.} \textbf{a,b} Raw coincidence counts for detector pairs CC$_{15}$ (dark blue) and CC$_{01}$ (green) show that signal for both Identity in \textbf{a} and Beamsplitter \textbf{b} is essentially overlapping with the noise floor (red and grey, respectively). This is further underlined by corrected coincidences in \textbf{c,d} where \textbf{c} corresponds to CC$_{15}$ solely, while CC$_{01}$ is depicted in \textbf{d}. Here, Identity and Beamsplitter cannot be discriminated, essentially rendering this operation day unreliable for further data analysis. The count rate issue could be resolved in the subsequent operation day 4 in Sec.~\ref{sec:operation_day_4} by adjusting the SPAD overvoltage.}
    \label{fig:op_day_3_counts}
\end{figure}

\begin{figure}[!tbp]
    \centering
    \includegraphics[width=1\textwidth,height=0.73\textheight,keepaspectratio]{ 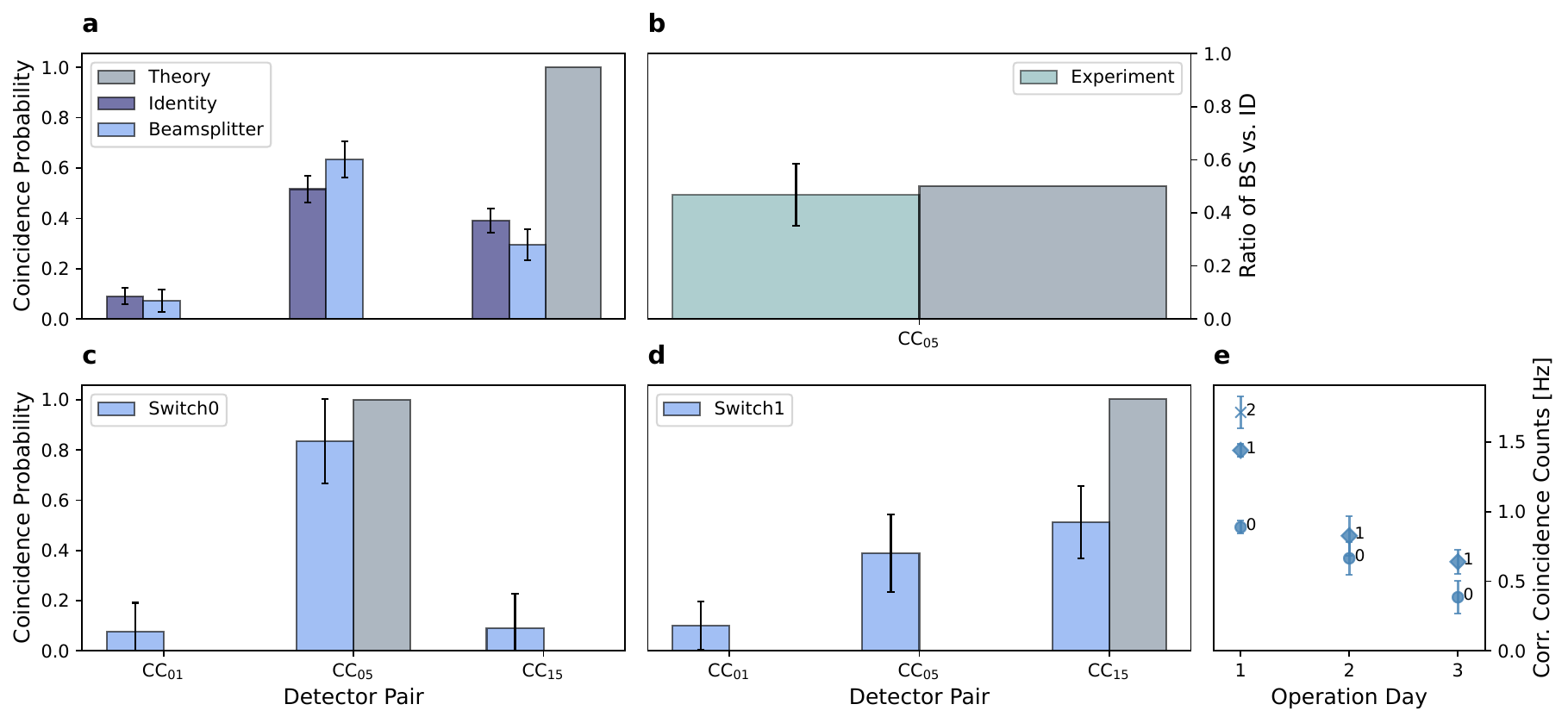}
    \caption[]{\textbf{Op. Day 3 - On-board programmable unitary performance and ratio.} \textbf{a} (Weighted) average coincidence probability from shadow measurements for Identity (dark blue) and Beamsplitter (light blue) shows that the count rate was too low to resolve any unitary. This is reflected in the strong deviation from theory (grey); ratio in \textbf{b} shows value around expected behaviour of $R_T=0.5$, and further underlines that ratio is in general more robust. However, the error is too high to consider it a valid measurement. \textbf{c,d} uPIC health test also shows that performance is negatively influenced by further count rate reduction in \textbf{e} of operation day $3$. Here, the gradual decrease indicates that a $\ensuremath{7\,\mathrm{V}}$ SPAD overvoltage is insufficient to conduct a meaningful experiment.}
    \label{fig:op_day_3_probabs}
\end{figure}

This operation day consisted of further HOM scans and health tests, similar to Sec.~\ref{sec:operation_day_2}, but with target detector pairs CC$_{15}$, CC$_{01}$. This was motivated by the experimental outcome of operation day $2$ in Sec.~\ref{sec:operation_day_2} that showed strongly reduced counts on detector pair CC$_{05}$. Despite this change, the count rate situation worsened compared with the former operation day, as shown in Fig.~\ref{fig:op_day_3_probabs}e. This problem can again be attributed to outgassing, which induced a further decrease in laser power (see Fig.~\ref{fig:laser_power_vac}) and relatively strongly increased dark counts (SPAD damage) after 52 days in orbit (see Fig.~3 of the main text or Fig.~\ref{fig:dark_counts_and_complete_rate}a). The problem could only be resolved in the subsequent operation day in Sec.~\ref{sec:operation_day_4} as here the SPAD overvoltage was increased from $\ensuremath{7\,\mathrm{V}}$ to $\ensuremath{12\,\mathrm{V}}$, effectively increasing the SPAD detection efficiency as shown in measurements on the EM in Fig.~\ref{fig:spads}. In the present operation day, the overvoltage was still set to $\ensuremath{7\,\mathrm{V}}$ and raw/corrected countrates of both detector pairs CC$_{15}$, CC$_{01}$ in Fig.~\ref{fig:op_day_3_counts} show that the observed counts are nearly overlapping with the noise floor. Moreover, the two configurations, Identity and Beamsplitter, cannot be distinguished; hence, this operation day also does not provide reliable insight into the experimental outcome and must be excluded from the data analysis.
\clearpage
\subsection{Operation Day 4 - 59 Days in Orbit}
\label{sec:operation_day_4}
\begin{figure}[!tbp]
    \centering
    \includegraphics[width=1\textwidth,height=0.73\textheight,keepaspectratio]{ 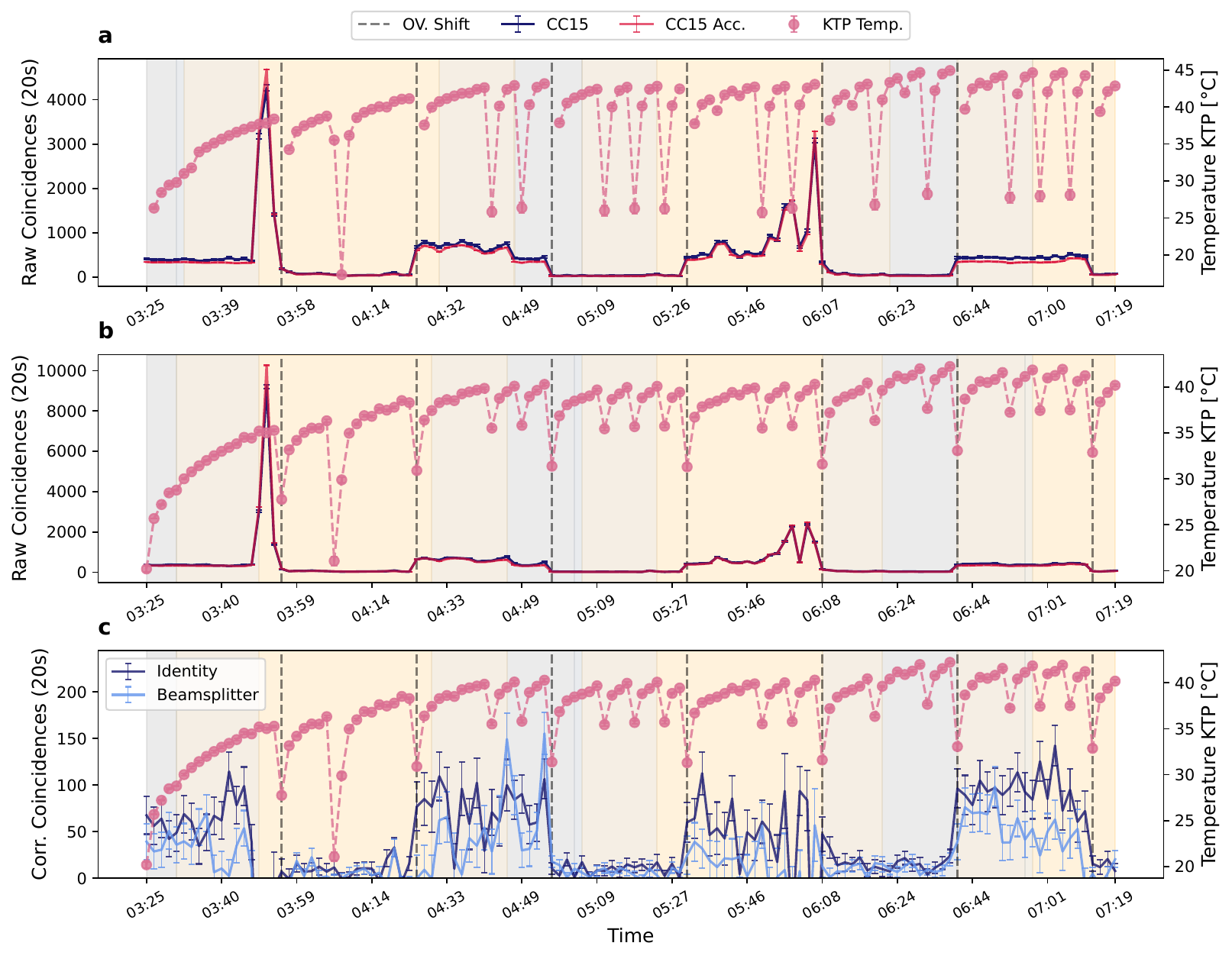}
    \caption[]{\textbf{Op. Day 4 - Coincidences overview.} \textbf{a} Raw Identity coincidence counts (dark blue) for detector pair CC$_{15}$ show severe spikes in countrate during sun measurements. SPAD overvoltage configuration cycles between $\ensuremath{7\,\mathrm{V}}$ and $\ensuremath{12\,\mathrm{V}}$ (vertical dashed lines), and a big difference in baseline countrate is visible. \textbf{b} Similar to \textbf{a}, but here Beamsplitter is measured and shows more drastic spikes in countrate; KTP temperature stability (red points, secondary $y$-axis) is significantly worsened with respect to operation day $1$ in Fig.~\ref{fig:op_day_1_counts} as measured temperature for Identity and Beamsplitter is not identical. \textbf{c} Corrected coincidence counts for Identity and Beamsplitter highlight the necessity to set the SPAD overvoltage to $\ensuremath{12\,\mathrm{V}}$ to regain some signal. KTP temperature is shown for the Beamsplitter configuration.}
    \label{fig:op_day_4_counts}
\end{figure}

\begin{figure}[!tbp]
    \centering
    \includegraphics[width=1\textwidth,height=0.82\textheight,keepaspectratio]{ 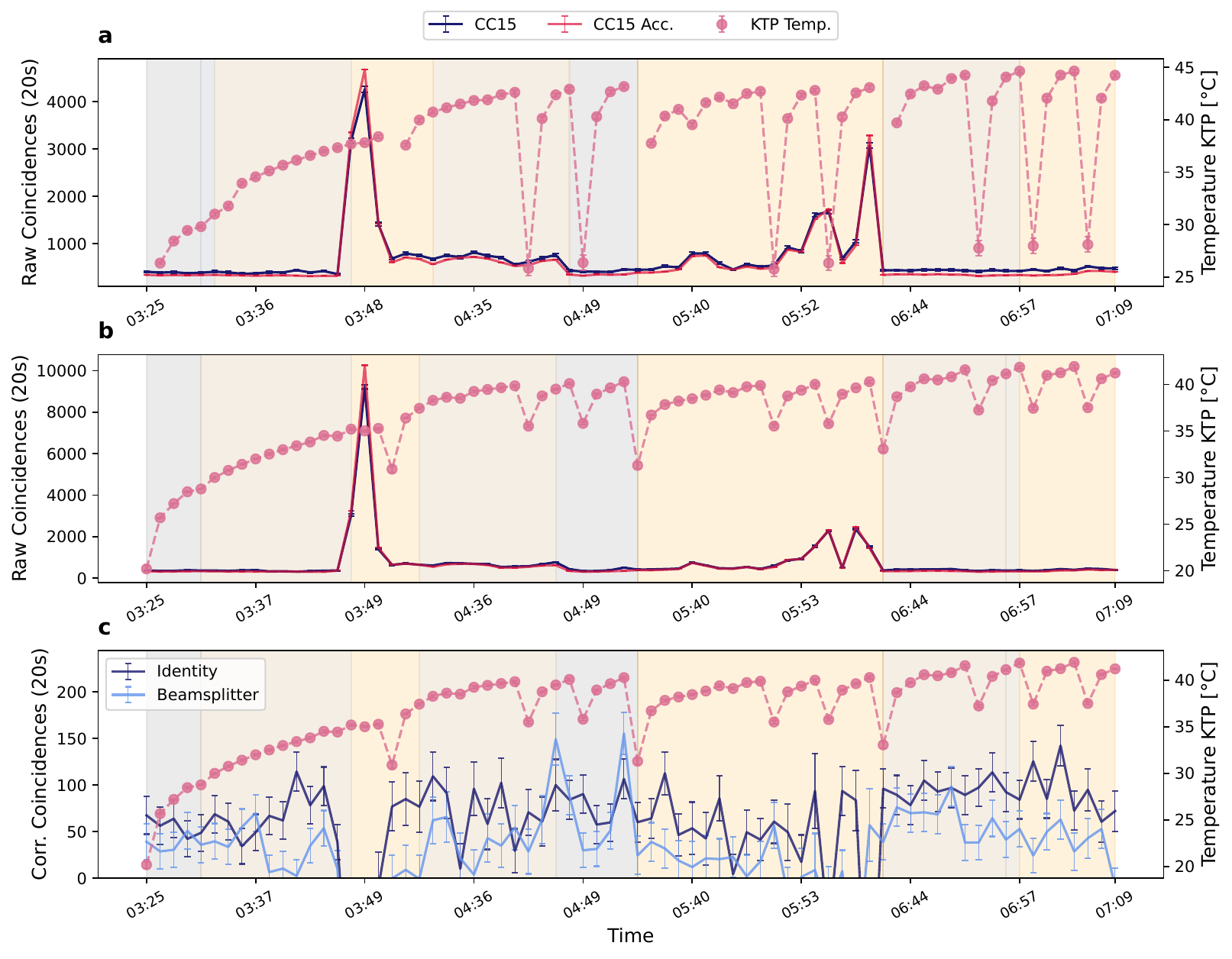}
    \caption[]{\textbf{Op. Day 4 - Coincidence Overview with $\mathbf{\ensuremath{12\,\mathrm{V}}}$ SPAD overvoltage.} \textbf{a,b,c} Identical to Fig.~\ref{fig:op_day_4_counts} but here only the case of $\ensuremath{12\,\mathrm{V}}$ SPAD overvoltage is shown. A clear signal above the noise floor is revealed in \textbf{c}, but due to strong spikes in \textbf{a,b}, the complete measurement has to be further divided into chunks of shadow- and sun measurements in Fig.~\ref{fig:op_day_4_counts_close_up} to reveal more information regarding the experimental outcome.}
    \label{fig:op_day_4_counts_1200}
\end{figure}

The task of this operation day was again to scan for Hong-Ou Mandel interference while also tuning the SPAD overvoltage from $\ensuremath{7\,\mathrm{V}}$ to $\ensuremath{12\,\mathrm{V}}$ to regain some of the lost signal by effectively increasing the SPAD detection efficiency (see measurements on the EM in Fig.~\ref{fig:spads}). The drastic change in overvoltage was essential, as the two previous operation days in Sec.~\ref{sec:operation_day_2} and Sec.~\ref{sec:operation_day_3} showed that the original overvoltage was no longer sufficient to conduct a meaningful experiment. At this project stage, the TEC temperature calibration was also not understood properly (see Sec.~\ref{sec:operation_day_1}), therefore the degeneracy temperature (HOM dip temperature) was thought to be around $T_{TEC}\sim \ensuremath{40\,{}^{\circ}\mathrm{C}}$. This also explains why the measured KTP temperature in Fig.~\ref{fig:op_day_4_counts} is mostly around $\ensuremath{40\,{}^{\circ}\mathrm{C}}$. For this operation day, the target detector pair is chosen to be CC$_{15}$ (dark blue), and a first observation of the measurement outcome in Fig.~\ref{fig:op_day_4_counts}a-b shows drastic spikes in detector signal for Identity and Beamsplitter unitary over the whole operation day. Nevertheless, the corrected signal\footnote{Extracted via Eq. \ref{eq:signal_extraction}.}in Fig.~\ref{fig:op_day_4_counts}c reveals that increasing the overvoltage is essential, as four regions with significantly higher coincidence counts can be identified here. In the case where the signal is compatible with zero, the overvoltage was still set to the original $\ensuremath{7\,\mathrm{V}}$. As the latter is excluded from data analysis, the subsequent Fig.~\ref{fig:op_day_4_counts_1200} only focuses on the coincidence counts with $\ensuremath{12\,\mathrm{V}}$ SPAD overvoltage.

\begin{figure}[!tbp]
    \centering
    \includegraphics[width=1\textwidth,height=0.69\textheight,keepaspectratio]{ 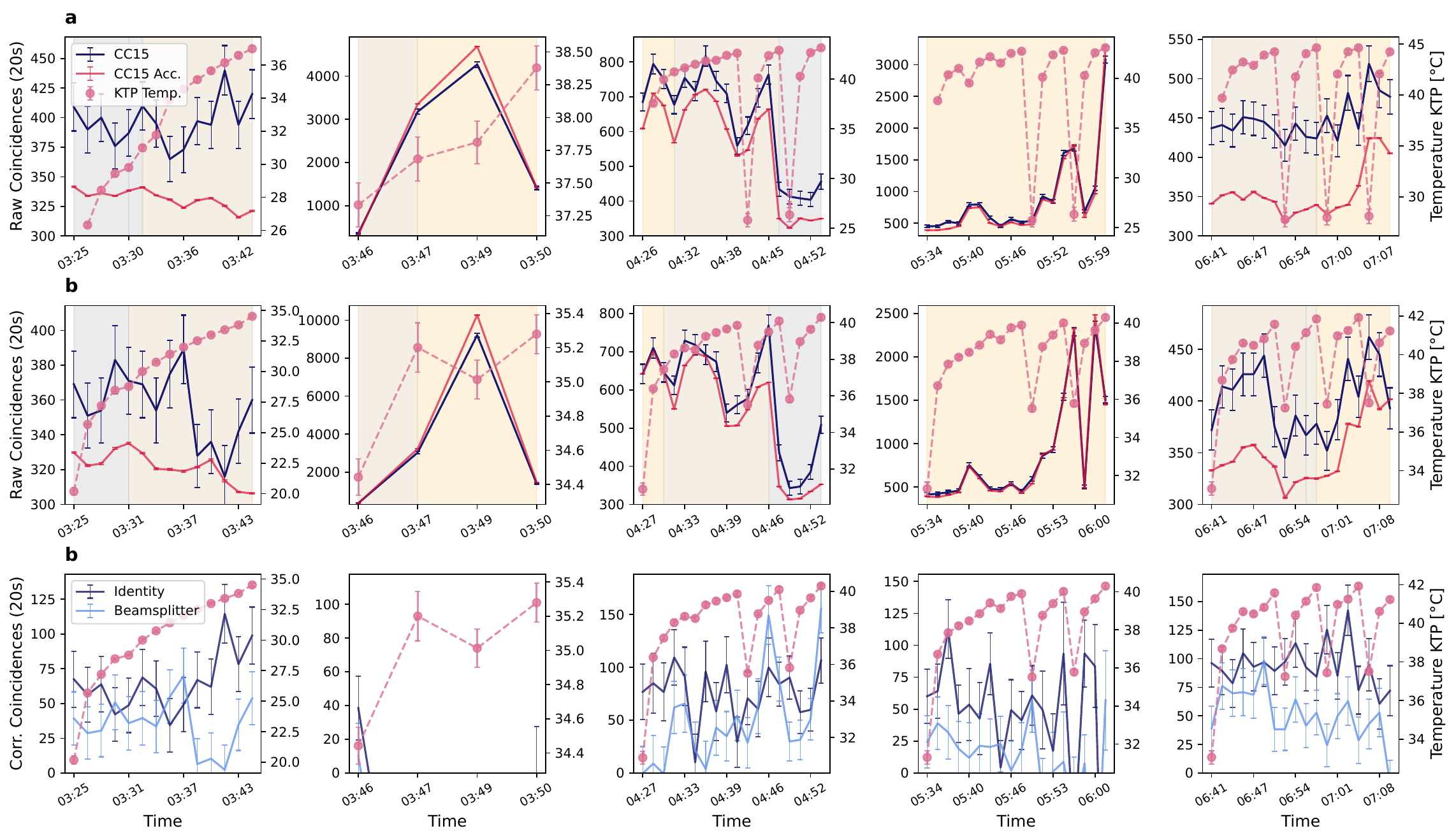}
    \caption[]{\textbf{Op. Day 4 - Grouped coincidence overview with $\mathbf{\ensuremath{12\,\mathrm{V}}}$ SPAD overvoltage.} \textbf{a (row)} Complete measurement of Fig.~\ref{fig:op_day_4_counts_1200}a is dissected in five individual regimes in order to identify shadow- and sun regimes in a more visible manner. Here, target detector pair CC$_{15}$ (dark blue) for Identity shows clear signal on top of noise floor (red) for left, partially center and right measurement; severe SPAD saturation can be seen in second and fourth section and highlights again that the host satellite's position has strong implications on the experimental outcome; yellow area marks when payload faces the sun while grey area means payload is in the shadow of the Earth. \textbf{b} Similar as \textbf{a} but here Beamsplitter unitary is applied. \textbf{c} Corrected signal shows that noise completely washed out the signal in, e.g., the second column for Identity- (dark blue) and Beamsplitter unitary (light blue); Count rate is more stable in shadow regimes and distinction between Identity and Beamsplitter is clearer in the first and last columns.}
    \label{fig:op_day_4_counts_close_up}
\end{figure}

\begin{figure}[!tbp]
    \centering
    \includegraphics[width=1\textwidth,height=0.65\textheight,keepaspectratio]{ 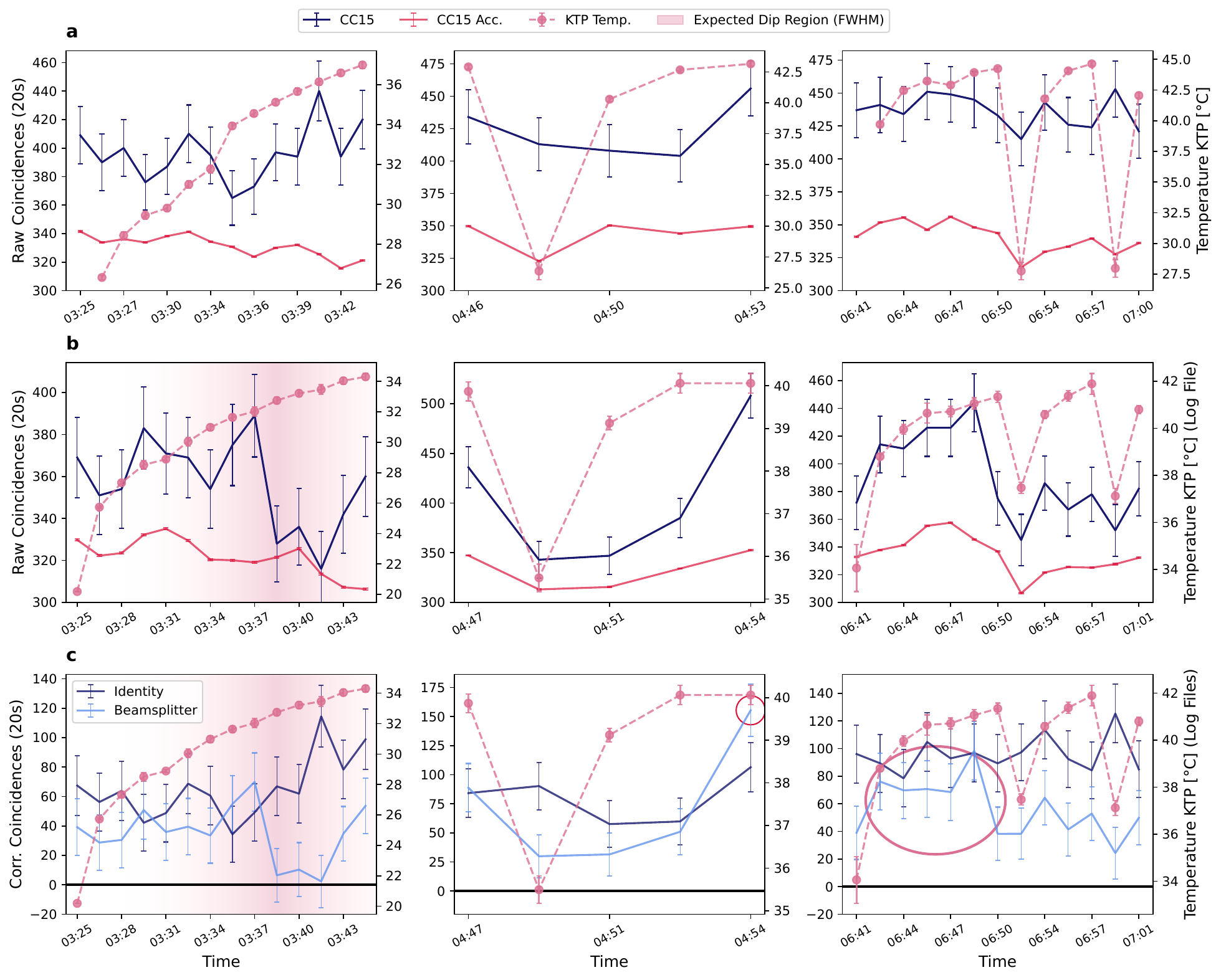}
    \caption[]{\textbf{Op. Day 4 - Coincidences in shade.} \textbf{a (row)} Raw Identity coincidence counts (dark blue) show a clear signal above the noise floor (red) for the target detector pair CC$_{15}$ in three shadow regimes. Overall count rate stability is worse than on operation day $1$ in Sec.~\ref{sec:operation_day_1}, as shown by the error bars. \textbf{b (row)} Similar setting as in \textbf{a}, but here Beamsplitter is measured, and KTP temperature (red points - secondary $y$-axis) is extracted from LOG file; significant drop in coincidences in regime of expected HOM dip (red-shaded area) can be seen, while also center and right figure show stronger fluctuations in countrate. Corrected signal for Identity (dark blue) and Beamsplitter (light blue) in \textbf{c (row)} shows that Beamsplitter is effectively zero within expected HOM dip KTP temperature and this is not the case for Identity; Beamsplitter rate regularly overlaps with Identity rate and this especially happens for five consecutive datapoints (red ellipse) in right figure; overlap between both unitaries can be explained by internal shifter jitter in Fig.~\ref{fig:op_day_1_mzi_jitter}. Strong outlier of Beamsplitter (red circle) can be identified in the center figure, as the data point significantly overshoots Identity.}
    \label{fig:op_day_4_counts_shadow}
\end{figure}

The severe spikes in the measured signal make the interpretation of Fig.~\ref{fig:op_day_4_counts_1200}a-b difficult, and it is therefore useful to further dissect the $\ensuremath{12\,\mathrm{V}}$ SPAD overvoltage count overview in Fig.~\ref{fig:op_day_4_counts_1200} into smaller chunks that divide the measurements into shadow- and sun measurements, respectively. This eventually helps in disentangling the noise contributions from the real signal, and Fig.~\ref{fig:op_day_4_counts_close_up} divides the complete $\ensuremath{12\,\mathrm{V}}$ measurement interval into five shadow- or sun regimes. Here, a general decay in measurement stability with respect to the first operation day in Sec.~\ref{sec:operation_day_1} is visible when observing the fluctuations of the noise floor (red) in the shadow regimes of Fig.~\ref{fig:op_day_4_counts_close_up}a (row, left, center, right). This is also directly reflected in the higher error bars of the signal for the target detector pair CC$_{15}$. This behaviour can be attributed to the increased dark counts (damage in SPADs) and relatively low integration time of $\ensuremath{20\,\mathrm{s}}$ per data point\footnote{This is increased to $\ensuremath{60\,\mathrm{s}}$ for upcoming operation days.}.  Nevertheless, in the shadow measurements, a clear signal above the noise floor is evident. The severity of the noise contributions in the sun is highlighted in the second, third, and fourth columns of Fig.~\ref{fig:op_day_4_counts_close_up} as here the countrate either fluctuates excessively or the signal is completely washed out by noise. Similar to the first operation day, we have to identify the shadow regimes by comparing the TLE data (colored background) with the stability and baseline count rate of the accidentals. With these boundary conditions in mind, we can restrict valid measurements to three shadow regimes in Fig.~\ref{fig:op_day_4_counts_shadow}, and here we are again within the KTP temperature regime (red points - secondary $y$-axis) of expected Hong-Ou Mandel interference (red-shaded area).

Similar to operation day $1$ in Sec.~\ref{sec:operation_day_1} we see a drop in coincidences around the expected HOM dip temperature regime of $T_{TEC}\sim \ensuremath{32.5\,{}^{\circ}\mathrm{C}}$ (red-shaded area) and as now the overall noise contributions are higher than on the first operation day in, e.g. Fig.~\ref{fig:op_day_1_umbra_first_look}, we have to carefully understand if this dip in coincidences is meaningful. Moreover, as the TEC stability is worsened on this operation day (compare the difference between Identity and Beamsplitter KTP temp.), the Beamsplitter results in Fig.~\ref{fig:op_day_4_counts_shadow}b-c contain KTP temperature information from an additional LOG file, instead of the otherwise used mean KTP temperature data. This LOG file records the current status of an experiment, including ADC readings, and also tracks the KTP temperature with a timestamp. This more frequent logging procedure for temperature data was implemented starting on operation day $2$ and records the temperature every $\sim \ensuremath{15\,\mathrm{s}}$. This is different from the usual case of extracting the KTP temperature, as in, e.g., Fig.~\ref{fig:op_day_4_counts}, where the mean KTP temperature is automatically evaluated within the experimental script. Using the LOG file's timing information and continuously updated temperature readings, a more precise KTP temperature can be extracted. This can be identified in the secondary $y$-axis of Fig.~\ref{fig:op_day_4_counts_shadow} as additional label ``Log File''.

\begin{figure}[!tbp]
    \centering
    \includegraphics[width=1\textwidth,height=0.63\textheight,keepaspectratio]{ 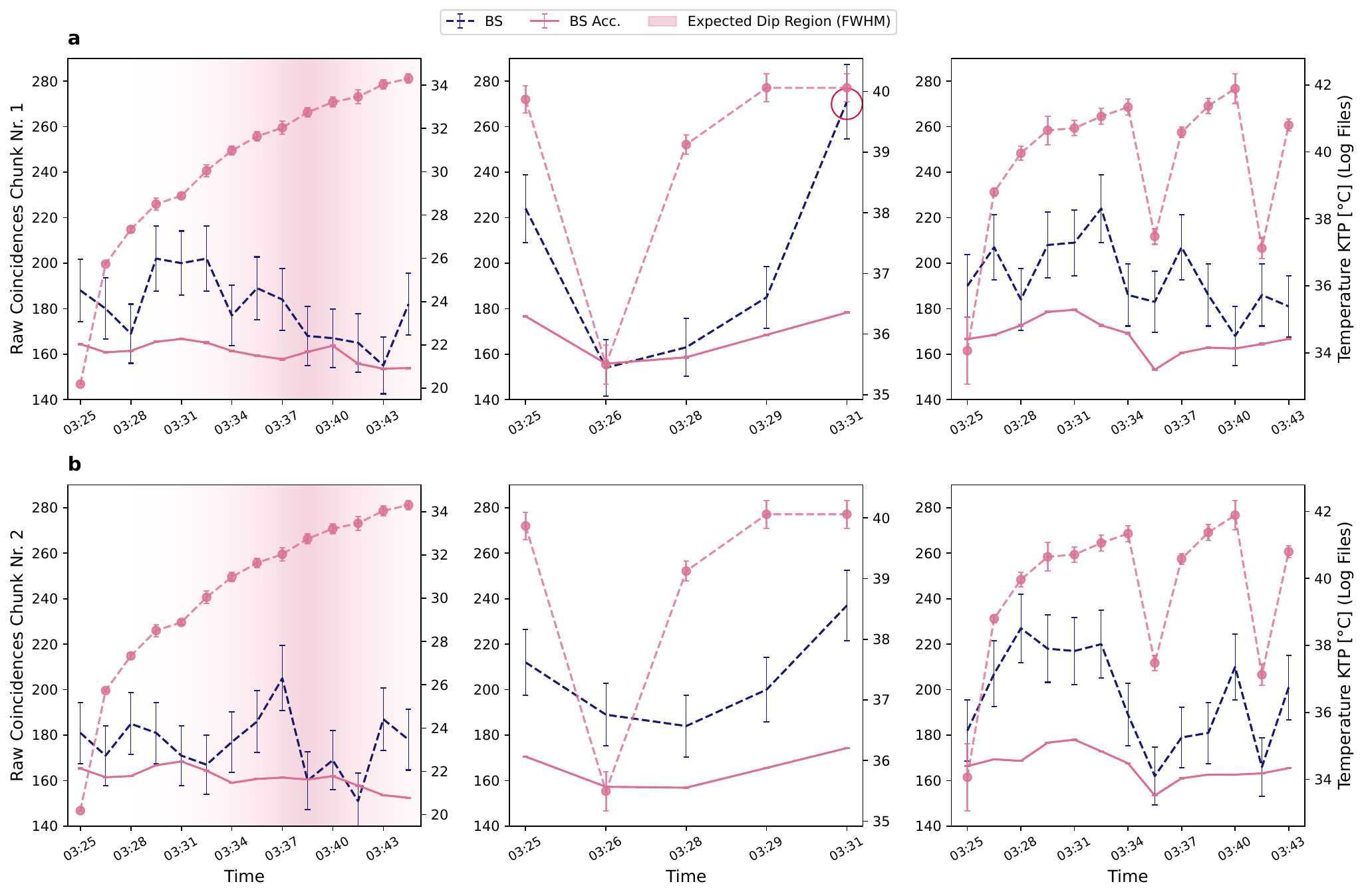}
    \caption[]{\textbf{Op. Day 4 - Beamsplitter outlier analysis.} \textbf{a,b (row)} Chunks of $\ensuremath{10\,\mathrm{s}}$ individual measurements reveal no significant outlier that might mimic a HOM dip, as seen in Fig.~\ref{fig:op_day_9_outlier_anomaly}, where a timetagger anomaly resulted in strong deviations between signal and noise floor; significant outlier can be identified for center figure in \textbf{a} where last datapoint (red circle) severely overshoots the other data points and must therefore be removed. In contrast to the outlier just before the dip on operation day $1$ in Sec.~\ref{sec:operation_day_1}, the spike in \textbf{b} (left) is not removed from data analysis. This is because the spike lies within the expected HOM-dip regime, and as fluctuations in count rate are also higher on this operation day, the spike is left untouched in this shadow measurement to stay more faithful to the data. High fluctuations in the count rate make general interpretation of individual $\ensuremath{10\,\mathrm{s}}$ chunks difficult; thus, individual chunks have to be summed to improve statistics.}
    \label{fig:op_day_4_outlier_analysis}
\end{figure}

\begin{figure}[!tbp]
    \centering
    \includegraphics[width=1\textwidth,height=0.65\textheight,keepaspectratio]{ 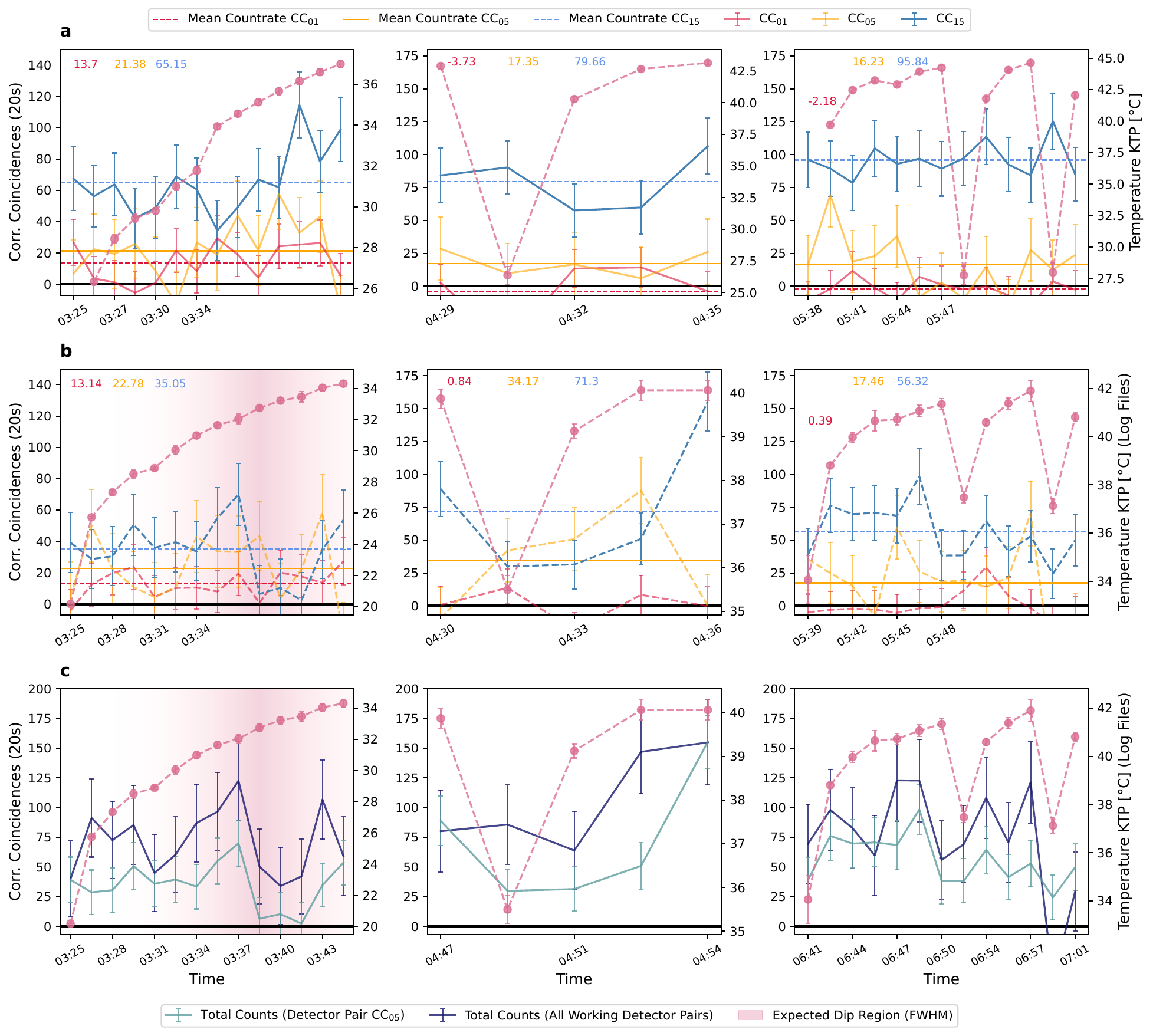}
    \caption[]{\textbf{Op. Day 4 - Total corrected coincidences.} \textbf{a (row)} Corrected Identity count rate for all three shadow measurements shows all working detector pairs CC$_{01}$ (red), CC$_{05}$ (yellow), and CC$_{15}$ (blue). In strong contrast to total count analysis of operation day $1$ in Fig.~\ref{fig:op_day_1_total_counts_anomaly_check}, non-target modes CC$_{01}$ (red) and CC$_{05}$ (yellow) show relative high countrate with respect to target pair CC$_{15}$ (blue); situation improves over span of operation day as beginning of operation (left) has more counts on unwanted modes than final measurement (right). \textbf{b} Beamsplitter unitary shows a similar fate as in \textbf{a} and suggests that the photon loss in unwanted detector modes is independent of the unitary. This is reflected in the nearly identical average countrate of CC$_{01}$ (red label) and CC$_{05}$ (yellow label) in \textbf{a,b}. Within the expected interference regime (red shaded area), counts from unwanted modes are not maximal, suggesting that the dip in coincidences of CC$_{15}$ cannot be attributed to erroneously routed photons. Outlier (red circle) in Fig.~\ref{fig:op_day_4_outlier_analysis} is also reflected in corrected Beamsplitter data in \textbf{b} (center) and needs to be removed for final analysis.\textbf{c (row)} Sum of all detector pairs (dark blue) is relatively strongly shifted due to constant photon loss in other modes.}
\label{fig:op_day_4_total_counts_anomaly_check}
\end{figure}

The analysis of the corrected signal in Fig.~\ref{fig:op_day_4_counts_shadow}c (row) reveals some similarities to the first operation day. First, the Beamsplitter regularly overlaps with the countrate of Identity, and this is especially true within the red ellipse of the last shadow measurement, where both countrates essentially overlap within their errors for five consecutive data points. This behaviour can again be attributed to the jitter of the internal phase shifter of MZI6. Second, and visible in the subsequent Fig.~\ref{fig:op_day_4_total_counts_anomaly_check}a-b, the mean countrate (dashed blue) gradually increases over the complete operation day due to thermalization of the payload. The present dip in coincidences in Fig.~\ref{fig:op_day_4_counts_shadow}c (left) is significant as three datapoints around the expected HOM dip KTP temperature (red-shaded area) are essentially zero, and this is clearly not the case for Identity within a similar KTP temperature regime. It is our task now to further test this dip in coincidences for potential anomalies. Identical to the procedure in Sec.~\ref{sec:operation_day_1}, the Beamsplitter data is divided into its two measurement chunks of $\ensuremath{10\,\mathrm{s}}$ of integration time, and Fig.~\ref{fig:op_day_4_outlier_analysis} shows that there is no significant outlier that could mimic a dip in coincidences within the three shadow regimes. In other words, the signal (dashed dark blue) is either above or compatible with the accidentals (solid red) and not significantly below (more than $2\sigma$) as in the case of the timetagger anomaly in Fig.~\ref{fig:op_day_9_outlier_anomaly}.

As a next step, and similar to the procedure of the first operation day in Sec. \ref{sec:operation_day_1}, we evaluate all detector pairs and the total counts in Fig.~\ref{fig:op_day_4_total_counts_anomaly_check}. Here, a large difference in programmable unitary performance relative to the first operation day is evident. The first shadow measurement of Fig.~\ref{fig:op_day_4_total_counts_anomaly_check}a-b shows that for Identity and Beamsplitter, a significant amount of photons also end up at the unwanted detector pairs CC$_{01}$ and CC$_{05}$. In both cases, the average rate of the former is $ \sim 13$ coincidences (red) and $\sim 21$ coincidences (yellow) for the latter. The situation improves over the course of the operation day: in the final shadow measurement, no photons reach CC$_{01}$, while approx. $16$ coincidences end up at CC$_{05}$. This can again be explained by a slow thermalization of the payload. If the payload is too cold in the beginning (compare the $\sim \ensuremath{7\,{}^{\circ}\mathrm{C}}$ in Fig.~\ref{fig:temperature-control} for a typical operation day), then also the uPIC is farther off its optimal operating temperature of $\sim \ensuremath{25\,{}^{\circ}\mathrm{C}}$. As the uPIC is calibrated at this specific temperature, a high operating-temperature offset might explain impaired unitary fidelity at the beginning of an operating day. A similar behaviour is also reflected in the thermalization performance of the TEC. In all the measurements of Fig.~\ref{fig:op_day_4_counts}, the temperature setpoint was around $\sim \ensuremath{40\,{}^{\circ}\mathrm{C}}$; however, due to the lower payload temperature in the beginning, the TEC only managed to reach $\sim \ensuremath{35\,{}^{\circ}\mathrm{C}}$ in the first measurements. Both observations suggest that the payload was significantly colder on this operation day when compared to the first operation day in Sec.~\ref{sec:operation_day_1}.

As the spurious counts on detector pairs CC$_{01}$, CC$_{05}$ are the same for both applied unitaries (see mean countrate as horizontal dashed lines in Fig.~\ref{fig:op_day_4_total_counts_anomaly_check}), they can be considered as a unitary independent photon loss in non-target modes. Moreover, in the regime of potential interference (red-shaded area) in Fig.~\ref{fig:op_day_4_total_counts_anomaly_check}b (left), the spurious detector counts on the other modes (dashed yellow/red) are not maximal, hence we can further exclude an artifact of wrongly routed photons into the other detector modes for this specific KTP temperature. Due to the photon routing problem of the uPIC, the sum of all detector pairs (dark blue) in Fig.~\ref{fig:op_day_4_total_counts_anomaly_check}c (row) is higher than the target detector pair CC$_{15}$, and this follows naturally from the spurious counts on the other modes, which form a constant baseline. This might seem problematic at first; however, measurements on subsequent operation days suggest that improperly routed photons in the Beamsplitter or Identity configuration do not affect the underlying interference capability. This can be seen consistently for the target detector pair CC$_{01}$ in Sec.~\ref{sec:operation_day_8} or Sec.~\ref{sec:operation_day_9}. Here, the overall unitary fidelity is very low, i.e., photons are lost to the unwanted detector pairs CC$_{05}$ and CC$_{15}$, but the ratio between Beamsplitter and Identity of target pair CC$_{01}$ is still $\sim 0.5$, consistent with MZI6 being set to Beamsplitter.

\begin{figure}[!tbp]
    \centering
    \includegraphics[width=1\textwidth,height=0.52\textheight,keepaspectratio]{ 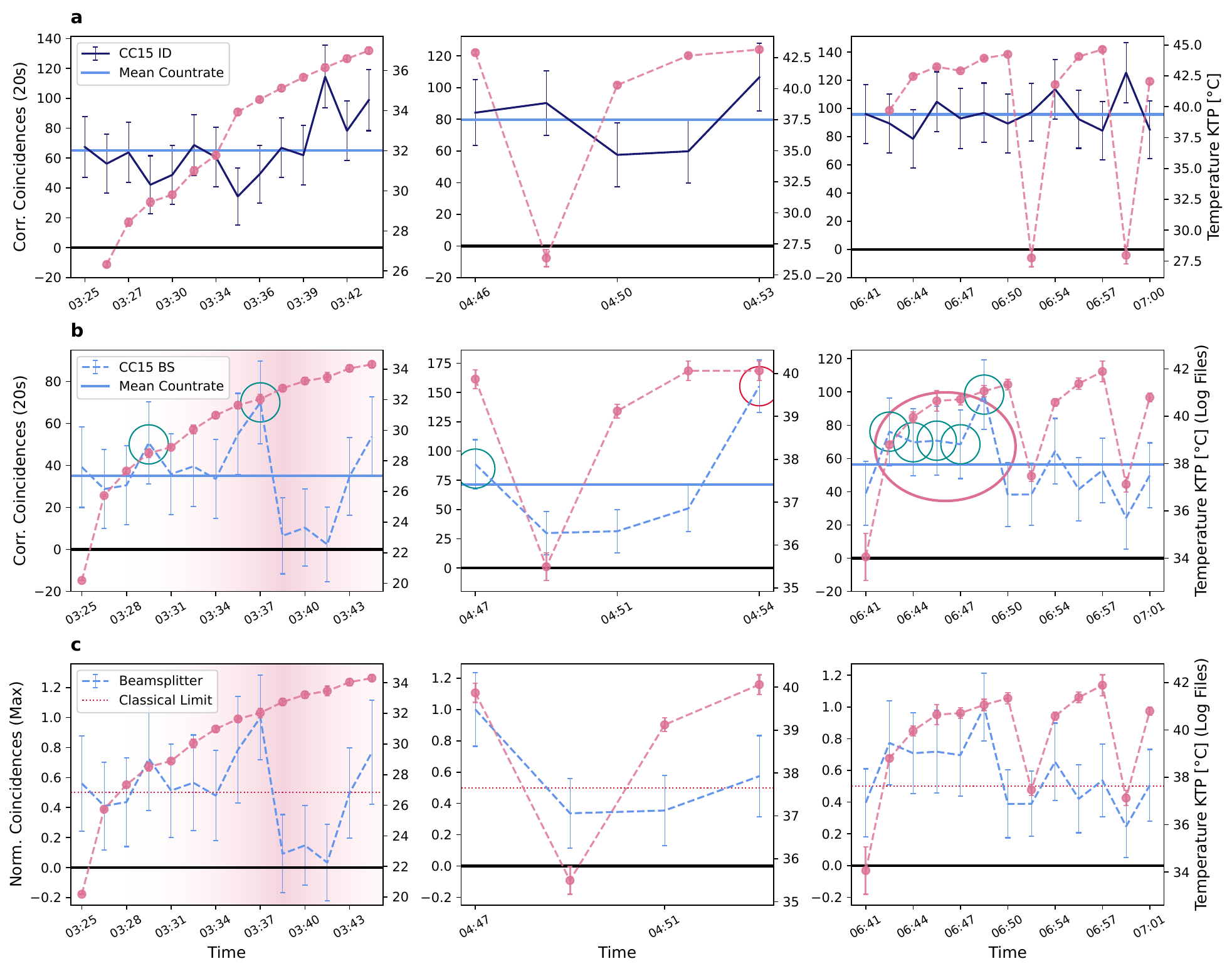}
    \caption[]{\textbf{Op. Day 4 - Corrected coincidences and HOM interference.} \textbf{a (row)} Corrected Identity coincidence counts for target detector pair CC$_{15}$ show a gradual increase in mean countrate (vertical blue line) and relatively stable rate over time for all shadow measurements; identity rate within expected HOM dip temperature regime (red-shaded area) does not show a dip in coincidences. \textbf{b (row)} Beamsplitter unitary shows a significant dip in the expected KTP temperature regime, where three datapoints are essentially zero. Overall count-rate stability is lower with respect to \textbf{a}, and this can be attributed to jitter in the internal shifter of MZI6; data points within green circles indicate where the Beamsplitter rate overlaps with the Identity rate (compare Fig.~\ref{fig:op_day_1_mzi_jitter}). This has a strong effect on the normalization with respect to the maximum in \textbf{c (row)}, as the last and middle shadow measurements also show data points below the classical limit (vertical dotted red line). Comparing the Beamsplitter rate and corresponding KTP temperature (red points - secondary $y$-axis) from all three shadow measurements shows that these points (center and right figure) cannot be attributed to two-photon interference as they are not minimal as within red-shaded area and happen at different KTP temperatures; behaviour is highlighted in last shadow measurement where local minima in coincidences occurs for $T_{TEC}=\{34,41,37\} \ \ensuremath{{}^{\circ}\mathrm{C}}$, respectively; mean normalization in Fig.~\ref{fig:op_day_4_bs_normalized_mean} further underlines that these datapoints correspond to fluctuations in Beamsplitter count rate. The outlier (red circle) in the second shadow measurement is removed in \textbf{c} because it strongly overshoots Identity; generally, the center-column measurement has to be validated with care, as it is only a partial shadow measurement and fluctuations are generally higher here (see Fig.~\ref{fig:op_day_4_counts_close_up} (center).}
    \label{fig:op_day_4_counts_close_up_3}
\end{figure}

The robustness of the Beamsplitter unitary in terms of photon loss to unwanted detector modes, together with the results of the artifact test in Fig.~\ref{fig:op_day_4_outlier_analysis} and Fig.~\ref{fig:op_day_4_total_counts_anomaly_check}, allows us to consider the data within the HOM dip region (red-shaded area) as meaningful. This is also shown in the final count rate analysis in Fig.~\ref{fig:op_day_4_counts_close_up_3}, where the individual corrected Identity and Beamsplitter rates are compared for all shadow measurements in the first two rows, and the normalized count rate is shown in the last row.

Within the expected HOM dip regime (red-shaded area) of Fig.~\ref{fig:op_day_4_counts_close_up_3}b (left), a clear dip in coincidences is only measured for Beamsplitter (dashed blue) but not for the corresponding Identity measurement. Here, the gradual increase in the count rate (vertical blue line) again underscores the need to treat each shadow measurement as an individual dataset for normalization. In the last row of Fig.~\ref{fig:op_day_4_counts_close_up_3}, the Beamsplitter counts of each shadow measurement are normalized with respect to their maximum as described in Eq. \ref{eq:maximum_normalization}. This time, the individual signal rates for a specific KTP temperature are extracted by Eq. \ref{eq:signal_extraction} (fixed coincidence window $\tau$) and plugged into the normalization routine in Eq. \ref{eq:maximum_normalization}. As already described in Sec.~\ref{sec:operation_day_1}, normalizing data to the maximum can be problematic due to jitter in the internal phase shifter of MZI6. Here, the normalized coincidence counts within the expected HOM dip temperature drop below the classical threshold of $p=0.5$ (vertical red dotted line); however, this is also true for the other two figures in which the significant outlier (red circle) in the second shadow regime of Fig.~\ref{fig:op_day_4_counts_close_up_3}b (center) is removed. The other two shadow areas also include data points that are not measured at the expected KTP temperature for two-photon interference but still fall slightly below the classical threshold. The latter can be explained by the measurements within the green circle in Fig.~\ref{fig:op_day_4_counts_close_up_3}b (row), which show a rate nearly identical to that of the Identity unitary. This phenomenon might mimic quantum interference below the classical threshold as the Beamsplitter rate is effectively normalized by the Identity rate.

\begin{figure}[!tbp]
    \centering
    \includegraphics[width=1\textwidth,height=0.73\textheight,keepaspectratio]{ 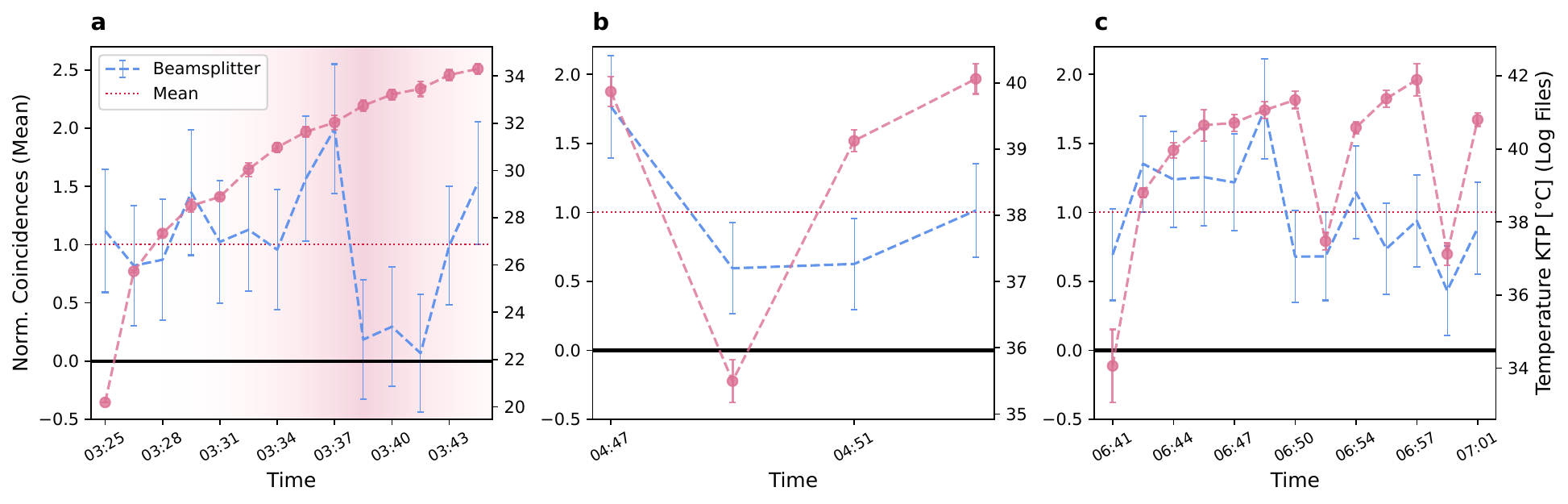}
    \caption[]{\textbf{Op. Day 4 - HOM interference - Mean normalization.} \textbf{a,b,c} Identical measurement configuration as in Fig.~\ref{fig:op_day_4_counts_close_up_3}c, but here the Beamsplitter counts are normalized with respect to the mean, and this procedure emphasizes that fluctuations in \textbf{b} and \textbf{c} can not be linked to two-photon interference as they essentially fluctuate around the mean. This is not the case in \textbf{a}, as the dip in coincidences within the expected HOM-dip KTP regime effectively overlaps with zero. The relatively large error bars come from the Gaussian error propagation routine in Eq. \ref{eq:mean_normalization}, and here the formula has a harsher effect on error bars compared with the maximum routine in Eq. \ref{eq:maximum_normalization}. Nevertheless, the result in \textbf{a} is critical as it shows the expected behaviour for two-photon interference.}
    \label{fig:op_day_4_bs_normalized_mean}
\end{figure}

This potential artifact is resolved by comparing the corrected Beamsplitter rate across all three shadow measurements and, as also done in the procedure on operation day $1$, by normalizing the raw data to their mean. As a start, the first shadow measurement in Fig.~\ref{fig:op_day_4_counts_close_up_3}b shows that the coincidence counts effectively drop to zero for three consecutive data points within the HOM dip KTP temperature regime. In the other two cases, the minimum countrate is around $\sim 25$ coincidences and moreover, the local minima of data points in, e.g. Fig.~\ref{fig:op_day_4_counts_close_up_3}b (right), happen for three completely different KTP temperatures $T_{TEC} \sim \{34,41,37\} \ \ensuremath{{}^{\circ}\mathrm{C}}$ in the same shadow measurement, hence they cannot be attributed to the phenomena of two-photon interference. In addition, the raw data points are normalized with respect to their mean in Fig.~\ref{fig:op_day_4_bs_normalized_mean}, and by comparing the individual shadow measurements, it becomes clear that the data of the second- and third shadow measurements is just fluctuating around the mean. This is not the case for the dip in coincidences within the expected HOM-dip temperature regime in Fig.~\ref{fig:op_day_4_bs_normalized_mean}a, as here the coincidence counts effectively drop to zero.

This dip in coincidences at the expected temperature of $T_{TEC}\sim \ensuremath{32.5\,{}^{\circ}\mathrm{C}}$ is a critical result as it coincides with the HOM dip measurement of operation day $1$ in Fig.~\ref{fig:op_day_1_bs_normalized} and also adds more datapoints as it occurs for three consecutive temperature points around the expected maximal HOM dip KTP temperature. In addition, this result could be obtained after $\sim 2$ months of the payload's exposure to the harsh environment of space, by which time its critical components have already sustained significant damage. The latter is reflected in the effectively measured damage in SPADs by the strongly increased dark counts in Fig.~\ref{fig:dark_counts_and_complete_rate} and the second critical damage can be attributed to the laser and its power loss in Sec.~\ref{sec:tvac} including operation day $2,3$ in Sec.~\ref{sec:operation_day_2} and Sec.~\ref{sec:operation_day_3}, respectively. The mean normalization used to evaluate the normalized counts is also used later in Sec.~\ref{sec:main_figures_hom}, where the individual operation days are combined to obtain a complete HOM dip scan as a function of temperature. This helps to further strengthen the result and minimize the effect of potential artefacts.

\begin{figure}[!tbp]
    \centering
    \includegraphics[width=1\textwidth,height=0.67\textheight,keepaspectratio]{ 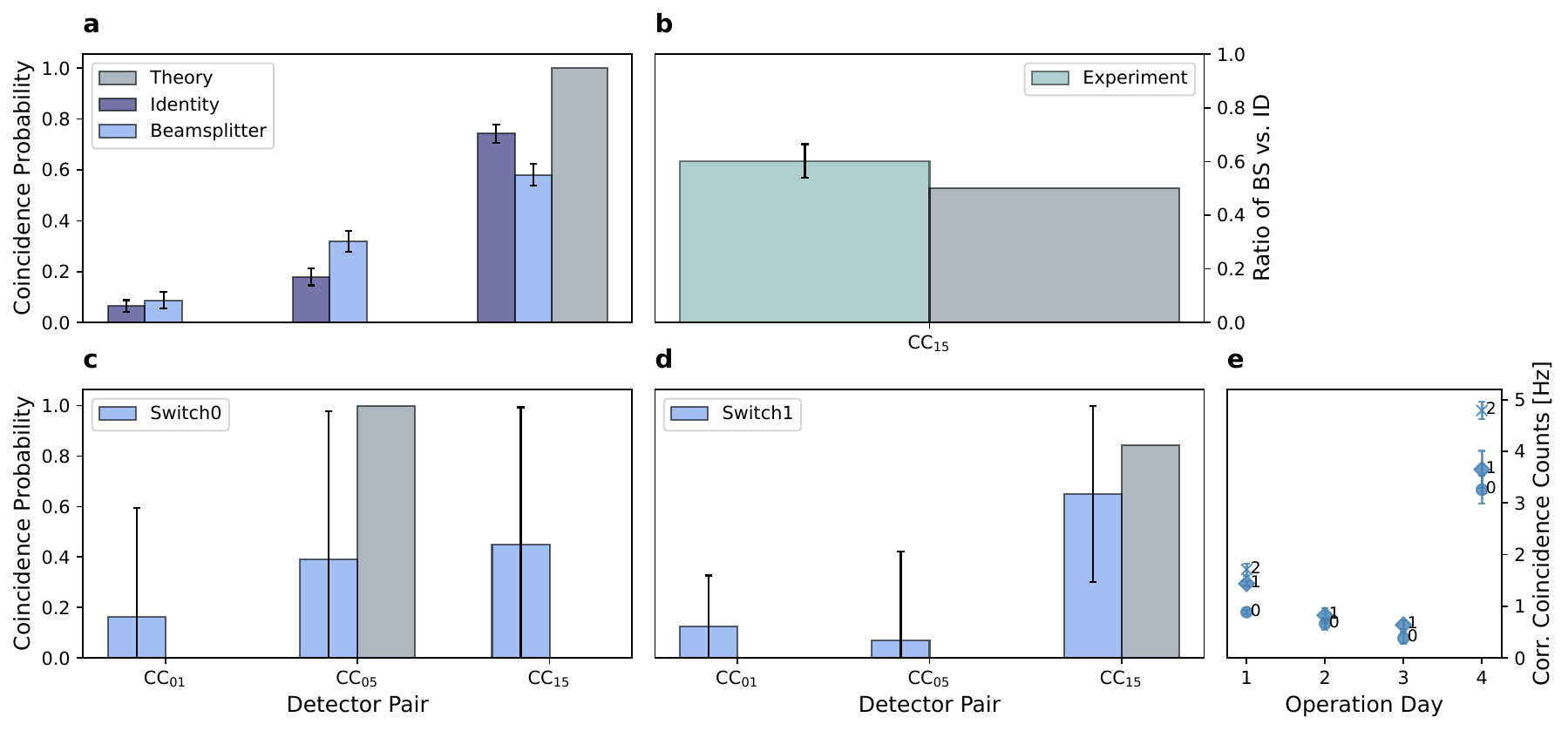}
    \caption[]{\textbf{Op. Day 4 - On-board programmable unitary performance and ratio.} \textbf{a} Significantly improved average shadow coincidence probability for Identity (dark blue) and Beamsplitter (light blue) with respect to the former operation day in Sec.~\ref{sec:operation_day_3} shows that the target detector pair CC$_{15}$ is now functioning; unitary fidelity could be improved in later operation days. \textbf{b} Measured ratio shows clear deviation with $R\sim 0.6$ (green bar) with respect to theory (grey bar) and can be attributed to problematic jitter in internal phase shifter of MZI6; data is solely taken from last shadow measurement as second shadow measurement is strongly unreliable due to fluctuations and outlier (see Fig.~\ref{fig:op_day_4_counts_close_up_3}) and first includes indistinguishable photons. \textbf{c,d} uPIC health test is negatively influenced due to data acquisition in the sun and relatively low integration time. \textbf{e} Most significant result is the overall improved coincidence rate due to the increased SPAD overvoltage, and this outcome enables meaningful experiments again; general unitary performance could be further improved on later operation days in Sec.~\ref{sec:operation_day_8} and Sec.~\ref{sec:operation_day_9}.}
    \label{fig:op_day_4_probabs}
\end{figure}

This chapter is concluded with some health-check unitaries performance in Fig.~\ref{fig:op_day_4_probabs} and shows that with respect to the former operation day in Sec.~\ref{sec:operation_day_3}, the average coincidence probability and coincidence rate improved significantly. Besides, the given ratio for target detector pair CC$_{15}$ clearly shows a higher ratio of $p\sim 0.6$ with respect to theory, and this aligns with the problematic jitter in Beamsplitter unitary performance. The latter could be improved in the upcoming operation days in Sec.~\ref{sec:operation_day_8} and Sec.~\ref{sec:operation_day_9}, and from the uPIC health test in Fig.~\ref{fig:op_day_4_probabs}c,d we must conclude that on operation day $4$ the general uPIC performance is less robust. This can be explained by considering the complete count rate overview in Fig.~\ref{fig:op_day_4_counts}, where the vertical dashed lines correspond to the shift in SPAD overvoltage. In our procedure, the uPIC health test is always measured after a HOM dip scan, and on this operation day, three of four $\ensuremath{12\,\mathrm{V}}$ measurements were completed with extremely high noise contributions from the sun. This means that the uPIC health test was mostly conducted in a high noise measurement setting and together with the relatively low integration time of $\ensuremath{20\,\mathrm{s}}$, the strong offset from theory in Fig.~\ref{fig:op_day_4_probabs}c,d can be explained\footnote{Note that the result of the uPIC health test consists of the sum of all individual health test measurements from each HOM dip scan for $\ensuremath{12\,\mathrm{V}}$ SPAD overvoltage in order to improve statistics.}.
\clearpage
\subsection{Operation Day 5 to 7 - 80 to 120 Days in Orbit}
\label{sec:operation_day_5_7}
\begin{figure}[!tbp]
    \centering
    \includegraphics[width=1\textwidth,height=0.84\textheight,keepaspectratio]{ 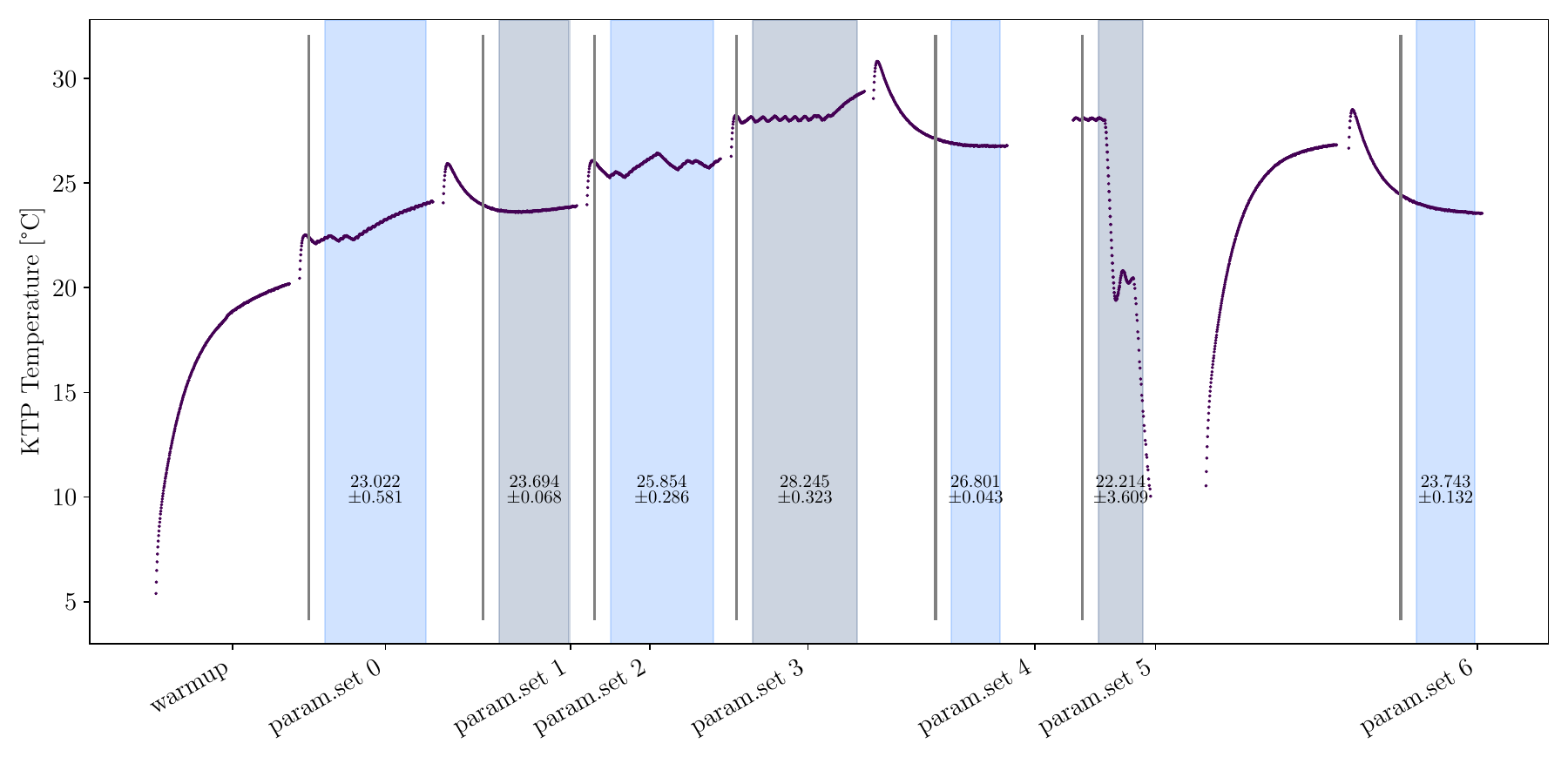}
    \caption[]{\textbf{KTP temperature control loop stability measurements.} For a number of different parameter sets (found in Table~\ref{tab:ktp_pid_stability_parameters}), temperature traces were recorded and temperature stability was evaluated once the loop had settled.}
    \label{fig:ktp_pid_stability}
\end{figure}

These operation days were designed to further conduct some health tests on the general uPIC and TEC performance. More specifically, the TEC performance tests included TEC stability measurements for different PID settings (see  Fig.~\ref{fig:ktp_pid_stability}, parameters in Table~\ref{tab:ktp_pid_stability_parameters}). Optical experiments were strongly hindered by a severe radiation event after $80$ days in orbit, as can be seen in the dark count rate in Fig.~\ref{fig:dark_counts_and_complete_rate}. Here, the dark counts for C$_5$ increased drastically, and the overall measurement quality, in terms of signal-to-noise ratio, was strongly impaired due to this damage to the SPADs. Despite this severe radiation event, SPAD C$_5$ eventually recovered, although its dark count rate settled to a higher level than before, and after $144$ days in orbit, optical experiments could be continued. As mentioned, the higher dark counts in C$_5$ had a general negative impact on countrate stability, hence the integration time per shot had to be increased from $\ensuremath{20\,\mathrm{s}}$ to $\ensuremath{60\,\mathrm{s}}$ to faithfully increase the statistics and reliability of a measurement.

\begin{table}
\centering
\caption[]{\textbf{Secondary (software) control loop parameters.} Parameter sets as used in the KTP-temperature stabilization measurements in Fig. \ref{fig:ktp_pid_stability}. The secondary control loop was either active from the beginning (\textbf{pid}), or engaged only after the hardware PID settled on some temperature (\textbf{pid\_after\_stable}). In the latter case, the secondary loop setpoint is given by the stable temperature reached by the hardware PID.
Temperature stability is evaluated over a rolling window, with a number of \textbf{stab\_w} measurements feeding into the evaluation, and stability accepted as soon as the rolling standard deviation fell below \textbf{accuracy}. Final mean and standard deviation are listed under \textbf{temp}. In Fig.~\ref{fig:ktp_pid_stability}, the respective datapoints contributing to these values are highlighted by colored regions.}
\label{tab:ktp_pid_stability_parameters}
\begin{tabular}{cccccc}
\hline
parameter set & pid & pid\_after\_stable & stab\_w & accuracy & temp \\
\hline
0 & False & True  & 20 & 0.100 & $23.02 \pm 0.58$ \\
1 & False & False & 10 & 0.025 & $23.694 \pm 0.068$ \\
2 & False & True  & 10 & 0.025 & $25.85 \pm 0.29$ \\
3 & False & True  & 10 & 0.100 & $28.24 \pm 0.32$ \\
4 & False & False & 40 & 0.100 & $26.801 \pm 0.043$ \\
5 & True  & False & 20 & 0.100 & $22.2 \pm 3.6$ \\
6 & False & False & 20 & 0.100 & $23.74 \pm 0.13$ \\
\hline
\end{tabular}

\end{table}

\subsection{Operation Day 8 - 144 Days in Orbit}
\label{sec:operation_day_8}
\begin{figure}[!tbp]
    \centering
    \includegraphics[width=1\textwidth,height=0.65\textheight,keepaspectratio]{ 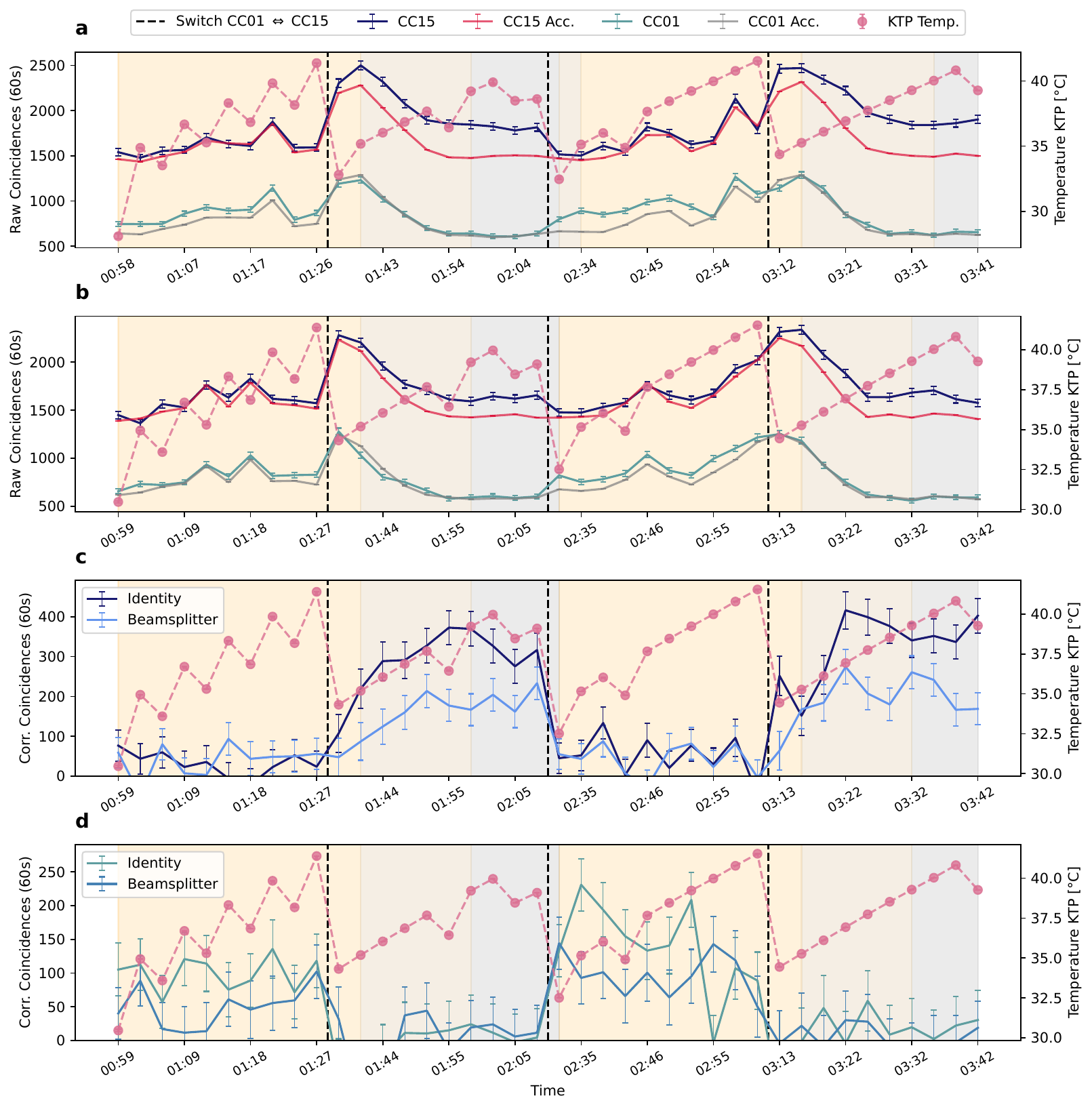}
    \caption[]{\textbf{Op. Day 8 - Coincidences overview.} \textbf{a} Raw Identity coincidence counts for both target detector pairs CC$_{01}$ (green) and CC$_{15}$ (dark blue) show working switch between both modes (vertical dashed lines); signal for CC$_{15}$ is very clear and above noise floor (red), while signal CC$_{01}$ is less prominent; KTP temperature (red points - secondary $y$-axis) is not within expected FWHM of HOM dip. \textbf{b} Beamsplitter instead of Identity shows similar count rate behaviour to \textbf{a}, and for both unitaries, the shade measurements show a very stable signal. \textbf{c} Detector pair CC$_{15}$ for Identity (dark blue) and Beamsplitter (light blue) highlights the uPIC switching performance. In regimes where CC$_{01}$ is targeted, the counts are essentially zero, whereas in the other regimes a clear signal can be extracted. Coincidence rate behaviour for Identity and Beamsplitter again proves that sun measurements are unreliable for an HOM dip measurement, as both unitaries are minimal during the sun. \textbf{d} Detector pair CC$_{01}$ for Identity (green) and Beamsplitter (blue) responds successfully for the first time in the complete mission and gives a clear signal when targeted; count rate is lower than for the other working detector pairs.}
    \label{fig:op_day_8_counts}
    \end{figure}

After regaining the functionality of SPAD C$_5$ (compare dark counts in Fig.~\ref{fig:dark_counts_and_complete_rate}) and increasing the integration time from $\ensuremath{20\,\mathrm{s}}$ to $\ensuremath{60\,\mathrm{s}}$ per shot, further optical experiments could be carried out, and for this operation day also HOM dip scans are continued. Similar to operation day $4$ in Sec.~\ref{sec:operation_day_4}, and as further explained in Sec.~\ref{sec:operation_day_1}, the TEC temperature calibration in Fig.~\ref{fig:arroyo_vs_tec_board} was not yet understood. For this reason, the KTP temperature data (red points - secondary $y$-axis) in Fig.~\ref{fig:op_day_8_counts}, is not at the expected HOM dip KTP temperature regime of $T_{TEC}\sim \ensuremath{32.5\,{}^{\circ}\mathrm{C}}$,\footnote{Except one single data point during a sun measurement for detector pair CC$_{01}$.} but still around $T_{TEC} \sim \ensuremath{40\,{}^{\circ}\mathrm{C}}$. In this sense, the operation day can be used to measure data points outside the HOM dip temperature regime.

Similar to operation day $3$ in Sec.~\ref{sec:operation_day_3}, target detector pairs are CC$_{01}$ and CC$_{15}$, but this time the resolution of both detector pairs is successful as can be seen in the unitary switching regimes (vertical dashed lines) of Fig.~\ref{fig:op_day_8_counts}a-b. Signal on CC$_{15}$ (dark blue) is more prominent than on CC$_{01}$ (green), and this is also evident in the corrected signal Fig.~\ref{fig:op_day_8_counts}c-d. The corrected signal for Identity (dark blue) and Beamsplitter (light blue) in Fig.~\ref{fig:op_day_8_counts}c reminds us that data from sun measurements (yellow area) is, in general, unreliable for Hong-Ou Mandel interference experiments. This is reflected in Identity (dark blue) and Beamsplitter (light blue) being both minimal in the sun regime. The decreased counts in Beamsplitter could eventually be explained by the KTP temperature being close to the FWHM of the expected KTP temperature ($T_{TEC}\sim \ensuremath{34\,{}^{\circ}\mathrm{C}}$); however, the Identity rate is also minimal here. On the other hand, the data points from CC$_{15}$ within the shade of Fig.~\ref{fig:op_day_8_counts}c show a very clear signal and can be used for completing the complete HOM dip in Sec.~\ref{sec:main_figures_hom}.

\begin{figure}[!tbp]
    \centering
    \includegraphics[width=1\textwidth,height=0.65\textheight,keepaspectratio]{ 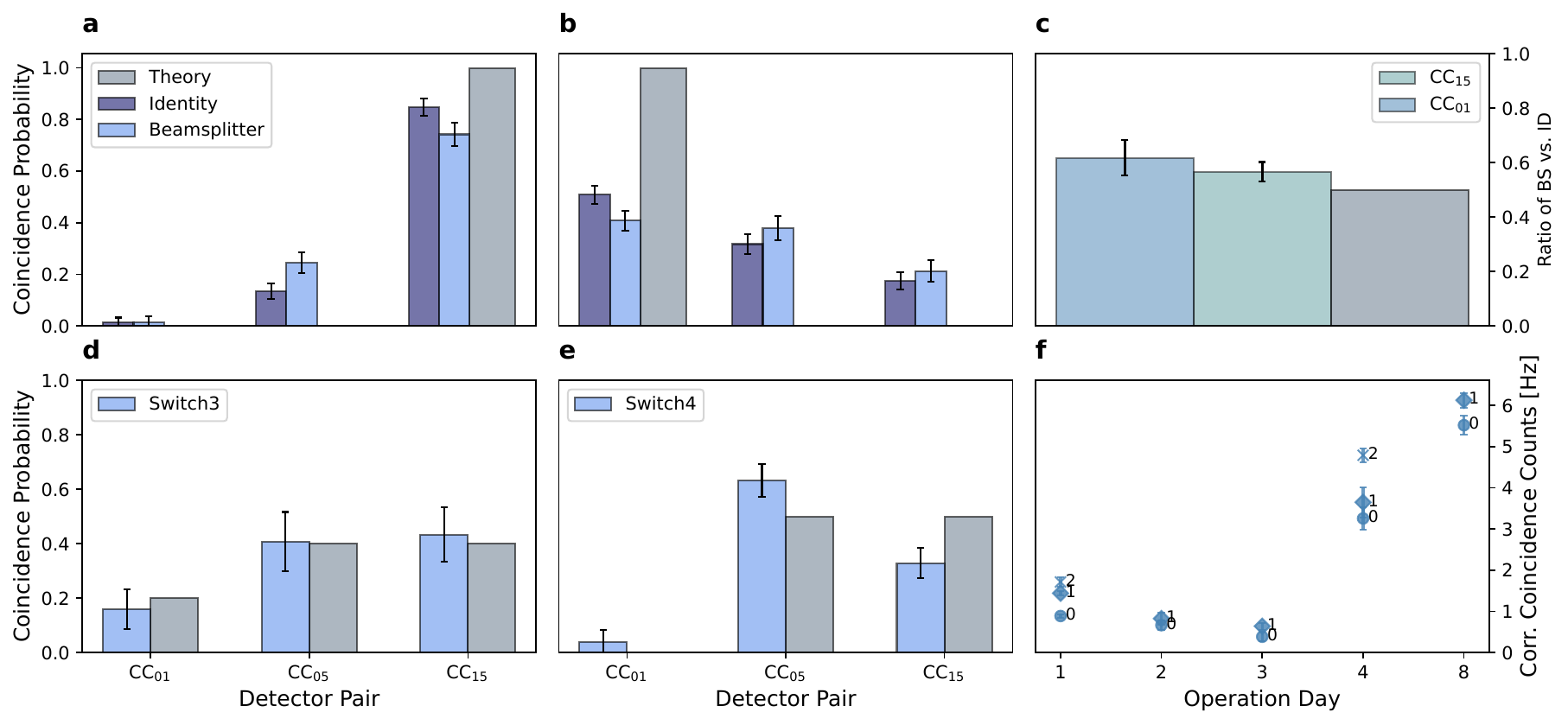}
    \caption[]{\textbf{Op. Day 8 - On-board programmable unitary performance and ratio.} \textbf{a} Average shadow coincidence probability of Identity (dark blue) and Beamsplitter (light blue) for target detector pair CC$_{15}$ shows improved unitary fidelity with respect to former operation days. \textbf{b} Average sun coincidence probability for target detector pair CC$_{01}$ differs strongly from theory (grey bar) and this can be explained by the thermal environment of uPIC; payload temperature is significantly colder than optimal operating temperature of uPIC and some unitary current configurations might be more sensitive to thermal influences (compare MZI jitter in Sec.~\ref{sec:operation_day_1}). Despite the low general unitary fidelity for CC$_{01}$, its ratio in \textbf{c} is not far off the theoretical value, and this is also true for target pair CC$_{15}$. \textbf{d,e} More complex target unitaries that include all three detector pairs in \textbf{d}, and here a good average fidelity is measured. Unbalanced Beamsplitter in \textbf{e} explains the offset from the theoretical value, and the reason can also be attributed to the aforementioned MZI jitter in the Beamsplitter setting. \textbf{f} Corrected coincidence rate could again be improved for detector pair CC$_{15}$ and this highlights the robustness of the payload after $144$ days in orbit.}
    \label{fig:op_day_8_probabs}
\end{figure}

The test for unitary performance in Fig.~\ref{fig:op_day_8_probabs} reveals new insights in the payload performance, and here the fidelity for target detector pair CC$_{15}$ and CC$_{01}$ shows a big difference in programmable unitary performance. While the fidelity of Identity (dark blue) and Beamsplitter (light blue) for CC$_{15}$ could be improved relative to former operation days, the measured coincidence probability for Identity and Beamsplitter for CC$_{01}$ is far from the expected behaviour (grey bars). In addition to these unitary switches, some more complex unitaries are also measured after each HOM dip scan, and Fig.~\ref{fig:op_day_8_probabs}d,e show that the obtained result agrees well with theory for switch $3$, while less so for switch $4$. Switch $4$ corresponds to a Beamsplitter between detector modes $0$ and $1$, heralded by mode $5$. The target configuration of this Beamsplitter should be balanced, such that an equal amount of photons is registered in both modes (grey bars). With the insight of Fig.~\ref{fig:op_day_1_mzi_jitter}, we can also explain the measured offset here, as due to the jitter in internal phase, it is in general harder to implement a balanced beamsplitter.

The low unitary fidelity for specific target unitaries, such as the one targeting CC$_{01}$, can in general be attributed to the thermal environment of the uPIC. As already mentioned in Sec.~\ref{sec:operation_day_4}, the uPIC is calibrated in vacuum but at room temperature. As our payload is significantly colder than $\sim \ensuremath{25\,{}^{\circ}\mathrm{C}}$ (see Sec.~\ref{sec:pay_temp}), this large temperature difference can negatively impact the performance of the thermal phase shifters. This was already shown for an individual phase shifter in Fig.~\ref{fig:op_day_1_mzi_jitter} to explain the jitter in the Beamsplitter configuration, but the explanation can directly be generalized to the complete set of thermal phase shifters. In this sense, some unitaries (e.g., CC$_{01}$) have corresponding current configurations that may be more sensitive to thermal effects and therefore exhibit reduced unitary fidelity.

Despite the low unitary fidelity of CC$_{01}$, its ratio between Beamsplitter and Identity is not far off from the expected value of $R_T=0.5$ in the ratio overview of Fig.~\ref{fig:op_day_8_probabs}c. This result is further strengthened on the next operation day in Sec.~\ref{sec:operation_day_9} and is important for the HOM dip analysis on the previous operation day in Sec.~\ref{sec:operation_day_4}. A significant increase in the corrected rate could also be measured for the present operation day in Fig.~\ref{fig:op_day_8_probabs}f, and this result highlights the robustness of the payload as it already had to endure $144$ days in orbit while facing continuous damage from radiation and other environmental hazards of space.

Let us note that, due to the improved count rate stability and coincidence rate, the coincidence probability and ratio results of Fig.~\ref{fig:op_day_8_probabs} also contain data from sun measurements for detector pair CC$_{01}$. This has to be done, as otherwise the entire set of measurements for CC$_{01}$ would have been excluded from the data analysis, since it was measured solely during the sun phase. As the coincidence probability and ratio are more robust with respect to fluctuations and outliers, this approach is meaningful and is also done for the measurements of target detector pair CC$_{15}$ for the upcoming operation day $9$ in Sec.~\ref{sec:operation_day_9}. Here, the mentioned target detector pair is also measured only during the sun phase, and, for robust unitary fidelity and ratio, these areas are included. This approach is considered only if shade measurements are unavailable.
\clearpage
\subsection{Operation Day 9 - 156 Days in Orbit}
\label{sec:operation_day_9}
\begin{figure}[!tbp]
    \centering
    \includegraphics[width=0.95\textwidth,height=0.75\textheight,keepaspectratio]{ 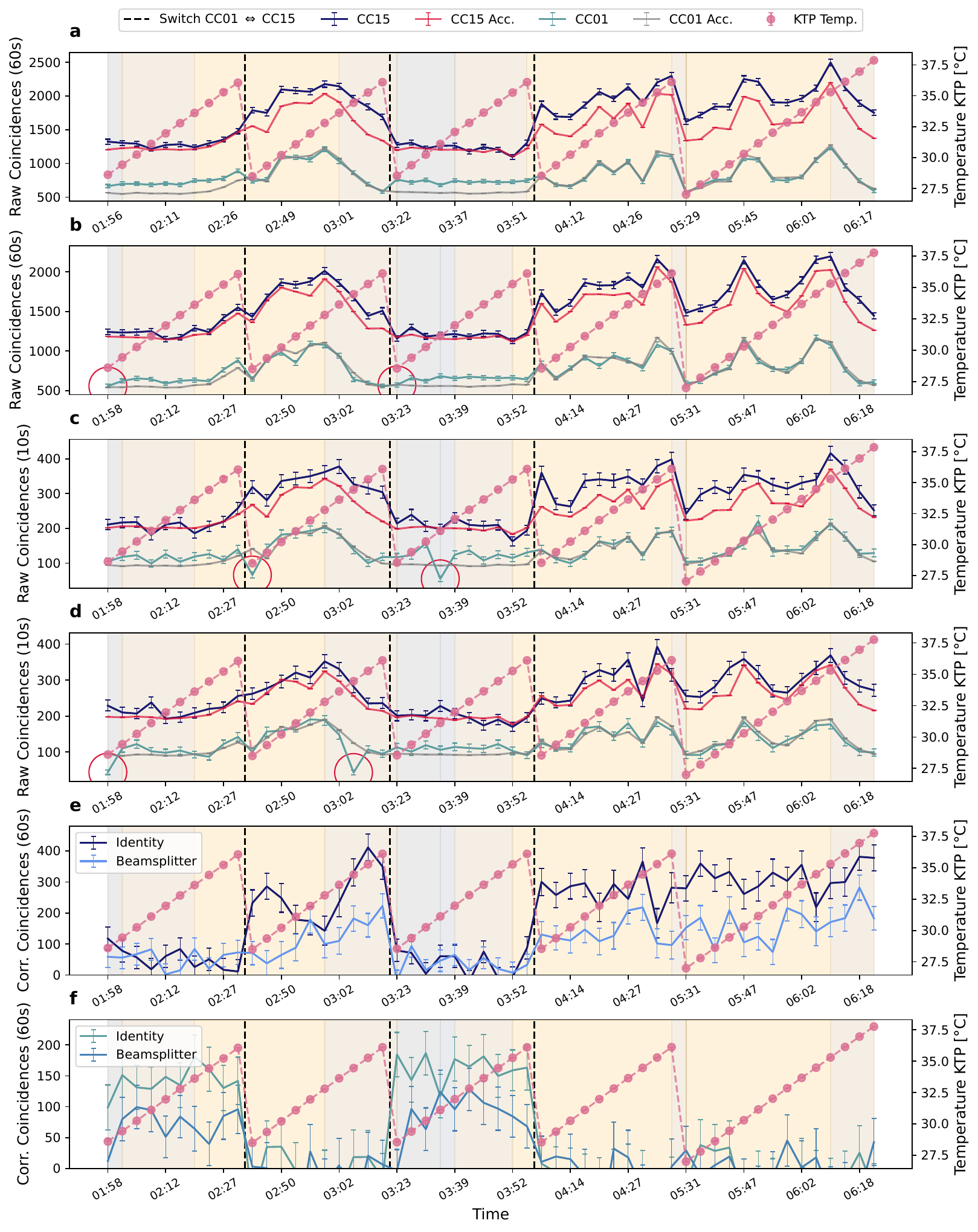}
    \caption[]{\textbf{Op. Day 9 - Coincidences overview.} \textbf{a,b} Identical configuration as in Fig.~\ref{fig:op_day_8_counts}a,b shows good switching performance between both target detector pairs (vertical dashed lines) but no dip in coincidences in shadow areas of \textbf{b} for expected KTP temperature regime (red points - secondary $y$-axis); potential reason is loss of laser SLM property as also shown for EM laser in Fig.~\ref{fig:laser_power_vac} that experienced similar damage as FM laser. \textbf{b,c,d} Red circles highlight the timetagger anomaly for the Identity (\textbf{c}) and Beamsplitter (\textbf{d}) configurations, in which counts drop significantly below the predicted noise floor. \textbf{e,f} Identical configuration and similar performance as in Fig.~\ref{fig:op_day_8_counts}c,d shows corrected signal for both detector pairs, respectively.}
    \label{fig:op_day_9_counts}
\end{figure}

The experimental protocol of this operation day is identical to the former one in Sec.~\ref{sec:operation_day_8}, but this time the TEC calibration in Fig.~\ref{fig:arroyo_vs_tec_board} was understood, hence the KTP temperature in Fig.~\ref{fig:op_day_9_counts} is scanned around $T_{TEC}\sim \ensuremath{32.5\,{}^{\circ}\mathrm{C}}$. Two shadow regions can be identified for the target detector pair CC$_{01}$ (grey shaded area). Despite reaching the HOM dip KTP temperature regime in these areas, no dip in coincidences was measured, consistent with the laser having lost its ability to operate in SLM mode as described in Sec.~\ref{sec:tvac}. The behaviour of the FM laser is, in this case, very similar to that of the EM laser on Earth, which experienced similar conditions, i.e., thermal vacuum, thermal distortion, and outgassing-induced power loss. The EM laser lost its coherence and ability to show a HOM dip after approx. $7$ months (see Fig.~\ref{fig:laser_power_vac}) and operation day $9$ suggests that the FM laser experienced a similar fate. We attribute this to multimode backreflections from the outgassing layer on the optical components, which would suppress the desired narrow-bandwidth backreflections of the VBG. In addition, ongoing thermal distortion might change the alignment of the VBG with respect to the laser diode output. A detailed analysis of coherence loss is presented in Sec.~\ref{sec:tvac}.

\begin{figure}[!tbp]
    \centering
    \includegraphics[width=1\textwidth,height=0.65\textheight,keepaspectratio]{ 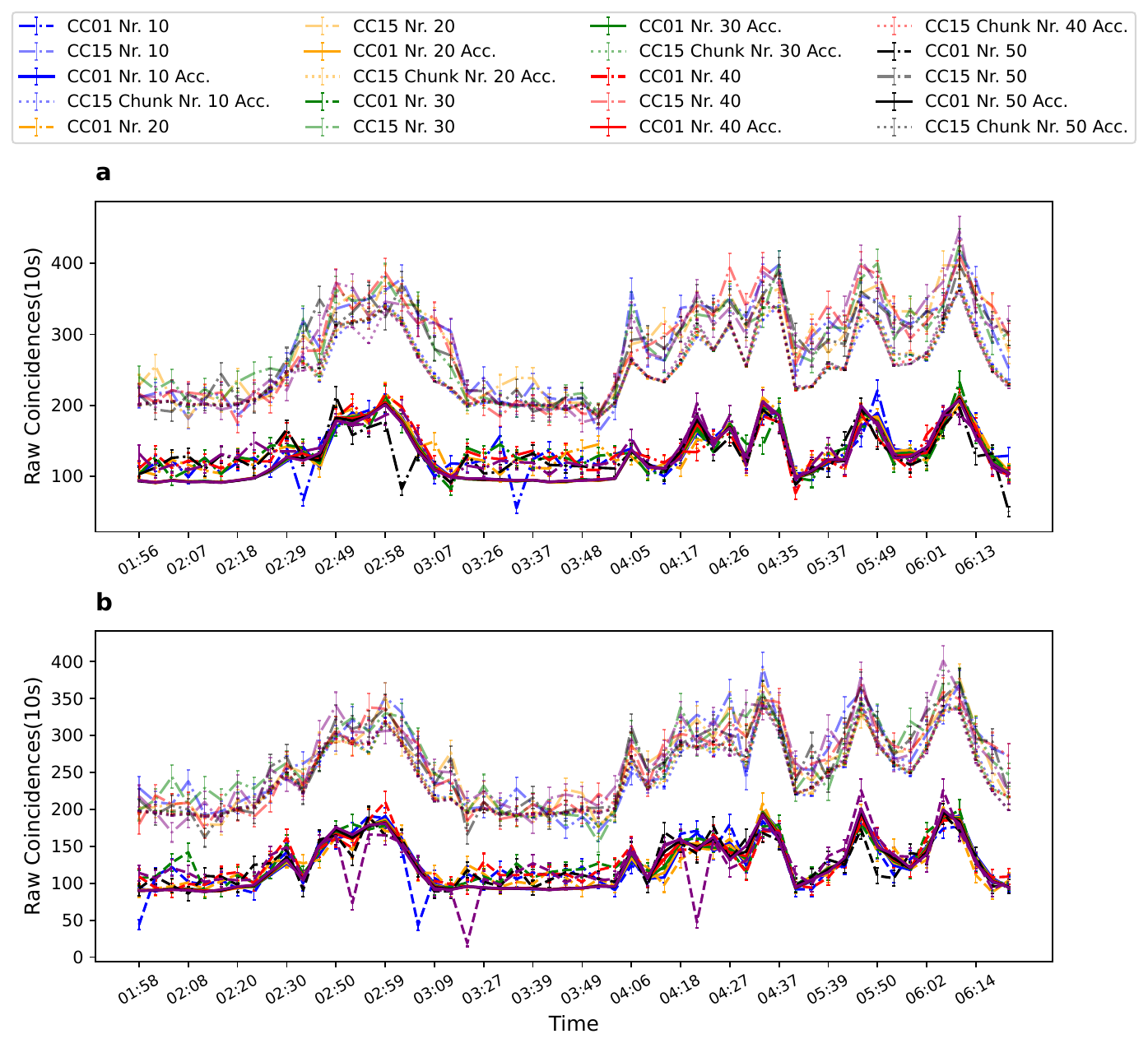}
    \caption[]{\textbf{Op. Day 9 - Timetagging anomaly.} \textbf{a} Raw Identity coincidence counts for all six measurement chunks of $\ensuremath{10\,\mathrm{s}}$ integration time and both target detector pairs CC$_{01}$ (high saturation),  CC$_{15}$ (low saturation) show that detector pair CC$_{01}$ has significant drops in countrate (dashed) below the predicted noise floor (solid). This is even more frequent in Beamsplitter setting in \textbf{b} where spikes in countrate are at least $2\sigma$ below the predicted noisefloor and must be considered outliers. Later operation days in Sec.~\ref{sec:operation_day_12} confirm this phenomenon for an overvoltage $>\ensuremath{12\,\mathrm{V}}$ but also for the standard $\ensuremath{12\,\mathrm{V}}$ configuration during measurements in the sun. This effect suggests that this anomaly must be related to SPAD saturation or a saturation of the counting logic in the passive quenching circuit (too high count rate). The anomaly has strong consequences for the sum of each measurement chunk, as the resulting datapoint with $\ensuremath{60\,\mathrm{s}}$ effective integration time may be strongly suppressed if saturation happens for one $\ensuremath{10\,\mathrm{s}}$ chunk. For this reason, each measured dip on other operation days is tested for this specific timetagger anomaly by examining its subsequent chunks of $\ensuremath{10\,\mathrm{s}}$ integration time.}
    \label{fig:op_day_9_outlier_anomaly}
\end{figure}

Besides these laser complications, detector pair CC$_{01}$ shows some additional timetagger anomalies that can mimic interference as highlighted in the red circles in Fig.~\ref{fig:op_day_9_counts}b-d for Identity and Beamsplitter configuration. These anomalies can be further identified in Fig.~\ref{fig:op_day_9_outlier_anomaly}, where the raw timetagger data (dashed) drops significantly (more than $2\sigma$) below the predicted noise floor (solid) for some $\ensuremath{10\,\mathrm{s}}$ integration chunks. This unphysical behaviour is detected only for CC$_{01}$ on this operation day and may be related to detector/quenching circuit saturation. This can be seen on later operation days, especially in Sec.~\ref{sec:operation_day_12}, as here, too, high overvoltage or measurements in the sun permanently saturate the counting logic or SPADs. In this sense, the measured anomalies correspond to the short-term effect of complete saturation of the detector/quenching circuit, encountered in later operation days. Due to this issue, every dip in coincidences from other operation days is tested for a similar anomaly to faithfully exclude artifacts.

\begin{figure}[!tbp]
    \centering
    \includegraphics[width=1\textwidth,height=0.71\textheight,keepaspectratio]{ 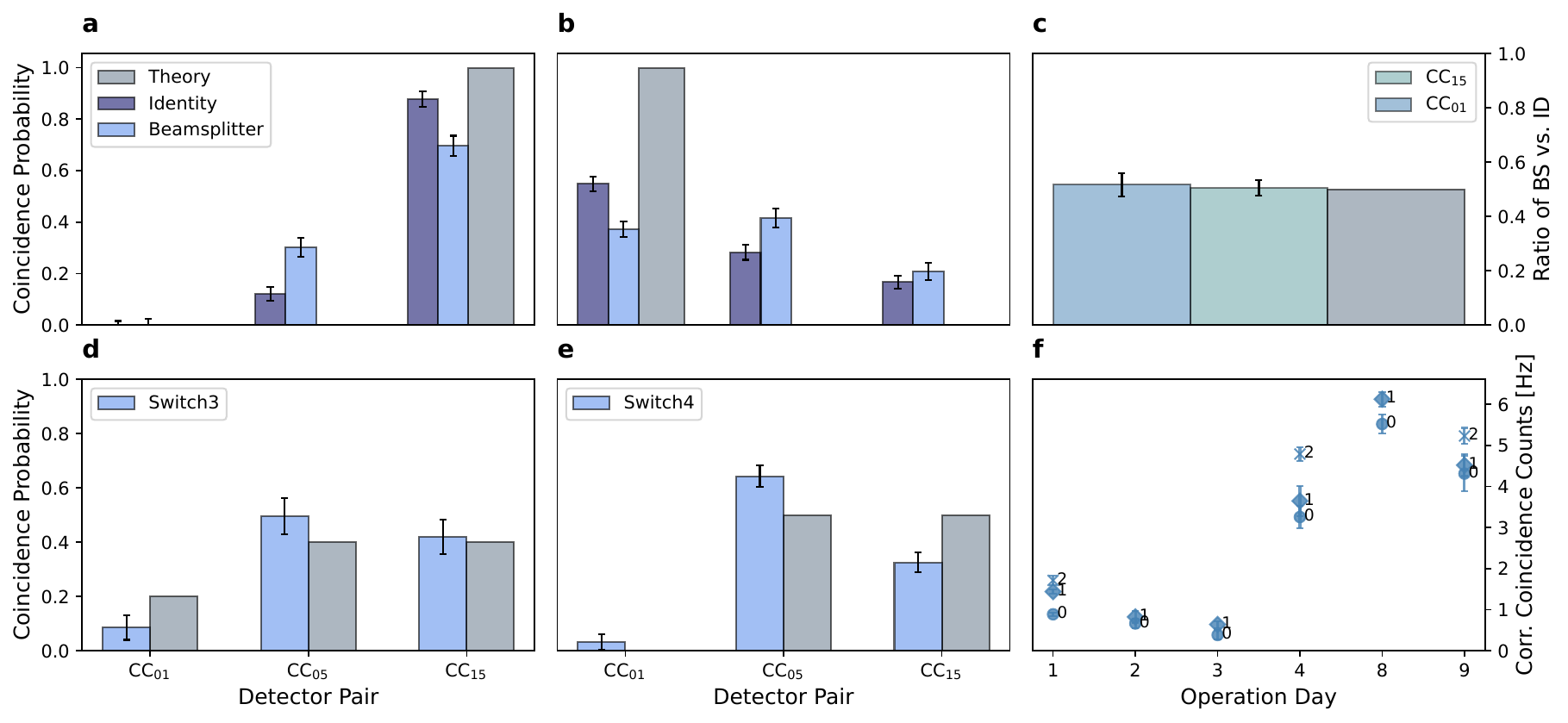}
    \caption[]{\textbf{Op. Day 9 - On-board programmable unitary performance and ratio.} \textbf{a} Average sun coincidence probability of Identity (dark blue) and Beamsplitter (light blue) for target detector pair CC$_{15}$ agrees well with theory and is also consistent with the result of the former operation day. \textbf{b} Average shadow coincidence probability for target detector pair CC$_{01}$ is also consistent with former operation day and emphasizes that this uPIC current configuration must be more sensitive to the thermal environment of space (compare Sec.~\ref{sec:operation_day_8}). Despite the low average fidelity for \textbf{b}, the ratio between Beamsplitter and Identity is nearly perfectly overlapping with theory, and this is also true for the case in \textbf{a}. \textbf{d,e} Average coincidence probability of more complex unitaries shows similar performance as in the former operation day, and a slightly decreased rate in \textbf{f} can be explained by CC$_{15}$ being in the sun for the whole operation day.}
    \label{fig:op_day_9_probabs}
\end{figure}

From a general programmable unitary performance operation day $9$ showed good results, as can be seen by the general improved target detector pair coincidence probability in Fig.~\ref{fig:op_day_9_probabs}a-b and a high overlap between measured and target ratio in Fig.~\ref{fig:op_day_9_probabs}c for both detector pairs. The latter again indicates that the unwanted photon loss in other detector modes is not problematic for the Beamsplitter configuration itself; hence, the result from operation day $4$ in Sec.~\ref{sec:operation_day_4} is further strengthened. A slight decrease in countrate is observed for detector pair CC$_{15}$, but this can be explained by the fact that this detector pair is only recorded during sun phase (see Sec.~\ref{sec:operation_day_8} for justification). The more complex unitaries in Fig.~\ref{fig:op_day_9_probabs}d,e show similar overlap with theory as in Sec.~\ref{sec:operation_day_8} and underline consistent uPIC performance.
\clearpage
\subsection{Operation Day 10 -  179 Days in Orbit}
\label{sec:operation_day_10}
\begin{figure}[!tbp]
    \centering
    \includegraphics[width=1\textwidth,height=0.82\textheight,keepaspectratio]{ 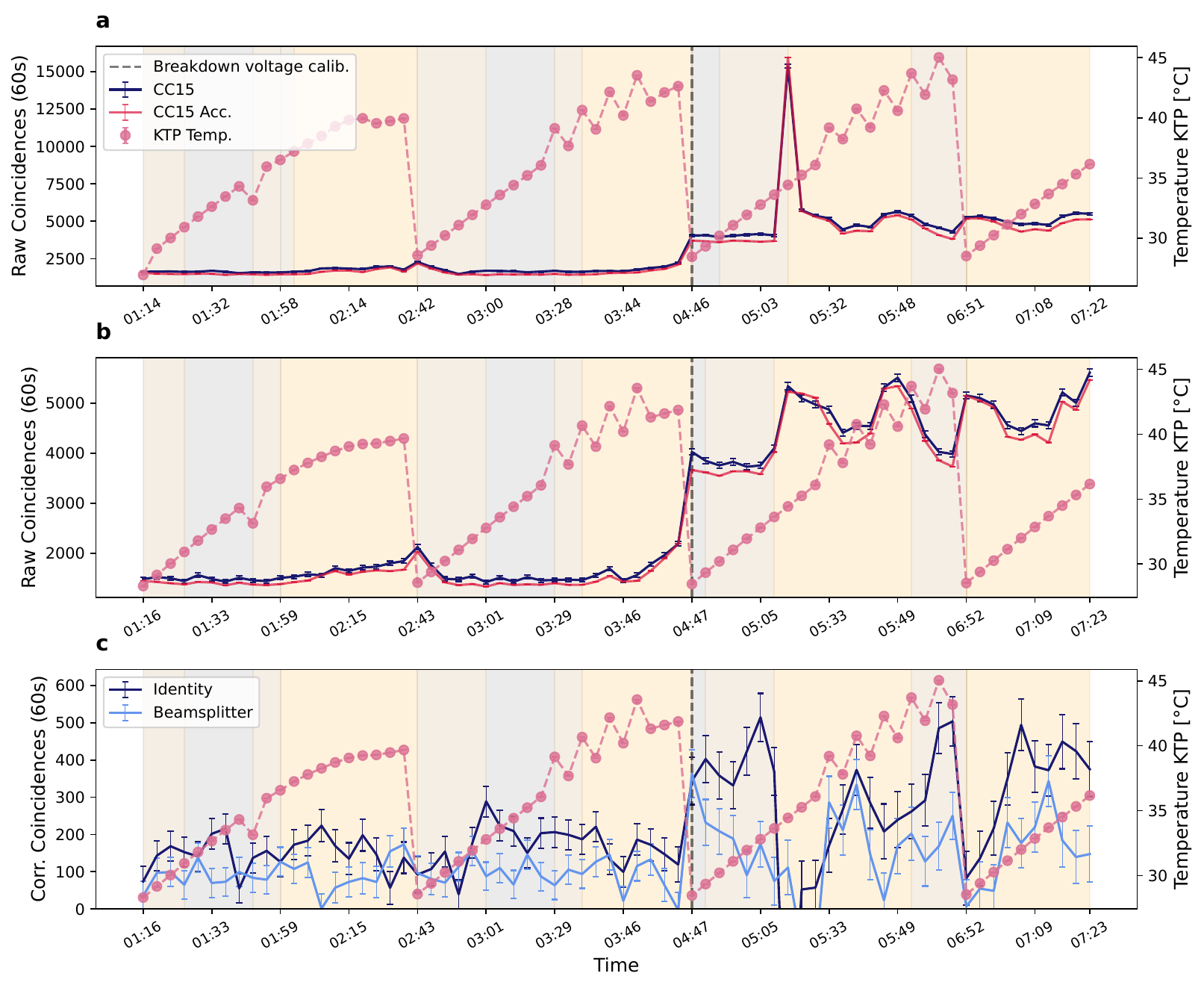}
    \caption[]{\textbf{Op. Day 10 - Coincidences overview.} \textbf{a,b} Raw coincidence counts for Identity (\textbf{a}) and Beamsplitter (\textbf{b}) for target detector pair CC$_{15}$ show strong increase in baseline countrate after breakdown voltage calibration (vertical dashed line); overall signal in corrected counts in \textbf{c} reveals that payload experienced a damage that cannot be attributed to a dark count increase on this day (compare Fig.~\ref{fig:dark_counts_and_complete_rate}).}
    \label{fig:op_day_10_counts}
\end{figure}

In a similar spirit to the former operation days, operation day $10$ included additional HOM dip scans despite having a damaged laser module in terms of its SLM property (see former operation day in Sec.~\ref{sec:operation_day_9}). Measurements on Earth (compare Fig.~\ref{fig:mode_hopping}) showed that one of the EM lasers randomly jumped back into a working SLM mode for its original diode temperature, and this was the main motivation to conduct further two-photon interference experiments. In general, as described in Sec.~\ref{sec:tvac}, the SLM feature could be deterministically regained with the diode temperature sweeping procedure described in Fig.~\ref{fig:laser_power_vac}. However, in the FM, the necessary UART was potentially damaged, so the only strategy we used was to power-cycle the laser before each HOM dip scan, as this was done successfully on EM, see Fig.~\ref{fig:mode_hopping}.  Together with these final attempts for additional HOM dip scans, a SPAD breakdown voltage calibration (see Fig.~\ref{fig:spads}) is performed and the effect of this recalibration can be seen by an increased overall countrate in Fig.~\ref{fig:op_day_10_counts} (vertical dashed line).

\begin{figure}[!tbp]
    \centering
    \includegraphics[width=1\textwidth,height=0.65\textheight,keepaspectratio]{ 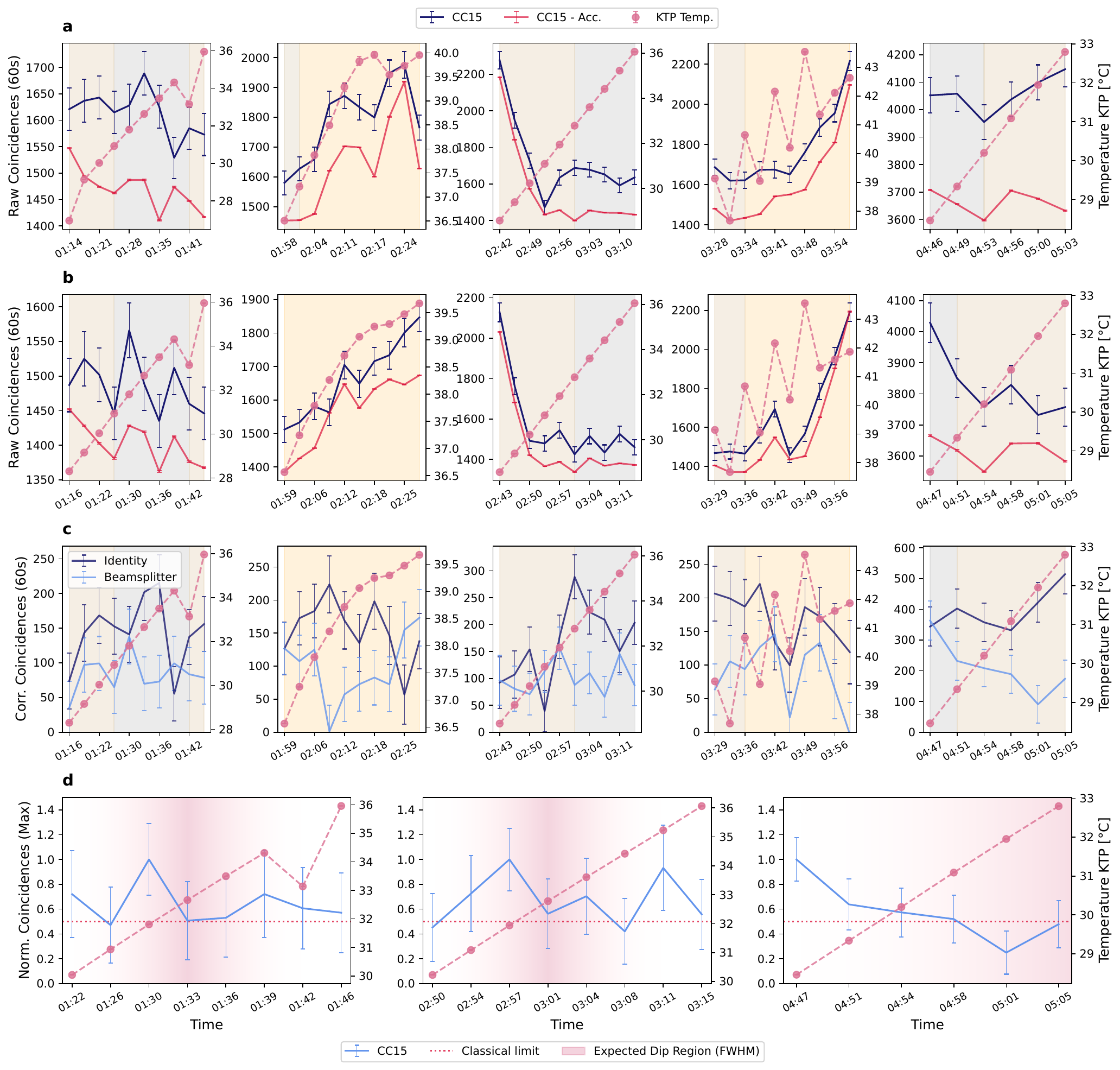}
    \caption[]{\textbf{Op. Day 10 - Grouped coincidence overview.} \textbf{a (row)} Similar to operation day $4$, countrate overview in Fig.~\ref{fig:op_day_10_counts} is dissected into smaller chunks; (sun) measurements after $5:05$ UTC are ignored (see $x$-axis) as highly unstable. Identity (\textbf{a}) and Beamsplitter configuration (\textbf{b}) show generally worse stability in noise floor (red) and signal overhead (dark blue), which might suggest additional damage to the payload. The corrected signal for both unitaries is also reduced in \textbf{c} and can only be restored with a new breakdown-voltage calibration (last column). Three shadow measurements can be identified in \textbf{d} where measurements are also within expected HOM dip temperature regime. Laser coherence damage in former operation days (see Sec.~\ref{sec:operation_day_9}) makes two-photon interference measurements extremely difficult as the laser lost its SLM property due to outgassing damage (more details in Sec.~\ref{sec:tvac}). Random frequency mode hopping into a stable SLM mode for the EM laser on Earth opened up the potential that the FM laser might manage the same (see Fig.~\ref{fig:mode_hopping}). The final shadow measurement shows signs of quantum interference, but due to lower visibility (similar to the case for the EM laser in Fig. \ref{fig:mode_hopping}), definitive statements about HOM interference are unreliable.}
    \label{fig:op_day_10_closeup}
\end{figure}

In general, the measurements prior to the new SPAD breakdown calibration reveal that the optical components must have suffered from a particular damaging event that cannot be attributed solely to a dark count increase (compare dark counts for this operation day in Fig.~\ref{fig:dark_counts_and_complete_rate}). This can be directly seen in the significantly reduced coincidence rate in Fig.~\ref{fig:op_day_10_probabs}e and the overall decreased countrate stability in Fig.~\ref{fig:op_day_10_closeup} (first four columns). TLE data and accidental stability analysis of Fig.~\ref{fig:op_day_10_counts} reveal three shadow regimes, and here also the KTP temperature (red points - secondary $y$-axis) is within the FWHM of the expected HOM dip temperature regime. Despite being in the shadow of the earth, the accidentals are more unstable with respect to the former operation days in Sec.~\ref{sec:operation_day_8} or Sec.~\ref{sec:operation_day_9} and identical measurement settings might suggest a different sort of damage than just increased dark counts. The row in Fig.~\ref{fig:op_day_10_closeup}d shows the normalized Beamsplitter counts for the three identified shadow phases, and greater attention should be given to the last measurement, as a different SPAD breakdown-voltage calibration is used here.

\begin{figure}[!tbp]
    \centering
    \includegraphics[width=1\textwidth,height=0.69\textheight,keepaspectratio]{ 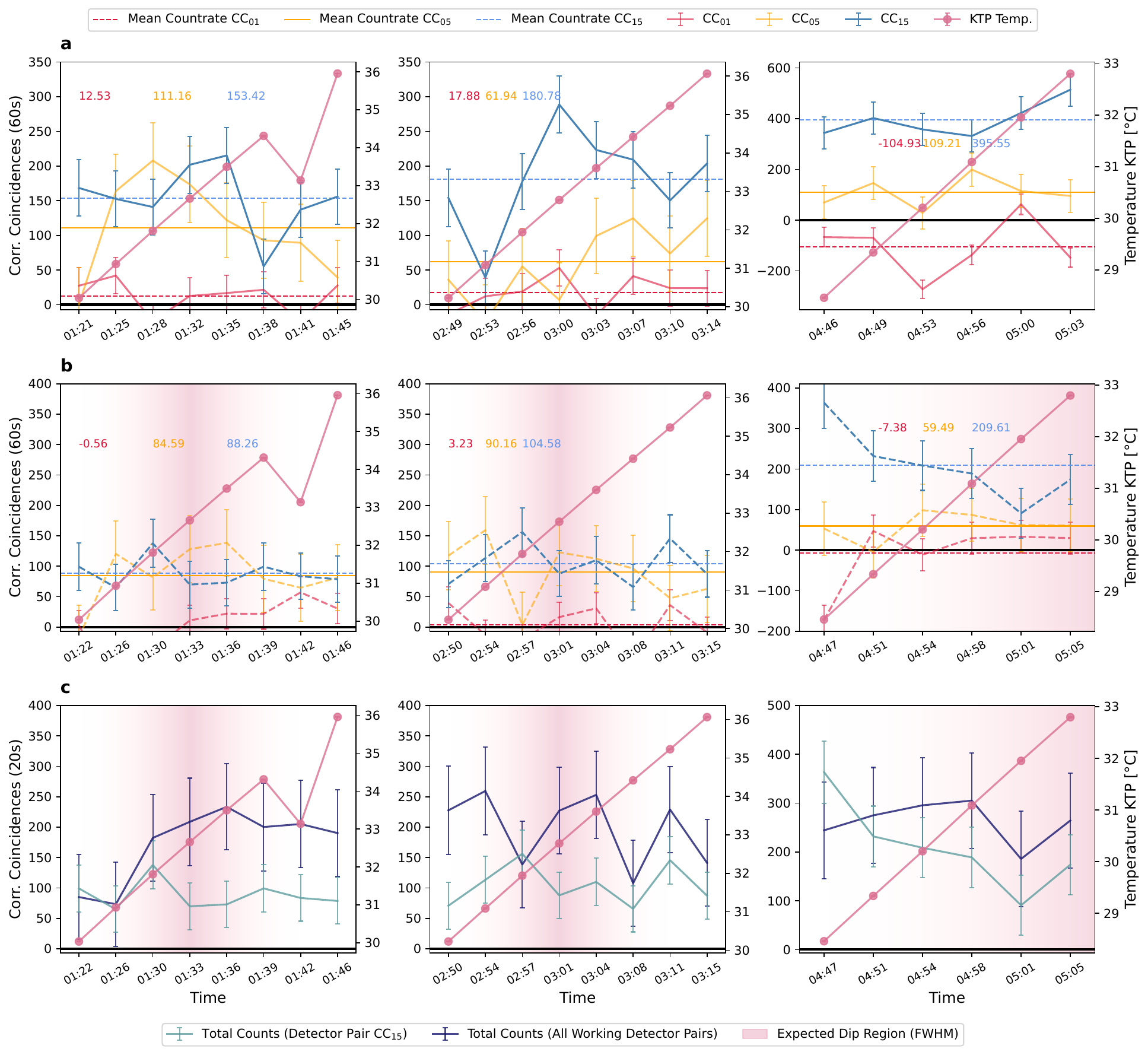}
    \caption[]{\textbf{Op. Day 10 - Total corrected coincidences.} \textbf{a (row)} Corrected coincidence counts for all detector pairs CC$_{01}$ (red), CC$_{05}$ (yellow) and CC$_{15}$ (blue) show highly unstable shadow countrate for Identity; comparison with former operation days (e.g. Sec.~\ref{sec:operation_day_8}) with identical measurement setting suggests that payload suffered a significant damage. The last column includes a new breakdown-voltage calibration that effectively increases the applied overvoltage; the detector pair CC$_{01}$ is strongly saturated, with corrected counts being negative. The latter is related to the timetagger anomaly in Fig.~\ref{fig:op_day_9_outlier_anomaly}, and this malfunction can be attributed to SPAD or quenching circuit saturation. \textbf{b (row)} Same as in \textbf{a} but Beamsplitter unitary, and the non-target detector pairs in the last column show no signs of potentially wrongly routed photons within the FWHM of the HOM dip temperature regime. The total rate in \textbf{c} also aligns with the worsened programmable unitary performance in Fig.~\ref{fig:op_day_10_probabs}, as a constant baseline on non-target modes is present.}
    \label{fig:op_day_10_total_counts_anomaly_check}
\end{figure}

In this last shadow measurement, the original SPAD breakdown calibration is replaced with a new calibration file, which has a slightly higher measured breakdown voltage for the working detectors. Applying the same $\ensuremath{12\,\mathrm{V}}$ of overvoltage will then have the same effect as increasing the overvoltage for the old calibration file, and the outcome of this new calibration is directly reflected in the observed higher countrate. While positive for the target detector pair CC$_{15}$, this configuration leads to a non-physical effect on detector pair CC$_{01}$ as can be seen in the corrected rate (red) in Fig.~\ref{fig:op_day_10_total_counts_anomaly_check}. Here, the noise floor significantly overshoots the measured signal, indicating that the timetagger anomaly in Fig.~\ref{fig:op_day_9_outlier_anomaly} is now evident across all data points. This can be explained by a sort of detector saturation that occurs for too high overvoltage or countrate, and this saturation is verified in additional tests on a later operation day in Sec.~\ref{sec:operation_day_12}.

\begin{figure}[!tbp]
    \centering
    \includegraphics[width=1\textwidth,height=0.82\textheight,keepaspectratio]{ 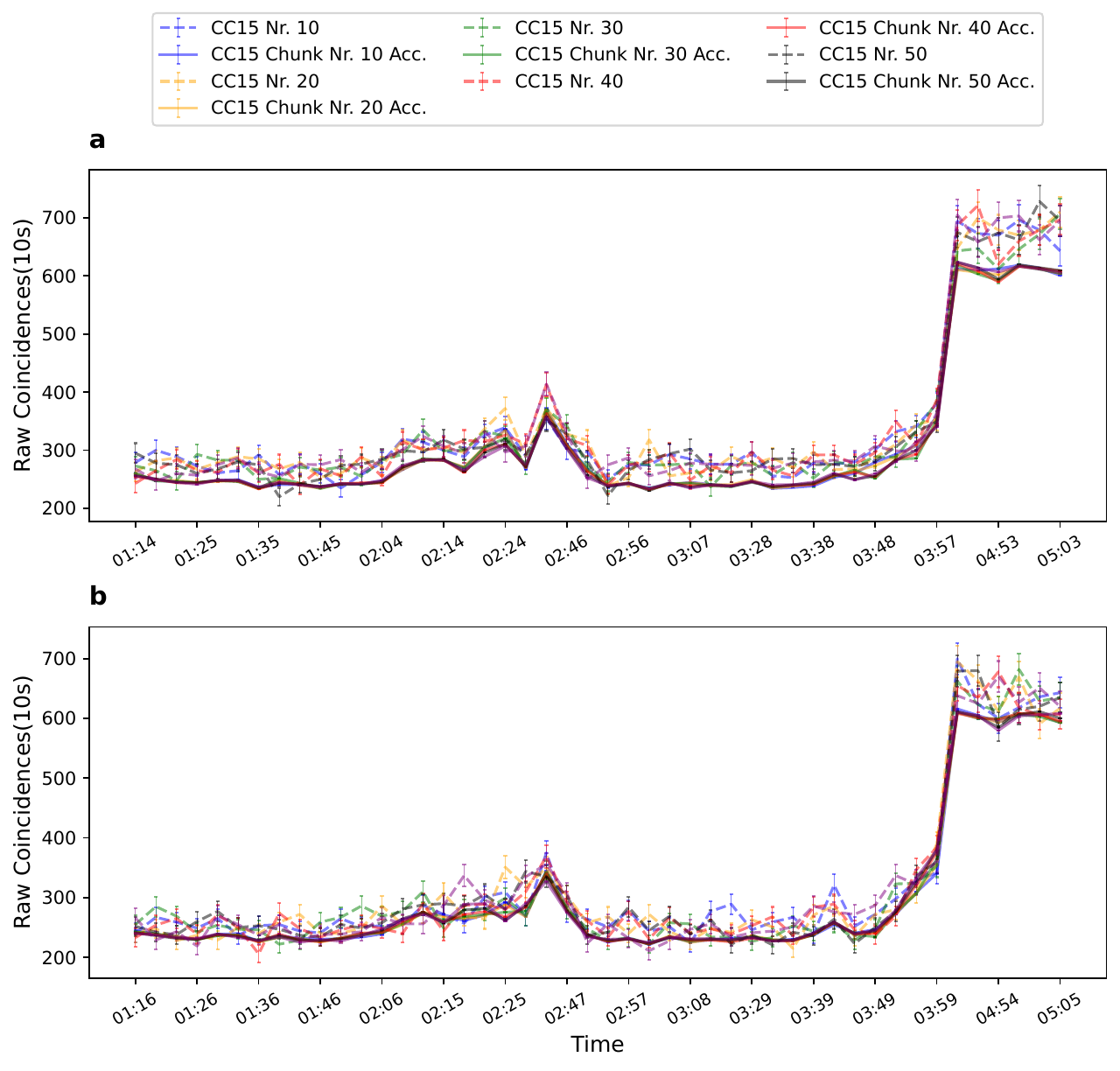}
    \caption[]{\textbf{Op. Day 10 - Beamsplitter outlier analysis.} \textbf{a} Raw Identity coincidence counts (dashed) with noisefloor (solid) for individual measurement chunks of $\ensuremath{10\,\mathrm{s}}$ reveal no timetagging anomaly, and this is also true for the Beamsplitter configuration in \textbf{b}. This test allows us to use the final shadow measurement as a meaningful dataset.}
    \label{fig:op_day_10_counts_outliers}
\end{figure}

\begin{figure}[!tbp]
    \centering
    \includegraphics[width=1\textwidth,height=0.75\textheight,keepaspectratio]{ 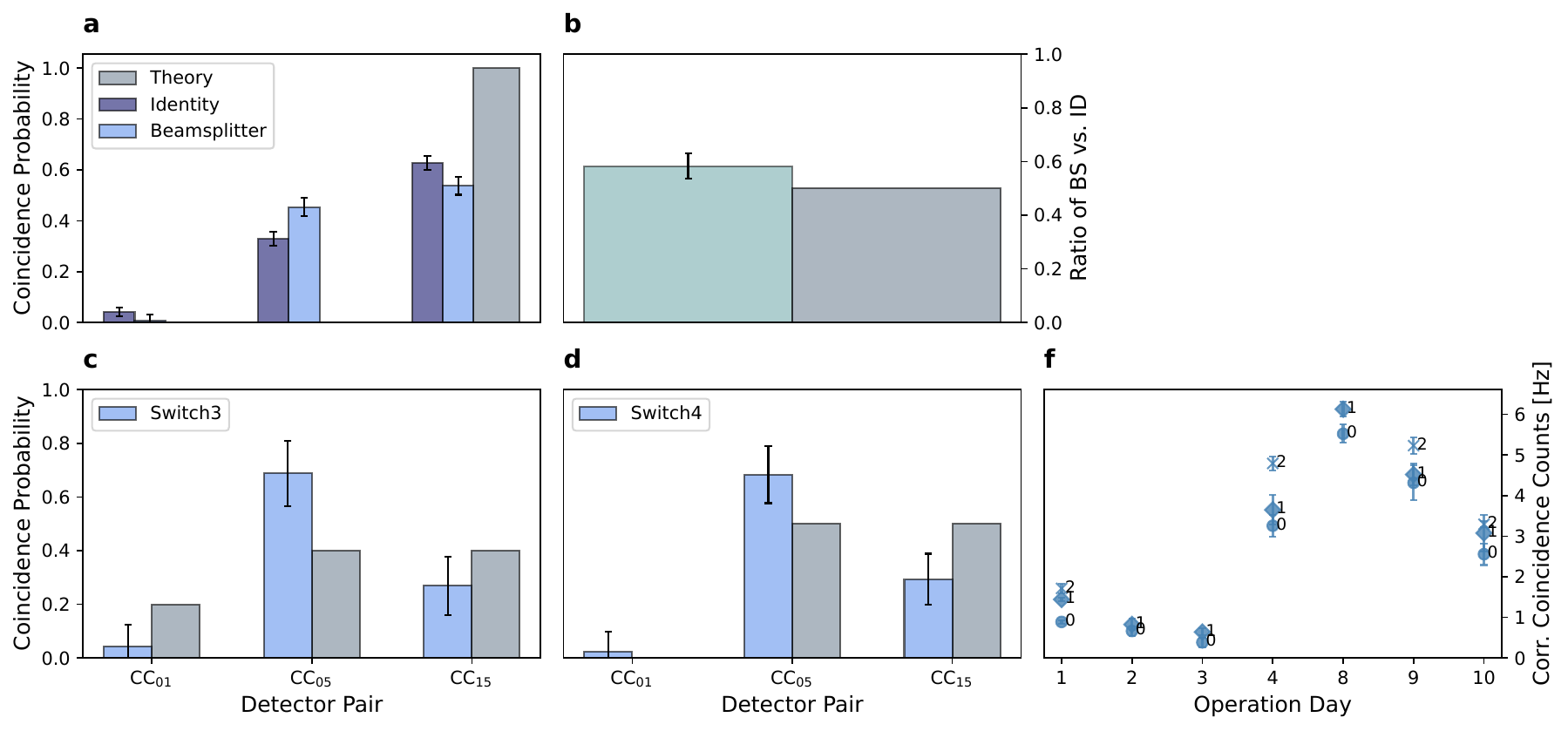}
    \caption[]{\textbf{Op. Day 10 - On-board programmable unitary performance and ratio.} \textbf{a} Average sun/shade coincidence probability for target detector pair CC$_{15}$ shows a clear decrease in photon routing performance with respect to the former operation days in Sec.~\ref{sec:operation_day_8} and Sec.\ref{sec:operation_day_9}. The ratio in \textbf{b} is not strongly affected, but a decrease in uPIC health is clearly observed in the obligatory health tests of \textbf{c,d}, as both complex unitaries cannot be resolved anymore. The overall trend is also clearly visible in the strongly reduced coincidence rate in \textbf{f}, and since this damage cannot be correlated with increased dark counts in Fig.~\ref{fig:dark_counts_and_complete_rate} or with changes in experimental parameters, this points to another influence on the payload that we could not identify.}
    \label{fig:op_day_10_probabs}
\end{figure}

Before each HOM dip scan, the laser is power cycled but for the first two shadow measurements in Fig.~\ref{fig:op_day_10_closeup}d no quantum interference can be observed. The latter suggests that the laser did not randomly ``hop'' into the right mode, and that only the last measurement, with a higher count rate, shows a drop below the classical threshold within the HOM dip temperature regime. The slightly shifted minima with respect to the expected dip temperature of $T_{TEC}\sim \ensuremath{32.5\,{}^{\circ}\mathrm{C}}$ and worsened interference visibility can be explained by the frequency mode hopping and visibility loss of the EM laser in Fig.~\ref{fig:mode_hopping}. Nevertheless, the observed dip in counts is within the FWHM of the expected KTP temperature. The overlap between the Beamsplitter and Identity coincidence rate can again be attributed to the internal phase shifter jitter in Fig.~\ref{fig:op_day_1_mzi_jitter}, and it is now our task to further dissect the measured drop in counts. The health-check unitaries, routing photons to CC$_{01}$ and CC$_{05}$ in the temperature regime of interference do not reveal any potential photon loss into other detector modes as given in Fig.~\ref{fig:op_day_10_total_counts_anomaly_check}b (right). The only significant anomaly is observed for CC$_{01}$ in the Identity configuration and for the first data point in Beamsplitter mode, as the canonical noise model from Eq. \ref{accidentals} breaks down here. The continuous offset between signal and noise is similar to the anomalies recorded in Fig.~\ref{fig:op_day_9_outlier_anomaly}, however, now the SPADs or quenching circuit face long-term saturation. The latter is further verified on subsequent operation days in Sec.~\ref{sec:operation_day_12}, suggesting that the SPADs, along with their corresponding quenching circuit, slowly begin to saturate. The analysis of the single $\ensuremath{10\,\mathrm{s}}$ measurement chunks for target detector pair CC$_{15}$ in Fig.~\ref{fig:op_day_10_counts_outliers} shows no anomaly between signal (dashed) and noise floor (solid), hence the final shadow measurement contains no unphysical artifacts.

Normalizing the raw data with respect to the maximum in Eq. \ref{eq:maximum_normalization} reveals potential quantum interference in the last shadow measurement of Fig.~\ref{fig:op_day_10_closeup}d, as here the measured counts drop below the classical threshold (horizontal dashed line) within the expected HOM-dip KTP temperature. Due to the random mode hopping of the laser, we cannot make a definitive statement about the quantum interference nature of this operation day, hence it is only considered optionally in the final HOM-dip analysis in Sec.~\ref{sec:main_figures_hom}. It is, however, important to note that this outcome can be related to the worsened overall interference capability of the laser as seen for measurements on Earth in Fig.~\ref{fig:mode_hopping} and even when normalizing it with the generally harsh mean normalization routine (see Fig.~\ref{fig:hom_interference_add_op_day_10}), it is consistent with the HOM-dip outcome of the complete mission.

This operation day also marked a turning point in terms of payload health, as can be seen in the overview of programmable unitary performance in Fig.~\ref{fig:op_day_10_probabs}. Here, the unitary fidelity for the target detector pair CC$_{15}$ is strongly reduced compared with the two previous operation days. Similarly, the outcome of the uPIC health-check shows that switches $3$ and $4$ could no longer be resolved. The decay in health is also reflected in a significant count rate drop in Fig.~\ref{fig:op_day_10_probabs}f, and this outcome is notable as the measurement parameters (integration time, overvoltage, etc.) are identical to the ones of the former operation days in Sec.~\ref{sec:operation_day_8} and Sec.~\ref{sec:operation_day_9}.
\clearpage
\subsection{Operation Day 11 -  221 Days in Orbit}
\label{sec:operation_day_11}
The former operation day in Sec.~\ref{sec:operation_day_10} marked a turning point in the health of the optical readout- and processing components, and this situation worsened by a severe radiation event after $221$ days in orbit, which further increased the dark counts of SPADs C$_1$ and C$_5$. This strong increase in Fig.~\ref{fig:dark_counts_and_complete_rate} coincides with the period of increased solar activity \cite{clette_silso_2015} within the operational timeline of the payload (summer 2025 to spring 2026). Similar to the earlier spikes in dark counts after $80$ days in orbit, both SPADs stabilised at a higher level, but system health tests on this operation day showed highly fluctuating and low signal levels. In the same spirit as on operation day $4$, it was believed that increasing the SPAD overvoltage  would eventually restore the signal. This is the task for the upcoming operation day in Sec.~\ref{sec:operation_day_12}.
\clearpage
\subsection{Operation Day 12 -  239 Days in Orbit}
\label{sec:operation_day_12}
\begin{figure}[!tbp]
    \centering
    \includegraphics[width=1\textwidth,height=0.72\textheight,keepaspectratio]{ 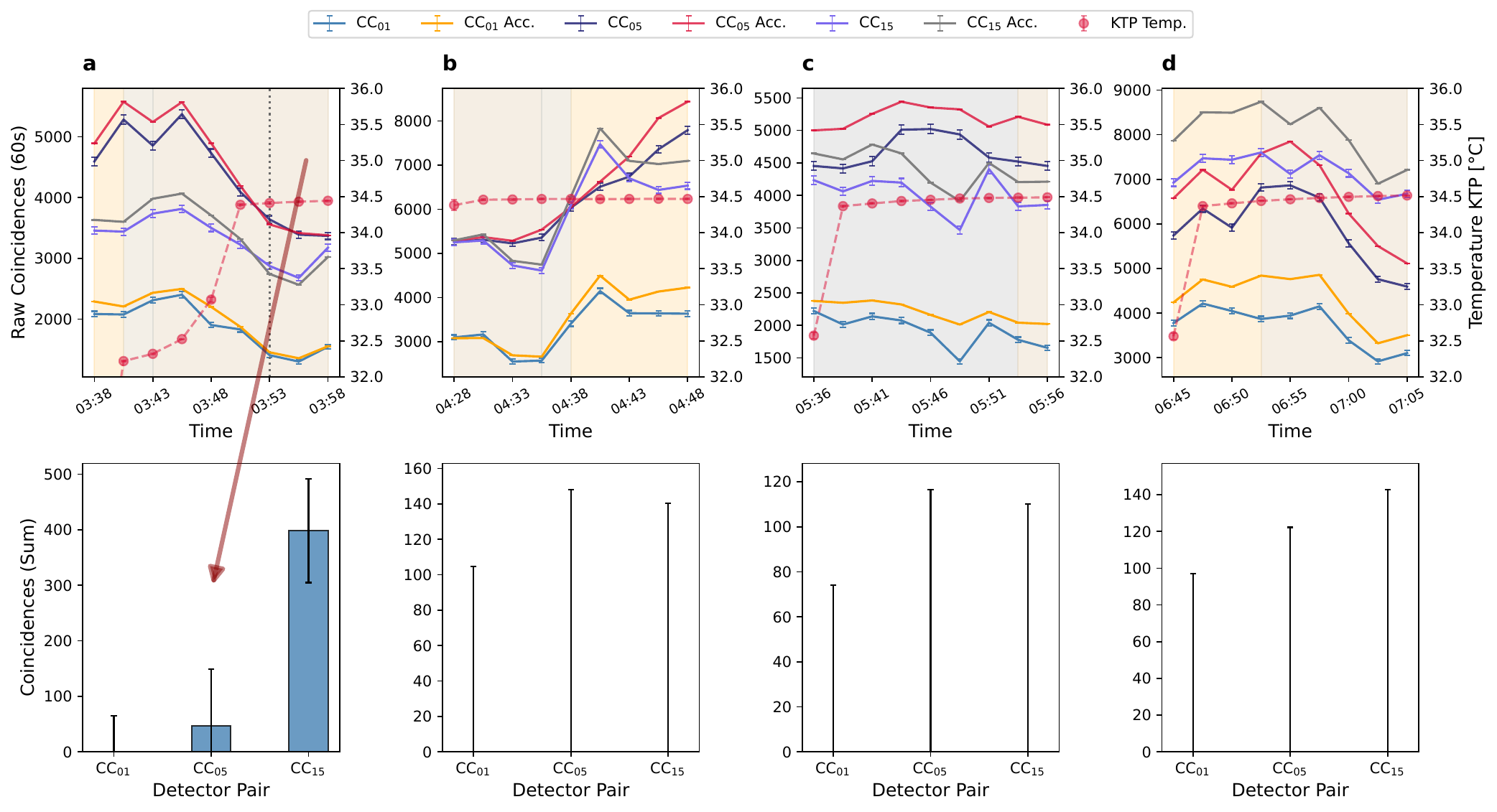}
    \caption[]{\textbf{Op. Day 12 - Identity Reference.} \textbf{a} Identity reference measurement that switches between CC$_{ij}$ with $ij=\{01,05,15\}$ and targets each pair for three data points. Original SPAD $\ensuremath{12\,\mathrm{V}}$ overvoltage shows already severe saturation when payload faces sun and only the last three data points (vertical dashed line) generate data from last identity switch CC$_{15}$; red arrow indicates sum of data points (blue bars). \textbf{b} Same setting as in \textbf{a}, but $\ensuremath{16\,\mathrm{V}}$ overvoltage shows clear saturation in SPADs or SPAD quenching circuit. No more data can be extracted with the canonical noise model in Eq. \ref{accidentals} as the noise floor overshoots the signal. \textbf{c} New breakdown voltage calibration file with higher breakdown voltage and SPAD $C_1$ at $\ensuremath{9\,\mathrm{V}}$, while others at $\ensuremath{12\,\mathrm{V}}$ effectively have the same outcome as \textbf{b}. \textbf{d} Same as in \textbf{c} but here SPAD $C_1$ is at $\ensuremath{15\,\mathrm{V}}$ and this even worsens the outcome.}
    \label{fig:op_day_12_id_check_0}
\end{figure}

In order to regain some faithful signal and eventually recover from the severe SPAD radiation damage after $221$ days in orbit (see Fig.~\ref{fig:dark_counts_and_complete_rate}), the SPAD overvoltage is further tuned. In order to test the effect of different SPAD overvoltages, three ``Identity switches'' are implemented that cycle through the working detector pairs CC$_{01}$, CC$_{05}$, and CC$_{15}$. In addition to these reference measurements, two special unitaries are measured, each encoded with the pixel information (RGB) from two images (testing unitaries). In this case, the images are from a building and water, and these unitaries are measured in a cyclic fashion with the Identity switches. In more detail, one measurement block consists of a total of $9$ datapoints for all the identity references that are grouped into $3$ datapoints for each target detector pair CC$_{ij}$ and $ij=\{01,05,15\}$, respectively. After recording an identity reference data point, the uPIC switches to a special unitary, and this cycling through the unitaries gives us a total of $9$ data points for a pixel unitary. Each data point has an integration time of $\ensuremath{60\,\mathrm{s}}$, and Fig.~\ref{fig:op_day_12_id_check_0}a-d provides an overview of the Identity switches from the SPAD overvoltage scan.

\begin{figure}[!tbp]
    \centering
    \includegraphics[width=1\textwidth,height=0.80\textheight,keepaspectratio]{ 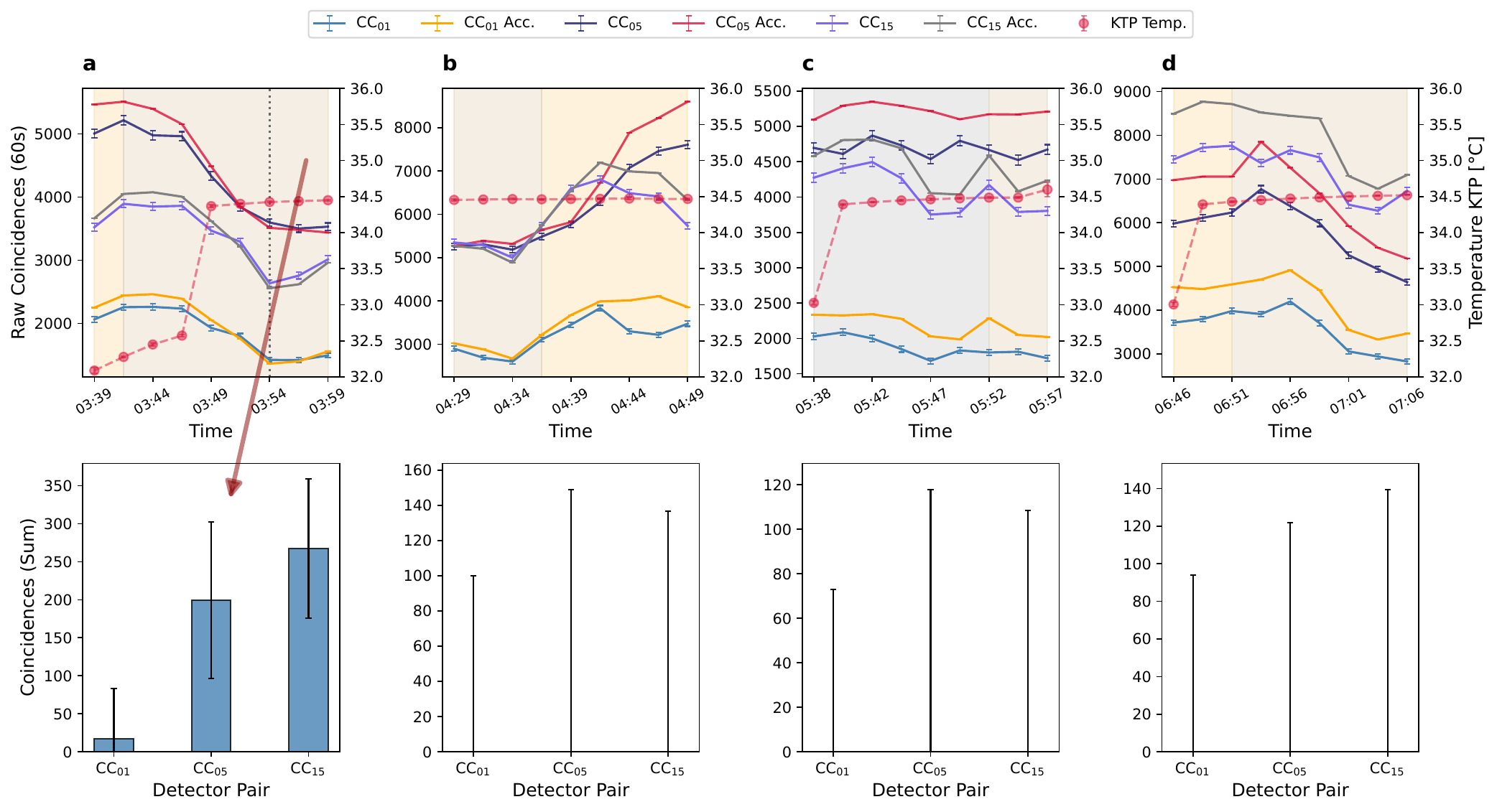}
    \caption[]{\textbf{Op. Day 12 - Unitary Check.} \textbf{a-d} Identical to Fig.~\ref{fig:op_day_12_id_check_0} but here the first pixel unitary of water is measured. Only the last three data points of the first original $\ensuremath{12\,\mathrm{V}}$ SPAD overvoltage (vertical dashed line) measurement are within the shaded area and thus generate a signal; the red arrow shows the sum of the individual data points (blue bars). Too low count rate (see Fig.~\ref{fig:dark_counts_and_complete_rate}b) and too high noise make the measurement unreliable for data analysis, as the measurement is only partially in shade.}
    \label{fig:op_day_12_id_check_1}
\end{figure}

From Fig.~\ref{fig:op_day_12_id_check_0}b-d, it is visible that any SPAD overvoltage above $\ensuremath{12\,\mathrm{V}}$ saturates the SPADs or SPAD board quenching circuit as the canonical noise model in Eq. \ref{accidentals} does not hold anymore. In all measurements, the recorded signal is below the predicted noise floor (accidentals), and this effect is the long-term manifestation of the timetagger anomaly of Fig.~\ref{fig:op_day_9_outlier_anomaly}. The original overvoltage in Fig.~\ref{fig:op_day_12_id_check_0}a, also shows the saturation/anomaly effect for regimes where the payload faces the sun (vertical dashed line), as here the accidentals overshoot the signal. Only the last $3$ datapoints are within a stable shadow regime, and here we can extract the last identity switch of CC$_{15}$ as indicated by the red arrow. This outcome indicates that our measurement capabilities are further constrained, as even with the original overvoltage, robust measurements in the sun are no longer reliable. This is in contrast to, e.g., operation day $9$ in Sec.~\ref{sec:operation_day_9}, where even in the sun we could extract robust measurements such as the ratio or unitary switches. In similar fashion, the measurements of the unitary in Fig.~\ref{fig:op_day_12_id_check_1} only generate reliable data for the original overvoltage and strictly within the shadow of the Earth where SPAD saturation does not occur.

\begin{figure}[!tbp]
    \centering
    \includegraphics[width=1\textwidth,height=0.84\textheight,keepaspectratio]{ 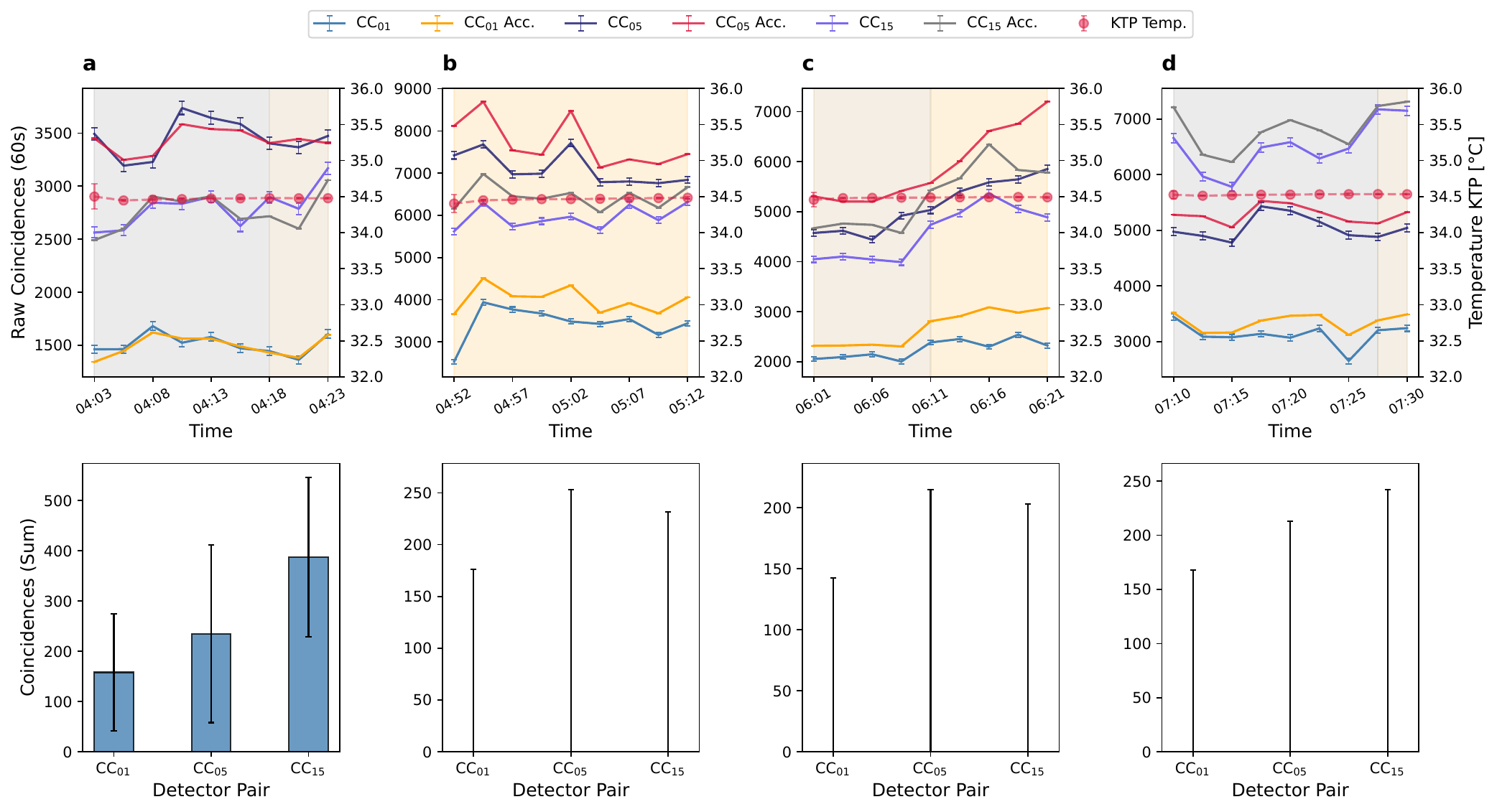}
    \caption[]{\textbf{Op. Day 12 - Routing Check 2.} \textbf{a-d} Identical to Fig.~\ref{fig:op_day_12_id_check_0} but fortunately identity switch measurement in \textbf{a (column)} is in complete shadow, hence all identity switches can be resolved for reference (blue bars). This indicates that the general measurement is more reliable and can be used for data analysis.}
    \label{fig:op_day_12_id_check_2}
\end{figure}

\begin{figure}[!tbp]
    \centering
    \includegraphics[width=1\textwidth,height=0.84\textheight,keepaspectratio]{ 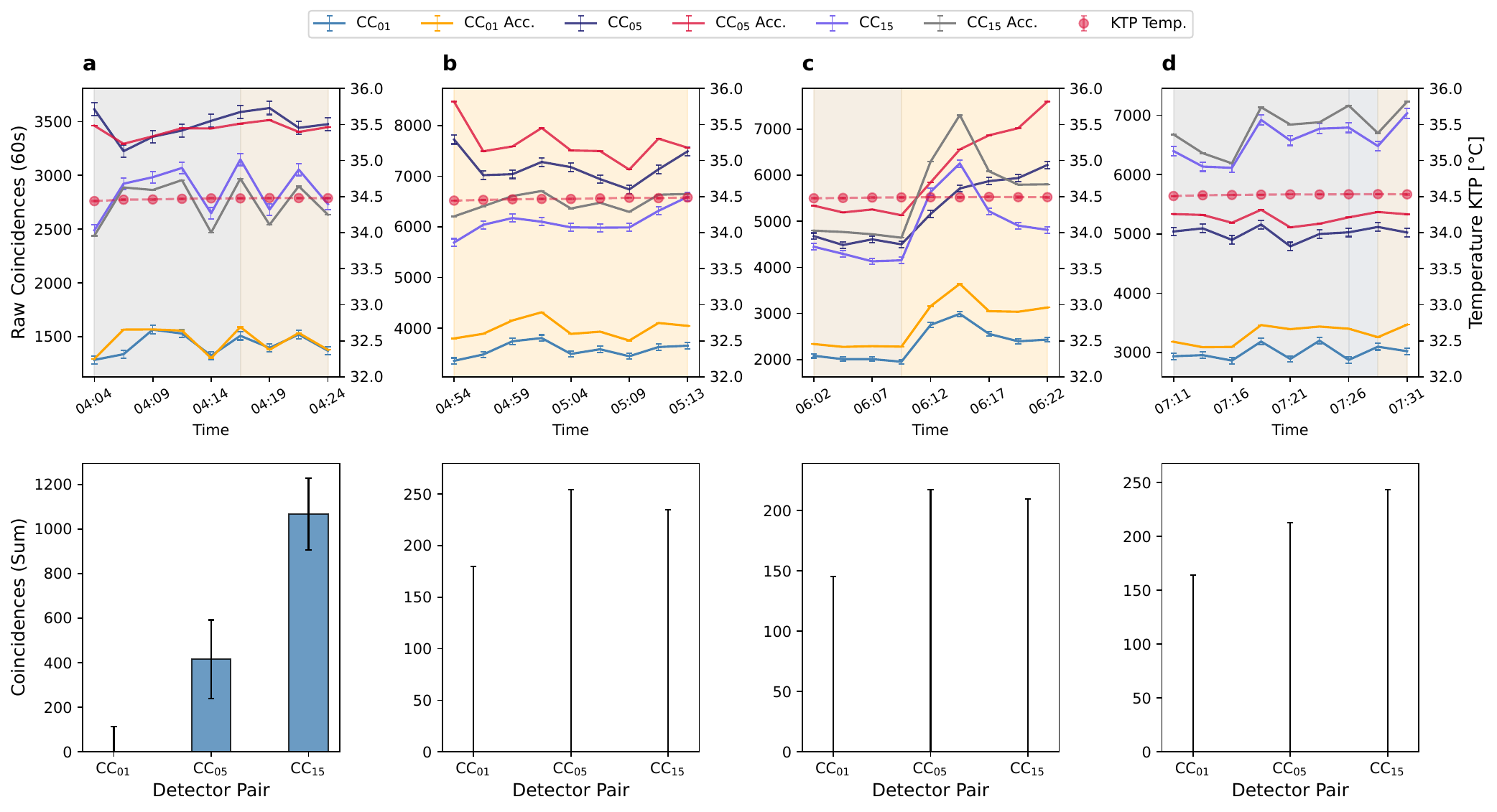}
    \caption[]{\textbf{Op. Day 12 - Unitary Check 2.} \textbf{a-d} Identical as in Fig.~\ref{fig:op_day_12_id_check_2}, but here building unitary is applied, and here the complete measurement is within the shadow and original SPAD overvoltage in \textbf{a (column)} generates reliable data. The individual data points are summed up (blue bars) to improve statistics.}
    \label{fig:op_day_12_id_check_3}
\end{figure}

\begin{figure}[!tbp]
    \centering
    \includegraphics[width=0.4\textwidth,height=0.84\textheight,keepaspectratio]{ 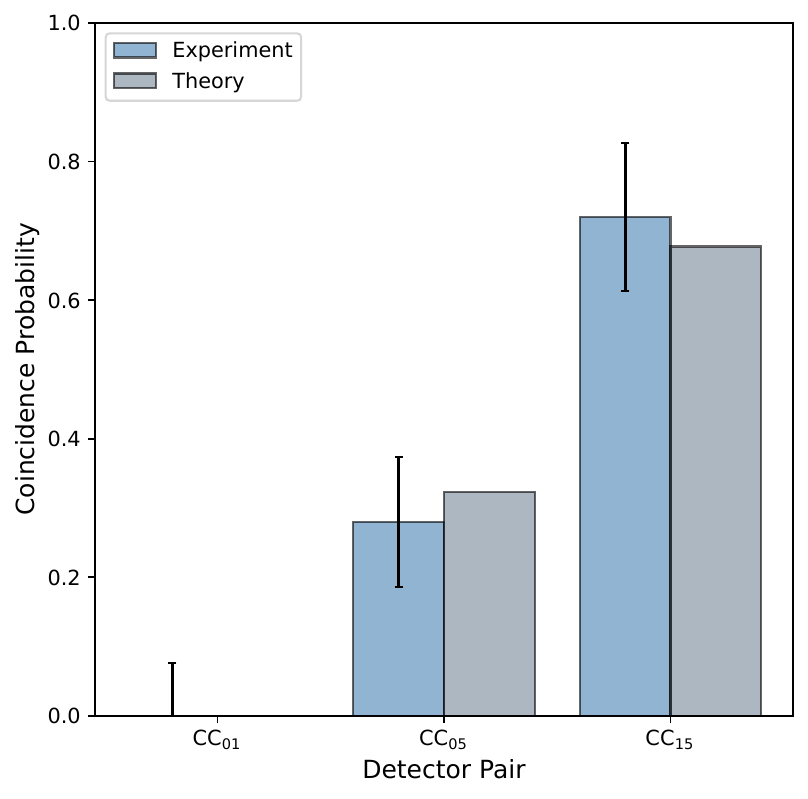}
    \caption[]{\textbf{Op. Day 12 - Building Unitary.} Coincidence probability (blue) of building unitary measured in Fig.~\ref{fig:op_day_12_id_check_3} shows good agreement with theory (grey) and further highlights the robustness of the payload, as even with the recurring saturation events in Fig.~\ref{fig:op_day_12_id_check_3}, meaningful data can still be extracted.}
    \label{fig:op_day_12_ml_unitary}
\end{figure}

In addition to these complications, the decay in overall countrate in Fig.~\ref{fig:dark_counts_and_complete_rate}b on this last operation day has implications on data acquisition, and the statistics from three data points in Fig.~\ref{fig:op_day_12_id_check_1} are too low to consider this dataset as reliable. In contrast, the secondary building unitary in Fig.~\ref{fig:op_day_12_id_check_3}a is meaningful as here the complete measurement is carried out in the shadow. The reliability of this measurement is also confirmed in the identity reference measurement in Fig.~\ref{fig:op_day_12_id_check_2}a, where all three switches can be resolved. Similarly, the other overvoltage settings no longer produce data due to potential saturation of the SPADs or the quenching circuit. The sum of the nine data points (average) from the special building unitary shows good agreement with theory in Fig.~\ref{fig:op_day_12_ml_unitary}, and this outcome again shows the robustness of the payload when considering the severe limitations of the mission at this project stage.

\clearpage
\subsection{Additional information on the results figures in main text}
\label{sec:main_figures_information}
\subsection{Figure~3 - In Orbit data acquisition}
\begin{figure}[!tbp]
    \centering
    \includegraphics[width=1\textwidth,height=0.52\textheight,keepaspectratio]{ 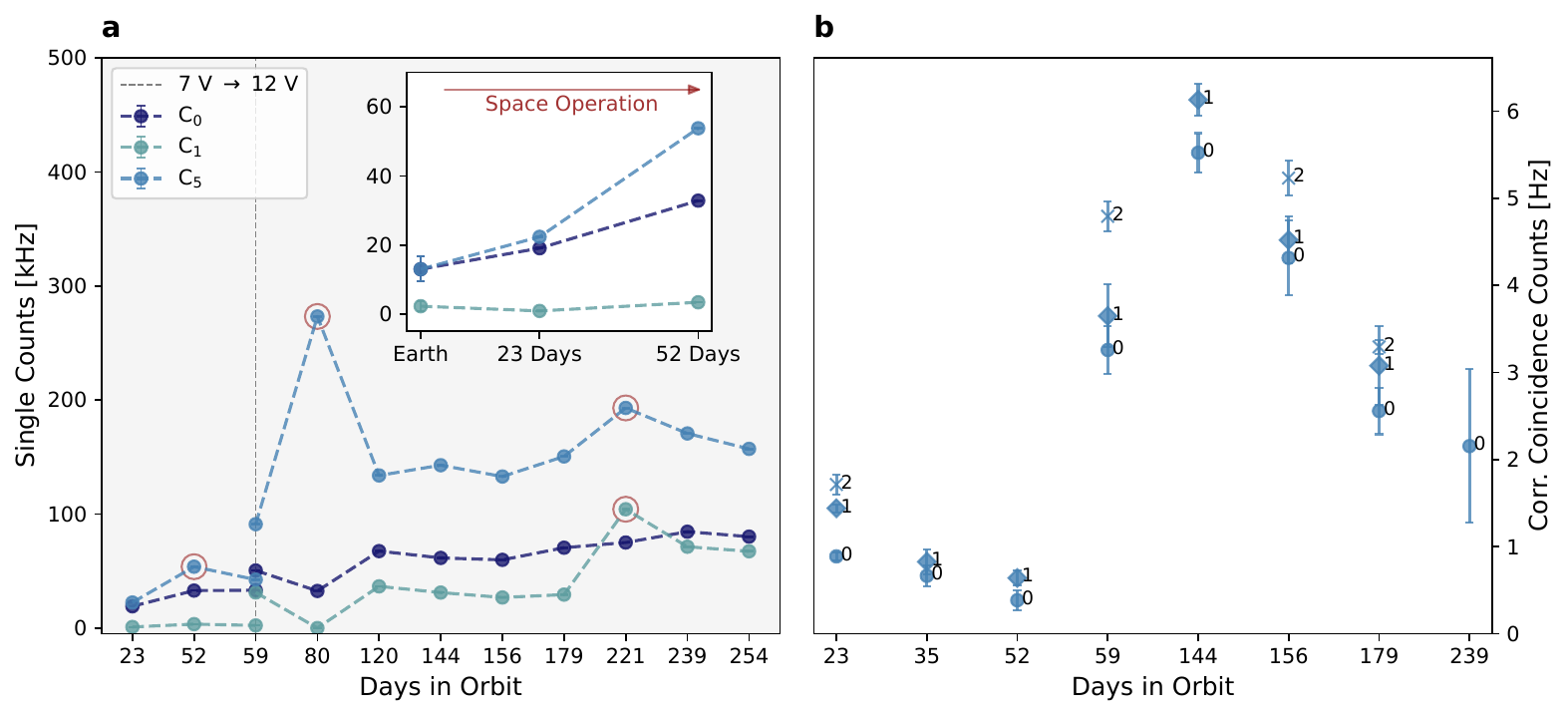}
    \caption[]{\textbf{Dark counts and corrected coincidence rate.} \textbf{a} Measured evolution of the dark count rate in the shadow for all three working detectors C$_0$, C$_1$, and C$_5$; small inset on the right shows a clear increase after $52$ days in orbit, and this can be correlated with radiation damage in SPADs as described in Sec.~\ref{sec:space_as_environment}. The pronounced spikes in count rate (red circles) can be linked to a severe radiation event. The rate eventually stabilises at a higher level than before, indicating damage from the prior radiation event. \textbf{b} Corrected coincidence rate in Hertz starts with $\ensuremath{7\,\mathrm{V}}$ SPAD overvoltage until $52$ days in orbit; decrease in laser output power and radiation damage in SPADs nearly washes out signal at original overvoltage, but overvoltage tuning to $\ensuremath{12\,\mathrm{V}}$ after $59$ days in orbit strongly increases coincidence signal due to optimised SPAD detection efficiency (see Fig.~\ref{fig:spads}). Small enumerated labelings indicate the individual shadow measurements (if available) and underscore how the rate improves over the course of an operation day (payload thermalization). The small decrease in rate after $156$ days in orbit can be explained by data acquisition in the sun (no shadow data available), but the drop in signal after $179$ days points to some other damage. The final measured rate after $239$ days is recorded shortly after the strong spikes in dark counts after $221$ days in orbit in \textbf{a} (red circles), and, together with the further decrease in signal, emphasizes the potential for radiation damage in SPADs. The parallel comparison of dark counts in \textbf{a} and the corrected signal in \textbf{b} highlights the payload's robustness to noise. Despite severe noise contributions, the signal could be improved and maintained over a relatively long time. More detail on individual rate and noise floor contributions is provided in the corresponding sections of the former operation days.}
    \label{fig:dark_counts_and_complete_rate}
\end{figure}

The SPAD dark count evolution in Fig.~3b of the main text includes data from nearly every operation day and is recorded using the same SPAD breakdown-voltage calibration at a fixed SPAD temperature of $T\sim -\ensuremath{15\,{}^{\circ}\mathrm{C}}$. Similar to the HOM dip scan, the dark counts are only meaningful when the payload faces the shadow, hence the overview in Fig.~\ref{fig:dark_counts_and_complete_rate}a and the figure of the main text includes only shadow measurements. There are significant spikes in dark counts (red circles), but all detectors partially heal after a severe radiation event. This is reflected in the measured decrease in count rate after a spike, and the rate stabilises at higher levels than before. The overall trend is a steady increase in dark counts, and the small inset in Fig.~\ref{fig:dark_counts_and_complete_rate}a shows in detail how strongly C$_0$ and C$_5$ were affected by radiation after only $52$ days in orbit. This behaviour agrees with the predicted radiation damage in SPADs, and the underlying concepts are described in more detail in Sec.~\ref{sec:space_as_environment}. Despite this gradual increase in noise and the generally high noise levels after $120$ days in orbit, the payload remained resilient, as meaningful optical experiments could be carried out up to $8$ months after launch. This is even more significant, as during the operational period (summer 2025 to spring 2026), there were peak levels of solar activity \cite{clette_silso_2015}.

The overview of the corrected coincidence rate in Hertz is shown in Fig.~\ref{fig:dark_counts_and_complete_rate}b, and here the strong coincidence count improvement between $52$ and $59$ days in orbit is immediately evident. This efficiency boost from $\sim \ensuremath{1.5\,\mathrm{Hz}}$ to $\sim \ensuremath{6\,\mathrm{Hz}}$ is linked to the SPAD overvoltage tuning on operation day $4$ in Sec.~\ref{sec:operation_day_4}. This boost in overvoltage eventually increases the dark counts as well, but it has such a positive impact on the corrected signal that it outweighs the increased uncorrelated
noise. The notable drop at the beginning of the mission ($35$ to $52$ days in orbit) can be explained by the loss of laser power (see the outgassing problem in Sec.~\ref{sec:tvac}) and by the generally increased SPAD radiation damage. The peak of the registered coincidence rate is measured between $59$ to $156$ days in orbit, and within this timeframe, the majority of data is generated. After $179$ days of space operation, the outgassing-related power loss of the laser should have saturated, so the loss in count rate points to a different source of damage. The last operation day in Sec.~\ref{sec:operation_day_12} further highlights that radiation events, such as the one after $221$ days in orbit (see the red circles in the dark count overview), have a strong negative influence on the measured signal rate.
\subsection{Figure~4 - On-board programmable unitary operations}
\begin{figure}[!tbp]
    \centering
    \includegraphics[width=1\textwidth,height=0.74\textheight,keepaspectratio]{ 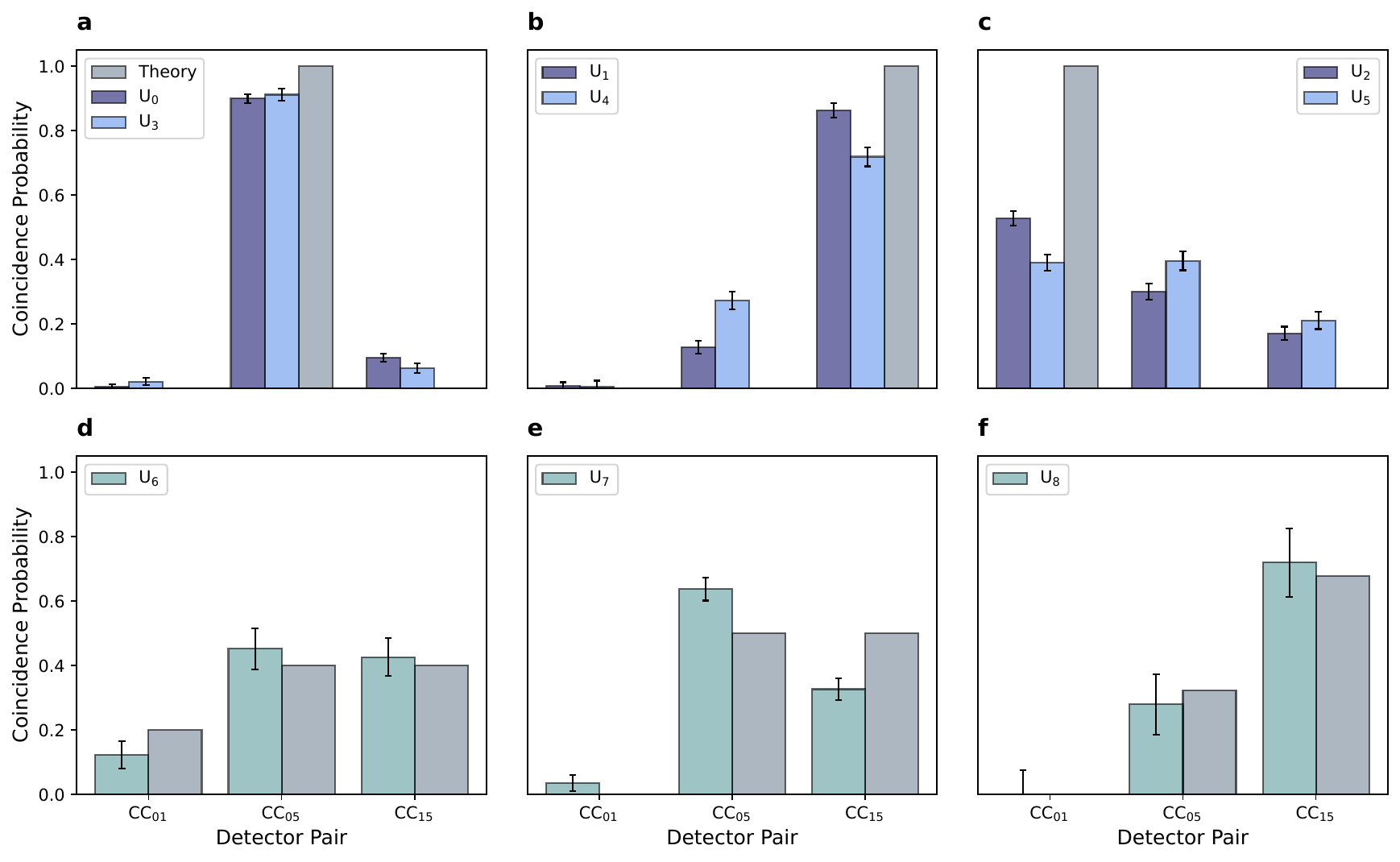}
    \caption[]{\textbf{Overview - On-board programmable unitary operations.} \textbf{a} Identical to the weighted average result of operation day $1$ in Fig.~\ref{fig:op_day_1_probabs}a, while \textbf{b-e} gives the average results for each unitary for the two operation days $8$ and $9$. The list concludes with the pixel-encoded building unitary from Sec.~\ref{sec:operation_day_12} in \textbf{f}. The overview shows, in general, good on-board programmable unitary performance, and it is also noteworthy that the target unitaries are very specific. The latter requires precise control of each MZI, and, given the harsh, uncontrollable conditions in space, these results demonstrate the payload's general robustness. More details on the results can be found in the individual operation days, and Table~\ref{tab:unitary_fidelities} corresponds to the relevant sections.}
    \label{fig:fm_unitary_fidelity_individual_unitaries}
\end{figure}

The overview in Fig.~4 of the main text provides insight into the on-board programmable unitary performance; more precisely, Fig.~4c compares the unitary fidelity for all measured unitaries $U_0$ to $U_8$. The chosen fidelity measure is related to the classical analog of the quantum state fidelity and is known as the ``Bhattacharyya coefficient,'' which quantifies the ``closeness'' of two random statistical ensembles \cite{dodge_oxford_2003}. In a formal way, the extracted average coincidence probabilities $\bar{p}_{ij}$ for all working detector pairs $ij \in \{01,05,15\}$ with its error $\sigma_{p_{ij}}$ from Eq. \ref{eq:weighted_average} is further processed by plugging it into the fidelity estimator $F$
\begin{equation}
    F=\sum_{ij} \sqrt{t_{ij}\cdot \bar{p}_{ij}} \quad \text{and} \quad \sigma_F=\frac{1}{2} \sqrt{\frac{t_{01} \cdot \sigma^2_{\bar{p}_{01}}}{\bar{p}_{01}}+\frac{t_{05} \cdot \sigma^2_{\bar{p}_{05}}}{\bar{p}_{05}}+\frac{t_{15} \cdot \sigma^2_{\bar{p}_{15}}}{\bar{p}_{15}}}
    \label{eq:fidelity_measure}
\end{equation}
In other words, $\bar{p}_{ij}$ is the experimental detection frequency for the $ij$th output, while $t_{ij}$ corresponds to the theoretical detection frequency for the $ij$th output \cite{yin2025experimental}. We include coincidence probability data from relevant operation days, and that is U$_0$, U$_3$ from operation day $1$ in Sec.~\ref{sec:operation_day_1}, which targets detector pair CC$_{05}$. The other two target detector pairs CC$_{01}$ (U$_2$, U$_5$) and CC$_{15}$ (U$_1$, U$_4$) are measured with higher precision ($\ensuremath{60\,\mathrm{s}}$ integration time per data point) and nearly identical measurement settings on operation day $8,9$ (see Sec.~\ref{sec:operation_day_8} and Sec.~\ref{sec:operation_day_9}, respectively) and also operation day $10$. The latter showed a strong decrease in overall system health, e.g., a significant drop in countrate, worsened unitary fidelity and no uPIC health test response (see Sec.~\ref{sec:operation_day_10}), hence it is excluded from this analysis. For the healthy operation days $8$ and $9$ we take the average of the obtained fidelities, including the special unitaries U$_6$ and U$_7$ that are acquired after each HOM dip scan. The final measured unitary corresponds to the pixel encoded building unitary from Sec.~\ref{sec:operation_day_12} and is labeled U$_8$. The following table in Table~\ref{tab:unitary_fidelities} gives an overview of all the measured unitary fidelities, and Fig.~\ref{fig:fm_unitary_fidelity_individual_unitaries} shows the corresponding measured average coincidence probabilities with the related theory (grey bars). The average unitary fidelity is $F= 0.888 \pm 0.011$ and improves to $F= 0.949 \pm 0.014$, once the two significant outliers U$_2$ and U$_5$ are removed. This outcome is significant and indicates that the uPIC and related optical components still operate well far from their optimal operating temperature and are resistant to continuous changes in environmental conditions (see, e.g., payload temperature in Fig.~\ref{fig:temperature-control}).
\begin{table}
        \centering
\caption[]{\textbf{Unitary fidelities and ratio.} An overview of all (averaged) unitary fidelities via Eq. \ref{eq:fidelity_measure} is given and shows overall good uPIC performance as the majority of measured unitaries is higher than $0.9$. This outcome is critical, as the in-principle thermally sensitive uPIC is continuously exposed to the harsh thermal environment of space. The latter manifests as relatively strong temperature drifts, as seen in the measured payload temperature data in Fig.~\ref{fig:temperature-control}. This again underlines that the uPIC (and all other optical components) still operate well far off from their typical optimal operating temperature (room temperature). The temperature offset might be responsible for the two significant fidelity outliers, U$_2$ and U$_5$, suggesting a uPIC calibration issue for these specific unitaries. The latter has no influence on the ratios of Beamsplitter unitaries (U$_3$, U$_4$, and U$_5$) vs. Identity unitaries (U$_0$, U$_1$, and U$_2$), as the measured average ratio is close to the theoretical one of $R_T=0.5$. The slight offset in ratio can be explained by the increased thermal sensitivity of the Beamsplitter unitary described in Sec.~\ref{sec:operation_day_1}. Note that the measured unitaries from operation day $8$ and $9$ are further averaged, and more details on the individual results can be extracted from the corresponding sections, which are referred to in the last column.}
\label{tab:unitary_fidelities}
        \begin{tabular}{|c c c c|}
            \hline
            Unitary & Fidelity & Ratio Beamsplitter vs. Identity & Operation Day\\ [0.5ex]
            \hline\hline
            U$_0$ & $0.949\pm 0.007$ & N.A. & $1$ - Sec.~\ref{sec:operation_day_1} \\
            \hline
            U$_1$ & $0.929 \pm0.012$ & N.A.  & $8$ and $9$ - Sec.~\ref{sec:operation_day_8} and Sec.~\ref{sec:operation_day_9}\\
            \hline
            U$_2$ & $0.727 \pm 0.016$ & N.A. & $8$ and $9$ - Sec.~\ref{sec:operation_day_8} and Sec.~\ref{sec:operation_day_9} \\
            \hline
            U$_3$ & $0.955\pm 0.010$ & $0.547 \pm 0.043$  & $1$ - Sec.~\ref{sec:operation_day_1}\\
            \hline
            U$_4$ & $0.848\pm0.018$ & $0.536 \pm 0.023$  & $8$ and $9$ - Sec.~\ref{sec:operation_day_8} and Sec.~\ref{sec:operation_day_9}\\
            \hline
            U$_5$ & $0.624\pm0.020$ & $0.567 \pm 0.026$ & $8$ and $9$ - Sec.~\ref{sec:operation_day_8} and Sec.~\ref{sec:operation_day_9} \\
            \hline
            U$_6$ & $0.969\pm0.026$ & N.A.   & $8$ and $9$ - \ref{sec:operation_day_8} and Sec.~\ref{sec:operation_day_9}\\
            \hline
            U$_7$ & $0.992\pm0.050$ &N.A.    & $8$ and $9$ - \ref{sec:operation_day_8} and Sec.~\ref{sec:operation_day_9}\\
            \hline
             U$_8$ & $0.999\pm0.072$ & N.A.  & $12$ - Sec.~\ref{sec:operation_day_12}\\
            \hline

        \end{tabular}

    \end{table}
\subsection{Figure~5 - Hong-Ou Mandel interference in space}
\label{sec:main_figures_hom}
This section concerns the result shown in Fig.~5a of the main text and explains the necessary steps to extract the Hong-Ou Mandel dip in coincidences. Data is considered from operation days where the laser is considered ``healthy'', i.e., when the laser still had its SLM property (see Sec.~\ref{sec:tvac} for more details on laser damage). The relevant operation days include operation day $1$ (Sec.~\ref{sec:operation_day_1}), operation day $4$ (Sec.~\ref{sec:operation_day_4}) and eventually operation day $8$ (Sec.~\ref{sec:operation_day_8}). These operation days also constitute consecutive ``healthy'' operation days in a historical context, as the intervening days did not produce reliable data due to the aforementioned laser power loss or radiation damage to SPADs (refer to individual operation days for more details). In this sense, the HOM dip data resembles a chronological picture of the overall mission and payload health, and to obtain a complete temperature scan, all shadow measurements from the relevant operation days are combined. As the count rate varies significantly within a single operation day (payload thermalization) and even more between different operation days (SPAD overvoltage tuning), the individual shadow measurements are normalized with respect to their own maximum/mean.

To account for uncertainties across individual operating days, we also take the inverse-variance-weighted arithmetic mean of the entire dataset to obtain a robust overview of the underlying raw data. More specifically, the data from the three different operation days measure the same physical quantity, i.e., Beamsplitter unitary for a specific KTP temperature $T_{TEC}$, but the noise contributions and also measurement settings differ between the individual days. This is reflected not only in the varying coincidence window, integration time, or SPAD overvoltage, but also in the changing environmental and noise conditions on a specific operation day. For these reasons, data has to be averaged with the weighted arithmetic mean in Eq. \ref{eq:weighted_average_hom} to faithfully gauge the different measurement conditions of the same physical quantity. Equivalent to Eq. \ref{eq:weighted_average}, we define the weighted average for the HOM-dip scan as
\begin{equation}
    \bar{x}=\sum_i \tilde{w}_ix_i \quad \text{with} \quad \tilde{w}_i=\frac{w_i}{\sum_jw_j} \quad \text{and} \quad  w_i=\frac{1}{\sigma_{x_i}^2} \quad \text{with} \quad \sigma_{\bar{x}}=\sqrt{\frac{1}{\sum_i w_i}}.
    \label{eq:weighted_average_hom}
\end{equation}
Here, $x_i$ and its error $\sigma_{x_i}$ correspond to the already normalized coincidence counts for the maximum in Eq. \ref{eq:maximum_normalization} or for the mean normalization routine in Eq. \ref{eq:mean_normalization}. In this sense, we first normalize the individual shadow data sets by their maximum or mean, and then compute a weighted average via Eq. \ref{eq:weighted_average_hom}. This approach gives a robust estimate of the underlying data and should show us general trends.

\begin{figure}[!tbp]
    \centering
    \includegraphics[width=1\textwidth,height=0.67\textheight,keepaspectratio]{ 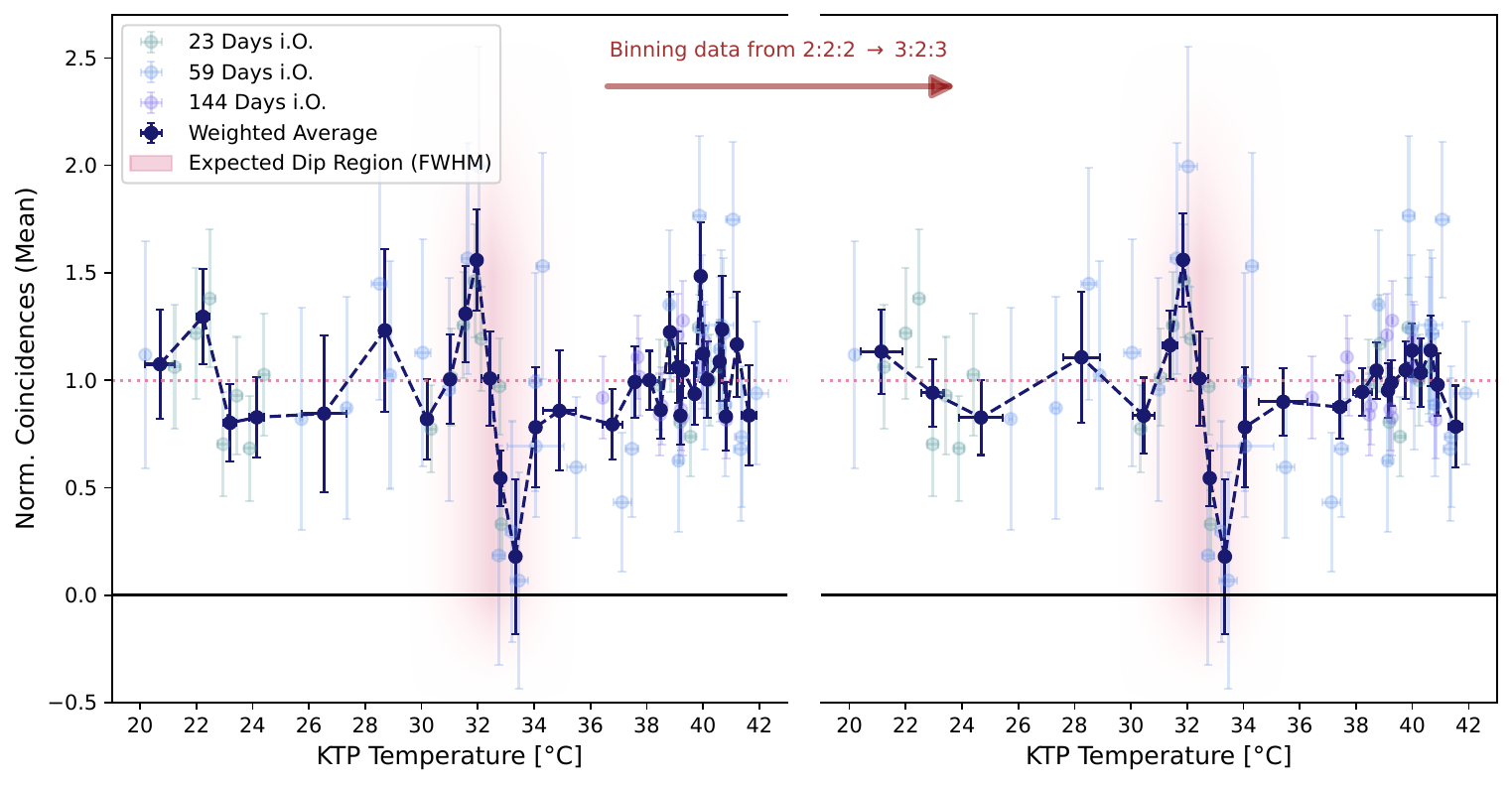}
    \caption[]{\textbf{HOM dip mean normalization and binning.} Weighted arithmetic mean of complete HOM dip data (dark blue) from all relevant operation days (shaded data points in the background refer to days in orbit) gives Hong-Ou Mandel interference overview for KTP temperatures from $T_{TEC}\sim \ensuremath{20\,{}^{\circ}\mathrm{C}}$ to $T_{TEC} \sim \ensuremath{42\,{}^{\circ}\mathrm{C}}$. KTP temperature data is binned in two (left), and individual data points are normalized with respect to the mean of the corresponding shadow measurement; mean normalization is helpful in showing a general trend for noisy data, and the weighted average shows already a significant dip in coincidences within the expected HOM dip KTP temperature regime (red-shaded area - FWHM). This result already highlights that the significant dip in coincidences is not just a fluctuation around the mean (red dashed line). The peak in counts right before the dip can be linked to a MZI jitter (see Fig.~\ref{fig:op_day_1_mzi_jitter}) and potential spectrum mismatch, which can lead to beating \cite{zhou2019second}. The red arrow indicates further processing of data outside the FWHM of the dip, and here data is grouped in bins of three to smooth out ``non-interference'' areas.}
    \label{fig:hom_dip_mean_binning}
\end{figure}

The list of measured KTP temperatures forms a relatively broad HOM scan from $T_{TEC}\sim \ensuremath{20\,{}^{\circ}\mathrm{C}}$ to $T_{TEC}\sim \ensuremath{42\,{}^{\circ}\mathrm{C}}$, and Fig.~\ref{fig:hom_dip_mean_binning} gives the first overview of the combined data, which utilizes the robust mean normalization in Eq. \ref{eq:mean_normalization}. As discussed in Sec.~\ref{sec:operation_day_1}, this normalization procedure helps in highlighting an overall trend in noisy data. Noise contributions in our data are generally high, mainly due to the high accidental noise floor. Error propagation in Eq. \ref{eq:corrected_signal} carries the error of the raw timetagger signal; hence, the error bars (one standard deviation) of the corrected signal are relatively large. The potential jitter in MZI performance (compare Fig.~\ref{fig:op_day_1_mzi_jitter}) also makes a mean normalization meaningful, as the countrate of the Beamsplitter unitary is sometimes close to that of Identity. In this case, the standard approach of normalizing to Beamsplitter coincidence counts to its maximum could mimic a dip in coincidences (see last shadow measurement in Sec.~\ref{sec:operation_day_4}), therefore, it is critical to first analyze HOM dip data with the mean normalization routine.

The weighted average of the individual operation days is emphasized as a dark blue graph in Fig.~\ref{fig:hom_dip_mean_binning}, while the underlying data points from the related operation days (days in orbit) are depicted in the background. Moreover, the weighted average is binned by KTP temperature, thereby summarizing similar data points. In Fig.~\ref{fig:hom_dip_mean_binning} (left), the KTP temperature data is binned in terms of two temperature steps, i.e., the temperature is grouped in bins of two. This approach already reveals a significant pattern of the raw data, and an evident dip in coincidences can be found at the expected HOM dip KTP temperature of $T_{TEC}\sim \ensuremath{32.5\,{}^{\circ}\mathrm{C}}$ (red-shaded area). To further smooth the shape of the weighted average, the KTP temperature is binned into three steps outside the FWHM dip region, while inside the dip region, the binning remains at two. The outcome of this procedure is shown in Fig.~\ref{fig:hom_dip_mean_binning} (right) and reduces the small fluctuations in coincidences outside the HOM dip regime.

Observing the overall behaviour also outlines the robustness of the weighted average with respect to fluctuations as can be seen in the specific KTP temperature steps of $T_{TEC} \sim \ensuremath{35\,{}^{\circ}\mathrm{C}}$ or $T_{TEC} \sim \ensuremath{37\,{}^{\circ}\mathrm{C}}$, where individual data points of the last shadow measurement of operation day $4$ are fluctuating strongly around the mean (see Fig.~\ref{fig:op_day_4_bs_normalized_mean}). These fluctuations do not affect the weighted average because additional measurements from operation day $8$ suppress them. The peak in the weighted average right before the dip might have two reasons. First, the Beamsplitter unitary could be closer to the Identity unitary due to jitter in MZI performance (see Sec.~\ref{sec:operation_day_1} for more details), which might explain the global maximum. Second, it is well known that just outside the dip region, an HOM interference experiment can exhibit beating behaviour if the interfering photons have different spectra \cite{zhou2019second}. This phenomenon could be validated for measurements on Earth, as the FM showed similar beating behaviour (see HOM dip before final integration in Fig.~\ref{fig:hom_interference_earth}a). It is important to note that no jitter in MZI performance can explain the dip in coincidences within the red-shaded area (see explanation in Fig.~\ref{fig:op_day_1_mzi_jitter}); therefore, the dip of the generally robust weighted average cannot be attributed to MZI jitter.

\begin{figure}[!tbp]
    \centering
    \includegraphics[width=1\textwidth,height=0.69\textheight,keepaspectratio]{ 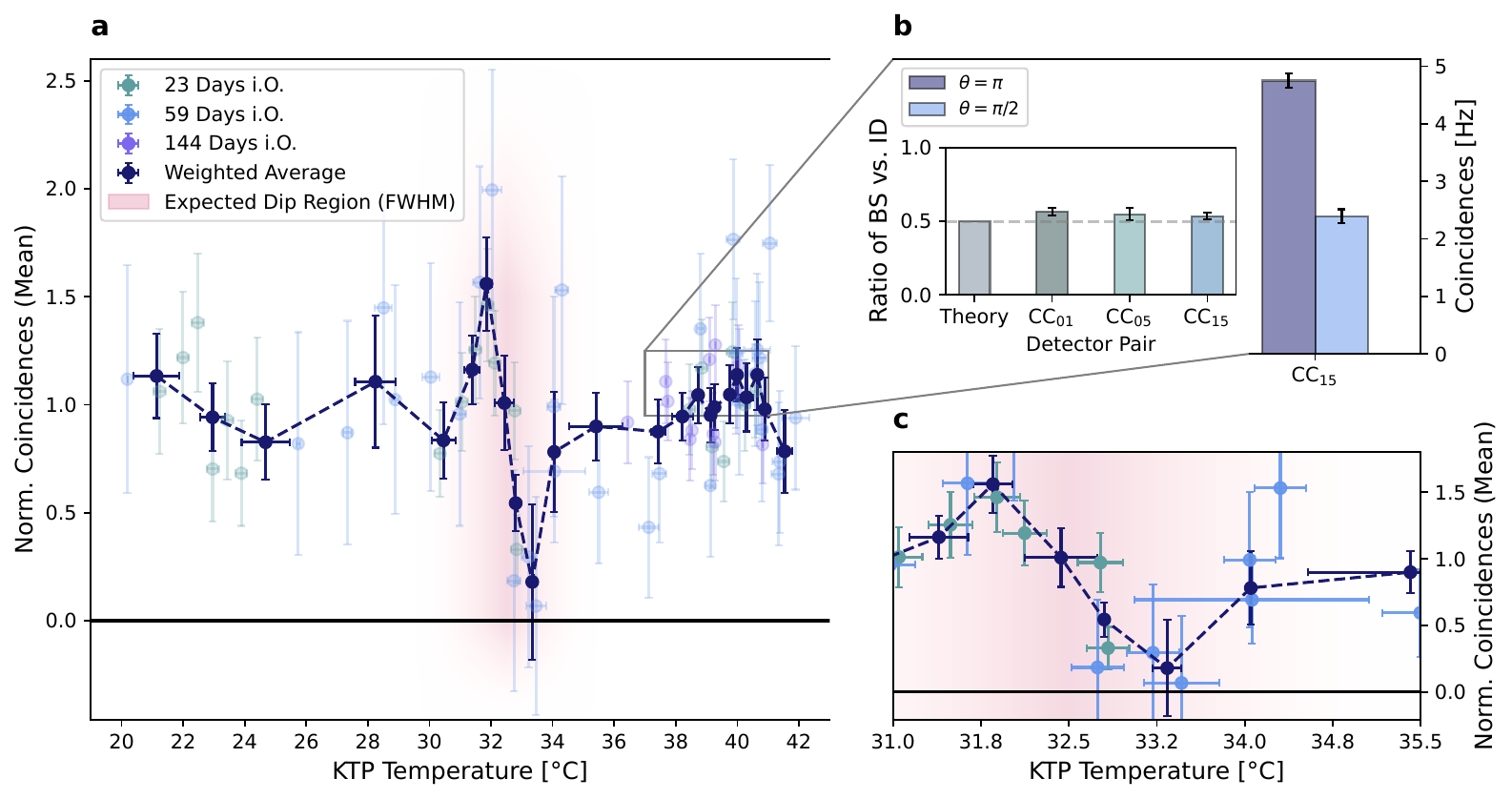}
    \caption[]{\textbf{HOM dip mean binning and ratio.} Alternative approach to Fig.~5 of the main text and \textbf{a} shows the weighted average (dark blue) of the individual shadow measurements from different operation days (transparent data points in the background); the number of days (e.g., $23$ days) corresponds to days in orbit. The normalization routine is with respect to the mean of individual shadow measurements, and this approach helps identify trends in highly noisy data. The overall trend of the weighted average in \textbf{a} and the close-up in \textbf{c} highlight the significant dip in coincidences within the expected HOM-dip KTP temperature regime (red-shaded area) and are consistent with quantum interference with a visibility of $V_{Max}= 0.885 \pm 0.231$. \textbf{b} Inset identifies behaviour of distinguishable photons, and here exemplatory Identity countrate (dark blue) is compared to Beamsplitter countrate (light blue); measured rate of the latter is half of the former. This outcome is consistent across all detector pairs (inset) and aligns well with theory (grey bar).}
    \label{fig:hom_interference_and_ratio_mean}
\end{figure}

The routine of normalizing HOM dip data to the mean is summarized in Fig.~\ref{fig:hom_interference_and_ratio_mean}, and this normalization method can be seen as an alternative approach to Fig.~5 of the main text. It further strengthens the overall two-photon interference result as it also shows that the dip in coincidences is not just a fluctuation around the mean. This is also highlighted in the close-up of the expected dip region in Fig.~\ref{fig:hom_interference_and_ratio_mean}c, where the dip in coincidences of the individual operation days and the combined weighted average is clear.

While the mean normalization is useful for showing general trends in noisy data, care is required when extracting additional information about Hong-Ou Mandel interference visibility. To see this, we first define the canonical approach to estimate the visibility of the HOM dip, which is extracted from the minimum and maximum coincidence probability as
 \begin{equation}
     V_{Max}=1-\frac{P_{min}}{P_{max}} \quad \text{and} \quad \sigma_{V_{Max}}=\sqrt{\frac{P^2_{min} \cdot \sigma^2_{P_{max}}+P^2_{max} \cdot \sigma^2_{P_{min}}}{{P^4_{max}}}}.
     \label{eq:visibility_max}
 \end{equation}
 Here, $P_{min}$ corresponds to our minimal value from the weighted average in Eq. \ref{eq:weighted_average_hom}, while $P_{max}$ relates to the maximum of the weighted average. Both of these expressions also have a related error $\sigma^2_{P_{min}}$ or $\sigma^2_{P_{max}}$. For the mean normalization, we have to modify this expression accordingly
 \begin{equation}
     V_{Mean}=1-\frac{P_{min}}{P_{mean}} \quad \text{and} \quad \sigma_{V_{Mean}}=\sqrt{\frac{P^2_{min} \cdot \sigma^2_{P_{mean}}+P^2_{mean} \cdot \sigma^2_{P_{min}}}{{P^4_{mean}}}} \quad \text{with} \quad \sigma_{P_{mean}}=\frac{1}{N}\sqrt{{\sum_i \sigma^2_{P_i}}}
     \label{eq:visibility_mean}.
 \end{equation}
Here, the error of the mean $\sigma_{P_{mean}}$ comes from standard error propagation on the individual errors of the weighted average data set $\sigma_{P_i}$. Note that this error estimation holds only for the uncorrelated weighted average in Eq. \ref{eq:weighted_average_hom}, but not for single data sets.

Despite the similarities between the two expressions for evaluating the Hong-Ou Mandel visibility in Eq. \ref{eq:visibility_max} and Eq. \ref{eq:visibility_mean}, the latter is strongly dependent on the choice of the experimental configuration. This can be seen directly in an example where most data points lie within the maximum dip in coincidences, but only one lies outside the dip. The mean visibility estimate in Eq. \ref{eq:visibility_mean} can then effectively wash out any interference signature, as the dip in coincidences is essentially normalized by the datapoints ``inside the dip''. The maximum normalization, on the other hand, is unaffected by this method, making it more canonical.

In the case of mean normalization in Fig.~\ref{fig:hom_interference_and_ratio_mean}, the weighted average essentially acts as a robust filter of the entire data set. In this spirit, it is meaningful to consider the maximum visibility estimate in Eq. \ref{eq:maximum_normalization} to faithfully extract the interference visibility of the already ``filtered'' mean data. The observed HOM dip visibility is $V_{Max}= 0.885 \pm 0.231$, consistent with quantum interference (1.7 standard deviations above the classical bound), and this clear sign of two-photon interference in the mean normalization routine also allows us to inspect the data with respect to the maximum normalization.

%Moreover, the mean visibility is way harsher on the error bars as can be directly seen in the error propagation of Eq. \ref{eq:visibility_mean} when compared to the maximum case in Eq. \ref{eq:visibility_max}. As $\sigma^2_{P_{mean}}$ is generally smaller than $\sigma^2_{P_{max}}$, the error bars are usually larger. This is also a general problem when normalizing the raw counts with respect to the mean as shown in Eq. \ref{eq:mean_normalization}. Besides these limitations, the classical threshold of $p=0.5$ is defined only for the maximum-visibility approach; for these reasons, the HOM-dip visibility with respect to the mean is provided only as an optional quantity. The resulting mean interference visibility from the HOM dip in Fig.~\ref{fig:hom_interference_and_ratio_mean} reads $V_{Mean}=(0.811 \pm 0.378$). The problematic error propagation under mean normalization is directly reflected in this visibility outcome, as the error is large. The visibility itself is sufficiently high, reflecting the strong HOM interference pattern in Fig.~\ref{fig:hom_interference_and_ratio_mean}; however, due to the difficulty of interpretation, the mean visibility should not be used as a faithful measure of interference visibility.\\[12pt]
\begin{figure}[!tbp]
    \centering
    \includegraphics[width=1\textwidth,height=0.71\textheight,keepaspectratio]{ 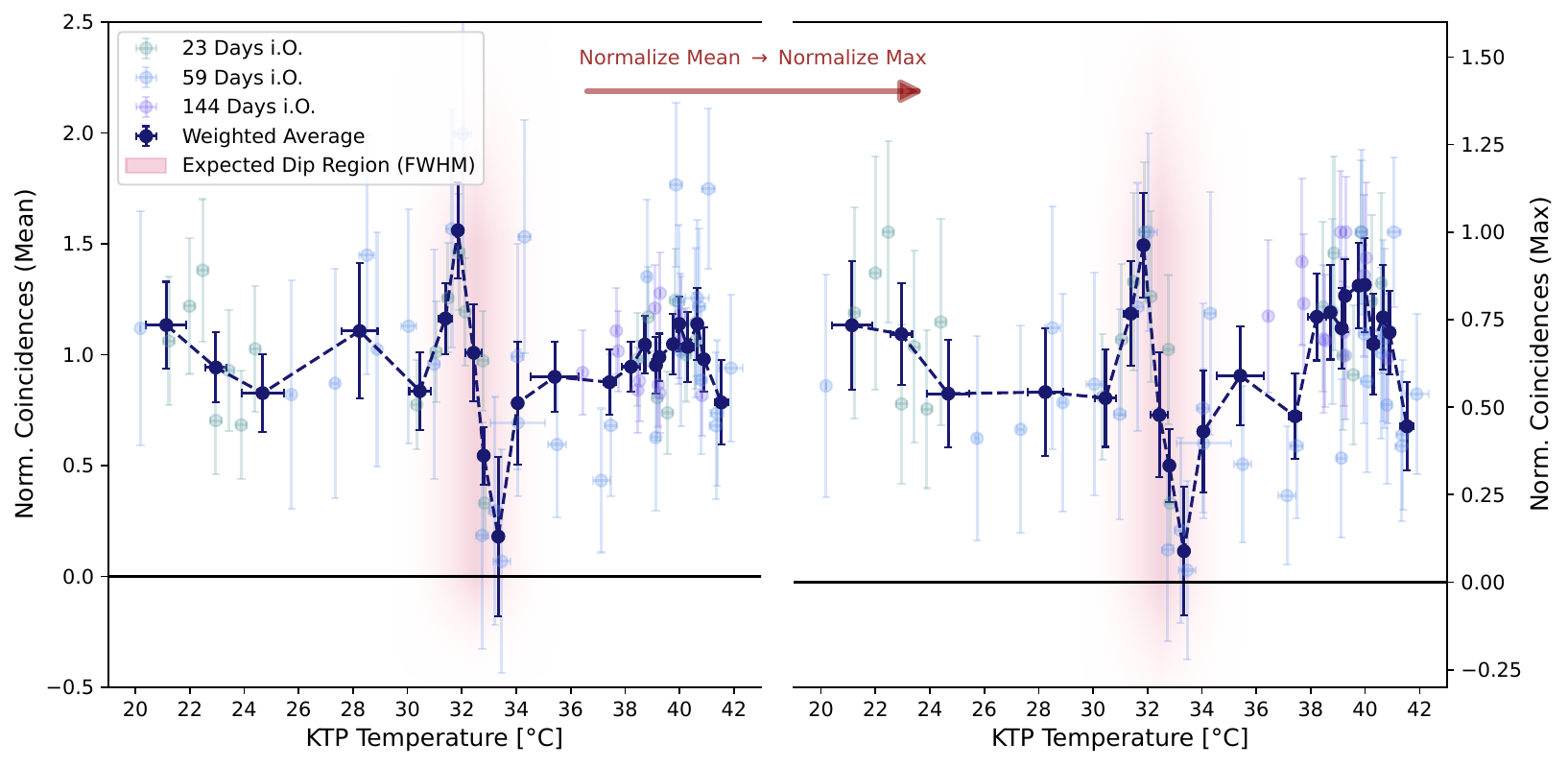}
    \caption[]{\textbf{HOM dip normalization transition.} HOM dip data normalized to the mean of data (left) is compared to the identical dataset but normalized to the maximum (right). The transition to the maximum normalization is more canonical, but due to the Beamsplitter unitary fluctuations (more details in Sec.~\ref{sec:operation_day_1}), some Beamsplitter data points are effectively normalized with respect to the Identity coincidence rate. This can mimic a slight dip in coincidences (blue data points at $T_{TEC}\sim \ensuremath{37\,{}^{\circ}\mathrm{C}}$). Further analysis of these data points in Sec.~\ref{sec:operation_day_4} reveals that they cannot be linked to two-photon interference but are more related to fluctuations in Beamsplitter stability; the weighted average (dark blue) is not strongly affected by these outliers, and minima in coincidences is evident within the red-shaded area, where HOM interference is expected.}
    \label{fig:hom_dip_mean_to_nax}
\end{figure}

The transition between both procedures is marked in Fig.~\ref{fig:hom_dip_mean_to_nax} and as expected, the weighted average of the maximum normalization is more sensitive to outliers as the fluctuations from the last shadow measurement of operation day $4$ ($T_{TEC} \sim \ensuremath{35\,{}^{\circ}\mathrm{C}}$ or $T_{TEC} \sim \ensuremath{37\,{}^{\circ}\mathrm{C}}$ - light blue data points from $59$ days in orbit) have a higher contribution. This slightly changes the shape of the weighted average; however, the overall trend stays the same, as the dip within the expected temperature regime (red-shaded area) is the distinct global minima.

\begin{figure}[!tbp]
    \centering
    \includegraphics[width=1\textwidth,height=0.75\textheight,keepaspectratio]{ 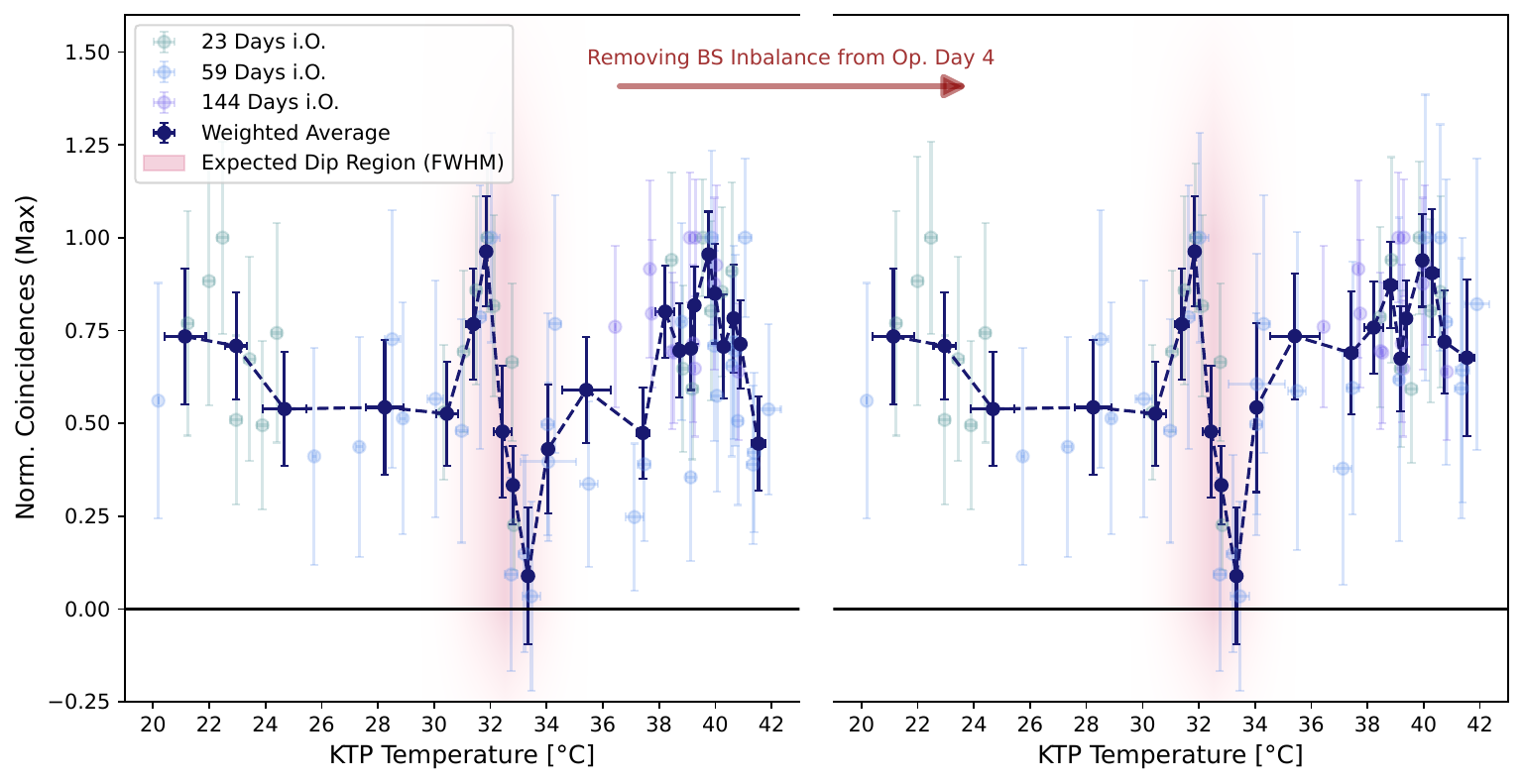}
    \caption[]{\textbf{HOM dip maximum normalization - outlier removal.} Transition to maximum normalization in Fig.~\ref{fig:hom_dip_mean_to_nax} has some visual implications on the shape of the weighted average. Outlier discussion in Sec.~\ref{sec:operation_day_4} and results from the mean normalization allow us to safely ignore these outliers as they can not be linked to interference; green circles from the second- and third shadow measurement in Fig.~\ref{fig:op_day_4_counts_close_up_3}b (row) are removed, as they essentially overlap with Identity. The red arrow marks the transition to outlier removal, smoothing the shape of the weighted average. It is important to note that this procedure does not affect the overall interference visibility of the weighted average.}
    \label{fig:hom_dip_max_outlier_consideration}
\end{figure}

From the discussions regarding the last two shadow measurements on operation day $4$, it is evident that these data points (light blue, $59$ days in orbit) are related to the strong fluctuations of the Beamsplitter unitary (MZI jitter). In more detail, the responsible datapoints correspond to the maximal Beamsplitter counts in the second- and third shadow measurement of Fig.~\ref{fig:op_day_4_counts_close_up_3}. As discussed in Sec.~\ref{sec:operation_day_4}, these fluctuations weigh stronger when normalized with respect to the maximum, but their effective count rate or mean normalization (see Fig.~\ref{fig:op_day_4_bs_normalized_mean}) does not show any meaningful dip in coincidences. Justified by the outcome of the robust mean normalization in Fig.~\ref{fig:hom_interference_and_ratio_mean}, we can safely exclude these five datapoints, and Fig.~\ref{fig:hom_dip_max_outlier_consideration} manifests this treatment. % TODO: Table~\ref{tab:binning_visibilities} lists 0.911 +- 0.184 (max norm., 2:2:2) and
% 0.908 +- 0.191 (max norm., 3:2:3, with and without outlier removal), but no 0.905 +- 0.197.
% Main text, this section and the table must quote the same number; the "2.1 standard deviations"
% statement in the main text depends on it (2.1 for 0.908, 2.2 for 0.911).
The interference visibility for this normalization scheme is $V= 0.905 \pm 0.197$, and alternatively, we can also evaluate the mean of the weighted average outside the FWHM of the dip (red-shaded area), and normalize the minima with respect to this averaged distinguishable count rate. Here, the visibility reads $V= 0.877 \pm 0.254$, so the two methods give consistent visibilities. The present approach of normalizing data and extracting the interference visibility with respect to the maximum is used in Fig.~5 of the main text, as its interpretation is more natural and frequent in the relevant literature \cite{bjurlin_versatile_2024}. Additionally, evaluating the HOM dip with respect to the mean normalization routine supports the maximum normalization routine, as the outcome is consistent in both approaches.
The interference visibility in Eq. \ref{eq:visibility_max} is summarized in Table~\ref{tab:binning_visibilities} for all relevant approaches from Fig. \ref{fig:hom_dip_mean_binning} to Fig. \ref{fig:hom_dip_max_outlier_consideration} and shows that the treatment of binning or normalization technique is equivalent. \begin{table}
        \centering
\caption[]{\textbf{HOM interference visibilities for different approaches.} The interference visibilities for all discussed approaches from Fig. \ref{fig:hom_dip_mean_binning} to Fig. \ref{fig:hom_dip_max_outlier_consideration} are shown for comparison and indicate that there is no significant difference between the approaches. This underlines that the transition from the mean binning $2:2:2$ to the final maximum binning $3:2:3$ and outlier removal stays faithful to the underlying two-photon interference phenomena.}
\label{tab:binning_visibilities}
        \begin{tabular}{|c c|}
            \hline
            Approach & Interference Visibility \\ [0.5ex]
            \hline\hline
            Mean norm. and binning $2:2:2$& $0.885\pm 0.232$  \\
            \hline
           Mean norm. and binning $3:2:3$& $0.885\pm 0.231$  \\
            \hline
     Max norm. and binning $2:2:2$ & $0.911\pm 0.184$  \\
            \hline
    Max norm. and binning $3:2:3$& $0.908\pm 0.191$  \\
            \hline
Max norm. and binning $3:2:3$ and outlier removal & $0.908\pm 0.191$  \\
            \hline
        \end{tabular}

    \end{table}

\begin{figure}[!tbp]
    \centering
    \includegraphics[width=1\textwidth,height=0.78\textheight,keepaspectratio]{ 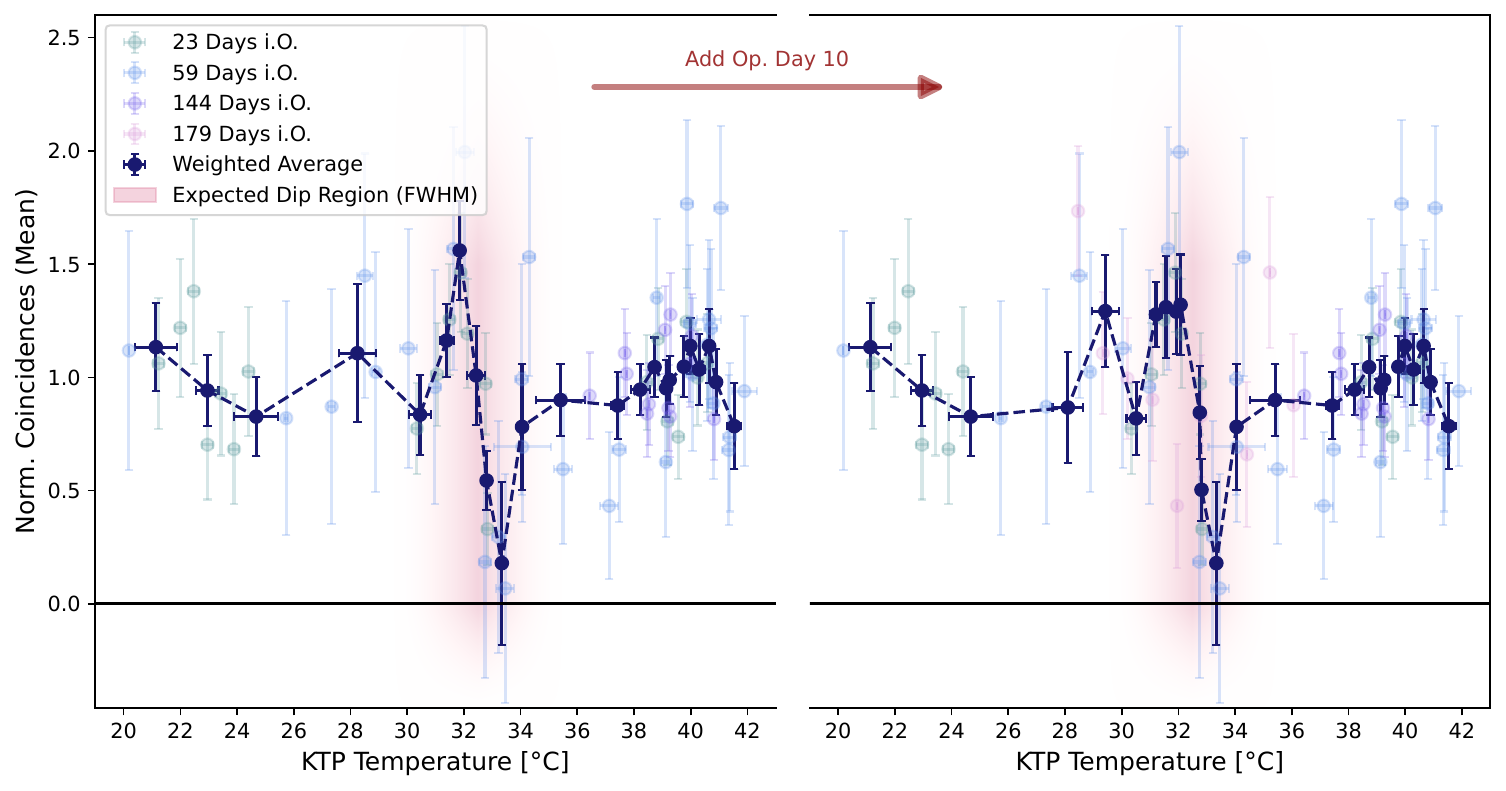}
    \caption[]{\textbf{HOM dip - inclusion of operation day 10.} Optional inclusion of HOM dip scan data ($179$ day in Orbit) from operation day $10$ to mean normalization dataset in Fig.~\ref{fig:hom_interference_and_ratio_mean}. Random laser mode hopping into a stable SLM mode makes interpretation of these datapoints difficult; however, their addition (red arrow) is consistent with the result of the main HOM dip analysis, as fluctuations outside the dip FWHM are further suppressed (especially the fluctuations of operation day $4$), and dip FWHM (red shaded area) is extended by additional interference data.}
    \label{fig:hom_interference_add_op_day_10}
\end{figure}

The HOM analysis is concluded with the optional consideration of the last shadow measurement from operation day $10$ in Sec.~\ref{sec:operation_day_10}. Here, the final shadow measurement eventually showed signs of interference; however, laser damage, in the form of random frequency-mode hopping, manifests itself as reduced interference visibility and a slightly shifted HOM dip at KTP temperature (see details on laser damage in Sec.~\ref{sec:tvac}). Moreover, mode hopping can happen completely randomly after power cycling the laser (compare Fig.~\ref{fig:mode_hopping}), and it did not occur for the first two shadow measurements on operation day $10$ or the two only shadow measurements on operation day $9$. In this sense, the result of the last shadow measurement from operation day $10$ in Sec.~\ref{sec:operation_day_10} is not clear, and for these reasons, they are excluded from the main HOM dip data analysis. Nevertheless, they can be included as supplementary information, and Fig.~\ref{fig:hom_interference_add_op_day_10} expands the mean normalization HOM dip in Fig.~\ref{fig:hom_interference_and_ratio_mean} with the additional data points. The weighted average is slightly shifted due to the shifted minima in coincidences of operation day $10$, but also the fluctuations from the second- and third shadow measurement of operation day $4$ ($T_{TEC} \sim \ensuremath{35\,{}^{\circ}\mathrm{C}}$ or $T_{TEC} \sim \ensuremath{37\,{}^{\circ}\mathrm{C}}$) are now further suppressed. In a sense, this outcome is consistent with the main result in Fig.~5 of the main text or the mean normalization in Fig.~\ref{fig:hom_interference_and_ratio_mean}, as fluctuations outside the dip are essentially washed away and the only additional ``dip'' in coincidences is within the expected HOM dip KTP temperature.

\clearpage
\subsection{Uncertainty estimation of data of operation day 1}
\label{sec:noise_model}
The uncertainty estimation procedure was slightly different for the data collected on operation day 1 (23 days in orbit). On that day, in addition to the temperature scan, several coincidence windows $\tau_i$ were used, scanning over five values: $4.05\times10^{-9}$s, $4.32\times10^{-9}$s, $4.59\times10^{-9}$s, $4.86\times10^{-9}$s and $5.13\times10^{-9}$s.

For a given temperature $T_i$, since the integration time was kept constant, the expected value of the corrected counts $CCc_{ij}^{\tau_{i},T_i}$ is the same across all coincidence-window measurements, while the number of accidentals grows linearly with the window size. For a coincidence window $\tau_{i}$ and temperature $T_i$, we have:

\begin{equation}
    CCc_{ij}^{\tau_{i},T_i}=CC_{ij}^{\tau_{i},T_i}-Acc_{ij}^{\tau_{i},T_i}
\end{equation}

where $CCc_{ij}^{\tau_{i},T_i}$ is the number of coincidence counts corrected for the accidental counts $Acc_{ij}^{\tau_{i},T_i}$, and $CC_{ij}^{\tau_{i},T_i}$ is the total number of registered coincidences.

Since $CCc_{ij}^{\tau_{i},T_i}=CCc_{ij}^{\tau_{i'},T_i}$ is expected for all $i, i'$, we can increase the precision of our signal estimate by averaging over the coincidence windows:

\begin{equation}
\label{eq:mean}
    N_{mean}^{T_m}=\sum_l \frac {CCc_{ij}^{\tau_{l},T_m}}{N_{\tau}}
\end{equation}

For the identity measurements, this averaged value is expected to remain constant across different temperatures $T_i$ as well, since no signal change is expected in this configuration. This property allows us to treat the observed fluctuations as pure noise and use them to characterize the overall distribution: we evaluated the mean and standard deviation of the collected measurements, obtaining $\mu=31.53$ and $\sigma=5.61\approx \sqrt{\mu}$.

\begin{figure}[!tbp]
    \centering
    \includegraphics[width=1\textwidth,height=0.73\textheight,keepaspectratio]{ 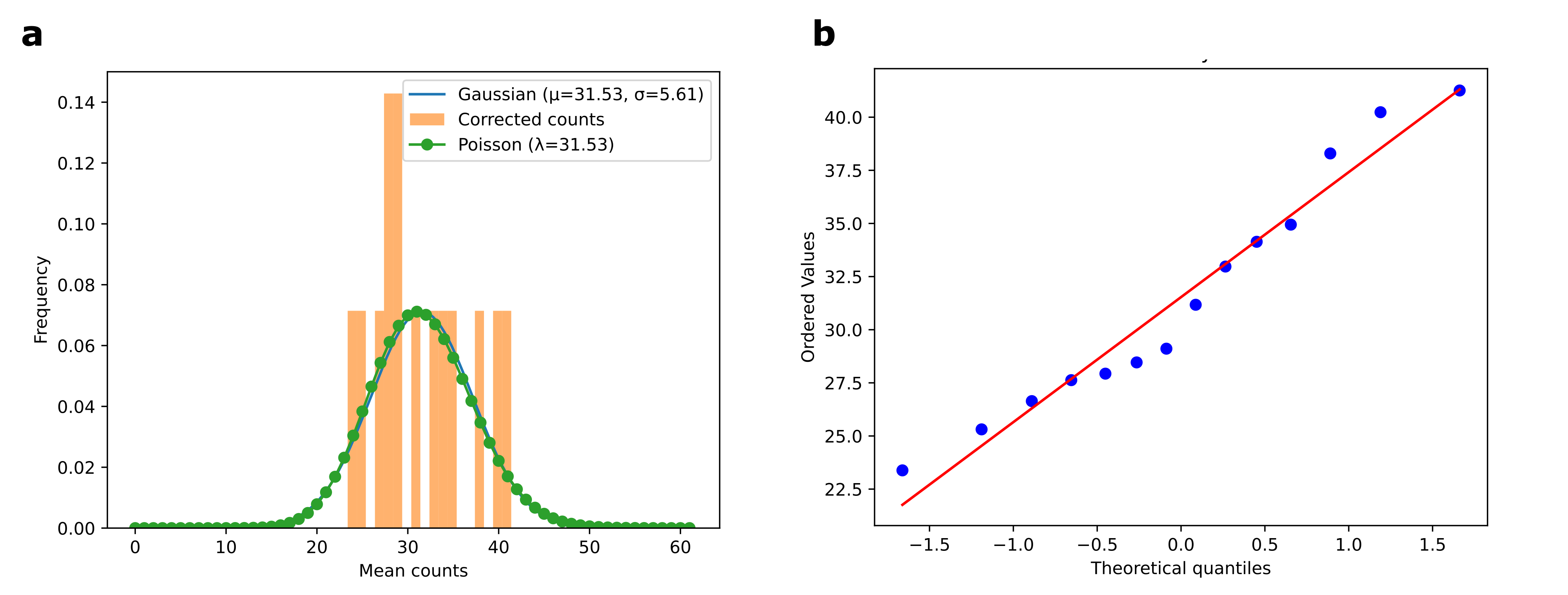}
    \caption[]{\textbf{Statistical distribution of coincidence counts in the identity setting.} To evaluate the correct uncertainty to be associated with the results obtained on the operation day 1, we take all of the coincidence counts, corrected for the accidentals, and take the mean over the different coincidence windows. In \textbf{a)} we report the histogram of the counts obtained at different temperatures, to check their distribution. The blue curve is a Gaussian with $\mu$ as the mean of the data and $\sigma$ is the standard deviation of the data. The green dots represent a Poissonian distribution, centered in $\mu$, which perfectly overlaps with the Gaussian distribution. From this histogram, it is visible how the Poissonian distribution approximates well the distribution of the data. This is further confirmed in \textbf{b)}, where it is visible that the Gaussian approximates well the data. }
    \label{fig:mean_1}
\end{figure}

As shown in Fig.~\ref{fig:mean_1}a, the distribution of this variable is well approximated by a Poissonian distribution centred on $N_{mean}=\sum_l \frac {CCc_{ij}^{\tau_{l},T_m}}{N_{\tau}}$, with $\sigma=\sqrt{N_{mean}}$.

To check whether our data are consistent with a Gaussian distribution, we used a quantile-quantile plot, comparing the data against the values expected under normality, together with a reference line, shown in Fig.~\ref{fig:mean_1}b. To further validate the compatibility of the data with the assumed model, we computed the reduced chi-square statistic, obtaining $\chi^2_{red}=1.41$ ($\chi^2=4.23$, with 3 degrees of freedom), corresponding to a p-value of 0.24. Since this value is well above standard significance thresholds (0.05), we find no evidence to reject the model, and the result is consistent with the Q-Q plot analysis, supporting the assumption of Gaussian/Poissonian statistics for the collected data.

\begin{figure}[!tbp]
    \centering
    \includegraphics[width=1\textwidth,height=0.72\textheight,keepaspectratio]{ 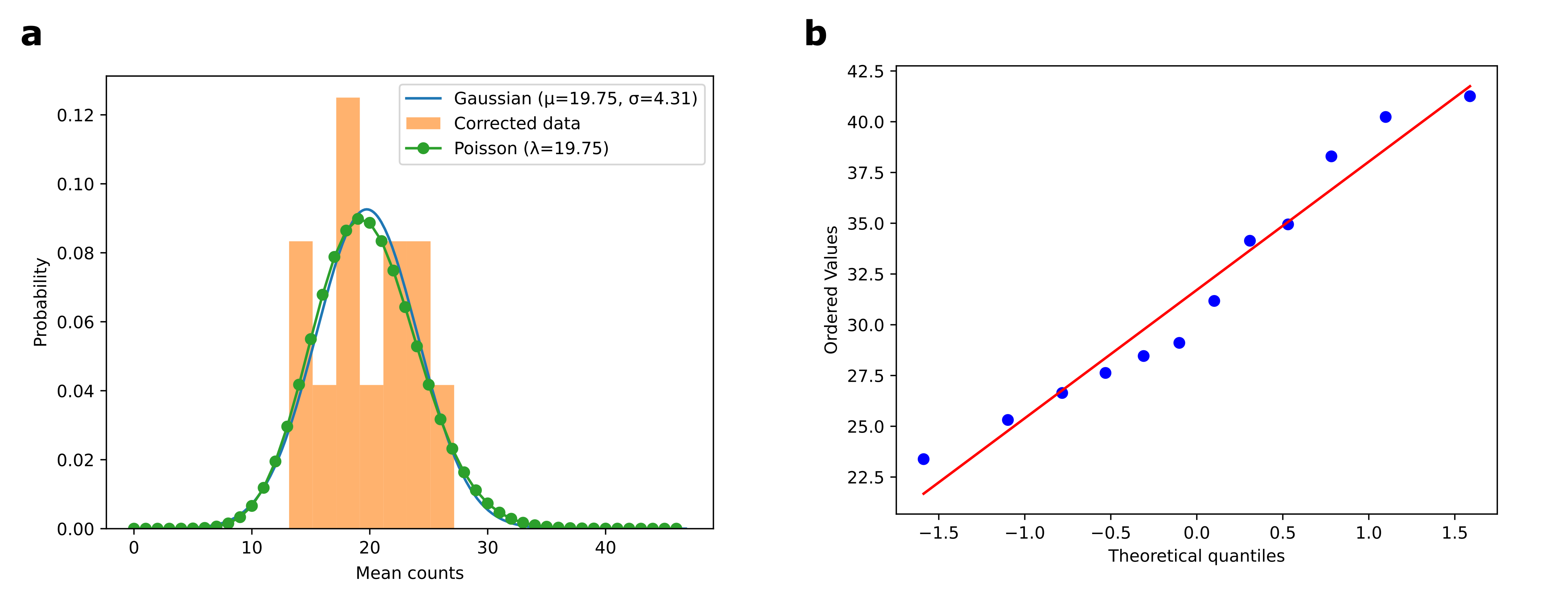}
    \caption[]{\textbf{Statistical distribution of coincidence counts in the beam-splitter setting.} To evaluate the correct uncertainty to be associated with the results obtained on the operation day 1, we take all of the coincidence counts, corrected for the accidentals, and take the mean over the different coincidence windows. In \textbf{a)} we report the histogram of the counts obtained at different temperatures, to check their distribution. The blue curve is a Gaussian with $\mu$ as the mean of the data and $\sigma$ is the standard deviation of the data. The green dots represent a Poissonian distribution, centered in $\mu$, which perfectly overlaps with the Gaussian distribution. From this histogram, it is visible how the Poissonian distribution approximates well the distribution of the data. This is further confirmed in \textbf{b)}, where it is visible that the Gaussian approximates well the data. }
    \label{fig:mean_2}
\end{figure}

The same behaviour is observed for the data obtained in the beam-splitter configuration, in the non-interference region, as shown in Fig.~\ref{fig:mean_2}. In this case, $\chi^2_{red}=1.82$ ($\chi^2=5.47$, with 3 degrees of freedom), giving a p-value of 0.14.

Given these results, we can associate each temperature with its estimated $N_{mean}$, as defined in Eq.~\ref{eq:mean}.

For the subsequent analysis, we normalize the counts with respect to the maximum value of the day:

\begin{equation}
    N_{mean}^{*T_m}=\sum_l \frac {CCc_{ij}^{\tau_{l},T_m}}{N_{\tau}}\Big/ \max_m\Big(\sum_l \frac {CCc_{ij}^{\tau_{l},T_m}}{N_{\tau}}\Big)=\sum_l CCc_{ij}^{\tau_{l},T_m}\Big/ \max_m\sum_l CCc_{ij}^{\tau_{l},T_m}=N_m \Big/ N_M.
\end{equation}

where $N_m$ is the sum of the registered counts for a temperature $T_m$ over the coincidence windows $\tau_l$ for $l \in (1,N_\tau =5)$ and $N_M$ is the maximum of $N_m$ over the temperatures.

The uncertainty on $N_{mean}^{*T_m}$ is given by $\sqrt{N_{mean}^{*T_m}}$; propagating the uncertainty to the normalized quantity yields:

\begin{equation}
    \sigma_{N_{mean}^{*T_m}}^2= \frac{\sigma_{N_m}^2}{N_M^2}+\frac{\sigma_{N_M}^2N_m^2}{N_M^4}=\frac{N_m N_\tau^2}{N_\tau N_M^2}+\frac{N_m^2N_M N_\tau^4}{N_\tau ^3N_M^4}
\end{equation}

where $N_\tau$ is the number of tested coincidence windows, i.e. 5.

The noise model of Eq. \ref{accidentals} is experimentally verified in Fig.~\ref{fig:noise_analysis} by switching the laser off and only validating the accidental coincidences of detector dark counts. As clearly illustrated in Fig.~\ref{fig:noise_analysis}a, the accidental noise model (yellow, green, violet) predicts the uncorrelated accidental signal (red, dark blue, light blue) well, as both graphs nearly perfectly overlap. This is also evident when averaging the data (small inset), as the corrected signal is consistent with zero across all working detector pairs.

\subsection{SPADs behaviour in FM}
\label{sec:spads_fm}
Our detection system is composed of six SPADs, however, as detailed in section~\ref{sec:detectors}, only three remained functional throughout the mission. In more detail, SPADs C$_2$, C$_3$ and C$_4$, were already faulty before
launch: system health tests of the FM, performed after the flight acceptance
test, showed no meaningful coincidence response for any pair involving
them.

\begin{figure}[!tbp]
    \centering
    \includegraphics[width=1\textwidth,height=0.67\textheight,keepaspectratio]{ 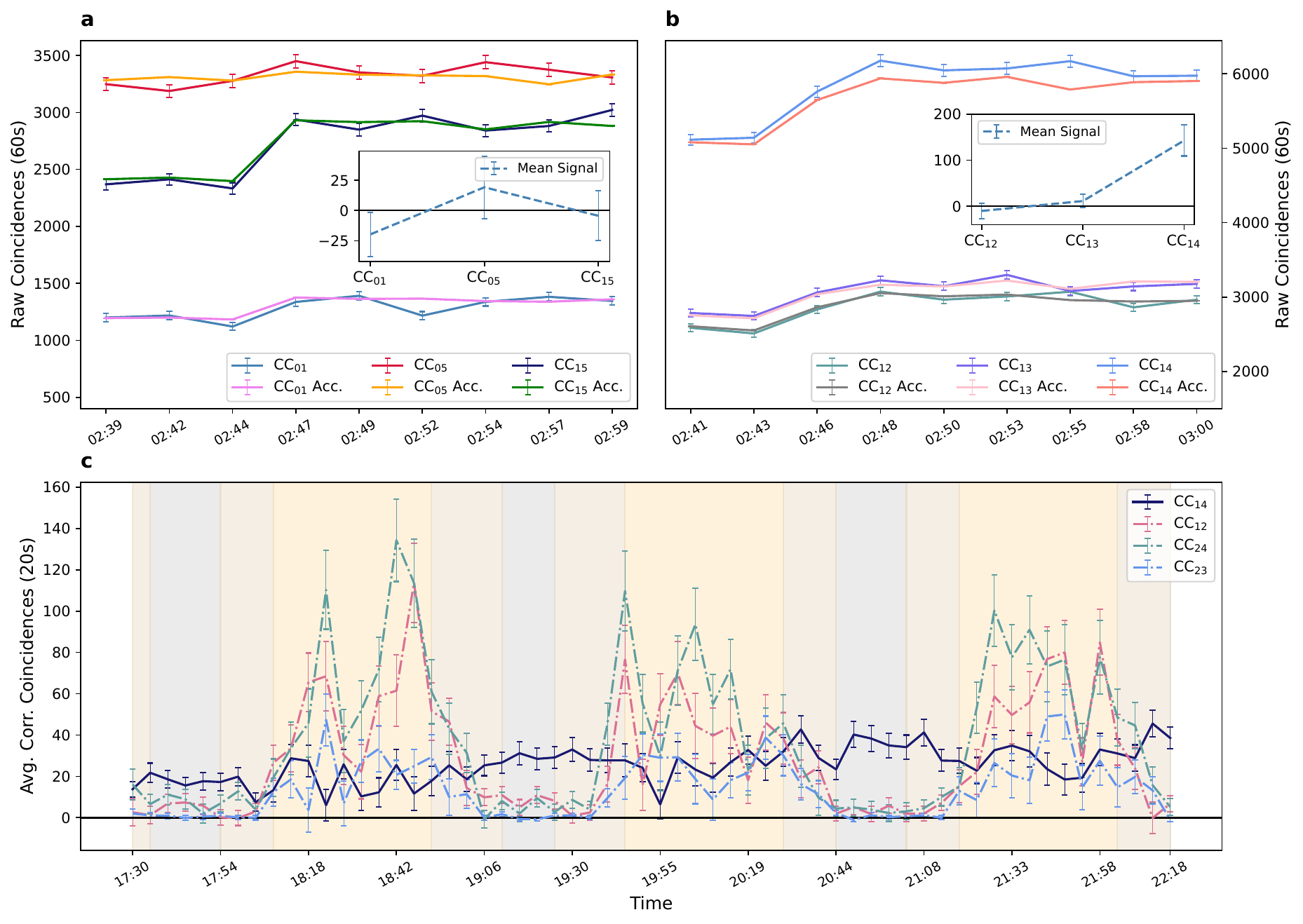}
    \caption[]{\textbf{Noise Model and exclusion of detectors.}
    \textbf{a} Measured timetagger coincidences counts for working detector pairs CC$_{ij}$ (red, dark blue, light blue) and $ij \in \{01,05,15\}$ for laser being switched off. The predicted (accidental) coincidence counts CC$_{ij}$ Acc. (orange, green, violet) in Eq. \ref{accidentals} show good agreement with the actually measured accidental counts and allow us to use the model of Eq. \ref{accidentals} as a proper accidentals estimation. \textbf{b} Excerpt of detector pairs from non working SPADs shows also good agreement of noise model including SPAD C$_1$, however, detector pair including SPAD C$_4$ is clearly indicating signal despite a switched off laser. This anomaly can only be linked to this specific SPAD and might arise from electronic noise. Detector pairs including SPAD C$_2$, C$_3$, and C$_4$ showed no meaningful coincidence response in tests on Earth and are thus excluded from the experimental pipeline. \textbf{c} Gives an overview of a working target detector pair CC$_{05}$ from operation day $1$ in Sec.~\ref{sec:operation_day_1} with respect to non working detector pairs including SPAD C$_2$. The noise behaviour is strongly different and suggests a potential saturation damage in SPAD C$_2$.}
    \label{fig:noise_analysis}
\end{figure}

\begin{figure}[!tbp]
	\centering
	\includegraphics[width=0.48\textwidth,height=0.78\textheight,keepaspectratio]{ 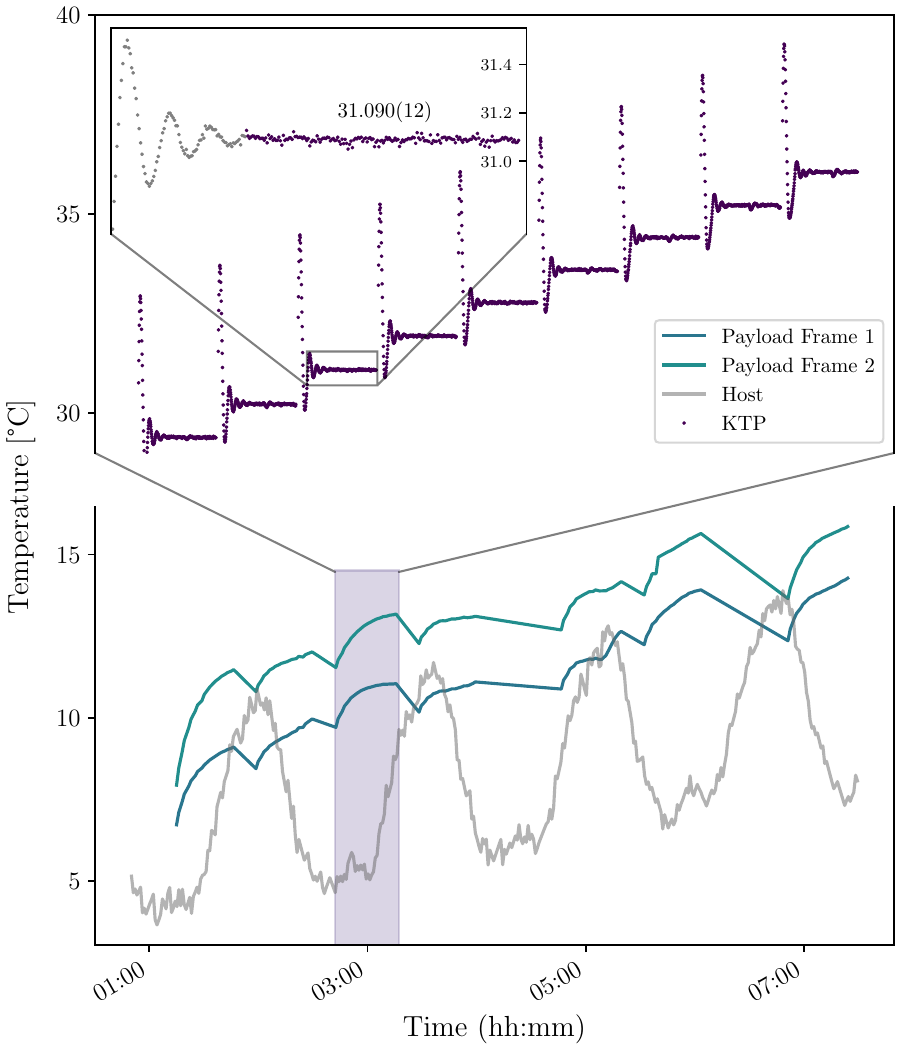}
	\caption[]{\textbf{Typical temperature profile of an operation day.} The payload's internal temperature (shades of green) is largely unaffected by the host's outside temperature (grey), which follows the orbital period. The shaded area marks the time frame of the top-panel scan. \textbf{Top: Temperature control of the KTP.} Stable regulation is obtained across a broad range of set temperatures, with temperature fluctuations below \ensuremath{0.1\,\mathrm{K}}. The inset shows the transient response and stability for one set point, where coloured data points contribute to the quoted mean and standard deviation.}
	\label{fig:temperature-control}
\end{figure}

SPAD C$_3$ did not respond to light at all. SPADs C$_2$ and C$_4$ both
registered singles counts, but each showed anomalous noise. C$_4$ exhibited
extremely high dark counts together with a constant electronic contribution
to the coincidences: as shown in Fig.~\ref{fig:noise_analysis}b, the pair
CC$_{14}$ (blue) sits above the noise floor with the laser off. We attribute
this to electronic correlations on the SPAD board affecting this specific
detector, since the same behaviour appears in the EM for a single detector
(the outlier C$_0$ in Fig.~\ref{fig:spads}) and in no other FM detector,
working or not (compare CC$_{12}$ and CC$_{13}$ in
Fig.~\ref{fig:noise_analysis}b). C$_2$ showed no such electronic
contribution, but its noise behaviour differed from that of all other pairs:
as shown by the dashed curves in Fig.~\ref{fig:noise_analysis}c, pairs
involving C$_2$ fluctuate strongly in the sunlit portions of the orbit and
markedly exceed the remaining pairs, suggesting a higher overall noise level
and closer proximity to saturation.
\clearpage
\subsection{Thermal Environment and KTP stability}

\subsection{Temperature-control architecture}
The temperature-control board (TCB) provides two thermoelectric-cooler (TEC) control channels for active thermal regulation within the payload. One channel was used to regulate the temperature of the periodically poled potassium titanyl phosphate (ppKTP) crystal oven. The controller is based on an LTC1923 pulse-width-modulation (PWM) TEC controller operated at a switching frequency of
$f_{\text{sw}} = \ensuremath{227\,\mathrm{kHz}}.$
The LTC1923 drives an external full-bridge, permitting bidirectional current flow through the TEC and, therefore, both active heating and active cooling of the crystal oven.

The oven temperature is measured using a negative-temperature-coefficient (NTC) thermistor (TDK B57550G1 8304) mounted on the source oven. The NTC forms the lower arm of a resistive divider. The divider centre tap is routed both to the temperature controller as the process variable and to an ADC for diagnostic observation. The divider's output voltage is
\begin{equation*}
    V_{\text{NTC}}(T)
    =
    V_{\text{exc}}
    \frac{R_{\text{NTC}}(T)}
         {R_{\text{b}} + R_{\text{NTC}}(T)},
\end{equation*}
where $V_{\text{exc}}=\ensuremath{5\,\mathrm{V}}$ is the divider excitation voltage and $R_{\text{b}}=\ensuremath{100\,\mathrm{k\Omega}}$ is the fixed bias resistor.

The thermistor resistance may be approximated using the beta model,
\begin{equation*}
    R_{\text{NTC}}(T)
    =
    R_0
    \exp\!\left[
        \beta
        \left(
            \frac{1}{T} - \frac{1}{T_0}
        \right)
    \right]
\end{equation*}
where temperatures are expressed in Kelvin, $R_0=\ensuremath{100\,\mathrm{k\Omega}}$ is the resistance at the reference temperature $T_0=\ensuremath{25\,{}^{\circ}\mathrm{C}}$, and $\beta=3507.82$ (fitted from the values in the datasheet) is the thermistor material constant.

The local voltage sensitivity is

\begin{equation*}
    \frac{\text{d}V_{\text{NTC}}}{\text{d}T}
    =
    -\frac{
        V_{\text{exc}}\,
        \beta R_{\text{b}}R_{\text{NTC}}
    }{
        T^2\left(R_{\text{b}} + R_{\text{NTC}}\right)^2
    }
    =
    -\frac{\beta}{T^2} V_{\text{NTC}} \left( 1- \frac{V_{\text{NTC}}}{V_{\text{exc}}}\right)
\end{equation*}

The setpoint is generated by an AD5669R DAC and applied to the negative input of the error amplifier. The voltage corresponding to DAC code $n$ may be written as

\begin{equation*}
    V_{\text{set}}(n)
    =
    V_{\text{ref}}
    \frac{n}{2^{16}-1},
    \qquad
    0 \leq n \leq 2^{16}-1,
\end{equation*}

A single DAC code, corresponding to $\approx$\ensuremath{76.3\,\mathrm{\mu{}V}}, therefore leads to a temperature setpoint shift of -\ensuremath{1.5\,\mathrm{mK}}, which is well below the targeted $\Delta T = \ensuremath{0.1\,\mathrm{K}}$, even with $\pm1$LSB DNL.

The controller drives the TEC until the thermistor-derived process voltage matches the setpoint voltage:
\begin{equation*}
    V_{\text{NTC}}\!\left(T_{\text{set}}\right)
    =
    V_{\text{set}}.
\end{equation*}

\subsection{Loop compensation}

The feedback network around the inverting error amplifier consists of a resistor $R_{\text{f}}=\ensuremath{909\,\mathrm{k\Omega}}$ in parallel with a capacitor $C_{\text{f}}=\ensuremath{4.7\,\mathrm{\mu{}F}}$. This produces the feedback impedance

\begin{equation*}
    Z_{\text{f}}(\omega)
    =
    R_{\text{f}}
    \parallel
    \frac{1}{j\omega C_{\text{f}}}
    =
    \frac{R_{\text{f}}}
         {1+j\omega R_{\text{f}}C_{\text{f}}}.
\end{equation*}

Together with an input resistance $R_{\text{i}}=\ensuremath{100\,\mathrm{k\Omega}}$, the transfer function is
\begin{equation*}
    C(\omega)
    =
    -\frac{Z_{\text{f}}(\omega)}{R_{\text{i}}}
    =
    -\frac{R_{\text{f}}/R_{\text{i}}}
          {1+j\omega R_{\text{f}}C_{\text{f}}}.
\end{equation*}

The implemented component values give a DC gain of

\begin{equation*}
    \left|C(0)\right|
    =
    \frac{R_{\text{f}}}{R_{\text{i}}}
    =
    9.09
    \label{eq:compensator-dc-gain}
\end{equation*}
and a compensator pole at
\begin{equation*}
    f_{\text{p}}
    =
    \frac{1}{2\pi R_{\text{f}}C_{\text{f}}}
    =
    \ensuremath{37\,\mathrm{mHz}}.
    \label{eq:compensator-pole}
\end{equation*}
The corresponding electrical time constant is
\begin{equation*}
    \tau_{\text{c}}
    =
    R_{\text{f}}C_{\text{f}}
    =
    \frac{1}{2\pi f_{\text{p}}}
    \simeq
    \ensuremath{4.27\,\mathrm{s}}.
\end{equation*}

Although a larger DC gain was initially estimated to be desirable, the implemented network was experimentally found to provide a satisfactory transient response and stability.

Ground verification was performed with a single-photon source installed inside a vacuum chamber maintained at approximately \ensuremath{22\,{}^{\circ}\mathrm{C}}. The ppKTP oven was regulated at a setpoint of \ensuremath{36\,{}^{\circ}\mathrm{C}}, and the temperature response was monitored during setpoint changes and steady-state operation.

\subsection{Full-bridge power stage}

The TEC voltage is supplied by a dedicated \ensuremath{3.8\,\mathrm{V}} rail generated by an LT8613 buck converter. This voltage was selected to remain below the specified \ensuremath{4\,\mathrm{V}} maximum operating voltage of the Laird OT24,31,F1,1010 TEC.

The output stage comprises two complementary MOSFET pairs (DMC3016) in a full-bridge configuration. The devices have maximum on-state resistances
\begin{equation*}
    R_{\text{DS,on,P}} < \ensuremath{0.038\,\mathrm{\Omega}},
    \qquad
    R_{\text{DS,on,N}} < \ensuremath{0.017\,\mathrm{\Omega}},
\end{equation*}

and total gate charges $Q_{\text{P}} = Q_{\text{N}} = \ensuremath{9.5\,\mathrm{nC}}.$

Let $D_A$ and $D_B$ denote the duty cycles of the two diagonal bridge states. Neglecting dead time and voltage drops, the average voltage across the TEC is
\begin{equation*}
    \overline{V}_{\text{TEC}}
    =
    V_{\text{bridge}}
    \left(D_A-D_B\right)
\end{equation*}

For complementary operation, $D_A+D_B=1$, and therefore

\begin{equation*}
    \overline{V}_{\text{TEC}}
    =
    V_{\text{bridge}}
    \left(2D_A-1\right).
\end{equation*}

A duty cycle $D_A=0.5$ produces approximately zero mean TEC voltage, whereas duty cycles above and below $0.5$ produce opposite current directions and hence heating and cooling.

\subsection{Output filtering and ripple-current}

Each bridge leg includes an inductor of $L = \ensuremath{15.4\,\mathrm{\mu{}H}}$
with an equivalent series resistance (ESR)
$ \text{ESR}_{\text{L}} = \ensuremath{0.0146\,\mathrm{\Omega}}.$

The peak-to-peak inductor ripple current can be determined by
\begin{equation}
    \Delta I_{\text{L,pp}}
    =
    \frac{
        V_{\text{bridge}}^2 - V_{\text{TEC}}^2
    }{
        4f_{\text{sw}}LV_{\text{bridge}}
    }.
    \label{eq:inductor-ripple-current}
\end{equation}
The largest ripple predicted by Eq.~\eqref{eq:inductor-ripple-current} occurs close to zero mean TEC voltage. With
\begin{equation*}
    V_{\text{bridge}} = \ensuremath{3.8\,\mathrm{V}},
    \qquad
    f_{\text{sw}} = \ensuremath{227\,\mathrm{kHz}},
    \qquad
    L = \ensuremath{15.4\,\mathrm{\mu{}H}},
    \label{eq:ripple-calculation-parameters}
\end{equation*}

The maximum estimated inductor ripple is
\begin{equation*}
    \Delta I_{\text{L,pp,max}}
    \simeq
    \frac{\ensuremath{3.8\,\mathrm{V}}}
         {4\left(\ensuremath{227\,\mathrm{kHz}}\right)
            \left(\ensuremath{15.4\,\mathrm{\mu{}H}}\right)}
    \simeq
    \ensuremath{272\,\mathrm{mA}}.
    \label{eq:maximum-inductor-ripple-current}
\end{equation*}

This means that to supply the TECs with their maximum rated current of \ensuremath{2.5\,\mathrm{A}}, the inductors (Würth 7443551151) must withstand $\ensuremath{2.5\,\mathrm{A}}+\Delta I_{\text{L,pp,max}}/2 = \ensuremath{2.636\,\mathrm{A}}$ - well within their rated maximum current.

Furthermore, the inductor ripple current influences the TEC ripple current, effectively derating the TECs maximum temperature differential. This effect can be summarised in

\begin{equation}
    \frac{dT}{dT_{\text{max}}} = \frac{1}{1+\text{N}^2}
    \label{equ:temp_differential_derationg}
\end{equation}
where $dT$ is the derated temperature differential, $dT_{\text{max}}$ the ideal maximum temperature differential, and N is the ratio of TEC ripple to DC current.

\begin{equation*}
    I_{\text{TEC,ripple}} = \frac{V_{\text{bridge}}^2-V_{\text{TEC}}^2}{16 f_{\text{sw}}^2 L C R_{\text{TEC}}V_{\text{bridge}}}
    +
    \frac{ \left( V_{\text{bridge}}^2-V_{\text{TEC}}^2 \right) \text{ESR}_{\text{C}}}{2 f_{\text{sw}} L V_{\text{bridge}} R_{\text{TEC}}}
\end{equation*}

The filtering capacitor's (Murata GCM31CR71A226KE02L) ESR of $\approx\ensuremath{4\,\mathrm{m\Omega}}$ (at \ensuremath{227\,\mathrm{kHz}} and \ensuremath{1.8\,\mathrm{V}} ($V_{\text{TEC}}/2$) DC bias and an estimated board temperature of \ensuremath{50\,{}^{\circ}\mathrm{C}} (where the capacitance of this multi layer ceramic capacitor is about \ensuremath{22\,\mathrm{\mu{}F}}, according to the manufacturers database) is therefore of specific importance. In the worst-case scenario, the TEC ripple current becomes $\approx\ensuremath{11\,\mathrm{mA}}$, or about $0.5\%$ of the maximal TEC current, practically not affecting the maximally achievable temperature differential, according to Eq. \ref{equ:temp_differential_derationg}.

\subsection{Electrical-loss model}

The dominant losses considered here are the MOSFET gate-drive loss and the resistive conduction loss. The current-independent gate-drive loss, arising from charging and discharging the MOSFET gates during each switching cycle, is
\begin{equation*}
    P_{\text{gate}}
    =
    2f_{\text{sw}}
    \left(Q_{\text{P}}+Q_{\text{N}}\right)
    V_{\text{DRV}},
    \label{eq:gate-drive-loss}
\end{equation*}
where $V_{\text{DRV}}$ is the gate-driver supply voltage.
For the DMC3016's $Q_{\text{P}} = Q_{\text{N}} = \ensuremath{9.5\,\mathrm{nC}}$ and $V_{\text{DRV}} = \ensuremath{5\,\mathrm{V}}$, $P_{\text{gate}}\approx\ensuremath{44\,\mathrm{mW}}.$

During conduction, the TEC current passes through one PMOS, one NMOS, both output inductors, and the current-sense resistor $R_{\text{sense}}=\ensuremath{10\,\mathrm{m\Omega}}$ (used by the PWM controller for current limiting purposes). The effective series resistance of the system is therefore
\begin{equation*}
    R_{\text{path}}
    =
    R_{\text{DS,on,P}}
    +
    R_{\text{DS,on,N}}
    +
    2R_{\text{L}}
    +
    R_{\text{sense}} = \ensuremath{94.2\,\mathrm{m\Omega}}.
    \label{eq:bridge-path-resistance}
\end{equation*}

If $I_{\text{TEC,ripple}} \ll I_{\text{TEC}}$, the corresponding conduction loss is

\begin{equation*}
    P_{\text{cond}}
    =
    I_{\text{TEC}}^2R_{\text{path}},
    \label{eq:bridge-conduction-loss}
\end{equation*}
which rises up to about \ensuremath{589\,\mathrm{mW}} at $I_{\text{TEC}}=\ensuremath{2.5\,\mathrm{A}}$, for a total bridge loss  of $\ensuremath{635\,\mathrm{mW}}$.

Additionally, the controller's quiescent consumption is \ensuremath{10\,\mathrm{mW}} under idle conditions.

\begin{figure}[!tbp]
	\centering
	\includegraphics[width=0.6\textwidth,height=0.80\textheight,keepaspectratio]{ 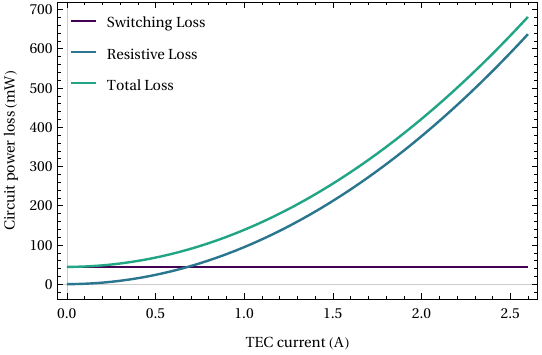}
	\caption[]{\textbf{Estimated losses of the TEC full-bridge power stage.} The modelled losses consist of an approximately current-independent gate-drive contribution and a conduction contribution proportional to $I_{\text{TEC}}^2$. The conduction term includes the on-state resistances of one PMOS and one NMOS, the series resistances of both filter inductors, and the current-sense resistor. The quiescent power consumption of the controller and associated circuitry, approximately \ensuremath{10\,\mathrm{mW}} under idle conditions, is not included.}
	\label{fig:tec-losses}
\end{figure}

\subsection{In-orbit observations}

During early in-orbit operation, stable regulation of the source oven could not be maintained under some operating conditions. Our initial guess of the internal payload's temperature being too low for stable operation could be strengthened by observing that continuously operating the camera throughout the day (starting from day 145) alleviated the issue. Yet, we cannot exclude that the temperature rise was caused by other or external effects, noting that the average frame temperature increased already on day 100.

\begin{figure}[!tbp]
    \centering
    \includegraphics[width=\textwidth,height=0.84\textheight,keepaspectratio]{ 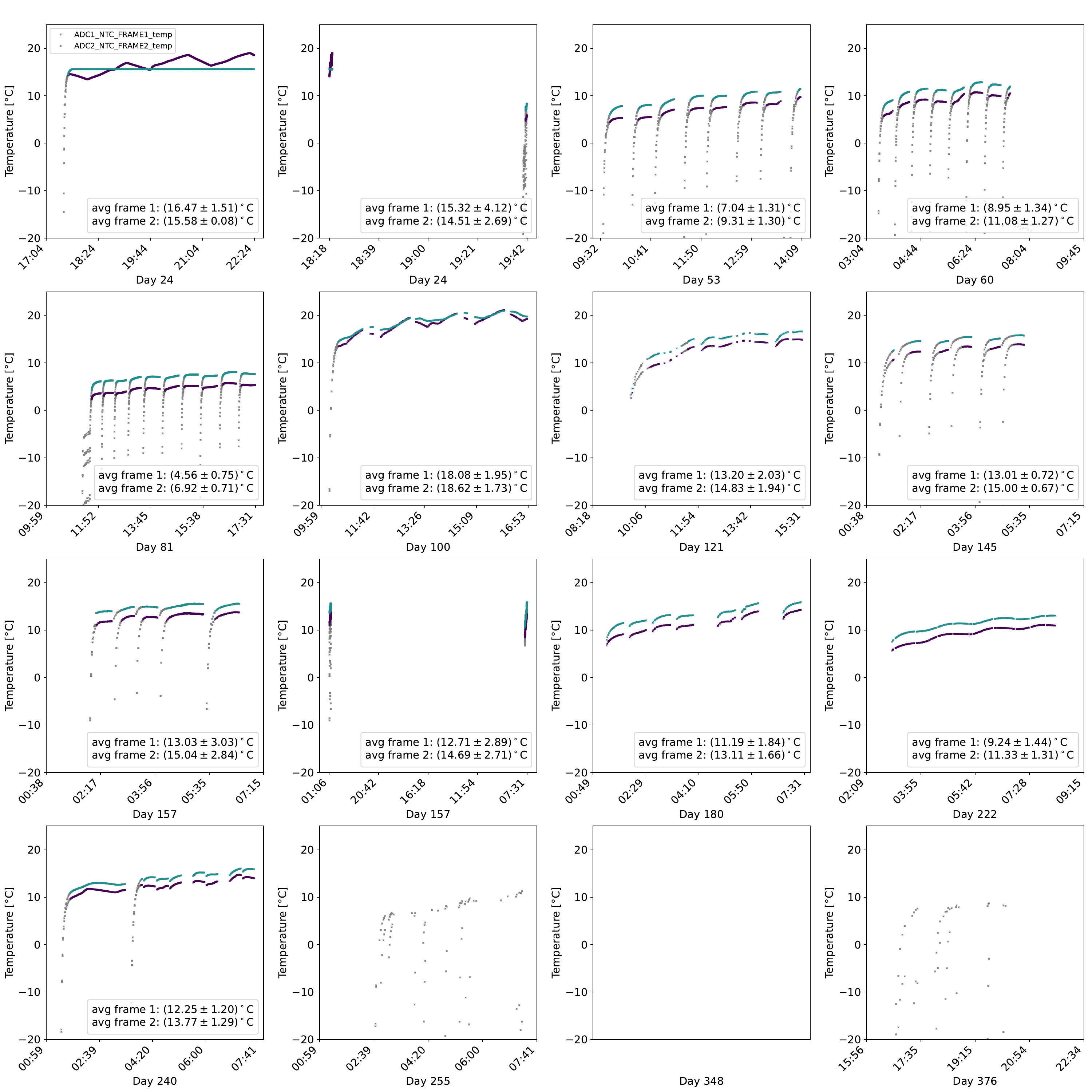}
    \caption[]{\textbf{Frame temperature throughout all operation days.} The camera was turned on all-day starting from day 145. Values where the ADC was idle are removed. Averages are taken over the colored datapoints only, excluding ramp-up behaviour caused by stabilising reference voltages.}
    \label{fig:frame_temperature}
\end{figure}

A reduction in the surrounding payload temperature increases the temperature difference that must be sustained and raises the required TEC current. Operating the camera increases the local ambient temperature and contributes parasitic heating, thereby reducing the required TEC power. Frame temperature traces and daily averages can be seen in Fig.~\ref{fig:frame_temperature}.

Starting from the first operation day, we decided to completely cut power to the experimental apparatus between each experiment (for example between individual temperature HOM-Dip scans), to start from a ``clean state'', avoiding the issue of stale parameters remaining set. Unfortunately, after each clean start, ADC/DAC reference voltages took in the order of a few minutes to stabilise, leading to artifacts in the temperature measurements manifesting as the ramps seen in Fig.~\ref{fig:frame_temperature}. To extract a somewhat meaningful average temperature of the day, we filtered the data by excluding datapoints whose 10-point rolling standard deviation is above \ensuremath{0.2\,\mathrm{K}}. The excluded points in Fig.~\ref{fig:frame_temperature} are plotted in grey.

\subsection{Additional Payload sensors}
\label{sec:pay_temp}

In addition to the thermistor used directly by the control loop, NTC probes were distributed throughout the payload to monitor temperature gradients, temperature-sensitive electronics, and support diagnosis of the thermal behaviour observed in orbit. Their locations are summarised in Table~\ref{tab:temperature-probes}.

The ADC inputs on the TCB and the sensor readout board (SRB) were also used to monitor the payload supply rails $3\text{V}3$, $5\text{V}0$, and $5\text{V}0\_2$, as well as the radiation-sensor output. Their read-out was distributed between the TCB and SRB board's ADC channels.

\begin{table}
    \centering
    \caption[]{Temperature-probe locations within the payload.}
    \label{tab:temperature-probes}
    \begin{tabular}{l c p{0.62\textwidth}}
        \hline
        \textbf{Location} &
        \textbf{Number} &
        \textbf{Placement and purpose}
        \\
        \hline
        Source oven &
        2 &
        Beaded NTCs attached to the top cover of the source oven.  One is used in the temperature control feedback loop.
        \\
        Payload frame &
        2 &
        Beaded NTCs bonded to diagonally opposite frame struts, approximately at the height of the rear of the camera. These channels monitor the payload's internal ambient temperature.
        \\
        uPIC assembly &
        2 &
        Beaded NTCs mounted on the titanium frame close to the V-groove-array on either side of the uPIC. These sensors were initially intended to support active regulation using the second TCB channel.
        \\
        TCB &
        4 &
        Two SMD NTCs were broken out through plated through-holes but became unavailable after conformal coating. Two additional NTCs were bonded to the rear side of the TCB, approximately opposite the bridge MOSFETs, to monitor local power-stage heating.
        \\
        SPAD board &
        3 &
        Three SMD NTCs were installed: one adjacent to the ADC and two between the SPADs, underneath the thermally conductive compound used to couple and mechanically stake the detectors.
        \\
        SRB &
        3 &
        Three SMD NTCs were installed: one on the rear side of each ADC and one adjacent to the radiation sensor.
        \\
        \hline
    \end{tabular}
\end{table}

% Do NOT include a reference list in the supplement (single list after the main text).
